\documentclass[11pt,superscriptaddress]{uopthesis}

\usepackage{bm}
\usepackage{amsfonts}
\usepackage{amssymb}

\newcommand{\mb}[1]{\mbox{\boldmath $#1$}}
\def \be {\begin{equation}}
\def \ee {\end{equation}}
\def \beq {\begin{eqnarray}}
\def \eeq {\end{eqnarray}}
\def \bean {\begin{eqnarray*}}
\def \eean {\end{eqnarray*}}
\def \D   {{\Delta}}
\def \al  {{\alpha}}
\def \ep  {{\epsilon}}
\def \la  {\lambda}
\def \La  {\Lambda}
\def \ga  {\gamma}
\def \om  {\omega}

\def \nn  {\nonumber}
\def \met {\mbox{g}}
\def \metb{\mbox{\bf g}}

\def \part {{\partial}}

\def \emLL  {e ^{-2 \, \Lambda }}

\def \pmr  { 4 \pi \,\left( p - \rho \right)}

\def \llcf  {l \left( l +1 \right) }
\def \Suo  { S^{(1,0)} }
\def \etuo  { \eta^{(1,0)} }
\def \F   { \textrm{\bf{F}} }
\def \U   { \textrm{\bf{u}} }
\def \S   { \textrm{\bf{S}} }
\def \AA   { \mathbf{A} }
\def \W   { \textrm{\bf{w}} }

\def \eor {{\left(1+\frac{p}{\rho}\right)}}
\def \cee {{C}}

\def \kbar {{c_s^2}}

\def \nat {Nature}
\def \apjs {Astrophys.J.Suppl.}
\def \apjl {Astrophys.J.Let.}
\def \apj {Astrophys.J.}
\def \prd {Phys.Rev. D}
\def \prl {Phys.Rev. Lett.}
\def \mnras {MNRAS}
\def \aap {A\&A}

\ifpdf%
  \hypersetup{
      pdftitle={A Java-based Extensible Operating System for Embedded Applications},
      pdfauthor={Hong Ong (hong.ong@port.ac.uk)},
      pdfsubject={PhD, Thesis},
      pdfkeywords={Operating Systems,Embedded},
      pdfpagemode=UseOutlines,
      pdfstartpage=1,
      pdfstartview=FitV,
      pdfview=FitV
  }
\fi

\title{ 
 Non-linear Oscillations of Compact Stars and  Gravitational Waves}

\ifpdf
  \author{\href{mailto:andrea.passamonti@port.ac.uk}{Andrea Passamonti}}
  \department{\href{http://www.tech.port.ac.uk/icg/}{Institute of Cosmology and Gravitation}}
  \university{\href{http://www.port.ac.uk}{University of Portsmouth, UK}}
  \principaladvisor{\href{}{Dr.~Marco Bruni}}
  \firstreader{First Reader Name}
  \secondreader{Second Reader Name}
\else
  \author{\Large Andrea Passamonti}
  \department{Institute of Cosmology and Gravitation}
  \university{University of Portsmouth, UK}
  \principaladvisor{Dr.~M.~Bruni}
  \firstreader{First Reader Name}
  \secondreader{Second Reader Name}
\fi

\submitdate{November, 2005}
\copyrightyear{2005}

\begin{document}

\begin{preliminary}
\begin{dedication}
  To my little flower Antonella
\end{dedication}

\newpage

\vspace*{-1.5cm}

\begin{center}
{ \huge  \bf Abstract}
\end{center}

\noindent This thesis investigates in the time domain a particular
class of second order perturbations of a perfect fluid non-rotating
compact star: those arising from the coupling between first order
radial and non-radial perturbations. Radial perturbations of a
non-rotating star, by themselves not emitting gravitational waves,
produce a peculiar gravitational signal at non-linear order through
the coupling with the non-radial perturbations.  The information
contained in this gravitational signal may be relevant for the
interpretation of the astrophysical systems, e.g. proto-neutron stars
and accreting matter on neutron stars, where both radial and
non-radial oscillations are excited. Expected non-linear effects in
these systems are resonances, composition harmonics, energy transfers
between various mode classes.

The coupling problem has been treated by developing a gauge invariant
formalism based on the 2-parameter perturbation theory~(Sopuerta,
Bruni and Gualtieri, 2004),
where the radial and non-radial perturbations have been separately
parameterized.  Our approach is based on the gauge invariant formalism
for non-radial perturbations on a time-dependent and spherically
symmetric background introduced in~Gerlach \& Sengupta (1979) and 
Gundlach \& M. Garc{\'\i}a (2000). 
It consists of further expanding the spherically
symmetric and time-dependent spacetime in a static background and
radial perturbations and working out the consequences of this
expansion for the non-radial perturbations.  As a result, the
non-linear perturbations are described by quantities which are gauge
invariant for second order gauge transformations where the radial
gauge has been fixed.  This method enables us to set up a boundary
initial-value problem for studying the coupling between the radial
pulsations and both the axial~(Passamonti et al.,2006) 
and polar~(Passamonti at el., 2004)
non-radial oscillations.  These non-linear perturbations obey
inhomogeneous partial differential equations, where the structure of
the differential operator is given by the previous perturbative orders
and the source terms are quadratic in the first order perturbations.
In the exterior spacetime the sources vanish, thus the gravitational
wave properties are completely described by the second order Zerilli
and Regge-Wheeler functions.

The dynamical and spectral properties of the non-linear
oscillations have been studied with a numerical code based on
finite differencing methods and standard explicit numerical
algorithms.  The main initial configuration we have considered is
that of a first order differentially rotating and radially
pulsating star, where the initial profile of the stationary axial
velocity has been derived by expanding in tensor harmonics the
relativistic j-constant rotation law.  For this case we have found
a new interesting gravitational signal, whose wave forms show a
periodic signal which is driven by the radial pulsations through
the sources. The spectra confirm this picture by showing that the
radial normal modes are precisely mirrored in the gravitational
signal at non-linear perturbative order. Moreover, a resonance
effect is present when the frequencies of the radial pulsations
are close to the first $w$-mode. For the stellar model considered
in this thesis the gravitational waves related to the fourth
radial overtone is about three orders of magnitude higher than
that associated with the fundamental mode.  We have also roughly
estimated the damping times of the radial pulsations due to the
non-linear gravitational emission. These values radically depend
on the presence of resonances.  For a $10~ms$ rotation period at
the axis and $15~km$ differential parameter, the fundamental mode
damps after about ten billion oscillation periods, while the
fourth overtone after ten only.



\begin{acknowledgements}

\indent Thessaloniki, Sunday 27th November 2005.

\vspace{0.2cm}
\noindent I am writing these acknowledgments in my partially furnished
flat in Thessaloniki, but instead of thanking you by writing words I
would like to organize a ``warming party'' and invite all you here.  I
do not know if the space is enough, but certainly there will be beer
and wine for everyone. I would like to invite my mamma Vanda, my
pap\`a Umberto, my sorella Anna and the Antonella's sweet eyes, who
have always encouraged me with their love. In particular, I am
grateful to Antonella for having shared and sustained my choices, even
though these led us to live in different countries.  It would be a
pleasure for me to invite my rabbit Sam{\'\i} and see him going around
the flat.
I would like to invite Roy Marteens for his kindness and availability,
and for having given me the possibility to work in his group with a
very friendly atmosphere. Of course all the members of the Institute
of Cosmology and Gravitation are welcomed, starting from the
cornerstone of the ICG, alias Chris Duncan, who received me every
morning with her smiles.  There is a glass ready for Rob Crittenden,
David Wands, David Matravers, Bruce Basset for their help and patience
with my initial desperate english.  For this aspect I want to thank
all the British people of the ICG.  They have really shown a
``British aplomb'' during my language mistakes and have tried to
correct me. I am grateful for this and for their friendship to Frances
White, Iain Brown, Kishore Ananda and Richard Brown.  I am very glad
to see at this party Marco Bruni, Carlos Sopuerta, Leonardo Gualtieri
and Alessandro Nagar, who shared with me the problems of my research
and gave me important suggestions. Of course Carlos, my invitation
includes Veronica and the little Sopuertino (Ariel).  It is a pleasure
to thank Nick Stergioulas, Kostas Kokkotas, Nils Andersson and Ian
Hawke for the interest they manifested in my research project and for
the fruitful discussions and suggestions.

The Italian crew in the ICG group has always had an important presence
during these years. I am very grateful to my old friends Andrea
Nerozzi, who first guided me in the Portsmouth life, to Christian
Cherubini ``like the angels'', who succeeded to go beyond the nuclear
densities by cooking a dense ``pasta e fagioli'', to Michele Ferrara
for his friendship and the intense football matches, to Marco
Cavaglia and Sante, for their help especially with the ``vecchia'',
to Fabrizio Tamburini for first guiding me in London and Venice.  It
is not possible a party without Federico Piazza and his ``ma che
meraviglia!'' or without Chris Clarkson who always tried to increase
my half pint of beer with ``Come on Andrea, Come on, another beer!''
With Federico I finally satisfied my child dream, i.e. going to
Wimbledon.  Unfortunately, I went only as a spectator but the day was
great anyway.  Hi Viviana, are you ready to come?  And you Mehri with
your ``polpette del tuo paese?''  Oh ``I am sorry to disturb you'' but
I want that also you Mariam will be here, ``Thank you very much''. I
would invite Ludovica Cotta-Ramusino (remenber to swich off the
mobile), Marta Roldo, Caterina, Shinji Tsujikawa (great giallorosso),
Nuria, Giacomo er lazialetto, Garry Smith and the wonderful Gretta
(but without the Italian shoes), Raz with his laughs and Rolando with
his guitar. I would like to thank Aamir Sharif for the good time
passed together and for making me love cricket. It is really a
pleasure to conclude this acknowledgment by thanking Hong Ong the
Great and his smiles, more than a friend he has been during these
three years a spiritual guide.

I will conclude to thank the magic Castle Road and London for its
multi-cultural atmosphere, for the free Museums that I could visit
many times. In particular, I am grateful to the astonishing Leonardo's
cartoon: ``The Virgin and Child with St. Anne and St. John the
Baptist'' for its intense beauty.

\end{acknowledgements}

\end{preliminary}

\newpage

\thispagestyle{empty}

\vspace*{1.5cm}
\begin{flushleft}
{ \Huge  \bf Notations and convenctions}
\end{flushleft}

\vspace{1cm}

\begin{itemize}

\item The signature of the metric is $\left( -,+,+,+ \right)$, thus time-like
4-vectors have negative norms.

\item The index notation of the tensor fields is the follows: Greek
 indices run from 0 to 3, capital Latin indices from 0 to 1, and small
 Latin indices from 3 to 4.

\item In a tensor expression we use the Einstein's sum convenction.

\item The Eulerian perturbation fields for the first and second order
perturbations are denoted with the notation of the 2-parameter
perturbation theory. Therefore, the upper index (1,0) denotes first
order radial perturbations, (0,1) the first order non-radial
perturbation and (1,1) their coupling.

\item In this thesis we have used the geometrical units in almost all
the expressions. Thus, the speed of light and the gravitational
constant are set $G=c=1$. Therefore, we have that:
\begin{eqnarray}
1~s & = & 2.9979 \times 10^{5}~km   \nn \qquad \qquad \qquad \\
1~g & = & 7.4237 \times 10^{-24}~km \nn \\
M_{\odot} & = & 1.4766~km \nn
\end{eqnarray}

\end{itemize}

\startthesis


\chapter{Introduction}
\label{ch:Introd}

Gravitational waves are the most elusive prediction of Einstein's
theory of gravity. The indirect evidence of their existence relies on
the observations of the binary pulsar PSR 1913+16 (Hulse and
Taylor~\cite{1975ApJ...195L..51H}), that shows a decay of the orbital
period consistent with the loss of angular momentum and energy due to
the emission of gravitational waves. The prospect of starting a new
astronomy based on gravitational radiation and providing a new
corroboration of General Relativity has motivated many theoretical and
experimental researches. As a result, the detection of gravitational
waves seems feasible in the next decade by an international network of
Earth-based laser interferometer detectors (LIGO, VIRGO, TAMA300, and
GEO600)~\cite{virgoetal}, bar resonant antennas (EXPLORER, AURIGA,
NAUTILUS, ALLEGRO)~\cite{bars} and by the Laser Interferometer Space
Antenna (LISA)~\cite{lisa}.  Three scientific runs have been so far
carried out by the LIGO detectors, in collaboration with GEO and TAMA
detectors for two of the three runs and with the bar detector ALLEGRO
for the last run.  The data analysis of the first and second science
run sets upper limits on the gravitational signal emitted by a number
of possible sources, such as stochastic background, coalescing binary
stars, pulsars~\cite{2004PhRvD..69l2004A, 2004PhRvD..69l2001A,
  2004PhRvD..69j2001A, 2004PhRvD..69h2004A,2005PhRvD..72f2001A,
  2005PhRvL..94r1103A}. The third science runs have been performed
with a higher sensitivity and the data analysis leads to a significant
improvement of the gravitational radiation upper
limits~\cite{Abbott:2005ez}.
Meanwhile, a second generation of detectors is already in the design
stage for the exploration of the high frequency band, up to several
kHz (advanced GEO600~\cite{schnabel-2004-21}, wide-band dual sphere
detectors~\cite{Cerdonio:2000bh}), with an improvement of sensitivity
up to two orders of magnitude with respect to the first-generation
instruments.

Gravitational radiation could provide new information about the nature
of astrophysical sources and help in the interpretation of the
dynamical evolution of many such systems.  Among the many sources of
gravitational waves, the oscillations of compact stars are considered
of great interest by astrophysicists and nuclear physicists. The
extreme conditions present in the core of compact stars make them a
unique laboratory, where nearly all the modern theoretical areas of
research in physics can be tested.

In many astrophysical scenarios, compact stars may undergo
oscillating phases. After violent events such as core collapse of
a massive star, an accretion-induced-collapse of a white dwarf, or
a binary white dwarf merger, the newly born protoneutron star is
expected to pulsate non-linearly before various dissipative
mechanisms damp the oscillations. Another system where pulsating
phases may occur is a massive meta-stable compact object, which is
born after the merger of a binary neutron star system.  The
gravitational signal emitted by the stellar oscillations lies in
the high frequency band ($\nu \gtrsim 1 kHz$) and strongly depends
on the structure and physics of the star, for instance on the
equation of state, rotation, crust, magnetic fields as well as on
the presence of dissipative effects such as viscosity, shock
formation, magnetic breaking, convective outer layers, etc. With a
detailed analysis of the gravitational wave spectrum emitted by
stellar pulsations we could infer through asteroseismology the
fundamental parameters of neutron stars, such as mass, radius and
rotation rate~\cite{1998MNRAS.299.1059A, Andersson:1998ak}.  This
information is necessary for the nuclear physicists as they can
test the equations of state proposed for the description of matter
at supra-nuclear densities.  However, the weakness of the
gravitational signal and the noise associated with the location
and technology of the detectors compels theorists to provide more
and more accurate models to predict the spectral and wave form
properties of the gravitational signal.  These templates are
indispensable for enhancing the chances of detection, by
extracting the signal from the noise with statistical methods.

The spectral and dynamical properties of the oscillations of
compact stars have been extensively investigated during the last
forty years in Newtonian and Einstein theories of gravity.  Linear
perturbative techniques are appropriate for the analysis of small
amplitude pulsations both in the frequency~\cite{Thorne:1967th,
kokkotas-1999-2} and the time domain approach~\cite{allen-1998-58,
Ruoff:2001ux, Seidel:1987in, Seidel:1990xb, Nagar:2004ns}. In
General Relativity, the oscillation spectrum of a compact object,
such as black holes and neutron stars, is characterized by a
discrete set of quasi-normal modes (QNM).  These modes have
complex eigenfrequencies whose real part describes the oscillation
frequency, and the imaginary part the damping time due to the
emission of gravitational waves.  The classification of QNM is
well known for a large set of stellar models and can be divided
schematically in fluid and spacetime modes.  The fluid modes have
a Newtonian counterpart and can be sub-classified by the nature of
the restoring force that acts on the perturbed fluid element.  The
spacetime modes are purely relativistic and are due to the
dynamical role assumed by the spacetime in General Relativity
(more details are given in chapter~\ref{ch:4Lin_Pert} and
reference therein).

Rotating stars in General Relativity can be described with various
approximations, such as the slow-rotation
approximation~\cite{Hartle:1967ha} or recently with codes
developed in numerical relativity~\cite{2003LRR.....6....4F}.  The
former approach is based on a perturbative expansion of the
equations in powers of the dimensionless rotation parameter
$\epsilon = \Omega / \Omega_{K}$, where $\Omega$ is the uniform
angular rotation and $\Omega_{K}$ is the Keplerian angular
velocity, which is defined as the frequency of a particle in
stable circular orbit at the circumference of a star.
The measured period of the fastest rotating pulsar corresponds to
a relatively small rotation parameter $\epsilon \sim 0.3$, which
may suggest that the slow rotation approximation provides an
accurate description of rotating stars even for high rotation
rates. However, in these cases the accuracy of this perturbative
approach is different for the various physical stellar quantities.
For instance, the quadrupole moment shows an accuracy to better
than twenty percent, while the radius of the corotating and
counterrotating innermost stable circular orbits is accurate to
better than one percent~\cite{2005MNRAS.358..923B}.
Nevertheless, a protoneutron star may be expected to have a higher
rotation rate that is not possible to describe with the slow rotation
approximation.  These regimes can be better addressed in numerical
relativity by evolving the full set of non-linear Einstein
equations~\cite{Dimmelmeier:2004prep, Stergioulas:2000vs,
  2003LRR.....6....3S}.  Furthermore, recent works on the core
collapse~\cite{Dimmelmeier:2002bk, Dimmelmeier:2002bm}, r-mode
instability~\cite{2000ApJ...531L.139R, 2001MNRAS.322..515L,
  2001PhRvL..86.1152L}, accretion from a
companion~\cite{1993ApJ...419..768F}, or supernova fall back
material~\cite{2002MNRAS.333..943W}, show that a neutron star
manifests a degree of differential rotation.  \\
These studies have clarified the effects of rotation on the
dynamics of the oscillations, as well as showed the importance of
using a relativistic treatment that takes into account the effects
of the dragging of inertial frames.  In particular, the presence
in rotating stars of instabilities due to emission of
gravitational waves has gained great attention. Almost all classes
of oscillations of rotating stars, such as the f- and r-modes, are
potentially unstable to the so-called Chandrasekhar-
Friedman-Schutz (CFS) instability~\cite{1970PhRvL..24..762C,
1978ApJ...222..281F}. This appears because beyond a critical value
of the stellar angular velocity, a mode that in a corotating frame
is retrograde and then has negative angular momentum, may appear
moving forward in the inertial frame of a distant observer. As a
result, this inertial observer will detect gravitational waves
with positive angular momentum emitted by this mode. Thus, the
gravitational radiation removes angular momentum by the retrograde
mode by making it increasingly negative and then leading to
instabilities.  The losses of the angular momentum through
gravitational waves slow down the star on secular timescales;
eventually the star rotates slower than a critical value and the
mode becomes stable. These instabilities could be strong sources
of gravitational radiation and also limit the rotation rate of
neutron stars, providing a possible explanation for the measured
rotation period of pulsars. Many studies are currently dedicated
to understand whether viscosity, magnetic fields, shock waves on
the stellar surface or non-linear dynamics of oscillations may
saturate this instability.

The non-linear analysis of stellar oscillations is more complex
and only recently are some investigations being carried out, due
to improvements achieved by the non-linear codes in numerical
relativity~\cite{2003LRR.....6....4F}.  Different methods have
been used to investigate the properties of non-linear
oscillations, such as for instance 3-dimensional general
relativistic hydrodynamics code in Cowling or conformal flatness
approximations~\cite{Stergioulas:2003ep, Dimmelmeier:2004prep}, a
combination of linear perturbative techniques with general
relativistic hydrodynamics simulations~\cite{2005MNRAS.356.1371Z},
or a new method where the non-linear dynamics is studied as a
deviation from a background, which is described by a stellar
equilibrium configuration~\cite{Sperhake:2001xi}.  These works,
which have been dedicated to investigate the non-linear dynamics
of different astrophysical systems: non-linear oscillations of a
torus orbiting a black hole~\cite{2005MNRAS.356.1371Z}, non-linear
axisymmetric pulsations of uniform and differential rotating
compact stars~\cite{Dimmelmeier:2004prep} and non-linear radial
oscillations of non-rotating relativistic
stars~\cite{Sperhake:2001xi}, have revealed a new phenomenology
associated with the non-linear regimes, the presence in the
spectrum of non-linear harmonics.  These harmonics arise from the
coupling between different classes of linear modes or from
non-linear self-couplings~\cite{Sperhake:2001xi}, and have a
characteristic that could be appealing for the detection of
gravitational waves: their frequencies appear as linear
combinations of the linear oscillation modes.
Therefore, some of these non-linear harmonics (sub-harmonics) can
emerge at lower frequencies than the related linear modes, and
then be within the frequency range where the detectors have higher
sensitivity. However, since the amplitude of non-linear
perturbations is usually the product of the amplitudes of first
order perturbations, in order to have a detectable gravitational
wave strain one needs non-linear effects that can enhance the
gravitational signal, such as
resonances, parameter amplifications or instabilities. \\
Strong non-linear regimes are adequately studied with a fully
non-perturbative approach.  However, many interesting physical
effects of  mild non-linear dynamics can be well addressed by
second order perturbative techniques. An example is given by the
analyses of black hole collisions in the so-called ``close limit
approximation'', where the second order perturbations of
Schwarzschild black holes~\cite{Gleiser:1995gx, Garat:2000gp} have
provided accurate results even in non-linear regimes where the
perturbative methods are expected to fail.  Non-linear
perturbative methods have been successfully used also for studying
linear perturbations of rotating stars, where the rotation is
treated perturbatively with the slow rotation approximation
~\cite{Hartle:1967ha, 2003LRR.....6....3S}. \\
\indent An important aspect of second order perturbative analyses
is that of providing an estimate of the error associated with the
first order treatment, as there is not an \emph{a priori} method
to determine the accuracy of the linear perturbative results.
Thus, the convergence and the corrections associated with any term
of the perturbative series can be determined only by investigating
higher perturbative orders. Furthermore, non-linear perturbative
equations are usually a system of partial differential equations,
thus their numerical integration is computationally less expensive
than the full Einstein equations which are treated in numerical
relativity. This relative simplicity of the perturbative approach
may then provide accurate results and can also be used to test the
full non-linear simulations.  However, an extension to second
order perturbative investigations is not always
straightforward~\cite{Gleiser:1995gx, Garat:2000gp}.  Some issues
may arise from the identification of the physical quantities among
the second order perturbative fields or from the movement of the
stellar surface in non-linear stellar
oscillations~\cite{Sperhake:2001si, Sperhake:2001xi}.  \\
\indent In physical systems where the perturbative analysis can be
described by more than a single parameter, as for stellar
oscillations of a slowly rotating star or mode coupling between
linear perturbations, the multi-parameter relativistic
perturbation theory~\cite{Bruni:2002sm, sopuerta-2004-70} can help
the interpretation of the gauge issues of non-linear
perturbations.  The identification of gauge invariant quantities
allows us to have direct information about the physical properties
of the system under consideration. The construction of such
quantities is not in general simple, but recent
works~\cite{Nakamura:2003wk, Nakamura:2004gi} show how to build
second order gauge-invariant perturbative fields from the
knowledge of the associated first order gauge invariant
perturbations.
For a specific class of astrophysical systems a gauge invariant and
coordinate independent formalism has been introduced nearly thirty
years ago ~\cite{Gerlach:1979rw, Gerlach:1980tx} for the analysis of
one-parameter non-radial perturbations on a time dependent and
spherically symmetric background. Recently, this formalism has been
further developed~\cite{Martin-Garcia:1998sk, Gundlach:1999bt}, and
has been used to study non-radial perturbations on a collapsing
star~\cite{2003PhRvD..68b4002H} and for linear perturbation on a
static star~\cite{Nagar:2004ns}.

The research project we have been working on aims to extend the
perturbative analysis of compact stars at non-linear orders, in
order to have a more comprehensive understanding of stellar
oscillations and the related gravitational radiation.  In
particular, this thesis presents a gauge invariant formalism and a
numerical code for studying the coupling between the radial and
non-radial perturbations of a perfect fluid spherical star.
The formalism for the polar perturbations has been worked out in a
first paper~\cite{Passamonti:2004je}. The formalism and applications
to axial perturbations are presented in~\cite{Passamonti:2005axial}.
Work in progress on the applications of the polar perturbative
formalism will be presented in a future work.
Radial and non-radial oscillations can be excited in the aftermath
of a core collapse or by accreting matter on a neutron star.
Radial perturbations of a non-rotating star are not damped by
emission of gravitational radiation, but they can emit
gravitational waves at non-linear order through the coupling with
the non-radial perturbations.  This picture changes in the
presence of rotation, where the radial pulsations become sources
of gravitational radiation and form a new class of modes, called
quasi-radial modes.
This coupling may be interesting for instance during a core
bounce, where it is expected that an excitation prevalently of the
quasi-radial and quadrupole modes. Even though the quadrupole
component provides the dominant contribution to the gravitational
radiation, the radial pulsations may store a considerable amount
of kinetic energy and transfer a part of it to the non-radial
perturbations. As a result, this non-linear interaction could
produce a damping of the radial pulsations and an interesting
gravitational signal. The strength of this signal depends
naturally on the efficiency of the coupling, which is an effect
worth exploring.  In this thesis, we start to investigate this
non-linear effect for small oscillations of a non-rotating star
with the aim of including rotation in future works.  \\
\indent The polar coupling, i.e. between the radial and the polar
sector of non-radial perturbations, is expected to be \emph{a
priori} more effective than the axial case. Indeed, the linear
polar modes have a richer spectrum than the axial sector and from
the values of the frequencies and the damping times of the fluid
QNMs, the resonances and composition harmonics should be more
probable between the polar fluid and the radial modes.  However,
for the purposes of this thesis we have implemented a numerical
code for the axial coupling for mainly two reasons: \emph{i)} the
axial coupling can have interesting physical effects, \emph{ii)}
the perturbative equations are simpler and enable us to understand
better the issues related to the numerical stability and accuracy
of the code as well as the effects of the low density regions
near the stellar surface on the non-linear simulations, etc. \\
\indent When we consider the first order axial non-radial
perturbations, we see that the only fluid perturbation is the
axial velocity, which can be interpreted to describe a stationary
differential rotation. Therefore, the linear axial gravitational
signal does not have any dependence on the dynamics of the stellar
matter. This picture changes at coupling order, where the
differential rotation and first order metric perturbations can
couple with the radial pulsations and source the axial
gravitational waves. We will see in chapter~\ref{sec:NumInt} that
this axial coupling produces a new class of quasi-radial modes,
which can exist only for differentially rotating stars.

This thesis is organized in seven chapters. In
chapter~\ref{ch3:Non-Lin_Per}, we introduce the perturbative
formalism used in this work, i.e. the multi-parameter relativistic
perturbation theory and the gauge invariant formalism introduced
by Gerlach and Sengupta and further developed by Gundlach and
Martin Garcia~(GSGM).  In chapter ~\ref{ch:4Lin_Pert}, we describe
the linear perturbations of a spherical star, i.e. the radial
pulsations, the polar and axial non-radial perturbations.  The
equations for describing the coupling between the radial and axial
and polar non-radial oscillations are presented in
chapter~\ref{ch:5_NL}, where in addition we discuss also the
boundary conditions.  In chapter~\ref{ch:GINLP}, we present the
proof of the gauge invariance of the perturbative tensor fields
that describe the non-linear perturbations for this coupling.
Chapter~\ref{sec:NumInt} is dedicated to the numerical code that
simulates in the time domain the evolution of the coupling between
the radial and axial non-radial perturbations. In this chapter we
give all the technical details relating to the code and the
results of the simulations. Finally in
chapter~\ref{ch:conclusions}, the conclusions and possible future
developments are discussed.

The appendix has seven sections. We have reported the source terms
of the equations derived by Gundlach and Mart\'{\i}n Garc\'{\i}a
in section~\ref{full-equations}. In section~\ref{AppSW11MG}, we
write the full expressions of the sound wave equation for the
fluid variable $H$, while in sections~\ref{AppSources}
and~\ref{AppSources_axial} we respectively present the source
terms of the perturbative equations that describe the coupling
between the radial pulsations and polar and axial non-radial
perturbations.  In addition, in section~\ref{sec:Tens_Harm} we
give the tensor harmonics, while some of the numerical methods
used in the numerical code are given in
sections~\ref{sec:finit_appr} and~\ref{sec:Num_meth}.


\chapter{Non-Linear Relativistic Perturbation Theory}
\label{ch3:Non-Lin_Per}

Exact solutions of the equations of physics may be obtained for
only a limited class of problems. This aspect is particularly
present in General Relativity, where the complexity of the
astrophysical systems and the non-linear Einstein field equations
allows us to describe exactly only simplified and highly symmetric
cases.
Among various approximation techniques, perturbation methods are
appropriate whenever the problem under consideration closely
resembles one which is exactly solvable. It assumes that the
difference from the exactly solvable configuration is small and
that one may deviate from it in a gradual fashion. Deviations of
the physical quantities from their exact solutions are referred to
as perturbations.  Analytically, this is expressed by requiring
that the perturbation be a continuous function of a parameter,
measuring the strength of the perturbation.
Although perturbative techniques are more appropriate for small
values of the perturbative parameter, sometimes they can give
reliable results also for mildly non-linear regimes as shown for
example in the analysis of black-hole collision
\cite{1999bhgr.conf..351P}. Hence, in many cases the validity
limit of perturbative methods cannot be determined \emph{a
priori}. A more accurate estimation can be reached by studying the
convergence of the perturbative series, which then involves the
analysis of the second or higher perturbative orders.

The gauge issue arises in General Relativity as in any other
theory based on a principle of general covariance. The
perturbative description of a physical system is not unique due to
the presence of unphysical degrees of freedom related to the
gauge, i.e. to the system of coordinates chosen for the analysis.
This ambiguity can be eliminated either by fixing a particular
gauge or by constructing perturbative variables which are
invariant for any gauge transformation. In the former case, the
properties and symmetries of the physical systems can help us to
decide an appropriate gauge.  In the latter approach, the
identification of the gauge invariant fields is the difficult
task.

In this section we review the perturbative framework we have used
for investigating the coupling between the radial and non-radial
perturbations.  In section~\ref{sec:3.1_MPPT} we report the main
results of the multi-parameter perturbation theory introduced by
Bruni et al. \cite{Bruni:2002sm} and Sopuerta et
al.~\cite{sopuerta-2004-70}. Section~\ref{sec:GSGM} is dedicated
to the formalism introduced by Gerlach and Sengupta
\cite{Gerlach:1979rw, Gerlach:1980tx}, which has been further
developed by Gundlach and M. Garcia \cite{Gundlach:1999bt}, while
in section~\ref{sec:3.3-NLFW} we outline the perturbative
structure of our work which is based on the 2-parameter expansion
of a static background.

\section{Multi-parameter perturbation theory}
\label{sec:3.1_MPPT}

Perturbation theory assumes the existence of two spacetimes,
namely the background and perturbed spacetimes.  The former is the
spacetime described by an exact solution of the field equations,
while the latter is the physical spacetime that the perturbation
theory attempts to describe through the perturbation fields. The
main requirement is that the physical description of the perturbed
spacetime slightly deviates from that of the background solution.

Let us call  $\mathcal{M}$ the spacetime manifold. The
multi-parameter relativistic perturbation formalism assumes the
existence of a smooth multi-parameter family of spacetime models
which are diffeomorphic to the physical spacetime:
\begin{equation}
M_{\vec{\lambda}} = \left( \mathcal{M}, \mathcal{T}_{\vec{\lambda}}
\right) \, . \label{multi-sptm}
\end{equation}
The quantity $\mathcal{T}_{\vec{\lambda}}$ represents a set of
analytic tensor fields which are defined on $\mathcal{M}$ and that
describe the physical and geometrical properties of the physical
spacetime.  The N-parameter vector $\vec{\lambda} \in \mathbb{R}^{N}$
labels any diffeomorphic representation of the physical spacetime
$M_{\vec{\lambda}}$, and controls the deviation from the background
quantities due to the contribution of the various perturbative
parameters of the system under consideration.  In this notation the
background manifold $M_{b}$ is then identified with the spacetime
model~$M_{b} \equiv M_{\vec{0}}$.
Furthermore, the validity of the Einstein field equations is assumed
on any manifold $M_{\vec{\lambda}}$:
\begin{equation}
 E \left( \mathcal{T}_{\vec{\lambda}}  \right) = 0 \, .
\end{equation}
In a perturbative approach, the comparison of perturbed and
background variables is crucial for determining the accuracy of
the perturbative description. In a physical theory based on a
principle of general covariance as in General Relativity, this
procedure requires a more precise definition that takes into
account the gauge issue.
Let us for instance consider a relation commonly used in perturbation
theory,
\begin{equation}
\mathcal{T}_{\vec{\lambda}} \left( q \right) = \mathcal{T}_{ b }
\left( p \right) + \delta \mathcal{T}_{\vec{\lambda}} \left( p \right)
\, , \qquad \textrm{where} \quad \delta \mathcal{T}_{\vec{\lambda}}
\ll \mathcal{T}_{ b} \, .
\label{pert_rel}
\end{equation}
Here,  $q$ and $\mathcal{T}_{\vec{\lambda}}$ are respectively a
point and a tensor field in the perturbed manifold
$M_{\vec{\lambda}}$, while the point $p$ and the tensors
$\mathcal{T}_{ b}$ and $\delta \mathcal{T}_{\vec{\lambda}}$ belong
to the background $M_{ b}$. From equation (\ref{pert_rel}) the
tensor $\mathcal{T}_{\vec{\lambda}} \left( q \right)$ can be
considered as a small deviation of the background value
$\mathcal{T}_{ b}$. However, we can notice that the perturbed and
the background tensors are applied at points belonging to
different manifolds.  Therefore, in order to have a well posed
relation it is necessary to establish a correspondence between
these two points $p$ and $q$ and consequently between the three
tensor fields in the expression (\ref{pert_rel}). This point
identification between the various representations of the physical
spacetime $M_{\vec{\lambda}}$ is provided by a N-parameter group
of diffeomorphisms~$\mathcal{D}_{\varphi} = \left\{ \varphi
_{\vec{\lambda}} : \vec{\lambda} \in \mathbb{R}^N \right\}$,
\begin{eqnarray}
\varphi : \mathcal{M} \times \mathbb{R}^N & \rightarrow &  \mathcal{M}  \label{Ndiff} \\
  (p,\vec{\lambda} )  & \mapsto & \varphi ( p,\vec{\lambda} ) \equiv
\varphi_{\vec{\lambda}} (p) \nn  \, ,
\end{eqnarray}
where the identity element corresponds to the null vector
$\vec{\lambda} = \vec{0}$, i.e. $\varphi_{\vec{\lambda}} (p) = p
$. The choice of the identification map $\varphi$ is completely
arbitrary and in perturbation theory this arbitrariness is called
``gauge freedom''. It is worth noticing that the gauge issue in
perturbation theory is in general independent on the gauge related
to the background spacetime, which fixes the system of coordinates
only on the background manifold $M_{ \, b}$.  In order to have a
correct description of a physical system the physical observables
have to be isolated from the gauge degrees of freedom. This can be
accomplished by fixing a particular gauge, where the variables
assume the correct physical meaning, or alternatively by
determining a set of gauge invariant quantities.  The latter
procedure can be more difficult to realize, but it provides
directly the physical quantities of the system.


\subsection{Taylor expansion}
\label{subsec:MLPT_Tay}

A perturbative solution of the Einstein equations is built as a
correction of the background solution. This property, expressed in
 equation (\ref{pert_rel}), allows us to approximate the physical
variables and the field equation by their Taylor series. However,
in order to define correctly a relativistic multi-parameter Taylor
expansion some concepts related to the properties of the
N-parameter group of diffeomorphisms have to be specified.

In general, a consistent perturbative scheme should not depend on
the order followed for performing two or more perturbations.  We
can then consider an Abelian group of diffeomorphisms which is
defined by equation~(\ref{Ndiff}) and the following composition
rule:
\begin{equation}
\varphi_{\vec{\lambda}} \circ \varphi_{\vec{\lambda '}} = \varphi
_{\vec{\lambda} + \vec{\lambda '}} \quad \textrm{for}  \quad
\forall \, \vec{\lambda}, \vec{\lambda '} \in \mathbb{R} ^N \, .
\end{equation}
Therefore, we can decompose every diffeomorphism $\varphi_{\vec{\lambda}}$
as a product of N one-parameter diffeormorphisms:
\begin{equation}
\varphi_{\vec{\lambda}} = \varphi_{(\lambda _1,0,..,0)} \circ \cdot
\cdot \cdot \cdot \circ ~\varphi_{(0,..,0,\lambda_N)} \, .
\label{dec_rule}
\end{equation}
It is also evident from the previous property and the commutation
rules that there are $N !$ equivalent decompositions of the
diffeomorphism~$\varphi_{\vec{\lambda}}$.
In any point of the perturbed manifold $M_{\vec{\lambda}}$, a
diffeomorphism $\varphi_{\vec{\lambda}}$ defines a N-parameter flow
$\varphi_{\vec{\lambda}} (p)$. This flow is generated by a vector
field $\vec{\zeta} \in \mathbb{R} ^N$ that acts on the tangent space
of $M_{\vec{\lambda}}$.  In an Abelian group the generators of two
different flows commute $\left[ \vec{\zeta}, \vec{\zeta}' \right] = 0$
and a N-dimensional basis with the N independent vectors can be
defined:
\begin{equation}
\left\{ \vec{\zeta}_{k}   \right\} _{k=1..N} \, , \qquad \textrm{where} \quad
\vec{\zeta}_{k} = \left(0,..,\zeta_k,..0  \right) \, .
\end{equation}
This basis generates the N one-parameter groups of diffeomorphisms
of equation~(\ref{dec_rule}) and can be used to decompose
the vector field $\vec{\zeta}$ in its N components
\begin{equation}
\vec{\zeta} = \sum_{k=1}^{N} \vec{\zeta}_{k} \quad \textrm{where}
\quad \vec{\zeta}_{k} = \left(0,..,\zeta_k,..0  \right)   \, ,
\end{equation}
and to define the Lie derivative of an arbitrary tensor field $\mathcal{T}$
\begin{equation}
 \pounds _{\vec{\zeta}_{k} } \mathcal{T} = \left. \frac{\partial
\varphi^{\ast} _{\vec{\lambda}_{k}} \mathcal{T} }{\partial \lambda_k}
\right| _{\lambda_k = 0 } \quad \textrm{where} \quad \vec{\lambda}_{k}
= (0,..,\lambda_k,..,0) \, .
\end{equation}
The operator $\varphi^{\ast} _{\vec{\lambda}_{k}}$ is the
\emph{pull-back} associated with the flow $\varphi_{\vec{\lambda}
_{k}}$.

\noindent The Taylor expansion of the pull-back~$\varphi^{\ast}
_{\vec{\lambda}_{k}}$ around the parameter $\vec{\lambda} =
\vec{0}$ is then defined as follows:
\begin{equation}
\varphi^{\ast}_{\vec{\lambda} _{k}} \mathcal{T} = \sum _{n \geq 0
} \frac{\lambda ^{n}_{k}}{n!} ~\pounds _{\vec{\zeta} _{k}} ^{n}
\mathcal{T} \equiv \exp{\left( \la _{k} \pounds _{\vec{\zeta}
_{k}}  \mathcal{T} \right)} \, . \label{Tay_pullback}
\end{equation}
The last equality in equation~(\ref{Tay_pullback}) is a formal
definition that will be very useful later for carrying out
calculations with the Baker-Campbell-Hausdorff (BCH) formula.  The
definition~(\ref{Tay_pullback}) and the diffeomorphism
decomposition expressed in  equation~(\ref{dec_rule}) allows us to
define the Taylor expansion associated with the diffeomorphism
$\varphi _{\vec{\la}}$:
\begin{equation}
\varphi^{\ast}_{\vec{\lambda} } \mathcal{T} = \sum _{n_1..n_N  \geq 0 }
\left(  \prod _{p=1}^{N}  \frac{\lambda ^{n_{p}}_{p}}{n_{p}!}
~\pounds _{\vec{\zeta}_{p}} ^{n_{p}} \right) \mathcal{T}  \equiv
\exp{\left( \sum_{p=1}^{N} \la _{p}~\pounds _{\vec{\zeta}_{p}}   \right) }
\mathcal{T}   \label{Tay_exp}
\end{equation}

\subsection{Perturbations}
\label{subsec:MLPT_Pert}

In a particular gauge,  the exact perturbations of a generic tensor field
$\mathcal{T}$ are  defined as follows:
\begin{equation}
\Delta \mathcal{T}_{\vec{\lambda}} ^{\varphi}   \equiv \varphi^{\ast}_{\vec{\lambda}}
\mathcal{T}_{\vec{\lambda}} - \mathcal{T}_{\vec{0}} \, ,  \label{Pert_def}
\end{equation}
where the tensors $\Delta \mathcal{T} _{\vec{\lambda}} ^{\varphi}$ and
$\varphi^{\ast}_{\vec{\lambda}} \mathcal{T}_{\vec{\lambda}}$ as
well as the background tensor $\mathcal{T}_{ b }$ are defined on the
background spacetime $M_{ b }$.  The definition~(\ref{Pert_def})
indicates that the background $M_{ b }$ is the fundamental spacetime
where all the perturbative fields are transported by the
N-parameter flows and then compared with the background fields.
The definition (\ref{Pert_def}) can be rewritten by using the Taylor
expansion (\ref{Tay_exp}) in the following form
\begin{equation}
\Delta \mathcal{T} _{\vec{\lambda}} ^{\varphi} \equiv \sum_{n_1,..,n_N \geq 0}
\left( \prod _{p=1}^{N} \frac{\lambda_{p}^{n_{p}} }{n_{p} !}  \right)
\delta _{\varphi} ^{ \vec{n}} \mathcal{T}
- \mathcal{T}_{ b } \, ,  \label{Pert_def_2}
\end{equation}
where the vector $\vec{n} = \left( n_1,..,n_N  \right)$ controls the
perturbation order of a tensor field with respect to the N-parameter,
\begin{equation}
\delta _{\varphi} ^{ \vec{n}} \mathcal{T} \equiv \frac{\partial ^{n_T}
\varphi^{\ast}_{\vec{\lambda}} T }{\partial \la _{1}^{n_1} . . \la
_{N}^{n_N} }  = \prod_{p=1}^{N} \pounds _{\zeta_{p} ^{n_{p}}} \mathcal{T} \, ,
\end{equation}
and $n_{T} = \sum _{p=1}^{N} n_{p}$ is the total perturbation order.

\subsection{Gauge transformations}
\label{subsec:MLPT_GT}

The representation of a perturbative field 
in the background manifold depends in general on the gauge choice
$\varphi _{\vec{\lambda}}$.  Let consider two generic gauges
represented by the diffeomorphisms $\varphi$ and $\psi$, which are
generated respectively by the vector fields $\left( {}^\varphi
\zeta _{1},.., {}^\varphi \zeta _{N} \right)$ and $\left( {}^\psi
\zeta _{1},.., {}^\psi \zeta _{N} \right)$. A gauge transformation
is then defined by a diffeomorphism
\begin{equation}
\Phi_{\vec{\lambda}} : \mathcal{M} \rightarrow
\mathcal{M}
\end{equation}
that for a given $\vec{\lambda}$, connects the physical descriptions
determined in the two gauges $\varphi$ and $\psi$ as follows:
\begin{equation}
 \Phi :   \equiv  \varphi_{\vec{\lambda}}^{-1}
\circ \psi_{\vec{\lambda}} = \varphi_{ - \vec{\lambda}}
\circ \psi_{\vec{\lambda}} \, .  \label{gauge_transf_diff}
\end{equation}
The family of all diffeomorphisms that relate two gauges does
not form in general a group,
\begin{eqnarray}
\Phi : \mathcal{M} \times \mathbb{R}^N & \rightarrow &  \mathcal{M} \\
  (p,\vec{\lambda} )  & \mapsto & \Phi ( p,\vec{\lambda} ) \equiv
\Phi_{\vec{\lambda}} (p) \nn \, .
\end{eqnarray}
Since the gauge transformation $\Phi_{\vec{\lambda}}$ is a
diffeomorphism we can extend to it some definitions used in the
previous sections for the identification maps. For instance, the
pull-back $\Phi_{\vec{\lambda}}^{\ast}$ of a generic tensor field
$\mathcal{T}$ induced by the gauge transformation
(\ref{gauge_transf_diff}) can be defined by using the expression
(\ref{Tay_exp}) in the following way:
\begin{eqnarray}
\Phi_{\vec{\lambda}} ^{\ast} \mathcal{T} & = & (\varphi_{-\vec{\lambda}}
\circ \psi_{\vec{\lambda}} )^{\ast} \mathcal{T} =
\psi_{\vec{\lambda}} ^{\ast} \circ \varphi_{-\vec{\lambda}} ^{\ast}
\mathcal{T}  \nn \\
& = & \exp{ \left(\sum_{P=1}^{N} \lambda _{P} \pounds _{{}^{\psi}
\zeta_{P}  }  \right) } \exp{\left( - \sum_{P=1}^{N} \lambda _{P}
\pounds _{{}^{\varphi} \zeta_{P} }  \right) }  \mathcal{T} \, .
\label{GT_pull}
\end{eqnarray}
The gauge transformation $\Phi_{\vec{\lambda}}$ is generated by the
vector field $\xi _{\vec{n}}$, which can be in general expressed in
terms of the two generators of the gauge transformation $\varphi$ and
$\psi$.  In \cite{sopuerta-2004-70}, the authors derive the relations
between these gauge generators as well as among the perturbation
fields by using the Baker-Campbell-Hausdorff (BCH) formula, which for
two linear operators $A,B$ is so defined:
\begin{equation}
e^{A} e^{B} = e^{f(A,B)}  \, ,
\end{equation}
where the functional $f(A,B)$ is given by,
\begin{eqnarray}
f(A,B) = \sum_{m\geq 1}^{\infty} \frac{(-1)^{m-1}}{m}
\hspace{-5mm}
\sum_{\begin{array}{c} p_i,q_i \\
p_i+q_i\geq 1\end{array}}
\hspace{-5mm}
\frac{[\overbrace{A\cdots A}^{p_1}
\overbrace{B\cdots B}^{q_1} \cdots \overbrace{A\cdots A}^{p_m}
\overbrace{B\cdots B}^{q_m}]}{\left[\sum_{\alpha=1}^{m}(p_j+q_j)\right]
\prod_{\alpha=1}^{m}p_\alpha!q_\alpha!} \,, \label{ch2}
\end{eqnarray}
and where in the previous expression the following notation for the
commutation operators has been used $\left[X_1 X_2 X_3 \cdot \cdot
X_n \right] \equiv \left[ \cdot \cdot \cdot \left[ \left[X_1, X_2
\right], X_3 \right], \cdot \cdot \cdot ,X_n \right]$.  By using
the BCH formula, the gauge transformation (\ref{GT_pull}) reduces
to a single exponential operator,
\begin{equation}
\Phi_{\vec{\lambda}} ^{\ast} \mathcal{T} = \exp{ \left\{  \sum_{n_1,..,n_N \geq 0}
\left(   \prod _{p=1}^{N} \frac{\la_p^{n_{p}}}{ n_p !}
\right) \pounds _{\xi_{\vec{n}}} - \mathcal{I}   \right\} } \mathcal{T} \, ,  \label{GT_BCH}
\end{equation}
where $\mathcal{I} \equiv \pounds _{\vec{0}}$ is the identity operator.

The gauge transformations at every perturbative order are then
determined
with the following procedure: \\
%
\emph{i)} by using the definition~(\ref{gauge_transf_diff}), one
defines a new relation between the pull-backs $\varphi_{\vec{\la}}
^{\ast} \mathcal{T}$ and $\psi _{\vec{\la }} ^{\ast} \mathcal{T}$:
\begin{equation}
\psi _{\vec{\la }} ^{\ast}
\mathcal{T}_{\vec{\la }} = \Phi _{\vec{\la }}  ^{\ast} \varphi
_{\vec{\la }} ^{\ast} \,  \mathcal{T}_{\vec{\la }} \, . \label{GT_rel}
\end{equation}
\emph{ii)} Thus, one can expand equation (\ref{GT_rel}) by using
the expressions (\ref{Tay_exp}) and (\ref{GT_BCH}). In doing that, one
can use the linearity of the functional $f(A,B)$ on the operators $A$
and $B$ and their commutators, and also the linearity of the operators
$A,B$ on the respective Lie derivatives.  \emph{iii)} At the end, the
desired relations can be determined by comparing the terms order by
order (for more details see~\cite{sopuerta-2004-70}).
%

In case of 2-parameter perturbations $N=2$, which is the parameter
number used in this thesis, the gauge transformations
at first order are the well known relations:
\begin{eqnarray}
\delta _{\psi} ^{(1,0)} \mathcal{T} & = & \delta _{\varphi} ^{(1,0)}
\mathcal{T} + \pounds _{\xi_{(1,0)}} \mathcal{T}_{b} \, ,
\label{gtr10_base} \\
\delta _{\psi} ^{(0,1)} \mathcal{T} & = & \delta
_{\varphi} ^{(0,1)} \mathcal{T} + \pounds _{\xi_{(0,1)}}
\mathcal{T}_{b} \, . \label{gtr01_base}
\end{eqnarray}
At second order, the perturbation fields in the two gauges are related
as follows:
\begin{eqnarray}
\delta _{\psi} ^{(2,0)} \mathcal{T} & = & \delta _{\varphi} ^{(2,0)}
\mathcal{T} + 2 \pounds _{\xi _{(1,0)}} \delta _{\varphi} ^{(1,0)}
\mathcal{T} + \left( \pounds _{\xi_{(2,0)}} + \pounds
_{\xi_{(1,0)}}^{2} \right)\mathcal{T}_{b} \, , \\
\delta _{\psi}
^{(0,2)} \mathcal{T} & = & \delta _{\varphi} ^{(0,2)} \mathcal{T} + 2
\pounds _{\xi _{(0,1)}} \delta _{\varphi} ^{(0,1)} \mathcal{T} +
\left( \pounds _{\xi_{(0,2)}} + \pounds _{\xi_{(0,1)}}^{2}
\right)\mathcal{T}_{b} \, , \\
\delta _{\psi} ^{(1,1)} \mathcal{T} & =
& \delta _{\varphi} ^{(1,1)} \mathcal{T} + \pounds _{\xi _{(1,0)}}
\delta _{\varphi} ^{(0,1)} \mathcal{T} + \pounds _{\xi _{(0,1)}}
\delta _{\varphi} ^{(1,0)} \mathcal{T} \nn  \label{gtr11_base} \\
&&{} + \left[ \pounds
_{\xi_{(1,1)}} + \left\{ \pounds _{\xi_{(1,0)}}, \pounds
_{\xi_{(0,1)}}\right\} \right] \mathcal{T}_{b} \, ,
\end{eqnarray}
where $\left\{ \cdot, \cdot \right\}$  is the anticommutator.
Gauge transformations for higher perturbative orders can be found
in reference~\cite{Bruni:2002sm}.

\section{Gauge invariant perturbative formalism (GSGM)}
\label{sec:GSGM}

Linear perturbations on a spherically symmetric background can be well
described by using the formalism of Gerlach and Sengupta
\cite{Gerlach:1979rw,Gerlach:1980tx}. With a 2+2 decomposition of the
spacetime, the authors set up a covariant formalism to study linear
non-radial perturbations on a time dependent and spherically symmetric
background. Gundlach and Mart\'{\i}n--Garc\'{\i}a have further
developed this formalism for a self-gravitating perfect fluid
\cite{Martin-Garcia:1998sk,Gundlach:1999bt,Martin-Garcia:2000ze}.  The
authors have specified a general fluid frame on which all the tensor
fields and perturbative equations can be decomposed.  This approach
leads to a set of scalar gauge invariant fields and equations, which
are easily adaptable to any coordinate system of the background.
Hereafter we refer to this formalism with the acronym GSGM.

\subsection{The time dependent background}
\label{sec:GSGM_Time_Bck}

The background manifold is a warped product $M^2\times S^2$
of a two dimensional Lorentzian manifold $M^2$ and the 2-sphere $S^2$.
The metric can be written as the semidirect product of a
general Lorentzian metric on $M^2$, $g_{AB}$, and the unit curvature
metric on $S^2$, that we call $\gamma_{ab}$:
\begin{equation}
\met_{\alpha\beta} = \left(\begin{array}{cc}
\met_{AB} & 0 \\
0 & r^2\gamma_{ab} \end{array} \right) \,.\label{met22}
\end{equation}
With Greek letters we denote the components defined in the 4-manifold,
whereas the capital and small latin letters describe respectively the
tensors in the $M^2$ and $S^2$ sub-manifolds. The scalar function
$r=r(x^A)$ is defined in $M^2$, and can be chosen as the invariantly
defined radial (area) coordinate of spherically-symmetric spacetimes.
Besides the covariant derivative in the four dimensional spacetime,
defined as usual
\begin{equation}
g_{\alpha \beta ; \mu } = 0 \, ,
\end{equation}
we can introduced in the two submanifolds two distinct
covariant derivatives
\begin{equation}
g_{AB|C}=0 \, , \qquad \gamma_{ab:c}=0\, ,
\end{equation}
where the vertical bar corresponds to the covariant derivative of
$M^2$ and the semicolon to that of the 2-sphere $S^2$. Moreover, we
can introduce the completely antisymmetric covariant unit tensors on
$M^2$ and on $S^2$, $\epsilon_{AB}$ and $\epsilon_{ab}$ respectively,
in such a way that they satisfy:
\begin{equation}
\epsilon_{AB|C}=\epsilon_{ab:c}=0 \, ,  \qquad
\epsilon_{AC}\epsilon^{BC}= -g_A^B \, , \qquad
\epsilon_{ac}\epsilon^{bc}= \gamma_a^b \,.
\end{equation}
The energy-momentum tensor in a spherically symmetric spacetime
has the same block diagonal structure as the metric,
\begin{equation}
T_{\alpha\beta}= \textrm{diag} \left( \
t_{AB}\;, \,r^2 Q(x^C) \gamma_{ab} \right)\,, \label{tblock}
\end{equation}
where $Q(x^C)$ is a function defined on $M^2$.
In this thesis we have used a perfect-fluid description of the stellar
matter, thus we specialize the GSGM formalism to this case.
Therefore, we have for $T_{\alpha \beta}$,
\begin{equation}
T_{\alpha\beta}=(\rho+p)u_{\alpha}u_{\beta} + p g_{\alpha\beta}\,,
\end{equation}
where $\rho$ and $p$ are the mass-energy density and pressure, and
$u_{\alpha}$ is the fluid velocity. The background fluid velocity
in spherical symmetry has vanishing tangential components,
\begin{equation}
u_{\alpha}=( u_A,0) \, . \label{u_bck}
\end{equation}
The velocity $u_A$ and the space-like vector
\begin{equation}
n_A\equiv-\epsilon_{AB}u^B~~~\Rightarrow~~~n_A u^A=0\, , \label{n_bck}
\end{equation}
provide an orthonormal two dimensional basis $\left\{ u_A, n_B \right\}$
for the submanifold $M^2$.

\noindent The metric $\met_{AB}$ and the completely antisymmetric
tensor $\epsilon_{AB}$ can be written
in terms of these frame vectors as follows
\begin{eqnarray}
g_{AB} =  -u_Au_B+n_An_B\,,~~~~ \qquad
\epsilon_{AB} = n_Au_B-u_An_B\,, \label{ge22}
\end{eqnarray}
while the energy-momentum tensor assumes the following form
\begin{eqnarray}
t_{AB}& = &\rho u_Au_B + p n_An_B\,,~~~~Q = p\,.
\end{eqnarray}
In any given coordinate system for $M^2$, $\{x^A\}\,,$ one can define the
following quantity:
\begin{equation}
v_A\equiv \frac{1}{r}r_{|A}\,.
\end{equation}
Then, any covariant derivative on the spacetime can be written in
terms of the covariant derivatives on $M^2$ and $S^2$, plus some terms
due to the warp factor $r^2$, which can be written in terms of
$v_A$.
The frame derivatives of a generic scalar function $f$ are defined by
\begin{equation}
\dot f \equiv u^Af_{|A}\,, \qquad f' \equiv n^Af_{|A}\,, \label{fr_der_bck}
\end{equation}
which obey the following commutative relation:
\begin{equation}
( \dot f ~)' - (f'~) \dot{}  = \mu f' - \nu \dot{f} \, .
\end{equation}
Furthermore, a set of background scalars are introduced in order to
write the background and perturbative equations in a scalar form:
\begin{equation}
\Omega \equiv \ln\rho, \quad U \equiv u^Av_A, \quad W \equiv n^Av_A, \quad \mu \equiv u^A_{~|A},
\quad \nu  \equiv n^A_{~|A}\,.   \label{backg_scal}
\end{equation}
Finally, the Einstein field equation for the background spacetime
\begin{equation}
R_{\al \beta} - \frac{R}{2} g_{\al \beta} = 8 \pi T_{\al \beta} \label{Ein_equations}
\end{equation}
in the 2+2 split are given by the following equations:
\begin{eqnarray}
-2 \left( v_{A \mid B} + v_{A}  v_{B} \right) + \left( 2 v_{C} ^{~\mid C}
+ 3 v_{C} v^{C} - \frac{1}{r^2}\right) g_{A B} & = &
8 \pi t_{A B} \, , \\
v_{C} ^{~\mid C}  + v_{C} v^{C} - R & = & 8 \pi Q \, ,
\end{eqnarray}
where $R = \frac{1}{2} R ^A _A$ is the Gauss curvature of $M^2$.
The conservation of the energy-momentum tensor
\begin{equation}
T ^{\al \beta} _{ ~~~~ ;\al} = 0 \,  \label{Cons_eqs}
\end{equation}
leads to the energy conservation equation and to the relativistic
generalization of the Euler equation:
\begin{eqnarray}
 \dot{\Omega} + \left( 1 + \frac{p}{\rho} \right) \left( \mu + 2U\right)
& = & 0 \, , \\
c_s^2 \Omega ' + \left( 1 + \frac{p}{\rho} \right) \nu & = & 0 \, ,
\end{eqnarray}
and $c_s^2$ is the speed of sound defined on the isentropic fluid
trajectories as follows:
\begin{equation}
c_s^2 = \left. \frac{\partial p }{\partial \rho } \right| _{s} \, .
\end{equation}

\subsection{Perturbations}
\label{sec:GSGM_pert}
Linear perturbations of a spherically-symmetric background can be
decomposed in scalar, vector and tensor spherical harmonics.  The
perturbative variables are then completely decoupled in a part
depending on the angular coordinates of the 2-sphere and a part
defined on the submanifold $M^2$. This expansion is really helpful
because the perturbative problem is reduced to a 2-dimensional
problem, usually a time and spatial coordinate.  The tensor harmonics
are decomposed in two different classes of basis, the so-called polar
(even) parity and axial (odd) parity tensor harmonics. These transform
differently under a parity transformation, namely as $(-1)^l$ for the
polar and as $(-1)^{l+1}$ for the axial.

The basis for scalar function is given by the scalar
spherical harmonics $Y^{lm}$, which are eigenfunctions of the
covariant Laplacian on the sphere:
\begin{equation}
\gamma^{ab}Y^{lm}_{:ab} = -l(l+1)Y^{lm}\,.   \label{Sph_H_cond}
\end{equation}
where the integers $(l,m)$ are respectively the multipole index
and the azimuthal number. For a given $l$, the azimuthal number
can assume the following $2l+1$ values:
$$   - l, \, - l+ 1,\,  ..0 .. \, l-1, \, l \, .$$
There is not any axial basis for the scalar case.
A basis of vector spherical harmonics (defined for $l\ge 1$) is
\begin{eqnarray}
Y^{lm}_a & \equiv & Y^{lm}_{:a} \qquad  \quad \textrm{polar} \, , \\
S^{lm}_a & \equiv & \epsilon_a^{~b} Y^{lm}_b \qquad \textrm{axial} \, ,
\end{eqnarray}
A basis of tensor spherical harmonics (defined for $l\ge 2$) for the
polar case is given by
\begin{equation}
Y_{ab}^{lm}\equiv Y^{lm}\gamma_{ab}\,, \qquad
Z^{lm}_{ab}\equiv Y^{lm}_{:ab}+\frac{l(l+1)}{2}Y^{lm}\gamma_{ab}\,,
\end{equation}
and for the axial class by the following definition:
\begin{equation}
S^{lm}_{ab}\equiv S^{lm}_{a:b}+S^{lm}_{b:a} \,,
\end{equation}
The explicit form of the tensor harmonics are given in
appendix~\ref{sec:Tens_Harm}.

The perturbations of the covariant metric and energy-momentum
tensors can be expanded in the polar basis as
\begin{eqnarray}
\delta g^{pol}_{\alpha\beta} & = & \left(\begin{array}{cc}
  h_{AB}^{lm}\, Y^{lm} &  h_A^{pol ~ lm} \, Y^{lm}_a \\
\\
  h_A^{pol ~ lm}  \, Y^{lm}_a    & \  r^2(K^{lm} \, \gamma_{ab} \, Y^{lm}+
  G^{lm} \, Y_{:ab}^{lm})\\  \label{gGSMG_pol}
\end{array}\right)\,,\\
\delta t_{\alpha\beta}^{pol} & = & \left(\begin{array}{cc}
\delta  t_{AB}^{lm} \, Y^{lm} & \delta t_A^{pol ~ lm} \, Y^{lm}_a \\ \\
  \delta t_A^{pol ~ lm} \, Y^{lm}_a & \ r^2 \, \delta t^{3\,lm} \,
  \gamma_{ab} \, Y^{lm}+ \delta t^{2\,lm} \, Z_{ab}^{lm} \\
\end{array}\right)\, , \label{talbe_pol}
\end{eqnarray}
and axial basis as
\begin{eqnarray}
\delta g_{\alpha\beta}^{ax} & = & \left(\begin{array}{cc}
   0  &  h_A^{ax ~ lm} \, S^{lm}_a \\
\\
  h_A^{ax ~ lm}  \, S^{lm}_a    & \  h \left( S_{a:b}^{lm} + S_{b:a}^{lm} \right)\\
\end{array}\right)\,,\\
\delta t_{\alpha\beta} ^{ax} &=&\left(\begin{array}{cc} 0 & \delta
t_A^{ax ~ lm} \, S^{lm}_a \\ \\ \delta t_A^{ax ~ lm} \, S^{lm}_a & \
\delta t^{lm} \,\left( S_{a:b}^{lm} + S_{b:a}^{lm} \right) \\
\end{array}\right)\,. \label{talbe_ax}
\end{eqnarray}
In a spherically symmetric background the axial and polar
perturbations are dynamically independent, and for a given
multipole index $l$ their dynamics does not depend on the value of
the azimuthal number $m$. For simplicity, we can then study the
non-radial perturbations on a spherical star by only considering
the axisymmetric case $m=0$. This approximation is not valid for
instance in a rotating configuration, where axisymmetric and
non-axisymmetric perturbations have different spectral and
dynamical features.

The true degrees of freedom on metric and matter perturbations can be
determined by a set of gauge-invariant variables.  In one parameter
perturbation theory, see section~\ref{subsec:MLPT_GT}, the first order
perturbation $\delta \mathcal{T}$ of a generic tensor field
$\mathcal{T}$ is gauge-invariant if and only if the following
condition is satisfied~\cite{1974RSPSA.341...49S}:
\begin{equation}
{\cal L}_{\xi}  \mathcal{T}_{b} = 0 \, , \label{GIFO}
\end{equation}
where $\mathcal{T}_{b}$ is the value of $\mathcal{T}$ on the
background spacetime and $\xi$ is an arbitrary vector field that
generates the gauge transformation (see
section~\ref{subsec:MLPT_GT} and
reference~\cite{1974RSPSA.341...49S}).
By combining separately the polar perturbation fields
$h_{AB},\,h^{pol}_A,\,K,\,G$, $ \delta t_{AB},\,\delta
t_A,\,\delta t^2,\,\delta t^3\,,$ and the axial ones $
h_A^{ax},\,h$, $ \delta t^{ax}_A,\,\delta t \, $, it is possible
to determined the following set of gauge-invariant variables
\cite{Gerlach:1979rw,Gerlach:1980tx},
where for clarity the harmonic indices $(l,m)$ are neglected, \\
\indent \emph{polar}
\begin{eqnarray}
k_{AB}&=&h_{AB}-(p_{A|B}+p_{B|A})\,, \label{pkab} \\
k&=&K-2v^A p_A\,, \label{pk} \\
T_{AB}&=&\delta t_{AB}-t_{AB|C}\, p^C-t_{AC} \,  p^C_{|B}-t_{BC} \, p^C_{|A}\,,
\label{pTAB}\\
T^3&=&\delta t^3-p^C(Q_{|C}+2Qv_C)+\frac{l(l+1)}{2} \, Q \, G\,, \\
T_A&=&\delta t_A-t_{AC} \, p^C-\frac{r^2}{2} \, Q \, G_{|A}\,, \\
T^2&=&\delta t^2-r^2\, Q\, G \,, \label{pt2}
\end{eqnarray}
where $T_A$ is defined for $l\ge 1\,,$  $T^2$ is defined for $l\ge 2$, and
\begin{equation}
p_A = h^{pol} _A - \frac{1}{2}r^2 G_{|A}\,. \label{ppa} \quad \quad \qquad \qquad  \qquad \qquad
\end{equation}
\indent \emph{axial}
\begin{eqnarray}
k_{A} &=& h_{A}^{ax} - h_{\mid A} + 2 h v_{A} \,, \label{pka} \qquad
\qquad \qquad \qquad \\
L_A &=& \delta t_{A}^{ax} - Q h_{A}^{ax} \,,
\label{pLa} \\
L &=& \delta t - Q h \,, \label{pL}
\end{eqnarray}
where $k_A$ and $L_A$ are defined for $l\ge 1$, and $L$ for $l\ge 2$.
Therefore, any linear perturbation of the spherically-symmetric background
(\ref{met22}) can be written as a linear combination of these gauge-invariant
quantities, which are tensor fields defined on the submanifold $M^2$.
The definition of the gauge invariant quantities of matter is
valid for any energy-momentum tensor and not only in the case of a
perfect fluid.

The perturbations of a perfect fluid are given
by four polar and one axial quantities.
The polar velocity perturbation can be written as follows:
\begin{equation}
\delta u_{\alpha}=\left(\left[\tilde \gamma^{lm} n_A+
\frac{1}{2}  h_{AB}^{lm}u^B\right]Y^{lm}\,,
\tilde \alpha^{lm} Y^{lm}_{:a}\right)\,, \label{deltaupol}
\end{equation}
where $\tilde\alpha$ is defined for $l\ge 1\,.$
The axial velocity perturbation is instead given by
\begin{equation}
\delta u_{\alpha}=\left( 0 \,, \tilde \beta^{lm} S^{lm}_{a}\right)\, .
\label{deltauax}
\end{equation}
The functions $ \tilde \gamma^{lm}, \, \tilde \alpha^{lm}, \,
\tilde \beta^{lm}$ are defined on $M^2$ and describe the rate of
the radial, tangential polar and tangential axial motion
respectively. Furthermore, the axial perturbation $\beta^{lm}$ is
gauge invariant for an odd-parity gauge transformation (see section~\ref{sub_Ax}).

The mass-energy density and pressure perturbations can be written in the following
form (using the barotropic equation of state)
\begin{equation}
\delta\rho=\tilde\omega^{lm} \rho Y^{lm} \,, \qquad \qquad \delta
p=c_s^2\delta\rho\,.  \label{pr}
\end{equation}
In terms of these quantities it is possible to define a
gauge-invariant set of fluid perturbations:
\begin{eqnarray}
\alpha&=&\tilde\alpha-p^Bu_B\,, \label{algi} \\
\gamma&=&\tilde\gamma-n^A\left[p^Bu_{A|B}
+\frac{1}{2}u^B(p_{B|A}-p_{A|B})\right]\,, \label{gamgi}  \\
\omega&=&\tilde\omega-p^A\Omega_{|A}\, ,   \label{omgi}
\end{eqnarray}
where in these expressions and in the remaining part of this section
we do not write explicitly the harmonic indices $(l,m)$.

The gauge-invariant tensors (\ref{pTAB})-(\ref{pt2}),
(\ref{pLa}) and \ref{pL})  for a generic energy-momentum
tensor can be written in terms of the perfect fluid
gauge invariant perturbations as follows:

\indent \emph{polar}
\begin{eqnarray}
T_{AB} & = & \left( \rho + p \right) \left[ \gamma \left( u_A n_B + n_A u_B \right)
 + \frac{1}{2} \left( k_{AC} u_{B} + u_{A} k_{BC} \right) u^{C} \right] {} \nn \\
{} && + \omega \rho \left( u_A u_B + c_s^2 n_A n_B \right) + p
k_{AB}
\, , \\
T_{A} & = & \al \left(\rho + p \right) u_{A}  \, ,\\
T^{3} & = & p k + c_s^2 \rho \omega \, , \\
T^2   & = &  0 \, .
\end{eqnarray}
\indent \emph{axial}
\begin{eqnarray}
L_A & = & \beta \left(\rho + p\right) u_A \, \qquad \qquad \qquad \qquad
\qquad \qquad \qquad \qquad \\
L   & = & 0 \, .
\end{eqnarray}

\subsection{Perturbative equations}
\label{sec:GSGM-peqs}
The dynamics of linear oscillations of a time dependent and
spherically symmetric spacetime is described by two independent
classes of oscillations: the axial and polar perturbations.  The
perturbative equations can be expressed in terms of the gauge
invariant GSGM quantities. In addition, when a decomposition with
respect to the vector basis~$\{u^A,n^A\}$ of the spacetime~$M_2$
is adopted they assume a scalar form~\cite{Gundlach:1999bt}.
Here, we report the main procedure; see~\cite{Gundlach:1999bt} for
details.

\noindent \emph{Polar sector:}  \\
\noindent The tensor $k_{AB}$ can be decomposed on the frame $\{u^A,n^A\}$:
\begin{equation}
k_{AB}=\eta(-u_Au_B+n_An_B)+\phi(u_Au_B+n_An_B)+\psi(u_An_B+n_Au_B)\,,
\label{kABdec}
\end{equation}
where $\eta$, $\phi$ and $\psi$ are scalars.  It is useful to consider
the following new scalar variable
\begin{equation}
\chi=\phi-k+\eta\,,  \label{chidef}
\end{equation}
in place of $\phi$.  Then, combining Einstein equations with the
energy-momentum equations we can obtain the following set of equations:
for $l\ge 2$,
\begin{equation}
\eta=0\,, \label{eta}
\end{equation}
for $l\ge 1$,
\begin{eqnarray}
-\ddot\chi+\chi''+2(\mu-U)\psi'&=&S_{\chi}\,, \label{chitt} \\
-\ddot k+c_s^2k''-2c_s^2U\psi'&=&S_k\,, \label{ktt} \\
-\dot\psi&=&S_{\psi}\,, \label{psit} \\
16\pi(\rho+p)\alpha&=&\psi'+C_{\alpha}\,, \label{psip}\\
-\dot\alpha&=&S_{\alpha}\,, \label{alphat}  \\
-\dot\omega-\left(1+\frac{p}{\rho}\right)\gamma'&=&\bar S_{\omega}\,,
\label{omegat} \\
\left(1+\frac{p}{\rho}\right)\dot\gamma+c_s^2\omega'&=&\bar S_{\gamma}
\label{gammat} \,.
\end{eqnarray}
And finally, for $l\ge 0$,
\begin{eqnarray}
8\pi(\rho+p)\gamma&=&(\dot k)'+C_{\gamma}\,, \label{ktp} \\
8\pi\rho\omega&=&-k''+2U\psi'+C_{\omega}\,, \label{kpp} \, .
\end{eqnarray}
where the expressions of
$S_{\chi},\,S_{\psi},\,C_{\alpha},\,S_{\alpha},\, \bar
S_{\omega},\,\bar S_{\gamma},\,C_{\gamma},\,C_{\omega}$ can be found
in Appendix~\ref{full-equations}. 

\noindent \emph{Axial sector:}  \\
\noindent The perturbed Einstein and hydrodynamics equations can be written as
\begin{eqnarray}
\label{maseq}
\left[\frac{1}{r^2} \left( r \Psi \right) _{\mid A }
\right] ^{|A } - \left( l-1 \right) \left( l+2 \right) r^{-3}
\Psi  &  = &  -16 \pi \ep^{AB} L_{A|B} \\
\label{traseq}
\dot{ \beta} - c_s^2 \left( \mu + 2 U\right) \, \beta & = &  0 \ ,
\end{eqnarray}
where, following references~\cite{Gerlach:1979rw,Gerlach:1980tx} we have
introduced the gauge-invariant odd-parity master function $\Psi$ as
\begin{equation}
\Psi \equiv
r^3 \, \ep ^{AB} \left( r^{-2} k_A \right) _{\mid B} \, . \label{Pidef}
\end{equation}
Once $\Psi$ is obtained as a solution of the odd-parity master
equation (\ref{maseq}), the metric components $k_A$ can be recovered
by means of the relation
\begin{equation}
\left( l-1 \right) \left( l+2 \right) k_A = 16 \pi \, r^2 \,
  L_A -
\ep _{AB} \, \left( r \, \Psi \right) ^{\mid B} \ . \label{mtr_form}
\end{equation}
The solutions are determined by specifying the initial values of
$\beta $, $\Psi $, and $\dot \Psi $ on a Cauchy surface. The fluid
conservation equation (\ref{traseq}) can be solved independently
from the odd-master equation, as it depends only on the fluid
perturbation $ \beta $. Its solution then provides a constant
value of axial velocity $\beta$ along the integral curves of $u^A$
\cite{Gundlach:1999bt}.  However, in the odd-master equation
(\ref{maseq}) the velocity perturbation $\beta$ couples with the
background quantities, which being time dependent can source the
non-radial oscillations.

\section{Non-linear perturbative framework}
\label{sec:3.3-NLFW}
The radial and non-radial perturbations are the two fundamental
families of stellar oscillations, which have different properties
with respect to the gravitational physics. In this section, we are
going to investigate the main characteristics of the non-linear
perturbations and their equations by adopting a two parameter
perturbative scheme which allows us to distinguish at any
perturbative order these two perturbation classes. The two
parameter perturbation theory, is the $N=2$ subcase of the multi
parameter theory reported in section~\ref{sec:3.1_MPPT}. The
parameter $\lambda$ denotes the family of radial perturbations,
namely the class of polar perturbations with vanishing harmonic
index $l=0$. On the other hand, the second parameter $\epsilon$
labels the class of non-radial perturbations with $l \ge 1$.  With
this notation the metric and energy-momentum tensors can be
expanded as follows:
\begin{eqnarray}
\! \! \! \! \! \! \! \! \! \met_{\alpha\beta} & = & \bar{\met} _{\alpha\beta} + \lambda \,
\met^{(1,0)}_{\alpha\beta}+ \epsilon \, \met^{(0,1)}_{\alpha\beta} +
\frac{ \lambda ^2}{2} \, \met^{(2,0)}_{\alpha\beta} + \lambda \epsilon \,
\met^{(1,1)}_{\alpha\beta} + \frac{ \epsilon^2 }{2} \, \met^{(0,2)}_{\alpha\beta}
+O(\lambda^n,\epsilon^k)\,, \label{initialmetric} \\
\! \! \! \! \! \! \! \! \!
T_{\alpha\beta} & = & \bar{T}_{\alpha\beta} + \lambda \, T^{(1,0)}_{\alpha\beta}+
\epsilon \,T^{(0,1)}_{\alpha\beta} + \frac{ \lambda ^2 }{2}\,
T^{(2,0)}_{\alpha\beta} + \lambda\epsilon \, T^{(1,1)}_{\alpha\beta} +
\frac{ \epsilon ^2 }{2}\, T^{(0,2)}_{\alpha\beta} +O(\lambda^n,\epsilon^k)
\,, \label{tmunu}
\end{eqnarray}
where the integers $(n,k)$ are such that $n + k > 2$. The background tensors have
been denoted with a bar, while the indices $(i,j)$ denote the
perturbations of order $i$ in  $\lambda$ and $j$ in $\epsilon$. A
similar expansion can be done for the other fluid perturbations,
i.e., velocity, mass-energy density and pressure.
The equations at any perturbative order can be determined with a
standard procedure: \emph{i)} one introduces the perturbative
expansions~(\ref{initialmetric}), (\ref{tmunu}) for the metric,
energy momentum tensors and those related to the other fluid
variables into the Einstein and conservation equations, \emph{ii)}
Taylor expand these equations with respect to the two perturbative
parameters $\lambda$ and $\epsilon$, and eventually \emph{iii)}
select the terms of the equation which refer order by order to the
same perturbative parameter $\lambda ^n \epsilon ^k$, where $n,k
\in \mathbb{N}$.
Let's carry out the analysis focusing on the Einstein equations, the
conservation equations can be addressed with the same method.  We can
start writing the full Einstein equations:
\begin{equation}
\mb{E}\left[\,\metb\,,\mb{\psi}_A\,\right] = \mb{G}\left[\,\metb\,\right] -
\mb{T}\left[\,\metb\,,\mb{\psi}_A\,\right] = 0 \,, \label{efes}
\end{equation}
where $\mb{G}$ is the Einstein tensor, and $\mb{\psi}_A$
(A=$1,\dots$)  the various fluid variables. After having
introduced the perturbative expressions~(\ref{initialmetric}) and
(\ref{tmunu}) into  Eq.~(\ref{efes}), we obtain the following
expression:
\begin{eqnarray}
\mb{E}  \left[\,\metb\,,\mb{\psi}_A\,\right]   & = &
 \mb{E}_{b} \left[\,\bar{\mb{\metb}} \,,\bar{\mb{\psi}}_A \,\right]
+ \lambda\,\mb{E}^{(1,0)}
\left[\,\mb{\metb}^{(1,0)}\,,\mb{\psi}_A^{(1,0)}\,\right]
+ \epsilon\,\mb{E}^{(0,1)}
\left[\,\mb{\metb}^{(0,1)}\,,\mb{\psi}_A^{(0,1)}\,\right]  \nn \\
& + & \frac{ \lambda^2 }{2}
\,\mb{E}^{(2,0)}\left[\,\mb{\metb}^{(2,0)}\,,\mb{\psi}_A^{(2,0)}
\,  , \mb{ J }^{(1,0)}\otimes\mb{ J }^{(1,0)}  \,\right]  \nn
\\
&+&\lambda\epsilon\,\mb{E}^{(1,1)}\left[\,\mb{\metb}^{(1,1)}\,,\mb{\psi}_A^{(1,1)}
\, , \mb{ J }^{(1,0)}\otimes\mb{ F}^{(0,1)}  \,\right]  \nn
\\
&+& \frac{ \epsilon ^2 }{2}
\,\mb{E}^{(0,2)}\left[\,\mb{\metb}^{(0,2)}\,,\mb{\psi}_A^{(0,2)}
\,  , \mb{ F}^{(0,1)}\otimes\mb{ F }^{(0,1)}  \,\right] +
O(\lambda^n,\epsilon^k)=0  \, ,  \label{total_EEq}
\end{eqnarray}
where the tensors $\mb{ J }^{(1,0)}$ and $\mb{ F}^{(0,1)}$ denote
the set of metric and fluid variables of the radial and non-radial
perturbations respectively. The linear differential operators
$\mb{ E }^{(i,j)}$ in the previous expression are defined as
follows:
\begin{equation}
\mb{ E }^{(i,j)} = \left. \frac{\rm{\partial} \,
^{i+j}}{\rm{\partial } \lambda ^{i} \rm{\partial} \epsilon^{j}}
\mb{E} \, \right| _{\lambda ^{i} = \epsilon^{j} = 0}  \, , 
\label{DiffE}
\end{equation}
and $\mb{E}_{b} \equiv \mb{ E }^{(0,0)}$.  They act linearly on the
perturbation of order $(i+j)$, and in general non-linearly on the
background quantities. \\
\indent An interesting aspect of the second order perturbative
equations is the presence in the expansion (\ref{total_EEq}) of
products between linear perturbations, which have been already
determined by solving the first order perturbations. Therefore, in the
non-linear perturbative equations they behave as source terms.
The equation of order $\lambda \epsilon$ can be then
written as follows:
\begin{equation} 
\mb{ E }^{(1,1)} \left[ \mb{\metb}^{(1,1)}\,,\mb{\psi}_A^{(1,1)}
 \right] = \mathcal{S} \left[ \mb{ J }^{(1,0)}\otimes\mb{ F}^{(0,1)}
 \right] \label{11exp} \, , 
\end{equation}
and a similar structure is also present in the $\lambda^2$ and
$\epsilon^2$ perturbative equations.
The iterative procedure of the perturbation techniques implies that
the part of the differential operators~(\ref{DiffE}) that acts
linearly on the perturbations $\lambda^i \epsilon^j$, as for instance
in equation~(\ref{11exp}), is equal at any perturbative order.
However, when the perturbative fields have different dependence on the
coordinates the resulting systems of perturbative equations are
different. This is the case for the radial perturbations, which unlike
the non-radial do not have any angular dependence.  In order to have 
more clear this distinction between the perturbative equations of the
radial and non-radial perturbations we redefine the first order
differential operators as follows:
\begin{equation}
\mb{E}^{(1,0)} \equiv \mb{L}_{R}  \, ,\qquad \qquad
\mb{E}^{(0,1)} \equiv \mb{L}_{NR}  \, ,  \label{LNRdef}
\end{equation}
where $R$ and $NR$ stand for ``radial'' and ``non-radial''
respectively.
%

\indent Equation~(\ref{total_EEq}) has to be satisfied for arbitrary and
relatively small values of the two perturbation parameters
$\left(\lambda, \epsilon \right)$. Therefore, each term of the
expansion has to vanish and provide an independent system of equations
associated with its perturbation parameters.
The equilibrium configuration in this thesis is a spherically symmetric and
perfect fluid star. The background spacetime is then determined by equations:
\begin{equation}
\mb{E}_{b} \left[\,\bar{\mb{\metb}} \,,\bar{\mb{\psi}}_A \,\right]
= 0 \, ,
\end{equation}
which represent the Tolman-Oppenheimer-Volkoff (TOV) equations (see,
e.g.,~\cite{Misner:1973cw}). \\ At first order we have two independent
systems of equations for the \emph{radial}
\begin{equation}
\mb{L}_{R} \left[ \mb{\metb}^{(1,0)}\,,\mb{\psi}_A^{(1,0)} \right]
 =  0  \, , \label{Taylor_R1} \\
\end{equation}
and for the \emph{non-radial} perturbations:
\begin{equation}
\mb{L}_{NR} \left[ \mb{\metb}^{(0,1)}\,,\mb{\psi}_A^{(0,1)} \right] =
  0 \, .\label{Taylor_NR1}
\end{equation}
The second order perturbative equations instead can be divided in
three independent systems of equations: the second order radial
perturbation, the coupling between the radial and non-radial
perturbations and the second order non-radial perturbations which
are respectively given  by the following expressions:
\begin{eqnarray}
\mb{L}_{R} \left[ \mb{\metb}^{(2,0)}\,,\mb{\psi}_A^{(2,0)} \right]
& = & \mathcal{S} \left[ \mb{ J }^{(1,0)}\otimes\mb{ J }^{(1,0)} \right]  \, , \label{Taylor_R2} \\
\mb{L}_{NR} \left[ \mb{\metb}^{(1,1)}\,,\mb{\psi}_A^{(1,1)}  \right]
& = & \mathcal{S} \left[ \mb{ J }^{(1,0)}\otimes\mb{ F}^{(0,1)}  \right]   \, , \label{Taylor_Cp2}\\
\mb{L}_{NR} \left[ \mb{\metb}^{(0,2)}\,,\mb{\psi}_A^{(0,2)} \right]
 & = & \mathcal{S} \left[ \mb{ F}^{(0,1)}\otimes\mb{ F }^{(0,1)}  \right]   \, .\label{Taylor_NR2}
\end{eqnarray}
As we discussed above, the differential part $\mb{L}_{R}$ of the
non-linear radial perturbative equations in~(\ref{Taylor_R2}) is
the same as in the first order equations~(\ref{Taylor_R1}), while
the differential part $\mb{L}_{NR}$ of linear non-radial
perturbative equations~(\ref{Taylor_NR1}) appears at second order
for non-radial and coupling perturbations, respectively in
equations~(\ref{Taylor_NR2}) and ~(\ref{Taylor_Cp2}). Perturbative
tensor fields on a spherically symmetric background can be
expanded in tensor harmonics (see section~\ref{sec:GSGM_pert}). As
a result, the angular dependence of the perturbations is decoupled
from the dependence on the two remaining coordinates, which
generally describe the time and a radial coordinate. Therefore,
any perturbative tensor field $\delta \mathcal{T}$ can be written
as follows:
\begin{equation}
\delta \mathcal{T} \left( t,r,\theta,\phi \right)= \sum _{lm}
\delta \mathcal{T}^{lm} (t,r) \, \mathcal{H}^{lm}(\theta,\phi) \, ,  \label{delT}
\end{equation}
where the quantity $\mathcal{H}^{lm}$ denotes the appropriate tensor
harmonics associated with the nature of the perturbative fields,
i.e. scalar, vector or tensor as well as even or odd parity
perturbation.  The tensor $\delta \mathcal{T}^{lm} (t,r)$ is the
harmonic component of this expansion related to the harmonic indices
$(l,m)$, which is determined by projecting the perturbation $\delta
\mathcal{T}$ on the related tensor harmonic
$\mathcal{H}^{lm}(\theta,\phi)$ through the internal product
associated with the 2-sphere $S^2$:
\begin{equation}
\delta \mathcal{T}^{lm} = \left( \delta \mathcal{T},\mathcal{H}^{lm}(\theta,\phi)   \right)
\equiv \int_{S^2} \mathcal{T} : \mathcal{H}^{lm} \rm{d} \Omega  \, ,
\end{equation}
where $\mathcal{T} : \mathcal{H}^{lm}$ has the following definition
for a 2-rank tensor field $\mathcal{T}$:
\begin{equation}
\mathcal{T} : \mathcal{H}^{lm} = \mathcal{T}_{ab}
\mathcal{H}^{lm}_{cd} \gamma^{ac}\gamma^{bd} \, ,
\end{equation}
and $\gamma_{ab}$ is the unit metric of the 2-sphere $S^2$.

\indent When we introduce the tensor harmonic expansion into the
linear perturbative equations~(\ref{Taylor_R1})
and~(\ref{Taylor_NR1}), they assume the following expressions for
the \emph{radial perturbations:}
\begin{equation}
\mb{L}_{R}
\left[\,\mb{\metb}^{(1,0)}_{00}\,,\mb{\psi}_{A,00}^{(1,0)}\,\right]
\mathcal{H} ^{00} = 0 \,  \label{1RR}
\end{equation}
and for the \emph{non-radial:}
\begin{equation}
\sum _{l,m} \mb{L}_{NR}
\left[\,\mb{\metb}^{(0,1)}_{lm}\,,\mb{\psi}_{A, \, lm}^{(0,1)}
\,\right] \mathcal{H} ^{lm}   = 0 \, . \label{1NR}
\end{equation}
Since the spherical harmonic $\mathcal{H} ^{00} = Y^{00}$ is a
constant, in this section we will consider its value implicitly
contained in the harmonic component $\mathcal{T}^{(i,0)}_{00}$ of
the radial perturbations.  The decomposition in tensor harmonics
allows us to describe independently the various harmonic
components $(l,m)$ of the linear non-radial perturbations, where
each component obeys the perturbative equation~(\ref{1NR}) related
to its indices $(l,m)$. As we will see later this property is not
in general valid in
a second perturbative analysis.  \\
Equations~(\ref{Taylor_R2})-(\ref{Taylor_NR2}) that describe the
non-linear perturbations assume the following form in terms of a
spherical harmonic expansion: \\ 
\emph{non-linear radial
perturbations}
\begin{equation}
\mb{L}_{R} \left[\,\mb{\metb}^{(2,0)}_{00}\,,\mb{\psi}_{A, \, 00}^{(2,0)}\,\right]   =  \mathcal{S}\left[ \mb{
J}^{(1,0)}_{00}\otimes\mb{ J }^{(1,0)}_{00} \right]   \label{2RR}
\end{equation}
\emph{coupling radial/non-radial perturbations}
\begin{equation}
\sum _{l,m} \mb{L}_{NR}
\left[\,\mb{\metb}^{(1,1)}_{lm}\,,\mb{\psi}_{A, \, lm}^{(1,1)}\,\right] \mathcal{H}
^{lm} = \sum _{l,m} \mathcal{S}\left[ \mb{
J}^{(1,0)}_{00}\otimes\mb{ F }^{(0,1)}_{lm} \right] \mathcal{H}^{lm}   \label{CP}
\end{equation}
\emph{non-linear non-radial perturbations}
\begin{equation}
\sum _{l,m} \mb{L}_{NR}
\left[\,\mb{\metb}^{(2,0)}_{lm}\,,\mb{\psi}_{A, \, lm}^{(2,0)}\,\right] \mathcal{H}
^{lm} = \sum _{l',m'} \sum _{l'',m''} \mathcal{S}\left[ \mb{
F}^{(0,1)}_{l'm'}\otimes\mb{ F }^{(0,1)}_{l''m''} \right] \mathcal{H}
^{l'm'} \mathcal{H}^{l''m''}   \label{2NR}
\end{equation}
The presence of the source terms in the non-radial perturbative
equations (\ref{2NR}) prevents us from completely decoupling the
perturbative components with different harmonic indices $(l,m)$.
In fact, the quadratic terms in the source couple different
spherical harmonics according to the familiar law for addition of
angular momentum in quantum mechanics.  For instance, a complete
analysis of the quadrupolar case ($l=2$) must take into account
the source terms provided by the coupling of the indices
$(l',l'')=(2,2)$ as well as in principle the indices
$(l',l'')=(200,198)$ and so on. Therefore, the dynamics of a
second order perturbation $\mathcal{T} ^{(2,0)}_{lm}$ depends in
principle on an infinite series of source terms
$\mb{F}^{(0,1)}_{l'm'}\otimes\mb{ F }^{(0,1)}_{l''m''}$, which
have to be solved by the related first order perturbative
equations. In a non-linear analysis it is then crucial to select
in the sources the dominant terms which provide the main
contributions to the non-linear dynamics. This selection is a
standard procedure which has been used for instance in the
perturbative analysis of the oscillations of a slowly rotating
star, where the rotation has been treated perturbatively with the
Hartle-Thorne slow rotation approximation~\cite{Hartle:1967ha,
Hartle:1968ht}.  In general, when the aim is the description of
the gravitational radiation emitted by a physical system, the
coupling between the quadrupole terms and the other moments with
$l$ close to $l=2$ are expected to give the dominant
contributions. In second order perturbations of Schwarzschild
black holes~\cite{1999bhgr.conf..351P, Gleiser:1995gx}, which have
been used by the authors also for describing the collision of two
BHs, the coupling between the quadupolar terms provides results
which show good accuracy with respect to the non-linear
simulations carried out in numerical relativity.

In this thesis, we investigate the coupling between the
radial/non-radial perturbations $\lambda \epsilon$, which obey the
perturbative equations of the form~(\ref{CP}). In this case the
equation can be easily decoupled, as for any harmonic component
$(l,m)$ of equation~(\ref{CP}) the source terms contribute only
with the following indices $(l',m') = (0,0)$ and $(l'',m'') =
(l,0)$.  The source terms are then determined by solving at first
order two system of equations, one for the radial
perturbations~(\ref{1RR}) and the other for the $(l,0)$ component
of non-radial perturbations~(\ref{1NR}).

\subsection{Time and frequency domain analysis}
The investigations of stellar oscillations are carried out in the
time and frequency domain. These two approaches provide
complementary information about the dynamics and the spectral
properties of the stellar perturbations.  \\
\indent In the \emph{frequency domain}, the time dependence of the
perturbative fields is separated by the spatial coordinate, by
assuming an harmonic dependence of the oscillations. Therefore,
the tensor fields~(\ref{delT}) can be written as
\begin{equation}
\delta \mathcal{T} \left( t,r,\theta,\phi \right)= \sum _{lm} \delta
\mathcal{T}^{lm}_{0}(r) e^{i\omega_{lm}t} \,
\mathcal{H}^{lm}(\theta,\phi) \, , \label{delTfd}
\end{equation}
where $\omega_{lm}$ is in general a complex frequency associated with
the harmonic indices $(l,m)$.  The different action of the radial and
non-radial perturbations on the quadrupole of a spherical star, is
also reflected on the mathematical nature of the frequencies
$\omega_{lm}$. A radially oscillating phase of a perfect fluid
spherically symmetric star can be described by a Sturm-Liouville
problem, whose solutions provide a complete set of normal modes with
frequecies $\omega_{lm}$, where $\omega_{lm} \in \mathbb{R}$.
On the other hand, a non-radially oscillating dynamics is also a
source of gravitational radiation, which damps the stellar
oscillations.  As a result, the non-radial spectrum is described
by a set of quasi normal modes (QNM), where $\omega_{lm}$ are
complex quantities whose real part describes the oscillating
frequency, and the imaginary part the damping time due to the
gravitational emission.  The spectrum of normal or quasi normal
modes for radial and non-radial perturbations can be determined by
eigenvalue problems, which can be set up by introducing the
expressions~(\ref{delTfd}) into the radial and non-radial
perturbative equations~(\ref{1RR}) and~(\ref{1NR}).  The numerical
methods to derive these results are for instance reviewed in
references~\cite{Bardeen:1966tm, kokkotas-1999-2}. \\
\indent In the \emph{time domain}, there is not any harmonic
assumption on the time dependence. The perturbative
equations~(\ref{1RR})-(\ref{1NR}) and~(\ref{2RR})-(\ref{2NR}) are
then integrated in a 1+1 numerical code, where one dimension
describes the time and the other the spatial coordinate.  This
approach provides information about the time evolution of the
perturbative variables of oscillating phases. In particular in
gravitational physics researches, this method gives the propeties
of the wave forms of the gravitational signal.  In addition, the
QNMs which have been excited in a time evolution can be determined
by means of a Fast Fourier Transformation (FFT) of the time
profiles.

\chapter{Linear Perturbations of Compact Stars}
\label{ch:4Lin_Pert}

Neutron stars oscillations have been extensively investigated with
linear perturbative techniques both in Newtonian and relativistic
approaches.  The classical analysis of the oscillating star
spectrum has revealed the presence of various classes of modes
which have been organized in a detailed classification.  Linear
perturbations are classified in two fundamental classes: the
\emph{radial} and \emph{non-radial oscillations}.  This definition
respectively discerns the perturbations that have or not an
angular dependence.  In a non-rotating and spherically symmetric
stellar model the adiabatic radial pulsations are not
damped by any dissipative or emitting mechanism. The single degree of
freedom of radial perturbations, which represents the radial movement
of the fluid, can be described by a Sturm-Liouville problem.
Therefore, the radial spectrum is formed by a discrete and complete
set of normal modes which provides a basis for decomposing the time
evolution of any radially oscillating quantity by Fourier
transformation.  On the other hand, in relativistic stars the
non-radial oscillations modify the stellar quadrupole and are damped
by gravitational emission.  The non-radial spectrum is then described
by quasi-normal modes (QNM), which have complex eigenfrequencies where
the real part provides the oscillation frequency of the modes, while
the imaginary part identifies the damping time of the oscillations.

The features of pulsation spectra are closely related to the
properties of the stellar model adopted.  The interpretation of
the relations between the gravitational radiation and the source
properties is the subject of the Astereoseismology, which is
already a prolific area of the electromagnetic astrophysics that
has revealed important aspects of the internal dynamics of the Sun
and non-compact stars.  The high densities and strong physical
conditions present in a relativistic stars prevent us from
studying the neutron stars properties directly in Earth's
laboratories. As a result, various equations of state have been
proposed for describing the matter at supranuclear densities. An
analysis of the gravitational spectrum related to these sources
can settle this uncertainty, by determining the neutron star
masses and radii with an accuracy sufficient to constrain the
parameters of the equations of state
proposed~\cite{Andersson:1996ak, 1998MNRAS.299.1059A}.

The general relativistic treatment of the radial pulsations
started in 1964 with the work of
Chandrasekhar~\cite{Chandrasekhar:1964tc}.  The aim of these first
studies was the stability issue of the stellar equilibrium
configuration under radial pulsations. Subsequently, the interest
moved to the investigation in the frequency domain of the spectrum
features for various stellar models that are described with more
realistic equations of state (see~\cite{Kokkotas:2000up} and
references therein).  The time evolution of the radial
perturbations has been addressed quite recently
in~\cite{Ruoff:2000nj} and~\cite{Sperhake:2001si}, in Eulerian and
Lagrangian gauges. Ruoff in~\cite{Ruoff:2000nj} has explored the
numerical stability of the radial and non-radial oscillations when
a polytropic equations of state of a star is replaced by a more
realistic equation of state. Sperhake in~\cite{Sperhake:2001si},
approaches the non-linear time evolution of radial pulsations of a
polytropic non-rotating star in Eulerian and Lagrangian gauges.

Non-radial oscillations of compact stars have been originally studied
with the Newtonian theory of
gravitation~\cite{1989nos..book.....U}. In this context, the
gravitational radiation is due exclusively to the oscillations of the
fluid, the emission rate is determined by the quadrupole
formula~\cite{Thorne:1969to, 1973ApJ...185..277O} and the damping time
by the expression $E / \dot E$~\cite{1982MNRAS.200P..43B}, where $E$
is the pulsation energy and here the dot denotes the time derivative.
Damping times of typical pulsation modes are very
low~\cite{1988ApJ...325..725M} due to the weak coupling between matter
and gravitational waves.  The two classes
of non-radial oscillations are the \emph{polar} and \emph{axial
perturbations}. The axial perturbations have a degenerate spectrum,
which is removed when the stellar model contains rotation, magnetic
fields or non-zero stresses~\cite{1988ApJ...325..725M}. For a perfect
fluid non-rotating star the axial perturbations can describe a
continuous differential rotation of the stellar fluid without any
oscillating character.  On the other hand, according to the nature of
the restoring force that governs their dynamics the polar
perturbations are classified in pressure~(\emph{p}),
gravity~(\emph{g}) and fundamental~(\emph{f}) modes.  They have the
following properties:
\begin{itemize}
\item[] \emph{f-mode}: the fundamental mode is nearly independent of
  the internal structure of relativistic and Newtonian stars. It is
  the only mode present in the simplest stellar model, i.e. a zero
  temperature non-rotating star whose density is uniform. There is a
  single $f$-mode for any harmonic index $l$, and for a cold NS its
  frequency depends on the average density of the star.  The f-mode
  reaches its maximum in amplitude at the stellar surface and does not
  have any node in the associated eigenfunction.  Typical values of the
  frequencies and damping times are in the range $1.5-3.5~kHz$
  and~$0.1-0.5~s$,~\cite{1983ApJS...53...73L, Detweiler:1985dl}.
\item[] \emph{p-modes}: they are associated with the acoustic waves
  that propagate inside the star, where the pressure gradients act as
  restoring forces. A polytropic perfect fluid star is the simplest
  stellar model which can sustain these modes.  The oscillating
  frequencies are higher than the $f$-mode, as they are related to the
  travel time of the acoustic wave across the star. The $p$-modes form
  for any harmonic index $l$ a countable infinite discrete set,
  where the first element $p_1$ has a typical frequency $5-6~kHz$,
  damping time of one or few seconds and one node in the associated
  eigenfunction.  The frequencies, damping time and the node number
  increase directly with the order of the mode.
\item[] \emph{g-modes}: These modes arise from temperature and
  composition gradients present inside the star. Gravity is the
  restoring force that acts through buoyancy forces. Like the
  $p$-modes,
  the $g$-modes form for any harmonic index $l$ a countable
  infinite discrete set, but their frequencies are lower than the
  $f$-mode frequency and are inversely proportional to the order of
  the mode.  The $g$-mode frequencies range from zero to a few hundred
  $Hz$, and in a perfect fluid star, which is the model adopted in
  this work, they are all degenerate at zero frequency.  The typical
  damping time has an order of magnitude of $10^{6}~s$.
\end{itemize}
For more details about the mode classification see the monographs
dedicated to this subject~\cite{1989nos..book.....U,
  1980tsp..book.....C}.

The first relativistic analyses of non-radial perturbations of
non-rotating stars is due to Thorne and his collaborators in a
series of papers \cite{Thorne:1967th, Thorne:1968tc,
Thorne:1969to, Thorne:1969th} that date back to 1967.  In General
Relativity, the spectral properties and damping times of the
stellar oscillations can be directly determined with eigenvalue
problems, which provide the stellar QNMs.  Subsequent researches
have been dedicated to have a more complete understanding of the
stellar QNMs, by extending the analysis to more realistic stellar
models.  For oscillations associated with the dynamics of the
matter variables (\emph{fluid-modes}), the relativistic analyses
provided some small corrections to the mode frequencies and more
correct values of the damping times than the Newtonian
approach~\cite{kokkotas-1999-2}.  However, the spacetime in
General Relativity is not a static and ``absolutum medium'' on
which the gravitational wave propagates, but has  its own
dynamical degree of freedom. This property adds to the Newtonian
picture a new class of oscillation modes, namely the gravitational
$w$ave modes ($w$-modes)~\cite{Chandrasekhar:1991fi,
Kokkotas:1992ks}, which are high frequency and strongly damped
modes that couple very weakly with the stellar fluid. This latter
characteristic implies that the axial and polar spacetime modes
have similar properties. These purely relativistic modes can be
separated in three classes.
\begin{itemize}
\item[] \emph{Curvature modes}: are the standard $w$-modes, which are
present in all relativistic stars. They are associated with the
``curvature bowl'' present inside the compact star.  The typical
first curvature mode has frequency $5-12~kHz$ and a damping rate
of tenth of milliseconds. For higher order $w$-modes the frequency
increases and the damping rate is shorter.
\item[] \emph{Trapped modes}: Some of the curvature modes for
increasingly compact stars ($R \leq 3~M$) can be trapped inside
the potential barrier, when the surface of the star is inside the
peak of the gravitational potential. The were first determined by
Chandrasekhar and Ferrari~\cite{Chandrasekhar:1991fi} for axial
stellar perturbations. They have frequencies between a few hundred
 Hz and a few kHz and are more slowly damped by gravitational radiation than
the curvature modes (few tenths of milliseconds).  The spectrum of
trapped modes is finite, the number of modes depends on the depth
of the potential well and then on the compactness of the star. The
main issue related to this class of modes is whether such an
ultra-compact star can exist in nature.
\item[] \emph{Interface} or $w_{II}$ \emph{modes}: were determined by
Leins et al.~\cite{1993PhRvD..48.3467L}.  There is a finite number
of $w_{II}$-modes for any multipole $l$, which have frequencies
that vary from 2 to 15 kHz and very short life (less than tenth of
milliseconds).  The existence of this family of spacetime modes
may be associated with the discontinuity at the surface of the
star.
\end{itemize}
More details about the numerical techniques and physical
properties of the QNM can be found in the
reviews~\cite{kokkotas-1999-2, Nollert_mio, kokkotas-2005-}.

In many works, the stellar fluid oscillations have been determined
by neglecting the quantities associated with the gravitational
field, i.e. the Newtonian gravitational potential or the metric
tensor of the spacetime. This method, known as the ``Cowling
approximation''~\cite{Cowling:1941co}, provides frequencies and
damping times of the fluid modes with an error usually less than
10 percent.

The investigations of relativistic stellar perturbations as an
initial value problem has been addressed quite recently for
spherical non-rotating stars in the context of gravitational
collapse by Seidel~\cite{Seidel:1990xb}, and for static stars by
Kind, Ehlers and Schmidt~\cite{Kind:1993kn}. This latter work has
determined the set of perturbative polar equations and the
appropriate boundary conditions for having a well posed Cauchy
problem, which determine a unique solution.  Subsequently, Allen
et al.~\cite{allen-1998-58} and Ruoff~\cite{Ruoff:2001ux,
Ruoff:2000nj}, the latter using the ADM formalism, explored in the
time domain the dynamics of linear polar non-radial oscillations
of a non-rotating star for polytropic and more realistic equations
of state. These works provided important information about the
numerical issues related to the numerical integration of the
perturbative equations, and about the initial configurations which
are able to excite the fluid and spacetime modes. By using the
GSGM formalism, Nagar et al~\cite{Nagar:2004ns} extended the time
domain analyses for investigating the non-radial perturbations of
non-rotating stars induced by external objects, like point
particles and accretion of matter from tori.

Linear perturbations have been studied also for rotating
relativistic stars with perturbative techniques and full
non-linear codes. In this thesis we will study the non-linear
oscillations of non-rotating stars, therefore we do not address
here this subject. The interested reader can find accurate and up
to  date information in the review by
Stergioulas~\cite{2003LRR.....6....3S}.

This chapter is organized in five sections.  The equilibrium
configuration is described in section~\ref{sec:4_Back}, while the
background quantities of the GSGM formalism in
section~\ref{sec:4.2GSGMLin}. The first order radial perturbations
are introduced in section~\ref{sec:Linear_Rad_pert_anal}, while
the polar non-radial perturbations are described in
section~\ref{sec:431_pol_non_rad} and the axial in
section~\ref{sec:Lin_Axial_per}.

\section{Background} \label{sec:4_Back}

The equilibrium configuration is a non-rotating spherically
symmetric star that is described by a static metric in
Schwarzschild coordinates:
\begin{equation}
\bar{\met}_{\alpha\beta}dx^\alpha dx^\beta =
-e^{2\Phi(r)}dt^2+e^{2\Lambda(r)}dr^2+
r^2(d\theta^2+\sin^2\theta d\phi^2)\,,
\end{equation}
where the functions $\Phi(r)$ and $\Lambda(r)$ are two unknown
functions that must be determined by the Einstein field equations.
The radial coordinate $r$ identifies for constant $t$ and $r$ a
2-dimensional sphere of area $4\pi r^2$.

The stellar matter is described by a single component perfect fluid,
where by definition viscosity, heat conduction and anisotropic
stresses are absent. This model, though simplistic, is suitable for a
first investigation of non-linear oscillations and for a correct
interpretation of the results. More realistic descriptions should
consider the presence of magnetic fields, viscosity, crust, details of
the stellar structure, superfluid and different particles, etc.  We
intend to take into account these specific elements in future works,
also in order to avoid possible numerical instabilities which can
always arise when the stellar model becomes more complex.  The perfect
fluid energy-momentum tensor is given by the following expression:
\begin{equation}
\bar{T}_{\alpha\beta} = (\bar{\rho}+\bar{p})\, \bar{u}_{\alpha}
\bar{u}_{\beta} + \bar{p}\, \bar{\met}_{\alpha\beta}\,,
\end{equation}
where $\bar \rho$ and $\bar p$ denote the mass-energy density and
the pressure in the rest-frame of fluid, and $\bar{u} _{\alpha}$
is the covariant velocity of the static background. The velocity
is a timelike vector, thus its components can be derived by the
normalization condition $u^{\alpha} u_{\alpha} = -1$. The
covariant velocity assumes the following form:
\begin{equation}
\bar{u}_{\alpha} = \left(-e^\Phi,0,0,0\right) \, .
\end{equation}
The metric variable $\Lambda$ can be related to a new function $M$
by the following definition,
\begin{equation}
M(r) \equiv \frac{r}{2} \left( 1 - e^{-2\Lambda(r)} \right) \, \label{M_def}
\end{equation}
In the stellar exterior this function assumes the constant value $M =
M(R_s)$, which is the gravitational mass of the star and $R_s$ is the
stellar radius.  In the Newtonian limit the functions $M(r)$ and
$\Phi(r)$ describe respectively the gravitational mass and the
gravitational potential of the star.

The Einstein~(\ref{Ein_equations}) and the fluid conservation
equations~(\ref{Cons_eqs}) form a system of ordinary differential
equations first derived by Tolman \cite{Tolman:1939jz} and Oppenheimer
and Volkoff \cite{Oppenheimer:1939ne} (TOV) in 1939:
\begin{eqnarray}
\Phi_{,r} & = & \left( 4 \pi \bar p r + \frac{M}{r^2} \right)
\frac{r}{r-2M} \label{Phi_r}\, \\
\Lambda_{,r} & = & \left( 4 \pi \bar
\rho r - \frac{M}{r^2} \right) \frac{r}{r-2M} \, \label{Lambda_r} \\
\bar{p}_{,r} & = &
- \left(\bar{\rho}+\bar{p} \right) \Phi_{,r} \label{p_r} \,, \\
M_{,r} & = & 4\pi\bar{\rho} \, r^2 \label{M_r} \,.
\end{eqnarray}
The integration of the TOV equations require the specification of
an equation of state for the stellar matter $p = p(\rho,s)$. We
consider a cold neutron star at zero temperature. This
approximation is certainly accurate for old isolated neutron stars
in absence of accretion. A few seconds after a core collapse the
temperature of a newly  born neutron star rapidly decreases, and
the thermal energy becomes much lower than the Fermi energy of the
degenerate neutron fluid.  The Fermi energy for nuclear densities
$\rho _{N} = 2 \times 10^{14} g cm^{-3}$ is about $E_{F} \sim 30
MeV = 3 \times 10^{11} K$ and increases for the supranuclear
densities of the neutron star core. As a result, the thermal
degrees of freedom can be considered frozen out.  In this thesis
we will investigate also the effects of coupling between radial
pulsations and differential rotation, which is present within the
first seconds of a proto-neutron star life. Therefore, the thermal
and dissipative effects due to convective zones and shock
formations near the stellar surface should have been included into
the physical model.  In order not to complicate our investigation
of the non-linear oscillations we neglect these effects with the
aim to include them in future works.

\indent In the present work, the star is then described by a
barotropic fluid $p=p(\rho)$, which is parameterized by a
polytropic equation of state (EOS):
\begin{equation}
p = k \, \rho ^{\Gamma} \, ,  \label{Poli_EOS}
\end{equation}
where $k$ is the adiabatic constant and $\Gamma$ the adiabatic index.
The background speed of sound in the fluid is then given by
\begin{equation}
\bar{c}_s^2= d\bar{p}/ d\bar{\rho}\, .
\end{equation}
The TOV and fluid equation of state provide a one-parameter family of
solutions that depend on the stellar central density $\rho _c$.  Its
numerical integration is described in section~\ref{sec:Num_Bac}.

The exterior of a non-rotating star is a Schwarzschild spacetime
represented by the following line-element in Schwarzschild coordinate:
\begin{equation}
ds^2 = - \left( 1-\frac{2M}{r} \right) dt^2 + \left(1-\frac{2M}{r}
\right)^{-1} dr^2 + r^2 d\theta^2 + r^2 \sin^2 \theta \,d \phi^2
\end{equation}
where $M$ is the gravitational mass of the star $M=M(R_s)$. The
internal and external solutions have to match on the stellar
surface. Therefore, the following condition must be satisfied by
the metric variable $\Phi$,
\begin{equation}
\Phi(R_s) = - \Lambda(R_s) = \frac{1}{2} \ln \left( 1-\frac{2M}{r}\right)
\label{Back_Srf_cond} \, .
\end{equation}

\section{GSGM background quantities for linear perturbations}
\label{sec:4.2GSGMLin}

The linear non-radial perturbations on a non-rotating star can be
studied with the gauge-invariant perturbative formalism set up by
Gerlach and Sengupta and further developed by Gundlach and
Mart\'{\i}n--Garc\'{\i}a, which has been introduced in
section~\ref{sec:GSGM}.  This formalism can be specialized to the case
of a static background by choosing the static frame vector basis
$\left\{ \bar u^A, \bar n^A \right\}$,
\begin{eqnarray}
\bar u^A = (e^{-\Phi},0)\,, \quad  \qquad \bar n^A = (0,e^{- \,
\Lambda})\,. \label{unst}
\end{eqnarray}
The associated background scalars (\ref{backg_scal}) assumes the
following expressions:
\begin{eqnarray}
\bar\mu = \bar U =0\,, \qquad  \bar\nu =
e^{-\Lambda}\Phi_{,\,r}\,, \qquad \bar W =
\frac{e^{-\Lambda}}{r}\, ,  \label{static-scalars}
\end{eqnarray}
while the frame derivatives of a generic scalar function
$f(x^{A})$ are given by:
\begin{equation}
\dot{f} = e^{-\Phi} f_{\, ,t}  \qquad \qquad f' = e^{-\Lambda}
f_{\, , r}  \label{static-frame-der}
\end{equation}

\section{Radial perturbations}  \label{sec:Linear_Rad_pert_anal}

The radial perturbations of non-rotating stars preserve the
spherical symmetry of their equilibrium configuration.  The
geometrical and physical quantities of a star  deviate from the
background values only along the radial coordinate.  Therefore,
the validity of  Birkhoff's theorem implies that for a
non-rotating star an external observer cannot receive any
gravitational information about the pulsating stellar dynamics.
Adiabatic radial oscillations can be described in terms of their
normal modes. This characteristic is more evident in the frequency
domain analysis, where the radial perturbative equations can be
set as a Sturm-Liouville problem, which provides a complete and
discrete set of the eigenfrequencies of the normal modes and their
associated eigenfunctions.

The time domain analysis of the radial oscillations has been
investigated in various gauges and with different set of variables and
equations \cite{Misner:1973cw, Ruoff:2000nj, Gundlach:1999bt}.  The
most common perturbative quantity used to describe the unique degree
of freedom of radial perturbations is the Lagrangian displacement $\xi
^{\mu} (x^{\alpha})$. This is a vector field that provides at any
instant of time the position of a fluid element with respect to its
equilibrium position. The Lagrangian displacement obeys a wave
equation that can be studied in the time or frequency domain.  In this
work, we describe the radial perturbations by using the GSGM
formalism. This choice allows us to use a more uniform set of
perturbative variables for both the radial and non-radial
oscillations.  The Lagrangian displacement and its eigenvalue equation
will be useful later for setting up the initial configuration
(\ref{sec:IV_Rad}) and for estimating the movement of the surface
along the evolution~(\ref{sec:Ch7_rad_sim}).  However, it is worth
noticing that the GSGM formalism fails to be gauge invariant for
radial perturbations, for instance some of the polar gauge-invariant
tensors~(\ref{pkab})-(\ref{ppa}) cannot be even defined. This property
does not produce any limitation to the approach of this thesis, as we
will prove later that the non-linear perturbations that describe the
coupling will be gauge-invariant only for a fixed radial gauge.

The metric of radial perturbations is given by the following expression:
\begin{eqnarray}
\delta \met_{\alpha \beta}^{(1,0)} & = & \left(\begin{array}{cc}
   h_{AB}^{(1,0)}  &  0  \\
\\
 0     & \  r^2 K^{(1,0)} \, \gamma_{ab}  \\
\end{array}\right)\,,
\end{eqnarray}
where the constant value of the harmonic scalar functions $Y_{00}$
is implicitly contained into the perturbative variables, while the
tensor $h_{AB}^{(1,0)}$ assumes the form:
\begin{equation}
h_{AB}^{(1,0)} = \eta^{(1,0)} \bar{\met}_{AB} + \phi^{(1,0)}
(\bar{u}_A\bar{u}_B+\bar{n}_A\bar{n}_B)
+ \psi^{(1,0)}
(\bar{u}_A\bar{n}_B+\bar{n}_A\bar{u}_B) \,.
\end{equation}
The four scalar quantities $\eta^{(1,0)}$,$\phi^{(1,0)}$,$\psi^{(1,0)}$ and
$K^{(1,0)}$ are functions of the coordinates~$(t,r)$ and describe the
metric radial perturbations.
A gauge choice that considerably simplifies the perturbative equations is
the following:
\begin{equation}
\psi^{(1,0)}=0\,,
\qquad k^{(1,0)}=0\,, \label{fixgauge}
\end{equation}
which makes the radial metric diagonal,
\begin{eqnarray}
\delta \met_{\alpha \beta}^{(1,0)} & = & \left(\begin{array}{cccc}
   e^{2\Phi}\left( \chi^{(1,0)} - 2\eta^{(1,0)} \right) & 0   &  0 & 0  \\
 0 & e^{2\Lambda}\,\chi^{(1,0)} & 0  & 0  \\
 0  & 0   & 0   & 0  \\
\end{array}\right) \, ,
\end{eqnarray}
where we have used the definition~(\ref{chidef}) for the new metric
variable $\chi^{(1,0)}$:
\begin{equation}
\chi^{(1,0)}= \eta^{(1,0)} + \phi^{(1,0)}.
\end{equation}
As shown in reference \cite{Martin-Garcia:2000ze}, the relations
(\ref{fixgauge}) do not fix completely the gauge of radial
perturbations. There is a residual gauge degree of freedom that can be
used in the boundary conditions to impose the vanishing of the metric
perturbation $\eta^{(1,0)}$ on the stellar surface.

\indent The radial perturbation of the fluid velocity is given by 
equation~(\ref{deltaupol}),
\begin{equation}
\delta u_{\alpha }^{(1,0)}= \left( \left( \frac{\chi^{(1,0)}}{2} -
\eta^{(1,0)} \right) e^{\Phi}\,, \, e^{\Lambda} \,\gamma^{(1,0)} , 0 ,
0 \right)\,,
\end{equation}
where the scalar function $\gamma^{(1,0)}$ depends on the $(t,r)$ coordinates
and describes the fluid element velocity along the radial coordinate.
The pressure and density perturbations are for a barotropic
fluid given by:
\begin{equation}
\delta \rho^{(1,0)}=\omega^{(1,0)}
\bar\rho\,,  \qquad \delta p^{(1,0)} = \bar{c}_s^2  \delta
\rho^{(1,0)} \,.
\end{equation}

\subsection{Radial perturbative equations}
\label{sec:Linear_Rad_pert_anal_Eqs}
The perturbations of Einstein and conservation equations lead to a
set of four partially differential equations for the four
variables $\chi^{(1,0)},
\eta^{(1,0)},\omega^{(1,0)},\gamma^{(1,0)}$, see
~\cite{Gundlach:1999bt}. Before writing the system of equations,
it is more convenient for calculation purposes to adopt a slightly
different set of radial perturbations. The density perturbation
$\omega^{(1,0)}$ will be replaced by the enthalpy perturbation
$H^{(1,0)}$, while the metric quantity $\chi^{(1,0)}$ will be
changed with the variable $S^{(1,0)}$, in order to use a set of
variables consistent with the one we will use for the non-radial
polar perturbations. These two new perturbations are defined as
follows:
\begin{equation}
H^{(1,0)} \equiv \frac{\delta p^{(1,0)}}{\bar{\rho}+\bar{p}} =
\frac{\bar{c}_s^2 \bar{\rho}}{\bar{\rho} +\bar{p}}\,\om^{(1,0)}\,,
\qquad \qquad
S^{(1,0)} \equiv \frac{\chi^{(1,0)}}{r} \,.
\end{equation}

The dynamics of the four radial perturbations $S^{(1,0)},
\eta^{(1,0)},H^{(1,0)},\gamma^{(1,0)}$ is then governed by the following
set of three partially differential equations that does not contain the
quantity $\eta^{(1,0)}$,
\begin{eqnarray}
 H_{, \, t}^{(1,0)} & = & - \bar{c}_s^2\, e^{\Phi -\La} \,
\ga^{(1,0)}_{, \, r} - \bar{c}_s^2 \,
\left[\left(1-\frac{1}{\bar{c}_s^2}\right) \left(4 \pi \bar p \, r +
\frac{M}{r^2} \right) + \frac{2}{r} \, e^{-2 \La}  \nn {} \right. \\
&&  \left.  - 4\pi
\left(\bar\rho + \bar p \right) r \right]\, e^{\Phi + \La} \,
\ga^{(1,0)}\,,
\label{eq:H10_ev}  \\
\ga_{, \, t}^{(1,0)} & = & - e^{\Phi-\La} \, H^{(1,0)}_{,\,r} - 4\pi
\left(\bar\rho + \bar p\right)\,r \, e^{\Phi + \La} \,H^{(1,0)} -
\left( 4 \pi \bar p \, r^2 + \frac{1}{2} \right) \, e^{\Phi + \La}
S^{(1,0)}\,, \label{eq:gam_t} \nn \\ \\
S_{, \, t}^{(1,0)} & = & -8 \pi
\left(\bar\rho + \bar p \right)\, e ^{\Phi + \La} \,
\ga^{(1,0)}\,. \label{chi_t}
\end{eqnarray}
Equation (\ref{chi_t}) has been derived by using equation (34) in
Ref.~\cite{Martin-Garcia:1998sk}.  The remaining metric variable $\eta
^{(1,0)}$ is then obtained by a constraint, which is the following
elliptic equation:
\begin{equation}
\eta^{(1,0)}_{, \, r} = 4\pi (\bar\rho+\bar p) \, r \,
\left[r\,S^{(1,0)}+\left(1+ \frac{1}{\bar{c}^2_s}\right)H^{(1,0)}
\right]  \, e^{2 \La} \,. \label{eq:eta_cn}
\end{equation}
The solutions of radial pulsations must satisfy the Hamiltonian constraint
on the initial Cauchy hypersurface and all along the evolution.
The equation of this constraint is given by:
\begin{equation}
S^{(1,0)}_{, r} = e^{2\Lambda}\biggl[ \left( 8 \pi \bar\rho r
-\frac{2}{r} + \frac{2 M}{r^2} \right) S^{(1,0)}+8 \pi \frac{\bar\rho
+ \bar p}{\bar c_s^2} H^{(1,0)}\biggr] \ . \label{eq:S10_cn}
\end{equation}
The system of equations (\ref{eq:H10_ev})-(\ref{eq:S10_cn}) is
equivalent to the one used by Ruoff~\cite{Ruoff:2000nj},
cf.~also~\cite{Misner:1973cw}. The evolution equations
(\ref{eq:H10_ev}) and~(\ref{eq:gam_t}) can be combined in order to
determine a wave equation for each of the variable involved
$H^{(1,0)}$ or $\gamma^{(1,0)}$. The resulting equation for the
enthalpy requires another equation to close the system, as it
contains the metric perturbation $S^{(1,0)}$ as one of its terms.
On the other hand, the radial velocity $\gamma^{(1,0)}$ satisfies
a single wave equation. Therefore, the radial velocity
$\gamma^{(1,0)}$ can be used to represent as well as the
Lagrangian displacement the single degree of freedom present in a
radially oscillating configuration.  This equation can be
determined by differentiating the two equations (\ref{eq:H10_ev})
and~(\ref{eq:gam_t}) with respect to the radial and time
coordinate respectively, then we can take an appropriate linear
combination of them and introduce some of the radial equations in
it. This procedure leads to the following wave equation:
\begin{equation}
\gamma_{, \, tt}^{(1,0)}    -    c_{s}^2 \,  e^{2 \, \left(\Phi
-\Lambda\right)  } \, \gamma_{, \, rr}^{(1,0)} + d_1(r) \,
\gamma_{, \, r} ^{(1,0)}  + d_2(r) \, \gamma ^{(1,0)} = 0 \label{ga10_WvEq} \, ,
\end{equation}
where the background coefficients $d_1(r)$ and $d_2(r)$ are the following:
\begin{eqnarray}
e^{ - 2 \, \Phi} \, d_1(r) & \equiv & \, \left\{ \left( \rho + p
\right) \left( 4 \pi p \, r + \frac{M}{r^2} \right)
\frac{1}{c_{s}^{2}} \frac{ \partial c_{s}^{2}}{\partial \rho} + \left[
4 \pi \left( \rho - 2 p \right) \, r + \frac{M}{r^2} - \frac{2}{r}
\right] c_{s}^{2} {} \nn \right. \\ & + & \left. 4 \pi p \, r +
\frac{M}{r^2} \right\} \, , {} \\
e^{ - 2 \, \Phi} \, d_2(r) & \equiv
& \left\{ \left( \rho + p \right) \left( 4 \pi p \, r + \frac{M}{r^2}
\right) \, \left[ \left( \frac{2}{r} - \left( 4 \pi \rho \, r -
\frac{M}{r^2}\right) \, e^{2 \,\Lambda} \right) \, \frac{1}{c_{s}^{2}}
\frac{ \partial c_{s}^{2}}{\partial \rho} - 8 \pi r e^{2 \,\Lambda}
\right] {} \right. \nn \\
& - & {} \left. \left[ 16 \pi p -
\frac{2}{r^2} + \frac{6 M}{r^3} - 8 \pi r \left( \rho + p \right)
\left( 4 \pi \rho r - \frac{M}{r^2} \right) e^{2 \,\Lambda} \right] \,
c_{s}^{2} - \frac{ 2 M}{r^2} \right\} \, .
\end{eqnarray}
Having solved equation~(\ref{ga10_WvEq}) for $\gamma ^{(1,0)}$, we
may use the first order evolution equations to determine the
enthalpy $H ^{(1,0)}$, and the metric $S^{(1,0)} = \chi^{(1,0)}/r$
and $\eta^{(1,0)}$ variables.

\subsection{Boundary conditions for radial perturbations}

The physical solutions of the radial perturbation problem have to
satisfy the boundary conditions at the origin and surface. The
origin of coordinates $r=0$ must be a regular point for the
perturbative fields and equations. The analysis of the Taylor
expansion of the perturbative fields around the centre leads to
the following expressions:
\begin{eqnarray}
S^{(1,0)} & = & S^{(1,0)}_o(t)\,r+ O(r^3)\, , \label{bc_rad_in}\\
\eta^{(1,0)} & = & \eta^{(1,0)}_o(t)+O(r^2)\,, \label{eta_bc}\\\
H^{(1,0)} & = & H^{(1,0)}_o(t)+O(r^2)\, , \label{H10_bc} \\
\gamma ^{(1,0)} & = & \gamma^{(1,0)}_o(t)\,r+ O(r^3) \, .
\label{bc_rad_ori}
\end{eqnarray}
The physical condition on the stellar surface is the vanishing of
the total pressure on the perturbed surface:
\begin{equation}
\Sigma = \left\{ x^{\mu} \in \mathcal{M} \quad |  \quad p\left(x^{\mu} \right) = 0   \right\} \label{Condtotpre}
\end{equation}
In a perturbative approach, the small movement of the surface from its
equilibrium position can be described by the Lagrangian displacement $
\xi _{(1,0)} = \xi ^{\,r} \left(x^{\alpha}\right)$ that gives the
position of the perturbed surface with respect to the background
surface.  Thus, the total pressure~(\ref{Condtotpre}) can be Taylor
expanded around the equilibrium surface
\begin{equation}
p \left( x_{R}^{\mu} + \lambda \, \xi^{\mu}  \right) = \bar{p} \left( x_{R} ^{\mu} \right) + \lambda
\Delta p ^{(1,0)} \left( x_{R} ^{\mu}  \right) + O ( \lambda ^{2} ) \, , \label{pTOT_exp}
\end{equation}
where $x_{R}^{\mu}$ are the coordinates of the background surface,
$\lambda$ controls the strength of the radial perturbations and
$\Delta p^{(1,0)}$ is the radial Lagrangian perturbation of the
pressure. This last quantity can be written in terms of the Eulerian
perturbation with the well known relation that connects the
perturbations in these two gauges~(see e.g.~\cite{Misner:1973cw}),
\begin{equation}
  \Delta p^{(1,0)} = \delta p^{(1,0)} + \pounds_{\xi_{(1,0)}}  \, \bar{p}  \, .
\end{equation}
The condition (\ref{Condtotpre}) and the perturbative expansion
(\ref{pTOT_exp}) leads to the vanishing of the background pressure
$\bar{p} \left( x_{R} ^{\mu} \right) = 0$ and its Lagrangian
perturbation $\Delta p^{(1,0)} = 0$ on the equilibrium surface of the star. \\
\indent The condition $\bar{p} \left( x_{R} ^{\mu} \right) = 0$
will be imposed for the numerical integration of the TOV
equations. The Lagrangian perturbation of the pressure can be
written in terms of the MTW {\em radial renormalized displacement
function} $\zeta$ as
\begin{equation}
r^{2} \D p^{(1,0)} = -\left( \bar \rho + \bar p  \right)  \bar c_{s}^{2} e^{-\Phi}
\frac{\partial \zeta }{\partial r}   \label{Surf_bc_zeta}
\end{equation}
where $\zeta$ is so defined:
\begin{equation}
\zeta \equiv r^2 \, e^{-\Phi} \, \xi_{(1,0)}\,.  \label{zeta_def}
\end{equation}
However, we cannot use directly the expression ~(\ref{zeta_def}), as
the Lagrangian displacement is not a dynamical variable in our set of
radial perturbations.  The rate of the surface movement is described
instead by the radial component of the fluid velocity, namely
$\gamma^{(1,0)}$ which is related to $\zeta$ as follows
\begin{equation}
\zeta_{,t} = r^2 e^{-\Lambda} \gamma^{(1,0)} \, . \label{zeta_gam_def}
\end{equation}
By differentiating 
equation~(\ref{Surf_bc_zeta}) with respect to the time one finds the
following boundary condition on the surface for $\gamma^{(1,0)}$:
\begin{equation}
\left. (\bar\rho + \bar p)\,\bar{c}^2_s\, e^{-\Phi}\left( r^2 e^{-\Lambda}
\gamma^{(1,0)}\right)_{,r} \right|_{r=R}= 0\,.  \label{Surf_bc_gam10}
\end{equation}
In a polytropic equation of state the pressure, mass-energy density
and the speed of sound vanishes on the static surface, thus the
equation~(\ref{Surf_bc_gam10}) implies that
the radial variable $\gamma^{(1,0)}$ and its spatial derivative
$\gamma^{(1,0)}_{, \, r}$ must be finite. On the other hand,
when the mass-energy density and the speed of sound are non-null
on the surface the following condition on $\gamma^{(1,0)}$ is required
\begin{equation}
\left. \left( r^2 e^{-\Lambda}
\gamma^{(1,0)}\right)_{,r} \right|_{r=R}= 0\,.  \label{Surf_bc_gam10_2}
\end{equation}
This is the only physical surface condition. The behaviour of the
metric perturbation $S^{(1,0)}$ and the enthalpy $H^{(1,0)}$ can be
directly deduced by the perturbative equations (\ref{chi_t})
and~(\ref{eq:H10_ev}) respectively. For the other metric perturbation
$\eta^{(1,0)}$ one can use the residual gauge degree of freedom that
has not been fixed by the radial gauge~(\ref{fixgauge}), see
reference~ \cite{Martin-Garcia:2000ze}. In accordance with the
physical properties of the radial perturbations, the more appropriate
choice is a null value of $\eta^{(1,0)}$ on the surface. This allows
us to eliminate all the gauge fields present in the external
spacetime~\cite{Martin-Garcia:2000ze}.

\subsection{Frequency domain analysis of radial perturbations}
\label{sec:Rad_freq}

The time domain integration of radial perturbative equations
(\ref{eq:H10_ev})-(\ref{chi_t}) requires the choice of the initial
values for the radial variables. A method used to determine the
initial profile of one of the radial perturbations is to provide the
eigenfunction associated with the particular radial mode. This method
allow us to select and excite the radial modes of our interest and
then simplify the interpretation of the non-linear effects due to the
coupling with the non-radial oscillations.
In order to determine the eigenfunctions for the radial variables
$\gamma^{(1,0)}$ we can elaborate the most common eigenvalue
equation used in literature \cite{Misner:1973cw, Kokkotas:2000up},
which is the wave equation of the Lagrangian fluid displacement
$\xi ^{(1,0)}$.
By using the renormalized Lagrangian displacement $\zeta^{(1,0)}$
defined in equation (\ref{zeta_def}) the wave equation
reads~\cite{Misner:1973cw}:
\begin{equation}
- W \, \zeta_{, \, tt}^{(1,0)} +
 \frac{d}{dr}\left(P \, \frac{d\zeta}{dr}^{(1,0)}\right)+Q \, \zeta^{(1,0)}
 =0\,, \label{eqzeta}
\end{equation}
where $W, P, Q$ are functions of the radial coordinate $r$ only. They
are defined as:
\begin{eqnarray}
r^2 \, W & \equiv & \left( \bar \rho + \bar p  \right) \,
e^{3 \Lambda  + \Phi },   \label{Wrad}\\
r^2 \, P & \equiv & \left( \bar \rho + \bar p  \right) \, \bar c _{s}^{2}
\, \bar p  \, e^{ \Lambda + 3 \Phi} \, , \\
r^2 \, Q & \equiv & \left( \bar \rho + \bar p  \right) \,
  \left[  \Phi _{,r} ^{2} - \frac{4}{r} \, \Phi_{,r} - 8 \pi
\, \bar p \, e^{2 \Lambda  } \right] \, e^{ \Lambda + 3 \Phi } \,
. \label{Qrad}
\end{eqnarray}
The radial component of the velocity perturbation~$\gamma
^{(1,0)}$ obeys the wave equation~(\ref{ga10_WvEq}). An eigenvalue
problem can be then set up by introducing the harmonic
ansatz~(\ref{delTfd}) in the variable~$\gamma ^{(1,0)}$. However,
in order to simplify the analysis the wave
equation~(\ref{ga10_WvEq}) can be written after some manipulations
with the same form of equation~(\ref{eqzeta}):
\begin{equation}
- W y _{, \, tt}^{(1,0)} + \frac{d}{dr}  \left(P \, \frac{d
  y}{dr}^{(1,0)}\right)+Q \, y^{(1,0)} = 0\,,
\label{eqyy}
\end{equation}
where we find it convenient to use the variable $y^{(1,0)}$, which
is related to $\zeta^{(1,0)}$ by
\begin{equation}
y^{(1,0)} = r^2 e^{-\Lambda} \gamma^{(1,0)} = \zeta^{(1,0)}_{, \, t }
=r^2 e^{-\Phi} \xi _{, \, t} \, , \label{yydef}
\end{equation}
and $W,P,Q$ are the same three functions defined in equations
(\ref{Wrad})-(\ref{Qrad}). We notice that the wave equation
(\ref{eqzeta}) can be also determined by time differentiating the
equation (\ref{eqzeta}) and then using the relation
(\ref{zeta_gam_def}) that connects the two variables $\zeta^{(1,0)}$
and $\gamma^{(1,0)}$. However, the definition (\ref{zeta_gam_def})
shows that the solution of the two equations (\ref{eqzeta})
and~(\ref{zeta_gam_def}) disagree at most for a function that depends
on the radial coordinate $r$ only. This function can always set to
zero with an appropriate choice of the initial conditions.

The eigenvalue problem can be set up by expressing the radial
variable $y^{(1,0)}$ in the following time harmonic form:
\begin{equation}
y^{(1,0)} = y_{0}(r) e^{i \omega t } \, \, \qquad \textrm{with} \quad
\omega \in \mathbb{R}\, \label{y_harm}
\end{equation}
where $\omega$ is the frequency of radial pulsations. With the
introduction of the harmonic functional dependence~(\ref{y_harm}) and
the boundary conditions that later we discuss, the wave
equation~(\ref{eqzeta}) becomes a Sturm-Liouville problem:
\begin{equation}
 \frac{d}{dr}  \left(P \, \frac{d
  y_{0}}{dr} \right)+ \left(Q + \omega^{2} W \right)\, y_{0} = 0 \, ,
\label{eqyy_fr}
\end{equation}
where the squared frequency $\omega^{2}$ is a free parameter. The
solutions of this linear ODE form a countable set of discrete
eigenvalues $\omega^2$.

\indent In order to solve numerically this equation, one can
transform it to a set of two first order ordinary differential
equations by defining the new variable $z^{(1,0)} = P y_{, \,
r}^{(1,0)}$ \cite{Ruoff:2000nj, Kokkotas:2000up},
\begin{eqnarray}
y_{,r}^{(1,0)}  & = & \frac{z}{P}^{(1,0)} \label{yeq} \, ,\\
z_{,r}^{(1,0)}  & = & - \left( \omega^2 W + Q \right) y ^{(1,0)}
\label{zeq}\, .
\end{eqnarray}
The boundary conditions associated with these equations are given
by the regularity of perturbative equations at the origin and the
vanishing of the Lagrangian perturbation of pressure on the
surface. At the origin the condition~(\ref{bc_rad_ori}) for
$\gamma^{(1,0)}$ leads to the following behaviour:
\begin{equation}
y^{(1,0)} = y_{0} r^3 + O(r^5)\, , \qquad \quad z^{(1,0)} =
z_{0}^{(1,0)} + O(r^3)\, ,
\end{equation}
that with equation~(\ref{yeq}) provides the following expression:
\begin{equation}
y_{0}  = \frac{z_{0}}{3 P} \, .
\end{equation}
On the other hand, the vanishing of the Lagrangian perturbation of
pressure on the stellar surface leads to the expression
(\ref{Surf_bc_gam10}), and then to the following condition:
\begin{equation}
\left. (\bar\rho + \bar p)\,\bar{c}^2_s\, e^{-\Phi}
y^{(1,0)} _{,r} \right|_{r=R}= 0\, . \label{BC10_srf_Fre}
\end{equation}
Various numerical methods can be adopted to integrate the system of
equations (\ref{yeq})-(\ref{zeq}). As we will see in section
(\ref{sec:IV_Rad}), we use the "relaxation method"~\cite{1992nrfa.book.....P}. 

\section{Polar non-radial perturbations}
\label{sec:431_pol_non_rad}

Linear non-radial perturbations of a non-rotating barotropic star
can be described in the GSGM formalism by a set of metric and
fluid  gauge-invariant perturbation fields:
\begin{equation}
\left\{ k_{AB}^{(0,1)}, k^{(0,1)}, \alpha^{(0,1)}, \gamma^{(0,1)},
\omega ^{(0,1)} \right\} \, ,
\end{equation}
where all these tensors live on the submanifold $M^2$ and their
definitions can be found in section~\ref{sec:GSGM}.
The metric perturbations are expressed by the two quantities
$k_{AB}^{(0,1)}$ and $k^{(0,1)}$, where the 2-symmetric tensor
$k_{AB}$ can be decomposed in the frame basis $\left\{ \bar{u}^A,
\bar{n}^A \right\}$ as in equation (\ref{kABdec}), by giving the
following expression:
\begin{eqnarray}
k_{AB}^{(0,1)} & = & \left(\begin{array}{cc}
  \left( \chi^{(0,1)} + k^{(0,1)} \right) e^{2\Phi}  &  - \psi^{(0,1)} e^{\Phi + \Lambda}   \\ \\
 - \psi^{(0,1)} e^{\Phi + \Lambda}      & \left( \chi^{(0,1)} + k^{(0,1)} \right) e^{2\Lambda}    \\
\end{array}\right)\, .
\end{eqnarray}
In the previous equation we have used the definition (\ref{chidef}) for
the gauge invariant scalar perturbation $\chi^{(0,1)}$ and the validity of the
Einstein equation (\ref{eta}), which, for $l \ge 2 $, sets a null
value for the metric perturbation $\eta^{(0,1)}$.
On the other hand, the fluid motion is described by the two
gauge-invariant variables $\gamma^{(0,1)}$ and $\alpha ^{(0,1)}$
defined in~(\ref{gamgi}) and~(\ref{algi}), which describe
respectively the velocity perturbation along the radial
coordinates $r$ and the latitude of the star. The perturbation of
the mass-energy density is instead described by the variable
$\omega^{(0,1)}$, see equation~(\ref{omgi}). The equations of
non-radial perturbations assume a simpler form when the enthalpy
perturbation replaces the density
perturbation~\cite{allen-1998-58, Ruoff:2000nj, Nagar:2004pr}. The
enthalpy perturbations is so defined:
\begin{equation}
H^{(0,1)} = \frac{\bar c _s ^2 \, \bar \rho  }{\bar \rho + \bar p
} \, \omega^{(0,1)} \label{H01def}  \, .
\end{equation}
The six metric and fluid perturbations
\begin{equation}
\chi^{(0,1)}, k^{(0,1)}, \psi^{(0,1)}, \gamma ^{(0,1)},
\alpha ^{(0,1)}, H^{(0,1)}  \, ,
\end{equation}
are not all independent variables. Linear non-radial oscillations
in the interior spacetime have a dynamics that can be described
with two degrees of freedom, as shown
in~\cite{Chandrasekhar:1991fi,Ipser:1991ip}. The external
spacetime instead is a perturbed Schwarzschild spacetime where the
single degree of freedom can be described by the Zerilli
function~\cite{Zerilli:1970fj}.
In the frequency domain approach, Chandrasekhar and
Ferrari~\cite{Chandrasekhar:1991fi} were able to describe stellar
oscillations in terms of pure metric perturbations. However, the fifht
order system of equations, which they derived in the ``diagonal
gauge'', contained also one spurious solution.  By using the
Regge-Wheeler gauge, Ipser and Price~\cite{Ipser:1991ip} succeeded to
determine a fourth order system of equations, which contains only
metric perturbations, and to clarify the origin of the additional
solution of the diagonal gauge, which can be eliminated by a
non-trivial gauge transformation.  These researches then suggested
that the fundamental information of the non-radial perturbations of
relativistic stars are contained in the dynamics of the spacetime, as
in the case for the black hole perturbations.
%
%
As a result, other authors investigated the time evolution of
non-radial oscillations with the same strategy, i.e. by setting up a
system of PDE for these two metric degrees of
freedom~\cite{Ruoff:2001ux, Seidel:1990xb}.

The GSGM polar perturbative equations that are reported in
section~\ref{sec:GSGM} have been determined following the same
method. The two independent perturbations are given by the metric
variables $\left\{ \chi^{(0,1)}, k^{(0,1)} \right\}$ that satisfy
the two coupled wave equations (\ref{chitt}) and~(\ref{ktt})
respectively, which are sketched here:
\begin{eqnarray}
- \chi^{(0,1)}_{, \, tt}  +  \chi^{(0,1)}_{, \, r^{\ast}r^{\ast}}   + ... & = &
0  \label{chi_sk}\\
- k^{(0,1)}_{, \, tt} +   \bar{c}_{s}^{\,2}  \,
k^{(0,1)}_{, \, r^{\ast}r^{\ast}} + ... & =  & 0 \label{k_sk}
\end{eqnarray}
The other four quantities can be then derived by using some of the
other seven Einstein and conservation 
equations~(\ref{psit})-(\ref{kpp}).

An analysis of the wave equation characteristics in the expressions
(\ref{chi_sk}) and~(\ref{k_sk}) allows us to interpret the physical
properties of these two quantities. The perturbation $\chi^{(0,1)}$
describes the gravitational degree of freedom that propagates
according to the wave equation in all the spacetime. The
interpretation of the variable $k^{(0,1)}$ is less clear as it is a
metric perturbations that propagates in the stellar interior at the
speed of sound.  In the exterior,  equation~(\ref{k_sk}) losses its
wave propagation character and another equation is required to evolve
the variable $k^{(0,1)}$, see reference~\cite{Ruoff:2001ux}. This
problem can be avoided by adopting outside the star the Zerilli
formulation.

The polar non-radial oscillations can be also studied with a different
set of independent variables. In the work of Allen et
al.~\cite{allen-1998-58} and Nagar et al.~\cite{Nagar:2004ns} for
instance, the enthalpy perturbations~$H^{(1,0)}$ has been evolved
together with the two independent metric perturbations in a system of
three partial differential equations. This system is well defined
inside and outside the star although the enthalpy perturbation
vanishes in the exterior. In reference \cite{Nagar:2004ns}, the
authors use the GSGM formalism to set up a system of two hyperbolic
and one elliptic equation for the three variables $\left\{
\chi^{(0,1)}, k^{(0,1)}, H^{(0,1)} \right\}$.  The two hyperbolic
equations are the gravitational wave equation (\ref{chitt}) for the
variable $\chi ^{(0,1)}$ and the sound speed equation for the enthalpy
$H^{(1,0)}$. The elliptic equation is the Hamiltonian constraint
(\ref{kpp}), which is used to update the metric variable
$k^{(0,1)}$. The choice of solving the Hamiltonian constraint for one
of the variables allowed them to have more control on the numerical
errors and stable long time evolutions.  \\
\indent In our work we have chosen to adopt this approach, as we
think that higher control of errors and stable evolutions are
crucial for having an accurate analysis of the non-linear coupling
between the radial and non-radial oscillations.  In fact as will
explain more in details in section~\ref{sec:NLP-Pol}, we argue
that with this system of equations we can reduce the numerical
errors also for the evolution of the second order perturbations.
In addition, longer simulation times will allows us to investigate
more accurately the interaction between the first order
perturbations.

\indent The three equations for the interior spacetime are then given by the following expressions: \\
\emph{Gravitational  wave equation:}
\begin{eqnarray}
- S^{(0,1)}_{,tt} & + & e^{2 (\Phi - \Lambda)} S^{(0,1)}_{,rr} +
e^{2 (\Phi - \Lambda)} \left[
 \left( 5 \Phi_{,r}-\Lambda_{,r}
\right) S^{(0,1)}_{,r} + \frac{4}{r}
 \left( {\frac{1-{e^{2 \Lambda}}}{{r}^2}}+\Phi_{,r}^2+
 \frac{\Lambda_{,r}}{r} \right)k^{(0,1)} \right. \nn \\
 {}  & + & \left. \frac{1}{r} \left( \Phi_{,r} \left( 5 +4 \Phi_{,r}  r \right)
+3 \Lambda_{,r}+{\frac{2- \left( l(l+1) + 2 \right)
e^{2\Lambda}}{r}} \right) S^{(0,1)} \right]  =  0 \, ,
\label{GW01}
\end{eqnarray}
\emph{sound  wave equation:}
\begin{eqnarray}
- H^{(0,1)}_{,tt}   & + & \bar{c}_s^2 e^{2 \left(\Phi -\Lambda
\right)} H^{(0,1)}_{,rr}   +  e^{2 \left(\Phi-\Lambda\right)}
\left\{\left[ \left(   \frac{2}{r} + 2\Phi_{,r} - \Lambda_{,r}
\right)
\bar{c}_s^2 - \Phi_{,r} \right] H^{(0,1)}_{,r}  \right. \nn \\
 & + &\left. \frac{1}{r} \left[ \left( 1 + 3\bar{c}_s^2\right)
 \left(\Lambda_{,r} + \Phi_{,r}  \right)-\bar{c}_s^2\frac{l(l+1)}{r}
 e^{2\Lambda} \right] H^{(0,1)} -  \frac{1-\bar{c}_s^2}{2}   \Phi_{,r}
\left[ \left(r S^{(0,1)}\right)_{,r} - k^{(0,1)}_{,r}\right] \right.  \nn \\
  & + & \left.
 \left[ -2 \Phi_{,r}^2 +\left[ \left(3\Phi_{,r} + \Lambda_{,r} \right) r
+ 1-  e^{2   \Lambda}  \right]  \frac{\bar{c}_s^2}{r^2} \right]
\left( r   S^{(0,1)} + k^{(0,1)} \right)  \right\} =  0 \,,
\label{SW01}
\end{eqnarray}
\emph{Hamiltonian constraint:}
\begin{eqnarray}
k^{(0,1)}_{,rr} & - & S^{(0,1)}_{,r}+ \left(\frac{2}{r} -
\Lambda_{,r} \right) k^{(0,1)}_{,r}+\frac{2}{r\bar{c}_s^2}
\left(\Lambda_{,r}+\Phi_{,r}\right) H^{(0,1)} + \frac{1}{r^2}
\left[ \left( 1- l(l+1)\right) e^{2\Lambda}
\right.   \nn \\
 && {} \left.
+ 2\Lambda_{,r}r-1 \right] k^{(0,1)}   
 -\frac{1}{2r} \left[ l(l+1) e^{2 \Lambda} + 4 - 4\Lambda_{,r} r
 \right] S^{(0,1)} = 0 \, . \label{Ham01}
\end{eqnarray}
The exterior spacetime is a perturbed Schwarzschild solution.  The
equations (\ref{GW01}) and~(\ref{Ham01}), where all the fluid
quantities present in the background coefficients and the enthalpy
perturbation vanish, remain well defined. On the other hand, the
sound wave (\ref{Ham01}) is not defined there and the system of
equations reduces to the two following equations:
\begin{eqnarray}
- S^{(0,1)}_{, \, tt}   +  e^{2(\Phi-\Lambda)} S^{(0,1)}_{, \,
rr} & + &
 e^{2  \Phi} \left[\frac{6M}{r^2} S^{(0,1)}_{,r} - \left[ \frac{2M}{r^3}
\left(1 - \frac{2M}{r} e^{2\Lambda} \right) + \frac{l(l+1)}{r^2}
\right] S^{(0,1)}  \right. \nn {} \\
{} & & - \left. \frac{4M}{r^4}\left(3 - \frac{M}{r}
e^{2\Lambda}\right) k^{(0,1)}  \right] = 0 \,, \\
\nn \\
e^{-2\Lambda} \left( k^{(0,1)}_{,rr} - S^{(0,1)}_{,r} \right) & + &
  \left( \frac{2}{r} - \frac{3M}{r^2} \right) k^{(0,1)}_{,r} -
  \frac{l(l+1)}{r^2} k^{(0,1)}
\nn \\
{} & & - \left(\frac{2}{r} -
  \frac{2M}{r^2} +\frac{l(l+1)}{2r}\right) S^{(0,1)} = 0 \,.
  \label{ham01_con}
 \end{eqnarray}
From the previous two equations we can deduce the existence of a
single degree of freedom for the exterior spacetime. In fact, the two
metric variables have to satisfy the Hamiltonian
constraint~(\ref{ham01_con}).
Zerilli showed in reference~\cite{Zerilli:1970fj} that the
propagation of the gravitational wave can be described by a single
wave equation, afterward known as the Zerilli equation:
\begin{equation}
 - Z_{,tt}^{(0,1)} + e^{2 \left( \Phi -\Lambda \right)}Z_{,rr}^{(0,1)}
+ \frac{M}{r^2}   e^{2   \Phi}   Z_{,r}^{(0,1)} - V_{Z} Z^{(0,1)} =0 \,.
\end{equation}
The Zerilli function, which represents the only degree of freedom of
the external spacetime, is a gauge-invariant quantity related as
follows to the GSGM metric variables $S^{(0,1)}, k^{(0,1)}$:
\begin{equation}
 Z^{(0,1)} = \frac{4 r^2 e^{-2\Lambda}}{ l \left(l+1\right) \left[(l+2)(l-1)r+6M \right]}
\left[r S^{(0,1)} +
 \frac{1}{2}\left(l\left(l+1\right) +\frac{2M}{r} \right)   e^{2\Lambda}k^{(0,1)}
 - r k_{,r}^{(0,1)}  \right] \,.
\end{equation}
The function $V_{Z}$ is the Zerilli potential:
\begin{equation}
V_{Z} = - \left(1-\frac{2M}{r}\right)\frac{n_{l}(n_{l}-2)^2 r^3
+6(n_{l}-2)^2 M r^2+ 36(n_{l} -2) M^2 r + 72M^3}{r^3[(n_{l}-2)r +
6M]^2} \,, \label{V_Zer}
\end{equation}
where $n_{l} = l(l+1)$.

The energy radiated at infinity in gravitational waves can be
determined though the Zerilli function \cite{Cunningham:1978cp}
with the following equation:
\begin{equation}
\frac{d E}{d t}^{(0,1)} = \frac{1}{64 \pi} \sum_{l\,, m} \,
\frac{\left( l + 2 \right)\, !}{\left(l-2\right)\, !} \,
|\dot{Z}_{lm}^{(0,1)}|^2\,,
\end{equation}
which is valid for $l\ge 2$. \\
\indent The boundary conditions for the polar non-radial
perturbations will be described in section~\ref{sec:BC01_ana} as a
particular case of the polar non-linear $\lambda \epsilon$
perturbations.

\section{Axial non-radial perturbations}
\label{sec:Lin_Axial_per}
The linear axial, non-radial perturbations of a non-rotating star
are described by the following metric perturbations:
\begin{equation}
ds^{2(0,1)} = h_A^{(0,1)} S_a ( dx^{A} dx ^{a} + dx^{a} dx ^{A} ) +
h^{(0,1)} \left( S_{a:b} + S_{a:b} \right) dx^{a} dx ^{b}
\label{metr01} \, ,
\end{equation}
and from the only fluid perturbations existent in the axial sector,
i.e. the velocity perturbation:
\begin{equation}
\delta u^{(0,1)}_{\alpha}=\left( 0 \,, \beta^{(0,1)
\,}S_{a}\right) \, ,
\end{equation}
where the scalar function $\beta^{(0,1)}$ is a gauge invariant
variable~(see section~\ref{ch:GINLP}), and in the previous two
equations we have used the axial basis of the tensor harmonics defined
in section~\ref{sec:GSGM_pert} \\
\indent In the GSGM formalism the dynamical information of the
spacetime is completely described by the gauge-invariant master
function $\Psi^{(0,1)}$ defined in equation~(\ref{Pidef}), that in
a static background is given by the following expression:
\begin{equation}
\Psi^{(0,1)}  =  \left[ r \left( k_{1 \, , \, t}^{(0,1)} - k_{0 \,
, \, r}^{(0,1)} \right) + 2 k_{0}^{(0,1)} \right] e^{-\left(\Phi +
\Lambda \right)}  \, ,  \label{Psi01def}
\end{equation}
while the perturbation $\beta^{(0,1)}$ accounts for the amount of
differential rotation present in the star~\cite{Thorne:1967th}.  The
system of perturbative equations can be determined by introducing the
static background quantities given in the expressions
(\ref{unst}),(\ref{static-scalars}) and~(\ref{static-frame-der}) into
equations~(\ref{maseq}) and~(\ref{traseq}):
\begin{eqnarray}
- \Psi_{, \, tt}^{(0,1)} & + & \Psi^{(0,1)} _{, \, r_*r_*} + \left[
 4 \pi \, \left( \bar p - \bar \rho  \right) + \frac{6 M}{r^3} -
 \frac{\llcf}{r^2} \right]  \, e^{2 \, \Phi } \, \Psi ^{(0,1)}  \nn \\
&& +
16 \pi \,  r \,  \left[  e^{-2 \La} \, \hat \beta_{, \, r }^{(0,1)}   +
 \left( 4 \pi \bar p \, r + \frac{M}{r^2} \right)  \hat \beta^{(0,1)} \right] e^{2\, \Phi+\La} =
0 \, ,
\label{Psi01maseq}   \\
\hat \beta^{(0,1)} _{, \, t} & = & 0 \label{traseq01} \, ,
\end{eqnarray}
where we have introduced the tortoise coordinate $dr_* \equiv e^{
\Phi - \La } dr $, in order to formally simplify equation
(\ref{Psi01maseq}). Furthermore, we have re-defined the axial
velocity $\beta^{(0,1)}$ with the following perturbation:
\begin{equation}
\hat \beta^{(0,1)} = \left( \bar \rho + \bar p \right) \, \beta^{(0,1)} \, .
\label{hatbeta_def}
\end{equation}
The introduction of this new function is motivated mainly by two
purposes: \emph{i)} the equation at linear and non-linear order are
simpler, and \emph{ii)} for a polytropic equation of state, $\hat
\beta^{(0,1)}$ vanishes on the surface of a static star. This property
will be very useful later in the numerical integration of non-linear
axial oscillations, where the axial perturbations appear in the source
terms.  The numerical integration seems cleaner and more reliable with
this new variable.  From equation~(\ref{traseq01}) emerges the
stationary character of the linear axial velocity perturbation $\hat
\beta^{(0,1)}$ and then of $\beta^{(0,1)}$, thus the amount of stellar
differential rotation is completely determined by the spatial profile
of the initial condition.  Furthermore, the time independence of $\hat
\beta^{(0,1)}$ allows us to divide the solution of the master
equation~(\ref{Psi01maseq}) in two parts: \emph{i)} the true dynamical
degree of freedom of the spacetime, namely the gravitational wave, and
\emph{ii)} a stationary solution which is related to the dragging of
the inertial frame due to the differential stellar rotation. Thus
mathematically the general solution is given by
\begin{equation}
\Psi^{(0,1)}  = \Psi^{(0,1)}_{hom} + \Psi^{(0,1)}_{p} \, ,  \label{sol_dec}
\end{equation}
where the propagation of the gravitational wave
$\Psi^{(0,1)}_{hom}$ is carried out by the homogeneous equation
associated with the PDE~(\ref{Psi01maseq}), which is:
\begin{equation}
- \Psi_{, \, tt}^{(0,1)}  +  \Psi^{(0,1)} _{, \, r_*r_*} + \left[
 4 \pi \, \left( \bar p - \bar \rho  \right) + \frac{6 M}{r^3} -
 \frac{\llcf}{r^2} \right]  \, e^{2 \, \Phi } \, \Psi ^{(0,1)}  = 0 \,  .
\label{Psi01_hom_eq}
\end{equation}
On the other hand, the stationary metric profile $\Psi^{(0,1)}_{p}$
can be determined by a particular time-independent solution of the
master equation. In this case, the system of equations is given by an
ordinary second order equation for $\Psi^{(0,1)}_{p}$ and a trivial
evolution equation $\beta^{(0,1)}$:
\begin{eqnarray}
\Psi^{(0,1)} _{p, \, rr} & + & \left[ 4 \pi \left( \bar p - \bar
\rho \right) \, r + \frac{2 M}{r^2} \right] e^{2\Lambda}\,
\Psi^{(0,1)} _{p, \, r} + \left[ 4 \pi \, \left( \bar p - \bar
\rho \right) + \frac{6 M}{r^3} - \frac{\llcf}{r^2} \right] \, e^{2
\,\Lambda}\, \Psi ^{(0,1)}_{p}  {} \nn     \\
&& {} + 16 \pi \, r \, \left[ e^{-2 \La} \, \hat \beta_{ , \,
r}^{(0,1)} + \left( 4 \pi \bar p \, r + \frac{M}{r^2} \right)
\hat \beta^{(0,1)} \right] e^{3\,\La} = 0 \, ,
\label{Psi01part}   \\
\hat \beta^{(0,1)} _{, \, t} & = & 0 \label{traseq01_B} \, ,
\end{eqnarray}
The particular solution $\Psi^{(0,1)}_{p}$ is related to the
component $k_0$ of the gauge invariant tensor $k_{A}$ through the
expression~(\ref{mtr_form}). The expression that connects this
metric variable with the frame dragging function
$\omega(r,\theta)$ is the following \cite{Thorne:1967th}:
\begin{equation}
-  r^2 \sin^2 \theta \, \omega(r,\theta) = \delta g_{t\phi}
^{(0,1)} = \sum_{lm} k_0 ^{lm} S_{\phi}^{lm}   \, .
\label{frmdr-rel}
\end{equation}
The gauge-invariant harmonic component $k_0 ^{lm}$ coincides with the
metric component $h_0 ^{lm}$ in the stationary configuration,
and more generally for time-dependent cases when
the Regge-Wheeler gauge is adopted.
Alternatively, the solution $k_0^{lm}$ and the related frame
dragging $\omega^{lm}$ can be determined directly by the following
ordinary differential equation:
\begin{equation}
e^{-2\La} k_{0\, , rr} ^{(0,1)}- 4 \pi \left(\rho + p\right) r
k_{0\, , r}^{(0,1)} + \left[ 8 \pi \left( \rho + p \right) +
\frac{4M}{r^3} - \frac{l \left( l+ 1 \right)}{r^2} \right] k_{0}
^{(0,1)} = 16 \pi
  e^{\Phi}  \hat{\beta} ^{(0,1)}  \, , \label{k0-fradr}
\end{equation}
which has been derived by introducing the
definition~(\ref{Psi01def}) into the expression~(\ref{mtr_form}).

The axial gravitational radiation in the exterior spacetime
produces small perturbations of the Schwarzschild solution.  The
propagation of these small ripples of spacetime is studied with
the solution of the Regge-Wheeler equation \cite{Regge:1957},
which is the equation one obtains by adapting the system of
equations~(\ref{Psi01maseq}) and~(\ref{traseq01}) to the exterior,
\begin{equation}
\label{RW01eq} \Psi^{(0,1)}_{, \, tt} - \Psi^{(0,1)}_{, \, r_*r_*
}+ V_l^{(\rm RW)} \Psi^{(0,1)} = 0 \, ,
\end{equation}
where the Regge-Wheeler potential is:
\begin{equation}
V^{(\rm RW)}_l =
\left(1-\frac{2M}{r}\right)\left(\frac{l(l+1)}{r^2}-\frac{6M}{r^3}\right)
\ ,
\end{equation}
and $r_*$ is the usual Regge-Wheeler tortoise coordinate $r_*\equiv
r+2M\ln[r/(2M)-1]$.

The boundary problem for the axial linear perturbations is complete
when we specify the boundary conditions at the origin, at the surface
and at infinity.  The requirement of regularity of perturbation fields
and equations at the origin gives
\beq \hat\beta ^{(0,1)} \sim r^{l+1}  \qquad \qquad
\Psi^{(0,1)}  \sim r^{l+1}  \, . \label{BC01_orig}\eeq
The junction conditions on the surface lead to the
continuity of metric variable $\Psi $, of its time derivatives,
and of the following expression \cite{Martin-Garcia:2000ze}:
\begin{equation}
 e^{-\La} \, \left( \Psi^{(0,1)} \, r^{-3} \right)_{,\,r} - 16 \pi \,
 r^{-2} \, \hat \beta^{(0,1)} \, . \label{SC01}
\end{equation}
In the case of a barotropic equation of state, the pressure and
mass-energy density vanish on the surface, and the condition
(\ref{SC01}) induces the continuity of $\Psi _{, \, r} ^{(0,1)}$.  At
infinity, we impose for equation~(\ref{Psi01_hom_eq}) the Sommerfeld
outgoing boundary condition in order to isolate the physical system
under consideration.  In addition, since at infinity the effects of
the dragging of the inertial frame disappear, we set for
equations~(\ref{Psi01part}) and~(\ref{k0-fradr}) a vanishing value for
$\Psi ^{(0,1)}_{p}$ and $k_{0}$ respectively.

For $l\ge 2$, The odd-parity master function $\Psi$  is related to the emitted
power in GWs at infinity as \cite{Nagar:2005ea}
\begin{equation}
\frac{d \, E}{d \, t} ^{(0,1)}= \frac{1}{16 \, \pi} \, \sum _{l,m} \,
\frac{ l \left( l + 1 \right) }{\left( l - 1 \right)\, \left( l + 2 \right)} \,
\left| \dot{{\Psi}} ^{(0,1)}_{lm}   \right|^{2}\label{gwpower} \ ,
\end{equation}
where we have explicitly restored the $(l,m)$ multipolar indices and
we have indicated with an overdot the derivative with respect to the
Schwarzschild coordinate time.

\chapter{Non-linear Oscillations of Compact Stars}
\label{ch:5_NL}

Non-linear oscillations of relativistic stars have recently
attracted  great interest. The achieved improvements in numerical
relativity now enable us to simulate non-linear evolutions of
various physical systems such as core
collapse~\cite{Dimmelmeier:2002bk, Dimmelmeier:2002bm}, non-linear
oscillations of differentially rotating
stars~\cite{Stergioulas:2003ep}, non-linear saturation of
$r$-modes~\cite{2003ApJ...591.1129A, 2001PhRvL..86.1152L,
Lindblom_mio, Stergioulas:2000vs}, non-linear radial
pulsations~\cite{Sperhake:2001xi}, etc.

In perturbation theory, the second perturbative order has been
developed for studying Schwarzschild black
holes~\cite{Gleiser:1995gx, Garat:2000gp}.  In particular, the
authors treated the polar and axial quadrupole perturbations and
applied this framework to the study of symmetric and asymmetric
collisions of two black holes. The perturbative results have been
compared with those obtained in numerical relativity, showing an
unforeseen accuracy even for non-linear regimes.

Perturbative techniques for studying non-linear oscillations have
been used also for studying linear radial and non-radial
perturbations of a slowly rotating star. In this case the rotating
configuration of the star is treated
perturbatively~\cite{Hartle:1967ha}, where the perturbative
parameter is given by the dimensionless ratio $\Omega /
\Omega_{K}$, where $\Omega$ is the uniform angular velocity
measured by an observer at infinity and $\Omega_{K}$ the Keplerian
angular velocity that determines the mass shedding limit.  Several
works have been carried out in the slow rotation approximation and
with numerical codes developed in numerical relativity for
determining the effects of the rotation on quasi-normal modes
(QNM).  The splitting of the non-axisymmetric QNM frequencies and
the presence of gravitational wave instabilities, i.e. the
Chandrasekhar-Friedman-Schutz
instabilities~\cite{1970PhRvL..24..762C, 1978ApJ...222..281F} are
two of the most relevant effects due to the rotation of the star
(see review~\cite{2003LRR.....6....3S}).

Another interesting property of the non-linear harmonics is that
their frequencies may come out as composition frequencies, i.e. as
linear combinations of the linear mode
frequencies~\cite{Stergioulas:2003ep, 2005MNRAS.356.1371Z,
1969mech.book.....L}. This aspect could be interesting if two
linear modes belonging to different families have nearly or
exactly the same frequencies. In this case the associated
non-linear harmonics could emerge at lower frequencies than the
respective linear modes and with an amplified signal due to
resonance effects. As long as they reach a high efficiency, these
simultaneous effects could produce non-linear gravitational
radiation in the sensitivity window of the new generation of Earth
based laser interferometers and mass resonant antennas.
In this chapter, we derive for the first time the perturbative
equations for studying the coupling between the radial and non-radial
perturbations of a non-rotating compact star, where the stellar matter
is described by a perfect and barotropic fluid.  In
section~\ref{sec:5.1_Cpl}, we describe the method used for getting the
non-linear perturbations, i.e.  the 2-parameter relativistic
perturbation theory on the GSGM gauge invariant formalism.  The
perturbations that describe the coupling between radial and polar and
axial non-radial perturbation are introduced in
sections~\ref{sec:NLP-Pol} and~\ref{sec:5.1.2AxNonL} respectively,
where we discuss also their boundary conditions.

The results obtained in these sections have been presented in a first
paper for the polar perturbations~\cite{Passamonti:2004je}, while the
axial sector is the subject of a second
paper~\cite{Passamonti:2005axial}.

\section{Coupling between radial and non-radial stellar oscillations}
\label{sec:5.1_Cpl}

The dynamics of the non-linear  oscillations that describe the
coupling between the radial and non-radial perturbations can be
studied with a system of equations similar to the linear
non-radial perturbations. In fact as explained in
section~\ref{sec:3.3-NLFW}, at second perturbative order the
equations preserve the differential operator of the first order
equations, and in addition
they have quadradic source terms made up of the first order
perturbations.  At this perturbative order, the first order
perturbations are already known by evolving the perturbative
equations introduced in the previous chapter, and then behave as
sources that drive the $\lambda \epsilon$ perturbations.

For the determination of the solutions of these inhomogeneous
partial differential equations the spherical symmetry of the
radial pulsations is very helpful. In fact as we see later, the
variables and equations of interest can be obtained by expanding
the gauge invariant formalism of GSGM with the 2-parameter
perturbation theory.  This approach allows us to split the
time-dependent and spherically symmetric spacetime of GSGM in a
static background and a radially pulsating spacetime. The
consequences of this splitting will then be
worked out also for the non-radial perturbations. \\

\indent The GSGM is valid for first order non-radial perturbations
on a general time-dependent and spherically symmetric spacetime
that we denote as follows:
\begin{equation}
M _{G} = \left( \mathcal{M}, \mathcal{T}_{G}
\right) \, , \label{M_GSGM}
\end{equation}
where $\mathcal{M}$ is a manifold and $\mathcal{T}_{G}$ is the
set of geometrical and physical tensor fields defined on this
spacetime. In this formalism, the perturbed spacetime is a
continuous one-parameter family of spacetimes diffeomorphic to the
physical one,
\begin{equation}
M _{\epsilon}  = \left( \mathcal{M}, \mathcal{T} _{\epsilon }
\right) \, ,
\end{equation}
where the strength of the non-radial perturbations is controlled by
the perturbative parameter $\epsilon$, and $\mathcal{T}_{\epsilon}$
are tensors that describe the geometrical and physical properties and
are defined on the physical spacetime. Any tensor field can then be
perturbatively expanded as follows:
\begin{equation}
\mathcal{T} = \mathcal{T}^{(0)} + \epsilon \mathcal{T}^{(1)} +
O(\epsilon ^2 )\, , \label{metGSGM}
\end{equation}
where $\mathcal{T}^{(0)}$ are tensor fields which describe the quantities
of the time-dependent and spherically symmetric spacetime, and  $\mathcal{T}^{(1)}$
their non-radial perturbations.

In our approach, the main point is the identification of the
time-dependent and spherically symmetric spacetime $M _{G}$ with a
radially pulsating spacetime. In particular, since we are interested
in pulsations of small amplitude around an equilibrium configuration
given by a non-rotating star, we can treat $M _{G}$ perturbatively.
As a result, the tensor fields associated with the $M _{G}$ spacetime
can be written in the following way:
\begin{equation}
\mathcal{T}^{(0)} =  \overline{\mathcal{T}}  + \lambda \,
\mathcal{T}^{(1,0)}  + O(\lambda ^2)\, . \label{bgsep_0}
\end{equation}
where $\overline{\mathcal{T}}$ and all the quantities with the upper
bar represent variables of the background that we consider in this
work, i.e.  a static star.  Consequently $\mathcal{T}^{(1,0)}$ denotes
the one-parameter radial perturbations of a static star,
where the strength of radial perturbations is controlled by the
perturbative parameter $\lambda$.

In accordance with the new interpretation of the GSGM background, we
can distinguish into the physical spacetime of non-radial
perturbations $M_{\epsilon}$ the quantities that are dependent and
independent on the radial perturbative parameter $\lambda$. The
independent part will be denoted with the symbol $\mathcal{T}^{(0,1)}$
and represents non-radial perturbations on a static stellar
background. The second part, which depends on the perturbative
parameter $\lambda$, describes the corrections to the non-radial
perturbations $\mathcal{T}^{(0,1)}$ due to the radial pulsations.  As
a result, it represents the non-linear perturbations which describe
the coupling between the linear radial and non-radial perturbations
that will be denoted with $\mathcal{T}^{(1,1)}$.  Any tensor
$\mathcal{T}^{(1)}$ can be then expanded as follows:
\begin{equation}
\mathcal{T}^{(1)} = \mathcal{T}^{(0,1)} + \lambda \,
\mathcal{T}^{(1,1)} + O(\lambda ^2)\, . \label{bgsep_1}
\end{equation}
Hence, when we introduce the expressions~(\ref{bgsep_0})
and~(\ref{bgsep_1}) into equation~(\ref{metGSGM}) we find the
following 2-parameter expansion:
\begin{equation}
\mathcal{T}=  \overline{\mathcal{T}}  + \lambda \,
\mathcal{T}^{(1,0)}  + \epsilon \mathcal{T}^{(0,1)}  + \lambda \epsilon \,
\mathcal{T}^{(1,1)}  + O(\lambda^2,\epsilon^2)\, .
\label{initialmetric_B}
\end{equation}
It is worthwhile to remark that according to the multi parameter perturbation
theory, the perturbative fields in the expression~(\ref{initialmetric_B})
are tensors defined on the background spacetime.

This strategy is very useful for saving calculations and setting
up a boundary initial-value problem for the description of these
non-linear perturbations. In fact, we can determine the non-linear
perturbative fields and the related systems of equations by
introducing into the GSGM objects the expansion in the second
parameter $\lambda$, as shown in equation~(\ref{bgsep_0}) and
(\ref{bgsep_1}). The desired quantities will then be selected by
virtue of their perturbative order $\lambda \epsilon$.  Therefore,
this approach is a shortcut that prevents us performing a
2-parameter expansion of the Einstein and conservation equations
directly from the static background quantities.

Another important aspect of this method is that we can identify the
gauge invariant variables for the metric and fluid non-linear coupling
perturbations. The method will be described in detail in
chapter~\ref{ch:GINLP}, we can give here only a qualitative analysis.
Let's assume that $\mathcal{T}^{(1)}$ of equation~(\ref{bgsep_1}) is a
GSGM gauge invariant quantity. Thus, we can immediately deduce that
also the variable $\mathcal{T}^{(0,1)}$ is gauge invariant. In fact,
it is a first order non-radial perturbation of a static background,
which can be treated as a subcase of the GSGM formalism. Let's now
focus our attention on the $\lambda \epsilon$ perturbation
$\mathcal{T}^{(1,1)}$. If the gauge for first order radial
perturbations is not fixed, $\mathcal{T}^{(1,1)}$ is not in general
gauge invariant at order $\lambda \epsilon$. On the other hand, for a
fixed radial perturbation gauge, equation~(\ref{bgsep_1}) may suggest
that the gauge invariance of $\mathcal{T}^{(1,1)}$ can be derived by
the gauge invariance of $\mathcal{T}^{(1)}$ and
$\mathcal{T}^{(0,1)}$. In chapter~\ref{ch:GINLP}, we will prove this
property in detail by studying the structure of $\mathcal{T}^{(1,1)}$,
which arises from the 2-parameter expansion of the GSGM formalism and
by using the $\lambda \epsilon$ gauge transformations for a fixed
radial perturbation gauge.

In the last part of this section, we will describe an alternative
procedure for studying the non-radial perturbations on a radially
pulsating star and will compare it with the 2-parameter perturbation
theory. 
In this alternative approach, the GSGM time dependent background can
be still treated perturbatively as in equation~(\ref{bgsep_0}),
whereas the quantity $\mathcal{T}^{(1)}$, i.e. a first order
non-radial perturbation of a radially pulsating background, is now
considered as an entire term without performing any further
perturbative expansion as in equation~(\ref{bgsep_1}).  The structure
of the Einstein equations is then:
 \begin{equation}
 \mb{L_{NR}} \left[\, \mathcal{T}^{(1)} \,\right] + \lambda
\mb{ E } \left[\, \mathcal{T}^{(1,0)} \otimes \mathcal{T}^{(1)}
\,\right]  + O \left( \lambda ^2 \right)  = 0 \,,
\label{kojima}
\end{equation}
where we have not explicitly written the non-radial perturbative
parameter $\epsilon$, and $\mathcal{T}$ represents both metric and
fluid variables. We can notice that a part of equation~(\ref{kojima})
is governed by $\mb{L_{NR}}$, which is the linear differential
operator~(\ref{LNRdef}) that acts on the first order non-radial
perturbations~(\ref{Taylor_NR1}), while $\mb{ E }$ is a linear
differential operator that describes the remaining part of
equation~(\ref{kojima}). In addition, it is worth noticing that the
error $O \left( \lambda ^2 \right)$ in equation~(\ref{kojima})
is given exclusively by the expansion of the time dependent
background~(\ref{bgsep_0}).

The perturbative equations~(\ref{kojima}) are now a system of
homogeneous partial differential equations, where $\mathcal{T}^{(1)}$
is the unknown and the radial pulsating part of the time dependent
background appears in the second term. Furthermore, it is worth
noticing that the radial perturbative parameter $\lambda$ explicitly
appears in the equation and controls the strength of the radial
pulsations. For $\lambda = 0$, equation~(\ref{kojima}) describes first
order non-radial perturbations on a static
background~(\ref{Taylor_NR1}). 
In order to understand whether this equation is equivalent to the
coupling equation~(\ref{Taylor_Cp2}), we introduce the
expansion~(\ref{bgsep_1}) in equation~(\ref{kojima}) obtaining:
 \begin{eqnarray}
 \mb{L_{NR}} \left[\, \mathcal{T}^{(0,1)} \,\right] & + & \lambda
 \left\{ \mb{L_{NR}} \left[\, \mathcal{T}^{(1,1)} \,\right] +
\mb{ E } \left[\,\mathcal{T}^{(1,0)} \otimes \mathcal{T}^{(0,1)} 
\,\right]  \right\}   \nn \\ 
& + & \lambda^2 \mb{ E }
\left[\, \mathcal{T} ^{(1,0)} \otimes \mathcal{T} ^{(1,1)} \,\right]  +
O  \left( \lambda ^2 \right) = 0 \,. \label{texp}
\end{eqnarray}
At first order in $\epsilon$ we have equation~(\ref{Taylor_NR1}):
\begin{equation}
 \mb{L_{NR}} \left[\, \mathcal{T}^{(0,1)} \,\right]= 0 \, , \label{firstterm}
\end{equation}
that describes the first order non-radial perturbation on a static
star. At $\lambda \epsilon$ order we get the perturbative
equations~(\ref{Taylor_Cp2}):
\begin{equation}
\mb{L_{NR}} \left[\, \mathcal{T}^{(1,1)} \,\right] + \mb{ E }
\left[\,\mathcal{T}^{(1,0)} \otimes \mathcal{T} ^{(0,1)} \,\right] = 0 \, ,
\label{secterm}
\end{equation}
which describe the coupling of first order radial and non-radial
perturbations on a static star.  In addition, a third part of higher
perturbative order $\lambda^2 \epsilon$ is implicitly contained in
equation~(\ref{kojima}), i.e.
\begin{equation}
\mb{ E } \left[\, \mathcal{T}^{(1,0)} \otimes \mathcal{T}^{(1,1)}
\,\right] = 0 \, .\label{thirdterm}
\end{equation}
If we neglect terms of order $\epsilon \lambda^2$ the solutions of
equations~(\ref{kojima}) and~(\ref{secterm}) are equivalent.  However,
we think that the 2-parameter perturbation theory can give cleaner
results. First, it directly gets rid of higher order information, as
for instance the extra terms of order $\lambda^2$ in
equation~(\ref{texp}). Even though this term is of order $\lambda^2
\epsilon$, it is formally contained in equation~(\ref{kojima}).
Second, it enables us to work with inhomogeneus partial differential
equations~(\ref{Taylor_Cp2}), where the linear differential operator
is defined on the static background. On the other hand, the
differential operator of the homogeneous equation~(\ref{kojima}) has a
part which is governed by radially oscillating quantities (see second
term of equation~(\ref{kojima})).  Therefore, we expect that the
simpler differential operator of equation~(\ref{secterm}) will produce
less numerical problems during the implementation of the code.

\subsection{GSGM formalism on a radially oscillating star}

In order to implement the GSGM formalism one has to specify all the
quantities (\ref{u_bck}), (\ref{n_bck}), (\ref{fr_der_bck})
and~(\ref{backg_scal}) that describe the time-dependent and
spherically symmetric spacetime $M_{G}$. In our approach this
spacetime is considered as a radially pulsating spacetime which is
treated perturbatively. Therefore, all the variables defined on it
must be expanded in a static and radial pulsating part as in the
equation~(\ref{bgsep_0}).

\indent The frame vector field basis $\left\{ u^{(0) \, A}, n^{(0)
\, A} \right\}$ of the submanifold $M_{G}$ assumes then the
following expressions:
\begin{eqnarray}
u^{(0) \, A} & = & \left( \left[ 1- \la  \left( \eta^{(1,0)} -
\frac{\chi^{(1,0)}}{2}
 \right)  \right]  e^{-\Phi} ,  \la \
e^{-\Lambda}   \ga^{(1,0)}  \right) \,,  \label{urdbg} \\
n^{(0) \, A} & = & \left( \la  e^{-\Phi}  \ga^{(1,0)} , \left(1-
\la \frac{\chi^{(1,0)}}{2} \,  \right) e^{-\Lambda}\right) \, .
\label{nrdbg}
\end{eqnarray}
These two  vectors satisfy the ortho-normalization up to
$\lambda^2$ perturbative order:
\begin{equation}
u^{(0) \, A} u^{(0)}_{A} = -1 + O(\lambda ^2) \, , \qquad  n^{(0) \, A}
n^{(0)}_{A} = 1 + O(\lambda ^2) \, , \qquad u^{(0) \, A} n^{(0)}_{A} =
 O(\lambda ^2) \, .
\end{equation}
With these definitions the metric and the completely antisymmetric
tensors of the spacetime $M_{G}$ take the following form:
\begin{equation}
g ^{(0)} _{A B} = - u^{(0)}_{A} u^{(0)}_{B} + n^{(0)}_{A}
n^{(0)}_{B} \, ,  \qquad \qquad  e^{(0)} _{A B} =  n^{(0)}_{A}
u^{(0)}_{B} - u^{(0)}_{A} n^{(0)}_{B} \, .
\end{equation}
The action of the frame derivatives on a scalar perturbation
$f^{(1)}=f^{(0,1)}+\lambda f^{(1,1)}$ defined on the radially
oscillating star  $M_{G}$ is defined by the following
expressions:
\begin{eqnarray}
\dot{f}^{(1)} & = & u^{(0) \, A} f^{(1)}_{, \, A} =  e^{-\Phi}
f^{(0,1)}_{, \, t} + \la \left\{e^{-\Phi} f^{(1,1)}_{, \, t} +
 e^{-\Lambda}  \ga^{(1,0)} f^{(0,1)}_{, \, r}   \qquad \qquad \right.  \nn \\
&& \left. -
e^{-\Phi}  \left( \eta^{(1,0)} -  \frac{\chi^{(1,0)}}{2} \right)
f^{(0,1)}_{, \, t} \, \right\}   + O(\lambda ^2)  \,,  \\
f^{(1)'} & = & n^{(0)
\, A} f^{(1)}_{, \, A}  = e^{-\Lambda} f^{(0,1)}_{, \, r} + \la
\left\{e^{-\Lambda} f^{(1,1)}_{, \,r} + e^{-\Phi} \ga^{(1,0)}
f^{(0,1)}_{, \, t}  \right. \nn \\
&& - \left. \frac{\chi^{(1,0)}}{2}
 \, e^{-\Lambda} f^{(0,1)}_{, \, r} \right\}  + O(\lambda ^2) \,.
\end{eqnarray}
Finally, we consider the remaining scalar functions defined in this
formalism, which are the components of the vector $v_{A} = r'/r$
with respect to the frame basis, i.e. $U,W$,  and the divergence of the
two basis vectors, i.e. $\mu, \nu$. The expansion of the manifold
$M_{G}$ leads to the following expressions:

\begin{eqnarray}
U ^{(0)}& = & u^{(0) \, A} v_A^{(0)} = \lambda \frac{e^{-\Lambda}}{r} \,   \ga ^{(1,0)}  + O(\lambda ^2) \,,   \label{Ubk} \\
W ^{(0)}& = & n^{(0) \, A} v_A^{(0)} = \left( \frac{1}{r}-\lambda
\frac{\chi^{(1,0)}}{2 \, r}
\right) \ e^{-\Lambda}  + O(\lambda ^2)\,,  \label{Wbk} \\
\mu ^{(0)} & = &  u^{(0) \, A}_{\, |A}  = \lambda \, \left(
\ga^{(1,0)} \, e^{-\Lambda}
\right)_{,\, r}  + O(\lambda ^2) \label{mubk} \,,\\
\nu ^{(0)}& = & n^{(0) \, A}_{~|A} = \Phi_{,r}  e^{-\Lambda} +
\lambda\,\left\{e^{-\Phi} \ga_{, \, t}^{(1,0)} +
e^{-\Lambda}\left[\left( \eta_{, \, r}^{(1,0)} -\frac{1}{2}\chi_{,
\, r}^{(1,0)}\right) -\frac{1}{2} \Phi_{, \, r}
\chi^{(1,0)} \right]\right\} + O(\lambda ^2) \, , \label{nubk} \nn \\
\end{eqnarray}
where in order to simplify the function (\ref{mubk}) we have used
the radial perturbative equation (\ref{chi_t}).

\section{Coupling of radial and polar non-radial perturbations}
\label{sec:NLP-Pol}
The non-linear perturbations and the perturbative equations at
order $\lambda \epsilon$ are determined with the 2-parameter
expansion of the GSGM perturbation fields.  The explicit
expressions of all the metric and fluid perturbations will be
given in chapter~\ref{ch:GINLP}, where we address the gauge
invariance issues. In this section, we focus only on the dynamical
variables formed by the two metric scalar fields $\chi^{(1,1)}$
and $k^{(1,1)}$ and the enthalpy $H^{(1,1)}$. In this way, we can
set up a system of perturbative equations with the same
differential operator as the first order
equations~(\ref{GW01})-(\ref{Ham01}).

In the GSGM formalism the even-parity metric perturbations are
described by the 2-rank symmetric tensor $k_{AB}^{(1)}$ and the scalar
$k^{(1)}$ that are respectively defined in equations~(\ref{pkab})
and~(\ref{pk}) on the radially oscillating spacetime $M_{G}$.  The
symmetric tensor $k_{AB}^{(1)}$ can be then decomposed with respect to
the vector frame $\left\{u^{(0)}_{A}, n_{A}^{(0)}\right\}$ of $M_{G}$,
and three gauge-invariant scalar functions $\eta^{(1)},\phi^{(1)}$ and
$\psi^{(1)}$ can be defined~(\ref{kABdec}).  The explicit expression
of the gauge-invariant tensor $k_{AB}^{(1,1)}$ components can be
determined by introducing the expansion of the basis
vectors~(\ref{urdbg}) and~(\ref{nrdbg}):
\begin{eqnarray}
\! \! \! \! \! \! \! \! \! \! \! \! \! \! \! \! k_{00}^{(1,1)} & = &
\left[ \chi^{(1,1)} + k^{(1,1)} + \left( 2 \, \eta ^{(1,0)} - \chi
^{(1,0)}\right) \left( \chi^{(0,1)} + k^{(0,1)} \right) + 2 \, \ga
^{(1,0)} \psi ^{(0,1)} \right] \, e ^{2 \, \Phi} ,
 \label{k00_11}   \\
\! \! \! \! \! \! \! \! \! \! \! \! \! \! \! \! k_{01}^{(1,1)}  &
= & - \left[ \psi^{(1,1)} + 2 \, \ga ^{(1,0)} \, \left(
\chi^{(0,1)} + k^{(0,1)}  \right) + \eta ^{(1,0)} \psi ^{(0,1)}
\right] \,e^{ \Phi + \Lambda }   \, , \\
\! \! \! \! \! \! \! \! \! \! \! \! \! \! \! \! k_{11}^{(1,1)} & =
& \left[ \chi^{(1,1)} + k^{(1,1)} + \chi^{(1,0)} \left(
\chi^{(0,1)} + k^{(0,1)}  \right) + 2 \, \ga ^{(1,0)} \psi
^{(0,1)}  \right] \, e ^{2 \, \Lambda} \, . \label{k11_11}
\end{eqnarray}
where we have used the definition (\ref{chidef}) for the $\chi^{(1)}$
perturbation
\begin{equation}
\chi^{(1)}=\phi^{(1)}-k^{(1)}+\eta^{(1)} \, \,
\end{equation}
and the Einstein's equation (\ref{eta}): $\left(\eta ^{(1)} = 0\right)$.

The other gauge invariant quantity $k^{(1)}$
assumes instead the following form at $\lambda \epsilon$ order:
\begin{eqnarray}
k^{(1,1)} & = & K^{(1,1)} -     \frac{e^{-2\,\Lambda}}{r}  \left(
  p_{1}^{(1,1)} -  \chi^{(1,0)} \, p_{1}^{(0,1)} \right) \, ,
\end{eqnarray}
where we have performed the further perturbative expansion in
$\lambda$ of the vector $p_{A}^{(1)}$,
\begin{equation}
p_{A}^{(1)} = p_{A}^{(0,1)} + \lambda p_{A}^{(1,1)} \, ,
\end{equation}
which according to its definition~(\ref{ppa}) is given by:
\begin{eqnarray}
p ^{(0,1)}_{A} & =  & h_{A}^{(0,1)} - \frac{r^2}{2} \,
G^{(0,1)}_{|A} \, ,
\\
p ^{(1,1)}_{A} & =  & h_{A}^{(1,1)} - \frac{r^2}{2} \,
G^{(1,1)}_{|A} \, .
\end{eqnarray}

\subsection{Perturbative equations for polar perturbations}
The explicit form of the perturbative equations that describe the
dynamical properties of the non-linear coupling can be derived
from equations~(\ref{chitt})-(\ref{kpp}) by introducing the
quantities~(\ref{urdbg})-(\ref{nubk}) of the radially oscillating
spacetime $M_{G}$ and the perturbative fields, which are further
decomposed as in equation (\ref{bgsep_1}). The desired equations
are given by the perturbative part associated with the
perturbative parameter $\lambda \epsilon$, and accordingly the
equation (\ref{CP}) will have the following structure for any
harmonic index $\left(l,m\right)$:
\begin{equation}
 \mb{L}_{NR} \left[\,\mb{\metb}^{(1,1)}_{lm}\,,\mb{\psi}_{A, \,
lm}^{(1,1)}\,\right] = \mathcal{S}\left[ \mb{
J}^{(1,0)}_{00}\otimes\mb{ F }^{(0,1)}_{lm} \right] \qquad \forall \,
l \geq 1
   \label{CPlm}
\end{equation}
This particular structure of the  $(1,1)$ equations  is quite
convenient in order to build a boundary initial-value problem and
solve it numerically by using time-domain methods.  The basic idea
is that given a numerical algorithm capable of evolving linear
non-radial perturbations, we can build an algorithm for our
$(1,1)$ perturbations by just adding source terms to the original
algorithm. The time evolution of non-radial perturbations of a
static star has been successfully analyzed by numerically
integrating different systems of perturbation
equations~\cite{allen-1998-58, Ruoff:2001ux, Nagar:2004ns}.
However, for the main features of our formulation, the scheme
introduced by Nagar et al.~\cite{Nagar:2004ns,Nagar:2004pr} seems
to be more appropriate for the implementation of a numerical code.
In their scheme the Hamiltonian constraint is not just an error
estimator for the evolution equations, as is usually done in many
free evolution schemes. Instead, it is part of the system of
equations and is solved at every time step for the perturbative
quantity $k$, equation~(\ref{pk}). This provides some control of
the errors induced by constraint violation.  As a consequence, the
resulting numerical code~\cite{Nagar:2004ns} is able to evolve
non-radial perturbations for long times and is capable of
estimating  the damping time and mode frequencies with an accuracy
comparable to frequency domain calculations.

Therefore, since the system of perturbative equations for the
linear and non-linear, non-radial perturbations have the same
differential structure we can expect to control the numerical
errors by solving the Hamiltonian constraint both at first and
second perturbative order. Otherwise, if we do not use this
hyperbolic-elliptic system of equations the errors accumulated
from constraint violation would be double that in a standard
computation of non-radial perturbations. Therefore, we expect that
this scheme allows us to obtain accurate long term evolutions.

In the stellar interior, we evolve a hyperbolic-elliptic system
of three partial differential equations for the three second order
variables:
\begin{equation}
\left\{ \chi^{(1,1)} , k^{(1,1)} , H^{(1,1)}  \right\} \, .
\end{equation}
The two metric variables $\chi^{(1,1)}$ and $k^{(1,1)}$ defined
above, obey respectively a wave-like equation that describes the
propagation of the gravitational radiation and the Hamiltonian
constraint. The Hamiltonian constraint is an elliptic equation and
is solved at any time-step to update the value of $k^{(1,1)}$. The
third variable is the second order fluid perturbations
$H^{(1,1)}$, whose expression can be derived by the perturbative
expansion of the analogous quantity on the radially oscillating
spacetime $H^{(1)}$:
\begin{equation}
 H^{(1)} \equiv \frac{c_s^{2 (0)} \rho^{(0)}}{\rho^{(0)} +
p^{(0)}}   \om^{(1)} \, . \label{En_11}
\end{equation}
This procedure leads to the following expression:
\begin{equation}
H^{(1,1)} = \frac{\bar{c}_s^2   \bar{\rho}}{\bar \rho + \bar p} \,
\omega^{(1,1)} + \left[ \bar c_s^2 + \bar \rho \left( \frac{d\bar
c_s^2}{d\bar\rho} - \left(1 + \bar{c}_s^2 \right)
\frac{\bar{c}_s^2}{\bar \rho + \bar p}\right)\right]
\frac{\bar{\rho}}{\bar \rho + \bar p}    \,  \omega^{(1,0)}
\omega^{(0,1)}  \, , \label{En_11_exp}
\end{equation}
where we have introduced the perturbative expansion
of the sound speed on the radially pulsating spacetime:
\begin{equation}
c_s^{2  {(0)}} = \bar{c}_s^2 + \la \,
\frac{d\bar{c}_s^2}{d\bar{\rho}}\,\delta\rho^{(1,0)}\,.
\end{equation}
In particular gauges, the Regge-Wheeler one for instance, the
gauge-invariant quantity $\om^{(1)}$ coincides with the gauge
dependent perturbation $\tilde{\om}^{(1)} $ [see
equation~(\ref{omgi})], and $H^{(1)}$ describes the enthalpy
perturbation,
\begin{equation}
H^{(1)} \equiv \frac{\delta p^{(1)}}{\rho^{(0)} + p^{(0)}} \, ,
\label{En_11_RW}
\end{equation}
where $\delta p^{(1)}$ is given by the definition (\ref{pr}).

\indent The other metric  ($\psi^{(1,1)}$) and fluid
($\ga^{(1,1)}$, $\alpha^{(1,1)}$) perturbations can be
successively derived by solving the perturbative equations
(\ref{psit}), (\ref{ktp}) and (\ref{psip}).

The wave equation for the variables $\chi^{(1,1)}$ and the Hamiltonian
constraint for $k^{(1,1)}$ are given by equations (\ref{chitt})
and (\ref{kpp}) respectively. On the other hand, the sound wave
equation for the perturbations $H^{(1,1)}$ must be determined by using
the perturbation of the conservation equations given in
\cite{Martin-Garcia:2000ze}. To this end, we prefer first to operate
within the GSGM framework and find the equation for the perturbation
$H^{(1)}$ on the radially pulsating spacetime $M_{G}$, and secondly
carry out the 2-parameter expansion and derive the equation for the
non-linear variable $H^{(1,1)}$. This equation is obtained as a linear
combination of the time frame derivative of equation (\ref{omegat})
and the spatial frame derivative of equation (\ref{gammat}). After
having introduced equations
(\ref{psit}),~(\ref{alphat}),~(\ref{gammat}),~(\ref{ktt}) and~
(\ref{kpp}) to reduce the number of perturbative unknowns and the
transformation (\ref{En_11}), we have the following wave equation
\begin{equation}
 -\ddot{H} ^{(1)} +  c_s^2 H^{(1) ''} +  \mathcal{F}_{H} = 0 \, ,
\label{En_eq_GG}
\end{equation}
where $\mathcal{F}_{H}$ contains all the remaining terms (with
derivatives of lower order). The complete equation  has been
written in Appendix \ref{AppSW11MG}. It is worthwhile to remark
that the wave equation (\ref{En_eq_GG}) is valid in the GSGM
framework for barotropic non-radial perturbations on a time
dependent background. In case of a static background, given the
introduction of the static quantities
(\ref{unst})-(\ref{static-frame-der}), it reduces to the equation
used in the literature (see references~\cite{allen-1998-58,
Ruoff:2001ux, Nagar:2004ns}).

We can now write the perturbative equations for the stellar
interior. We consider instead of the perturbative quantity
$\chi^{(1,1)}$, which diverges like $r$ as we approach spatial
infinity, the perturbation variable $S^{(1,1)} = \chi^{(1,1)}/r$
which of course is well behaved at infinity. This quantity
satisfies the following \emph{gravitational wave} equation:
\begin{eqnarray}
- S^{(1,1)}_{,tt} & + & e^{2 (\Phi - \Lambda)}
S^{(1,1)}_{,rr} + e^{2 (\Phi - \Lambda)} \left[
 \left( 5 \Phi_{,r}-\Lambda_{,r}
\right) S^{(1,1)}_{,r} + \frac{4}{r}
 \left( {\frac{1-{e^{2 \Lambda}}}{{r}^2}}+\Phi_{,r}^2+
 \frac{\Lambda_{,r}}{r} \right)k^{(1,1)} \right. \nn \\
 {}  & + & \left. \frac{1}{r} \left( \Phi_{,r} \left( 5 +4 \Phi_{,r}  r \right)
+3 \Lambda_{,r}+{\frac{2- \left( l(l+1) + 2 \right)
e^{2\Lambda}}{r}} \right) S^{(1,1)} \right]  =  e^{2  \Phi}{\cal
S}_{S}\, , \label{GW11}
\end{eqnarray}
where ${\cal S}_{S}$ denotes the source term for this wave
equation. In particular, the source term in the gravitational-wave
equation, ${\cal S}_S$, has the following form
\begin{eqnarray}
{\cal S}_{S} & = & a_1  S_{,rr}^{(0,1)} + a_2  S_{,r}^{(0,1)} +
a_3 S_{,t}^{(0,1)} + a_4  S^{(0,1)}  +  a_5  \left(
\psi_{,r}^{(0,1)} - 2 e^{\Lambda-\Phi} k_{,t}^{(0,1)} \right)  +
a_6  k^{(0,1)}   {} \nn \\
{} && + \, a_7 \psi^{(0,1)} \,, \label{Sgw11}
\end{eqnarray}
where the coefficients $a_i$ are just linear combinations of
radial perturbations with coefficients constructed from background
quantities.  Their explicit form is given in
Appendix~\ref{AppSources}.

The perturbative fluid variable $H^{(1,1)}$ also satisfies a wave equation,
but with a different propagation speed.  We call this equation the
\emph{sound  wave equation}.  It has the following form:
\begin{eqnarray}
- H^{(1,1)}_{,tt}   & + & \bar{c}_s^2 e^{2 \left(\Phi -\Lambda \right)}
H^{(1,1)}_{,rr}   +  e^{2 \left(\Phi-\Lambda\right)}
\left\{\left[ \left(   \frac{2}{r} + 2\Phi_{,r} - \Lambda_{,r} \right)
\bar{c}_s^2 - \Phi_{,r} \right] H^{(1,1)}_{,r}  \right. \nn \\
 & + &\left. \frac{1}{r} \left[ \left( 1 + 3\bar{c}_s^2\right)
 \left(\Lambda_{,r} + \Phi_{,r}  \right)-\bar{c}_s^2\frac{l(l+1)}{r}
 e^{2\Lambda} \right] H^{(1,1)} -  \frac{1-\bar{c}_s^2}{2}   \Phi_{,r}
\left[ \left(r S^{(1,1)}\right)_{,r} - k^{(1,1)}_{,r}\right] \right.  \nn \\
  & + & \left.
 \left[ -2 \Phi_{,r}^2 +\left[ \left(3\Phi_{,r} + \Lambda_{,r} \right) r
+ 1-  e^{2   \Lambda}  \right]  \frac{\bar{c}_s^2}{r^2}
\right]   \left( r   S^{(1,1)} + k^{(1,1)} \right)  \right\}
= e^{2 \Phi}   {\cal S}_{H}\,, \label{SW11}
\end{eqnarray}
and the source term can written as \beq {\cal S}_{H} & =  &   b_1
H^{(0,1)}_{,rr} + b_2  H^{(0,1)}_{,tr} + b_3 H^{(0,1)}_{,t} + b_4
H^{(0,1)}_{,r} + b_5   H^{(0,1)}  + b_6  k^{(0,1)}_{,t} + b_7   r
S^{(0,1)}_{,t} \nn \\
 {} & +  & b_8 \left[ k^{(0,1)}_{,r} - \left(r S^{(0,1)}  \right)_{,r}\right]
+  b_9  \left( r   S^{(0,1)} +  k^{(0,1)} \right) + b_{10}
\ga^{(0,1)}_{,r} + b_{11} \ga^{(0,1)} \nn \\
 {} & +  & b_{12} \psi^{(0,1)}_{,r} + b_{13}  \psi^{(0,1)} + b_{14}
\alpha^{(0,1)} \,, \label{Ssw11} \eeq where the coefficients $b_i$
have the same structure as the $a_i$ coefficients
in~(\ref{Sgw11}). Their explicit expressions can be found in
Appendix~\ref{AppSources}.

For the last perturbative variable, the metric perturbation
$k^{(1,1)}\,,$ we will use the \emph{Hamiltonian constraint}
instead of an evolution equation. After some calculations we get:
\begin{eqnarray}
k^{(1,1)}_{,rr} & - & S^{(1,1)}_{,r}+ \left(\frac{2}{r} -
\Lambda_{,r} \right) k^{(1,1)}_{,r}+\frac{2}{r\bar{c}_s^2}
\left(\Lambda_{,r}+\Phi_{,r}\right) H^{(1,1)}
 -  \frac{1}{2r} \left[ l(l+1) e^{2 \Lambda} + 4 - 4\Lambda_{,r} r
 \right] S^{(1,1)} \nn \\
 & + & {} \frac{1}{r^2} \left[ \left( 1- l(l+1)\right) e^{2\Lambda}
+ 2\Lambda_{,r}r-1 \right] k^{(1,1)}    = {\cal S}_{Hamil} \,,
\label{Ham11}
\end{eqnarray}
where ${\cal S}_{Hamil}$ is the source term for the Hamiltonian
constraint. As in the previous equations the precise form of ${\cal S}_{Hamil}$
is: \beq {\cal S}_{Hamil} & = &    c_1   \left( k^{(0,1)}_{,rr} -
S^{(0,1)}_{,r} \right) + c_2 k^{(0,1)}_{,r} + c_3 k^{(0,1)}_{,t} +
c_4     S^{(0,1)} + c_5    k^{(0,1)}
+ c_6 H^{(0,1)} + c_7 \psi^{(0,1)}_{,r}     \nn \\
  & + & {} c_8   \psi^{(0,1)} + c_9   \ga^{(0,1)}\,. \label{Sham11}
\eeq The coefficients $c_{i}$, in the same way as the coefficients
$a_i$ and $b_i$ only contain radial perturbations $g^{(1,0)}$ and
quantities associated with the static background. They are also
given in Appendix~\ref{AppSources}. It is worth remarking that the
polar non-radial perturbation equations on a static background are
obtained from equations (\ref{GW11}),~(\ref{SW11})
and~(\ref{Ham11}) by discarding the source terms and replacing all
the $(1,1)$ perturbations with the corresponding non-radial
$(0,1)$ terms. The sources are determined from first order
perturbations. The radial perturbations from equations
(\ref{eq:H10_ev})-(\ref{eq:eta_cn}), and the non-radial
perturbations (described by the quantities $S^{(0,1)}\,,$
$k^{(0,1)}\,,$ and $H^{(0,1)}$) from the first order analogous of
the above system (see reference~\cite{Nagar:2004ns}), and the
equations (\ref{psit}),~(\ref{ktp}) and~(\ref{psip}) adapted to a
static background to get $\psi^{(0,1)}$, $\ga^{(0,1)}$  and
$\alpha^{(0,1)}$.

The \emph{stellar exterior} is  described by a Schwarzschild
spacetime on which gravitational waves carry away some energy of
the stellar oscillations.   All fluid perturbations vanish
outside the star and the radial perturbations do the same
because of Birkhoff's theorem.   Therefore,  the source terms in
our perturbation equations disappear. Only the metric perturbations
survive, and they satisfy the gravitational wave equation
(\ref{GW11}) and the Hamiltonian constraint (\ref{Ham11}), which
take the following form:
\begin{eqnarray}
- S^{(1,1)}_{,tt}\! +  e^{2(\Phi-\Lambda)} S^{(1,1)}_{,rr} & + &
 e^{2  \Phi} \left[\frac{6M}{r^2} S^{(1,1)}_{,r} - \left[ \frac{2M}{r^3}
\left(1 - \frac{2M}{r} e^{2\Lambda} \right) + \frac{l(l+1)}{r^2}
\right] S^{(1,1)} \right. \nn {} \\
 & - &  \left. \frac{4M}{r^4}\left(3 - \frac{M}{r}
e^{2\Lambda}\right) k^{(1,1)}  \right] = 0 \,, \\
 e^{-2\Lambda}   \left( k^{(1,1)}_{,rr} -  S^{(1,1)}_{,r} \right)
 & + & \left( \frac{2}{r}  -  \frac{3M}{r^2} \right) k^{(1,1)}_{,r}
 -
 \frac{l(l+1)}{r^2} k^{(1,1)} - \left(\frac{2}{r} - \frac{2M}{r^2}
 \right. \nn \\
 & + & \left. \frac{l(l+1)}{2r}\right) S^{(1,1)} = 0 \,.
\end{eqnarray}
It is worth mentioning that the above equations  coincide with the
equations for non-radial perturbations of a static stellar
background outside the star, as expected.

On the other hand, Zerilli showed that the even-parity perturbations
of a Schwarzschild background have just one degree of freedom, and
therefore can be described by just one variable, the Zerilli function, satisfying
a wave equation. At order $(1,1)$ the  Zerilli function can be built from the
two metric perturbations $S^{(1,1)}$ and $k^{(1,1)}$ and their derivatives, as
at first order~\cite{Moncrief:1974vm, Moncrief:1974vmII}, and is given by
\begin{equation}
 Z^{(1,1)} = \frac{4 r^2 e^{-2\Lambda}}{l\left(l+1\right) \left[ (l+2)(l-1)r+6M \right]}
 \left[r S^{(1,1)} +
 \frac{1}{2}\left(l\left(l+1\right) +\frac{2M}{r} \right)   e^{2\Lambda} k^{(1,1)}
 - r k_{,r}^{(1,1)}  \right] \,.
\end{equation}
It satisfies the Zerilli equation~\cite{Zerilli:1970la,Zerilli:1970fj}
\begin{equation}
 - Z_{,tt}^{(1,1)} + e^{2 \left( \Phi -\Lambda \right)}Z_{,rr}^{(1,1)}
+ \frac{M}{r^2}   e^{2   \Phi}   Z_{,r}^{(1,1)} - V_Z Z^{(1,1)} =0\,,
\end{equation}
where $V_Z$ is the Zerilli potential(~\ref{V_Zer}). 

Finally, we can determine the power of the gravitational radiation
emission at infinity by using the following
expression~\cite{Cunningham:1978cp}
\begin{equation}
\frac{d E}{d t}^{(1,1)} = \frac{1}{64 \pi} \sum_{l\,, m} \,
\frac{\left( l + 2 \right)\, !}{\left(l-2\right)\, !} \,
|\dot{Z}_{lm}^{(1,1)}|^2\, ,
\end{equation}
for $l\ge 2$.

\subsection{Boundary conditions for polar perturbations}
\label{sec:BC01_ana}

In this Section we discuss the boundary conditions at the origin,
infinity and at the stellar surface for the $\lambda \epsilon$
perturbations describing the coupling of radial and polar
non-radial modes.  With regard to the outer boundary, we locate it
far enough from the star and we impose the well-known Sommerfeld
outgoing boundary conditions on our perturbative fields.

At the origin, the boundary conditions are just regularity conditions
on the perturbative fields, which can be obtained by a careful
analysis of the equations that they satisfy.  The analysis of Taylor
expansions of the differential operators that appear in our equations
near the origin leads to the following behaviour for the non-radial
perturbations $\mathcal{T}^{(1)}$~\cite{Gundlach:1999bt}:
\begin{eqnarray}
l & \ge & 0 \, , \quad S^{(1)} \sim \, r^{l+1} \quad k^{(1)} \sim \,
r^l \quad \psi^{(1)} \sim \, r^{l+1} \,, \label{01_or_condA}\\
l & \ge & 1 \, ,  \quad \ga^{(1)} \sim  \, r^{l-1} \quad H^{(1)}
\sim  \, r^l \, \quad  \alpha \sim \, r^l \,,  \label{01_or_condB}
\end{eqnarray}
where the upper index $(1)$ in the previous quantities denotes the
linear $\epsilon$ and non-linear $\lambda \epsilon$ perturbations as
in equation~(\ref{bgsep_1}).  The
conditions~(\ref{01_or_condA})-(\ref{01_or_condB}) and the
expressions~(\ref{bc_rad_in})-(\ref{bc_rad_ori}) for the radial
perturbations leads also to the regularity of the source terms.

The stellar surface in spherically symmetric spacetime is a
1-dimensional manifold embedded in the 2-dimensional manifold $M^2
\subset \mathcal{M}$~(section~\ref{sec:GSGM_Time_Bck}). In order to
prevent $\delta$ discontinuities on the energy momentum tensor the two
fundamental forms, i.e. the induced metric tensor and the extrinsic
curvature, have to be continuous on this hypersurface. In particular,
when the perturbative approach is used, these continuity conditions
have to be imposed at any perturbative order considered.  A boundary
condition must be imposed at the surface also for the matter variable
$H$, which vanishes outside the star.

\noindent Let $\bar\Sigma$ be the surface of the static unperturbed star
(i.e. $r=R_s$).  The surface of the perturbed star can then be defined as
\begin{equation}
\Sigma\equiv\left\{ x(t) = x+\lambda\xi^{(1,0)}+\epsilon\xi^{(0,1)}
+\lambda\epsilon\xi^{(1,1)}\,:\,x\in\bar\Sigma\right\} \,, \label{Srf_per}
\end{equation}
where $\xi^{(i\,, j)}$ is a vector field that denotes the
Lagrangian displacement of a fluid element due to the action of
perturbations of order $(i\,, j)$. A physical requirement that
follows from matching conditions is the vanishing of the
unperturbed pressure $\bar{p}$ at the unperturbed surface
$\bar\Sigma$.  In the same way, the corresponding boundary
condition for the perturbed spacetime is the vanishing of the
total pressure $p = \bar{p} +\lambda \delta p^{(1,0)}+\epsilon\delta
p^{(0,1)}+\lambda\epsilon\delta p^{(1,1)}$ at the perturbed
surface $\Sigma$.  This condition turns out to be equivalent to
the vanishing of the Lagrangian pressure perturbations on
$\bar\Sigma$, the unperturbed surface, at every order.  The
Lagrangian pressure perturbations are given by: \beq \Delta \,
p^{(1,0)}  &=& \delta \, p^{(1,0)}  + \pounds_{\xi_{(1,0)}} \,
\bar{p} \,,
\label{LPp10} \\
\Delta \, p^{(0,1)}  &=& \delta \, p^{(0,1)}  + \pounds_{\xi_{(0,1)}} \, \bar{p} \,,
\label{LPp01} \\
\Delta \, p^{(1,1)}  &=&
\delta \, p^{(1,1)}  + \left(\pounds_{\xi_{(1,1)}} +\frac{1}{2}
\left\{\pounds_{\xi_{(1,0)}}\,,\,\pounds_{\xi_{(0,1)}}
\right\} \right) \bar{p} + \pounds_{\xi_{(0,1)}}\delta p^{(1,0)}
+ \pounds_{\xi_{(1,0)}}\delta p^{(0,1)} \nn\\
&=& \delta \, p^{(1,1)}  + \left( \pounds_{\xi_{(1,1)}}
-\frac{1}{2}
\left\{\pounds_{\xi_{(1,0)}}\,,\,\pounds_{\xi_{(0,1)}} \right\}
\right) \bar{p}\,, \label{p11atsf}
\eeq
where $\delta$ and $\Delta$ denote the Eulerian and Lagrangian
perturbations respectively, and we have used the lower order
boundary conditions $\Delta p^{(1,0)}=\Delta p^{(0,1)}=0$ in order
to simplify the condition (\ref{p11atsf}).

We can then conclude that the boundary conditions for the fluid
perturbations are described by the set of expressions given
in~(\ref{LPp10})-(\ref{p11atsf}).  However, in practice, in many
applications of first order perturbation theory, {\em dynamical
boundary conditions} either for density or enthalpy perturbations
have been considered.  This alternative boundary condition follow
from the analysis of the time derivative of the condition
(\ref{LPp01}) (see~\cite{allen-1998-58} for more details).  In our
current development of the numerical implementation of the
perturbative equations we are considering both types of boundary
conditions with the perspective of analyzing which type works best
for our formulation.

Finally, the junction conditions for the metric perturbations can be
determined by imposing continuity of first and second fundamental
differential forms and their perturbations at the
surface~\cite{Darmois:1927gd, Lichnerowicz:1971al, Obrien:1952bs,
  Israel:1966nc}.  The explicit form of these conditions has been
presented in~\cite{Gerlach:1979ih, Martin-Garcia:2000ze} for first
order perturbations of a time-dependent stellar background.
According to our interpretation of the GSGM formalism~(see
section~\ref{sec:5.1_Cpl}) the time dependent and spherical
spacetime is identified with a radially pulsating spacetime, which
is treated perturbatively. Therefore, in order to determine the
correct expressions of the junction conditions for linear
$\epsilon$ and non-linear $\lambda \epsilon$ perturbations on the
static surface we have to perform an expansion of the GSGM
matching relations~\cite{Martin-Garcia:2000ze} similar to that
describe in equations~(\ref{LPp01})-(\ref{p11atsf}) for the
pressure.  However, since we are carrying out the analysis in an
Eulerian gauge, the implementation of the junction conditions for
the non-linear perturbations requires some
approximations~\cite{Sperhake:2001si, Sperhake:2001xi}. In fact,
due to the movement of the stellar surface some perturbations can
take unphysical values near the surface during the contraction
phases of the star. This is the case for instance of the total
density $\rho$, which can become negative. Furthermore, the low
densities which are present at the outermost layers of the star
can produce also some numerical errors in the
simulations~\cite{2003PhRvD..68b4002H}. These problems can be
avoided by imposing the matching conditions for the non-linear
perturbations not on the static surface, as we do for the linear
perturbations, but slightly
inside~\cite{Sperhake:2001si,2003PhRvD..68b4002H}. This
approximation corresponds to neglecting less than one percent of
the stellar mass, which does not produce significant changes in
the wave forms and
spectra of the gravitational signal.  \\
\indent For first order polar non-radial perturbations on a static
star, the junction conditions can be determined at the static
surface $r=R_s$ as a subcase of the expressions given in
reference~\cite{Martin-Garcia:2000ze}. These relations provide the
continuity of the following $l\geq 2$ scalar fields:
\begin{equation}
k^{(0,1)} \, , \qquad  S^{(0,1)} \, ,\qquad \psi^{(0,1)} \, ,\qquad   k^{(0,1)~'} \, ,\qquad
S^{(0,1)~'} \, , \label{BC01}
\end{equation}
where we have also used the vanishing of the static pressure, density
and speed of sound at the stellar surface.  \\
\indent At second perturbative order, the non-linear perturbations
are matched on the following hypersurface that during the
evolution is always inside the star,
\begin{equation}
\Sigma_{jc} \equiv \left\{ r = R_{jc} : \quad R_{jc} < x(t) \right\}  \, , \label{hyp_sigjc}
\end{equation}
where $x(t)$ describes the position of the perturbed surface as
defined in equation~(\ref{Srf_per}).  Therefore, we can perform a
2-parameter expansion of the GSGM junction
conditions~\cite{Martin-Garcia:2000ze} at $\Sigma_{jc}$ as well as we
have done for the pressure in
equations~(\ref{LPp01})-(\ref{p11atsf}). By using the continuity
conditions for the first order perturbations on the surface
$\Sigma_{jc}$, we obtain the following continuity conditions at
$\lambda \epsilon$ non-linear order:
\begin{equation}
k^{(1,1)} \, , \qquad  S^{(1,1)} \, ,\qquad \psi^{(1,1)} \, ,\qquad
e^{(- \Lambda)} k^{(1,1)}_{\, ,r} + \frac{8 \pi \bar \rho}{\Phi_{, \, r}} e^{\Lambda} H^{(1,1)} \, ,\qquad
S^{(1,1)}_{, \, r} \, . \label{BC11}
\end{equation}

Alternatively, one may use the ``extraction
formulas"~\cite{Martin-Garcia:2000ze} that relate the Zerilli function
with metric perturbations at the stellar boundary.  For linear polar
perturbations on a static star this relation reads for  $l \geq 2$:
\begin{equation}
Z^{(0,1)} = r k^{(0,1)} + \frac{2 r^4}{\left( l + 2 \right) \left( l
-1 \right) r + 6m } \left[ \frac{e^{-2\Lambda}}{r^2} \left( r
S^{(0,1)} + k^{(0,1)} \right) - \frac{e^{-2 \Lambda}}{r} k_{, \,
r}^{(0,1)} \right] \, .
\end{equation}
The extraction formula for the non-linear $\lambda \epsilon$
perturbations can be imposed on the hypersurface $\Sigma_{jc}$. By
using the first order continuity conditions we obtain:
\begin{eqnarray}
Z^{(1,1)} & = & r k^{(1,1)} + \frac{2 r^4}{\left( l + 2 \right) \left( l
  -1 \right) r + 6m } \left[ \frac{e^{-2\Lambda}}{r^2} \left( r
  S^{(1,1)} + k^{(1,1)} \right) - \frac{e^{-2 \Lambda}}{r} \left( k_{,
    \, r}^{(1,1)} \right. \right.  \nn \\
    && \left. \left.  + \frac{8
    \pi \bar \rho}{\Phi_{, \, r}} e^{\Lambda} H^{(1,1)} \right)
    \right] \, .
\end{eqnarray}

\section{Coupling of radial and axial non-radial perturbations}
\label{sec:5.1.2AxNonL}

In this section we derive the perturbative equations that govern the
non-linear coupling between first order radial and axial non-radial
oscillations.  Following the same method of the polar case in
section~\ref{sec:NLP-Pol}, we perform a 2-parameter expansion of the
perturbation fields in the odd-parity perturbative equations
(\ref{maseq}) and (\ref{traseq}).

The odd-parity metric perturbation of order $(1,1)$ can be expanded in
tensor spherical harmonics as
\begin{equation}
ds^{2(1,1)} = h_A^{(1,1)} S_a ( dx^{A} dx ^{a} + dx^{a} dx ^{A} ) +
h^{(1,1)} \left( S_{a:b} + S_{a:b} \right) dx^{a} dx ^{b}
\label{metr11} \, .
\end{equation}
while the axial velocity perturbation of the fluid is given by
\begin{equation}
\delta u^{(1,1)}_{\alpha}=\left( 0 \,, \beta^{(1,1) \,}S_{a} \right)
\, .
\end{equation}
The information of the dynamical properties of the spacetime is
completely contained even at this perturbative order in a scalar
function~$\Psi^{(1,1)}$. This axial master function can be
determined by expanding equation~(\ref{Pidef}) as we showed
for the metric in equation~(\ref{bgsep_1}), leading to the
following expression at $\lambda \epsilon$ order:
\begin{equation}
\Psi^{(1,1)}  =  \left[ r \left( k_{1 \, ,t}^{(1,1)} -
k_{0 \, ,r}^{(1,1)} \right) + 2 k_{0}^{(1,1)} \right] e^{-\left(\Phi +
\Lambda \right)} -  \eta^{(1,0)} \Psi^{(0,1)}   \label{Psi11_mast} \, .
\end{equation}
In the previous equation the $\lambda \epsilon$ perturbative
components of the gauge invariant vector $k_{A}^{(1)}$ appear, see
equation~(\ref{ka_11_exp}).

The system of perturbative equations is formed by the master wave
equation~(\ref{maseq}) and by the conservation 
equation~(\ref{traseq}), which assumes the following form after a
2-parameter expansion:
\begin{eqnarray}
  - \Psi^{(1,1)} _{, \, tt} & + & \Psi^{(1,1)}_{, \, r^{\ast}r^{\ast}}
+ \left[ \pmr + \frac{6 m}{r^3} - \frac{\llcf}{r^2} \right] e ^{2
\Phi}\Psi^{(1,1)} {} \nn \\
{} & + & 16 \pi \, r \, \left[ e^{-2 \La}
\, \hat \beta_{, \, r }^{(1,1)} + \left( 4 \pi \bar p \, r +
\frac{M}{r^2} \right) \hat \beta^{(1,1)} \right] e^{2\, \Phi+\La} = e
^{2 \Phi} \, \Sigma_{\Psi} \, , \label{Psi11maseq} \\
\hat \beta ^{(1,1)}_{, \, t}  & = & e^{ \Phi} \, \Sigma_{\beta}  \, ,
\label{traseq11}
\end{eqnarray}
where the explicit expressions of the source terms~$\Sigma_{\Psi}$
and~$\Sigma_{\beta}$ are given in Appendix~\ref{AppSources}.

In the previous perturbative equations, the axial velocity
perturbation~$\beta ^{(1,1)}$ has been replaced, as in the case of
the first order perturbations, with the variable~$\hat \beta
^{(1,1)}$. In this way we have a uniform set of perturbative
variables for any perturbative order. In order to get its explicit
expression we can first define this variable  for a generic
time-dependent spacetime:
\begin{equation}
\hat \beta^{(1)} = \left( \rho^{(0)} + p^{(0)}   \right) \, \beta^{(1)} \, ,
\end{equation}
and then introduce the perturbative expansion in two parameters.  At
order $\epsilon$ the definition (\ref{hatbeta_def}) valid for linear
axial velocity is restored, while at non-linear order, i.e. $\lambda
\epsilon$, the following expression is determined:
\begin{equation}
\hat \beta^{(1,1)} = \left( \bar \rho + \bar p  \right) \, \left(  \beta^{(1,1)} +
\frac{ 1 + \bar c_s^2 }{\bar c_s^2} \, H^{(1,0)} \, \beta^{(0,1)} \right) \label{hatbet11}
\end{equation}

It is worth noticing that due to Birkhoff's Theorem, the source terms
are present only in the stellar interior. Hence, the exterior is a
Schwarzschild space-time perturbed by the gravitational waves, which
can still be described by the Regge-Wheeler equation of $\lambda \epsilon$ order:
\begin{equation}
\Psi^{(1,1)}_{, \, tt} - \Psi^{(1,1)}_{, \, r_*r_*} + V_l^{(\rm o)}\Psi^{(1,1)} = 0
\label{RW11eq} \, .  \\
\end{equation}
Consequently, the emitted power in GWs at infinity reads:
\begin{equation}
\frac{d \, E}{d \, t} ^{(1,1)}= \frac{1}{16 \, \pi} \, \sum _{l,m} \,
\frac{ l \left( l + 1 \right) }{\left( l - 1 \right)\, \left( l + 2 \right)} \,
\left| \dot{\Psi} ^{(1,1)}_{lm}   \right|^{2} \, ,\nn  \label{Power11ax}
\end{equation}
where we have explicitly written the multipolar indices $(l,m)$ with
$l\ge 2$ and we have indicated with an overdot the derivative with
respect to Schwarzschild coordinate time.

\subsection{Boundary conditions for axial perturbations}
\label{sec:BC11_axil}
We still need to impose the correct boundary conditions to
Eqs.~(\ref{Psi11maseq}) and (\ref{traseq11}) at the origin,
at the stellar surface and at infinity. At the origin a regularity
analysis shows that
\begin{eqnarray}
\hat \beta^{(1,1)} \sim r^{l+1}
\qquad \qquad \Psi^{(1,1)} \sim r^{l+1} \, .
\label{BC11_orig}\
\end{eqnarray}
The junction conditions at the stellar surface for the non-linear
axial perturbations can be determined by using the GSGM equations for
a time dependent and spherically symmetric
spacetime~\cite{Gerlach:1979ze, Martin-Garcia:2000ze}, as we have
already illustrated for the polar case in section~\ref{sec:BC01_ana}.
In the axial case the movement of the surface is due uniquely to the
radial perturbations, as the axial perturbations at $\epsilon$ and
$\lambda \epsilon$ perturbative orders can only produce a fluid motion
in the tangential direction.  The continuity of the induced metric,
extrinsic curvature and their perturbations leads to the following
results: \emph{i)} continuity of~$\psi^{(1)}$ and its time
derivatives~$\left( \psi^{(1)} \right)\dot{}$~, and \emph{ii)}
continuity at the stellar surface of the following expression:
\begin{equation}
\left(r^{-3} \, \Psi^{(1)} \right) ' + 16 \pi r^{-2} \hat \beta^{(1)} \, . \label{psi11_r_bc}
\end{equation}
According to our interpretation of the GSGM formalism~(see
section~\ref{sec:5.1_Cpl}),
the correct expressions for the junction conditions on a hypersurface
$\Sigma_{jc}$ placed inside the star can be determined with the
procedure used in section~\ref{sec:BC01_ana}.  At first perturbative
order in $\epsilon$, we get the same conditions for the linear axial
perturbations expressed in section~\ref{sec:Lin_Axial_per}. For the
non-linear coupling, the continuity of the master non-linear
function~$\Psi^{(1,1)}$ and its time derivatives must be imposed at
$\Sigma_{jc}$. In addition, the condition~(\ref{psi11_r_bc}) leads to
the continuity of the following quantity
\begin{equation}
e^{-\Lambda} \Psi^{(1,1)}_{, \, r} + 16 \pi \, r \, \hat \beta ^{(1,1)} \, ,
\label{Sfbc11}
\end{equation}
where we have used the continuity of $\Psi^{(1,1)}$ and the linear
junction conditions.  
At infinity, we impose the outgoing Sommerfeld boundary condition.

\chapter{Gauge Invariance of Non-linear Perturbations}
\label{ch:GINLP}

This chapter is dedicated to the gauge invariance of non-linear
perturbations of relativistic stars. The discussion is specialized to
a particular class of second order perturbations, namely those that
describe the coupling between the radial and non-radial oscillations.
The construction of gauge-invariant quantities for these non-linear
perturbations will be based on the same strategy that we have used for the
determination of the perturbative equations, i.e. a 2-parameter
perturbative expansion of the GSGM formalism, see section~\ref{sec:5.1_Cpl}.
This approach will be very helpful for identifying and building the
non-linear perturbations with gauge invariant character directly
from the gauge-invariant quantities of the GSGM formalism.

The gauge invariance of relativistic linear and non-linear
perturbations has been investigated in many works
\cite{1974RSPSA.341...49S, Bruni:1996im, Bruni:2002sm,
  sopuerta-2004-70, Nakamura:2003wk}, where transformation rules have
been presented for studying one or multi-parameter perturbative
problems. In addition, within the multi-parameter perturbative
framework, a technique for the construction of gauge invariant,
non-linear perturbative fields has been introduced by
Nakamura~\cite{Nakamura:2003wk, Nakamura:2004gi}.  In the literature,
there are different ways to define the gauge invariance of non-linear
perturbations~\cite{Bruni:1996im, Cunningham:1980cp}.  A perturbative
tensor field of $n$ order can be defined as gauge invariant ``at n
perturbative order'' or ``up to n perturbative order''.  For the
former definition a perturbative tensor field is invariant only for
the gauge transformations relative to the $n$ perturbative order
considered.  Therefore, all the gauges of the previous $1,..,n-1$
orders must be fixed.  The second definition is more restrictive and
requires that the perturbative field is gauge invariant at any order
up to the desired perturbative order. Thus in this case, all the
gauges from the first to the $n^{th}$ orders are completely arbitrary.

The non-linear perturbations that we adopt for the analysis of the
coupling between the radial and non-radial oscillations will be gauge
invariant for gauge transformations associated with the $\epsilon$ and
$\lambda \epsilon$ non-radial perturbations with the restriction that
the first order gauge for the radial perturbations must be fixed.
This restriction is due to the absence of a gauge invariant
formulation for the radial perturbations. In fact, as we reported in
section~\ref{sec:Linear_Rad_pert_anal}, the GSGM formalism also fails
to provide a gauge invariant description for this class of
oscillations.  However, their physical properties have been well
described in both Eulerian and Lagrangian
gauges~\cite{Kokkotas:2000up, Sperhake:2001si}. In particular, the
latter is more appropriate for describing the regions near the stellar
surface~\cite{Sperhake:2001si}.

In section~\ref{sec:ConstGI}, we present the method to build the gauge
invariant perturbations at $\lambda \epsilon$ order. In
section~\ref{sub_Ax}, we illustrate this method for the axial
perturbations while the polar sector is discussed in
section~\ref{sub_Pol}. This method has been presented in the
paper~\cite{Passamonti:2004je} for the polar perturbations, while the
axial case will be treated in~\cite{Passamonti:2005axial}.

\section{Construction of gauge-invariant non-linear perturbations}
\label{sec:ConstGI}

Gauge transformations and gauge invariance in 2-parameter perturbation
theory have been studied in reference~\cite{Bruni:2002sm,
sopuerta-2004-70}.  In section \ref{sec:3.1_MPPT}, we have reported
the procedure based on the Baker-Campbell-Hausdorff (BCH) formula
given in the work of Sopuerta et al.~\cite{sopuerta-2004-70}, where
the gauge transformation rules for linear and non-linear perturbative
fields have been derived in a more general way.  Let ${\cal T}$ be a
generic tensorial quantity and ${\cal T}^{(i,j)}$ its perturbation at
$\lambda ^i \epsilon^j$ order.

The gauge transformations of linear radial and non-radial
perturbations ${\cal T}^{(1,0)}$ and ${\cal T}^{(0,1)}$ are
respectively given by the formulae (\ref{gtr10_base}) and
(\ref{gtr01_base}),
\begin{eqnarray}
\widetilde{\cal T}^{(1,0)}& =& {\cal T}^{(1,0)}
+ \pounds_{\xi_{(1,0)}} {\cal T}_{b}\,, \label{gaf10} \\
 \widetilde{\cal T}^{(0,1)}& =& {\cal T}^{(0,1)}
+ \pounds_{\xi_{(0,1)}} {\cal T}_{b}\,, \label{gaf01}
\end{eqnarray}
where ${\cal T}_{b}$ is the background value of the tensor field
${\cal T}$, and the tilde denotes the perturbative fields transformed
by the gauge transformation. The vector fields $\xi_{(1,0)}$ and
$\xi_{(0,1)}$ are respectively the generators of the gauge
transformations for the first order radial and non-radial
perturbations.  \\
At second $\lambda \epsilon$ order, the tensor fields ${\cal T}$
transforms according to equation (\ref{gtr11_base}):
\begin{eqnarray}
\widetilde{\cal T}^{(1,1)} = {\cal T}^{(1,1)} +
\pounds_{\xi_{(0,1)}}\,{\cal T}^{(1,0)}
+\pounds_{\xi_{(1,0)}}\,{\cal T}^{(0,1)} +
\left(\pounds_{\xi_{(1,1)}} +
\left\{\pounds_{\xi_{(1,0)}}\,,\,\pounds_{\xi_{(0,1)}}
\right\} \right) {\cal T}_{b} \,, \label{gtr11}
\end{eqnarray}
where $\{\,,\}$ stands for the anti-commutator $\{a,b\}= a\, b + b\,
a$, and $\xi_{(1,1)}$ is the generator of the $\lambda \epsilon$ gauge
transformations.  In the transformation rule (\ref{gtr11}), the
presence of non-linear terms, which contain perturbative fields and
gauge transformation generators of the previous order, make really
hard the construction of non-linear perturbations that are completely
''gauge invariant up to the second order''. A method to reach this
purpose has been presented by Nakamura in
reference~\cite{Nakamura:2003wk}, but here, we do not follow his
approach as the calculations will be too laborious for the aim of this
work.  Furthermore, as long as a gauge invariant description for the
radial pulsations is unknown we have to study these linear
perturbations by choosing a gauge, for instance as we did in
section~\ref{sec:Linear_Rad_pert_anal}.  Therefore, we can derive
non-linear perturbative fields which are invariant under a sub-class
of second order gauge transformations, namely those where the first
order gauge for radial perturbations is fixed. With this assumption
the general gauge transformation~(\ref{gtr11}) reduces to this simpler
expression:
\begin{equation}
\widetilde{\cal T}^{(1,1)} = {\cal T}^{(1,1)} +
\pounds_{\xi_{(0,1)}}\,{\cal T}^{(1,0)}
+\pounds_{\xi_{(1,1)}}\,{\cal T}_{b} \, , \label{ga_tran}
\end{equation}
which is determined by setting to zero the radial generator
$\xi_{(1,0)}$ in the expression~(\ref{gtr11}).

Now, we are going to explore the structure of the non-linear
perturbations that we have used in chapter~\ref{ch:5_NL} and as these
perturbations change under the gauge transformation~(\ref{ga_tran}).
The main idea behind our method is to build the $\lambda \epsilon$
gauge-invariant variables starting from the gauge-invariant quantities
of the GSGM formalism.  \\ 
\indent Let ${\cal G}^{(1)}$ be any of the
gauge-invariant metric or fluid quantities in
equations~(\ref{pkab})-(\ref{pt2}) and~(\ref{algi})-(\ref{omgi}) for the
polar sector or in equations~(\ref{pka})-(\ref{pL})
and~(\ref{deltauax}) for the axial case. In our perspective, these 
quantities can be considered as non-radial perturbations of a radially
pulsating spacetime, which is itself described as a perturbation of a
static background
(see section~\ref{sec:5.1_Cpl} for more details).  This perturbative
expansion then allows us to perform a splitting of the perturbation
${\cal G}^{(1)}$ in $\lambda$ to get
\begin{equation}
{\cal G}^{(1)} = {\cal G}^{(0,1)} +\lambda{\cal G}^{(1,1)}\, , \label{gp1p}
\end{equation}
where ${\cal G}^{(0,1)}$ is a first order, gauge invariant non-radial
perturbation on a static spacetime, while ${\cal G}^{(1,1)}$ describes
the corrective effects on the non-radial oscillations brought by the
radial pulsations. The non-linear perturbations ${\cal G}^{(1,1)}$ are
the quantities we have used to investigate the coupling between the
radial and non-radial perturbations. In this chapter, we explore in more
details their structure and prove their gauge invariance with
respect to a spherical static background.  It is important to remark
that the $(1,1)$ superscript refers not only to quantities constructed
from the $\metb^{(1,1)}$ perturbations, but in general to any
perturbative quantity of order $\lambda\epsilon$. This structure is
for instance evident in equations (\ref{k00_11})-(\ref{k11_11}),
(\ref{En_11_exp}) and in (\ref{Psi11_mast}) for some of the non-linear
variables used in this work.  Therefore, ${\cal G}^{(0,1)}$ and ${\cal
G}^{(1,1)}$ can be written in general as:
\begin{eqnarray}
{\cal G}^{(0,1)} & = & {\cal H}^{(0,1)} \,, \label{g01}\\
{\cal G}^{(1,1)} & = & {\cal H}^{(1,1)} + \sum_\sigma{\cal I}^{(1,0)}_\sigma
{\cal J}^{(0,1)}_\sigma\,,\label{g11}
\end{eqnarray}
where the objects ${\cal H}^{(0,1)}$ and ${\cal J}^{(0,1)}_\sigma$ are
linear in the ${(0,1)}$ perturbations, while ${\cal I}^{(1,0)}_\sigma$
and ${\cal H}^{(1,1)}$ are respectively linear in the $\lambda$ and
the $\lambda \epsilon$ variables.  \\
\indent By definition, the quantity ${\cal H}^{(0,1)}$ represents
any of the gauge-invariant quantities given in the expressions
(\ref{pkab})-(\ref{pt2}) and (\ref{algi})-(\ref{omgi}) for the polar
perturbations, and in~(\ref{pka})-(\ref{pL}) and~(\ref{deltauax}) for
the axial, which have been derived from first order non-radial
perturbations of a static background.  It is also clear that the
quantities ${\cal H}^{(1,1)}$ have the same formal structure as the
${\cal H}^{(0,1)}$, but are now constructed from second order
$\lambda\epsilon$ quantities. We will show later that these objects
are not gauge-invariant at $\lambda \epsilon$ order. In fact at this
second perturbative order the gauge invariant quantities must contain
some extra terms formed by the product of first order perturbations,
as in equation (\ref{g11}).

Let us now consider a class of non-radial gauge transformations of
first and second perturbative order where we have fixed the gauge for
the linear radial perturbations. This transformation is given by the
two expressions~(\ref{gaf01}) and~(\ref{ga_tran}).
The linear perturbations ${\cal G}^{(0,1)}$ are gauge-invariant by
definition at order $\epsilon$, thus from equation~(\ref{gaf01})
we have that
\begin{equation}
\widetilde{\cal G}^{(0,1)}-{\cal G}^{(0,1)} = \widetilde{\cal H}^{(0,1)}-{\cal
H}^{(0,1)} =0\,. \label{agi01}
\end{equation}
For the non-linear perturbations ${\cal G}^{(1,1)}$ we have
from equation~(\ref{g11}) and the fact that we have fixed the gauge for
radial perturbations the following transformation:
\begin{eqnarray}
\widetilde{\cal G}^{(1,1)}-{\cal G}^{(1,1)} & = & \widetilde{\cal H}^{(1,1)}
 -{\cal H}^{(1,1)} + \sum_\sigma {\cal I}^{(1,0)}_\sigma \,
 \left(\widetilde{\cal J}^{(0,1)}_\sigma-{\cal J}^{(0,1)}_\sigma
 \right)\,. \label{agi11}
\end{eqnarray}
This expression can be further elaborated if we note that every ${\cal
H}^{(1,1)}$ and ${\cal J}^{(0,1)}_{\sigma}$ can be expressed as
follows:
\begin{equation}
{\cal H}^{(1,1)} = {\cal A}\left[\metb^{(1,1)} \right]\,,~~~~
{\cal J}^{(0,1)}_\sigma = {\cal B}_\sigma\left[\metb^{(0,1)} \right]\,,
\label{hjlin}
\end{equation}
where ${\cal A}$ and ${\cal B}_\sigma$ are linear operators involving
differentiation with respect to the coordinates of $M^2$ and
integration on $S^2$.  These operators act on spacetime objects and
return objects with indices on $M^2$.  In equations~(\ref{hjlin})
and in the rest of this section we consider for simplicity only
the metric perturbations. For the energy-momentum tensor perturbations and
fluid variables the procedure follows along the same lines and is given 
in two later subsections of this chapter.
Using the gauge
transformations (\ref{gaf01}) and (\ref{ga_tran}), the transformation
rules for ${\cal H}^{(1,1)}$ and ${\cal J}^{(0,1)}_\sigma$ are given
by the following expressions:
\begin{eqnarray}
\widetilde{\cal H}^{(1,1)} &=&{\cal H}^{(1,1)}
 + {\cal A}\left[\pounds_{\xi_{(0,1)}}\metb^{(1,0)} +
\pounds_{\xi_{(1,1)}} \bar{\metb} \right]  \,, \label{ha11}\\
\widetilde{\cal J}^{(0,1)}_\sigma&=&{\cal J}^{(0,1)}_\sigma
 + {\cal B}_\sigma\left[ \pounds_{\xi_{(0,1)}} \bar{\metb}
\right]\,. \label{J01}
\end{eqnarray}
In addition, we can notice that the term ${\cal A}\left[ \,
\pounds_{\xi} \bar{\metb} \, \right]$ must vanish in equation
(\ref{ha11}) for any vector field $\xi$. This result is due to the
fact that the perturbative fields ${\cal H}^{(1,1)}$ and ${\cal
H}^{(0,1)}$ share the same functional structure, and that ${\cal
H}^{(0,1)}$ are first order gauge-invariant perturbations.  Thus, they
satisfy equation (\ref{agi01}) for any gauge transformation
(\ref{gaf01}) generated by a generic vector field $\xi$.
As a result, the quantity (\ref{ha11}) becomes
\begin{equation}
\widetilde{\cal H}^{(1,1)}={\cal H}^{(1,1)}
 + {\cal A}\left[\pounds_{\xi_{(0,1)}} \metb^{(1,0)}\right] \,, \label{ha11_B}
\end{equation}
and the gauge transformation (\ref{agi11}) reduces to the following
\begin{equation}
\widetilde{\cal G}^{(1,1)} - {\cal G}^{(1,1)} = {\cal
A}\left[\pounds_{\xi_{(0,1)}}\metb^{(1,0)} \right] + \sum_\sigma {\cal
I}^{(1,0)}_\sigma {\cal B}_\sigma \, \left[ \,
\pounds_{\xi_{(0,1)}} \bar{\metb} \,  \right] \,. \label{agi11_red}
\end{equation}
The gauge transformation (\ref{agi11_red}) is valid for $\lambda
\epsilon$ non-radial perturbations, when the gauge of linear radial
perturbations is fixed. This formula will be used in the following
sections in order to prove the gauge invariance of the ${\cal
G}^{(1,1)}$ variables. This will be achieved by showing that the right
hand side of equation~(\ref{agi11_red}) vanishes.

In order to prove the invariance of our non-linear polar and
axial perturbations ${\cal G}^{(1,1)}$ for a gauge transformation
defined at first order by the expression~(\ref{gaf01}) and at second
order by the transformation~(\ref{ga_tran}), where the radial gauge is
fixed, we proceed as follows:
\begin{itemize}
\item[1)] we expand the GSGM variables ${\cal G}^{(1)}$ as in equation
(\ref{gp1p}) and determine the second order quantities ${\cal
G}^{(1,1)}$.  Then, we identify in ${\cal G}^{(1,1)}$ the fields
${\cal H}^{(1,1)}$, ${\cal I}^{(1,0)}_\sigma$ and ${\cal
J}^{(0,1)}_\sigma$ as in equation (\ref{g11}).

\item[2)] We perform the gauge transformations of the fields ${\cal
 H}^{(1,1)}$ and ${\cal J}^{(0,1)}_\sigma$ by using the
 rules~(\ref{ha11_B}) and (\ref{J01}).

\item[3)] Finally, we introduce the results in the gauge
   transformation~(\ref{agi11_red}) and prove that it is satisfied,
   namely that the right hand side vanishes.
\end{itemize}

\subsection{Axial perturbations}
\label{sub_Ax}
Gauge invariant non-radial perturbations on a radially pulsating
spacetime are expressed for the axial sector by the tensor
fields~(\ref{pka})-(\ref{pL}), (\ref{deltauax}) and (\ref{Pidef}).
The perturbative expansion of these metric and fluid variables, which
has been illustrated in the previous section, leads to the following
quantities at first order in $\epsilon$:
\begin{eqnarray}
k_{A}^{(0,1)} & = & h_{A}^{(0,1)} - h^{(0,1)}_{\mid A } + 2\,
h^{(0,1)} \, \bar{v}_A \label{ka_01_exp}  \, ,\\
\Psi^{(0,1)} & = & \left[ r \left( k_{1 \, ,t}^{(0,1)} - k_{0 \,
,r}^{(0,1)} \right) + 2 k_{0}^{(0,1)} \right] e^{-\left(\Phi + \Lambda
\right)}  \label{Psi01_exp} \, \\
L_A ^{(0,1)} & = & \delta t_A ^{(0,1)} -
\bar{Q} \, h_A ^{(0,1)} \, , \label{LA_01_exp}  \\
L ^{(0,1)
}& = & \delta t ^{(0,1)} - \bar{Q} \, h ^{(0,1)} \, ,  \label{L_01_exp}
\end{eqnarray}
and at second order in $\lambda \epsilon $:
\begin{eqnarray}
k_{A}^{(1,1)} & = & h_{A}^{(1,1)} - h^{(1,1)}_{\mid A } + 2\,
h^{(1,1)} \, \bar{v}_A \label{ka_11_exp}  \, ,\\
\Psi^{(1,1)} & = & \left[ r \left( k_{1 \, ,t}^{(1,1)} -
k_{0 \, ,r}^{(1,1)} \right) + 2 k_{0}^{(1,1)} \right] e^{-\left(\Phi +
\Lambda \right)} {} 
-  \eta^{(1,0)} \Psi^{(0,1)}   \label{Psi11_exp} \\
L_A ^{(1,1)} & = & \delta t_A ^{(1,1)} -
\bar{Q} \, h_A ^{(1,1)} - Q^{(1,0)} \, h_A ^{(0,1)} \, , \label{LA_11_exp} \\
L ^{(1,1)
}& = & \delta t ^{(1,1)} - \bar{Q} \, h ^{(1,1)} - Q^{(1,0)} \, h
^{(0,1)} \, .  \label{L_11_exp}
\end{eqnarray}
Comparing the expressions (\ref{Psi01_exp})-(\ref{L_01_exp}) and
(\ref{Psi11_exp})-(\ref{L_11_exp}), it is easy to identify for any
second order perturbation the terms ${\cal H}^{(1,1)}$:
\begin{eqnarray}
k_A^{(1,1)} & : \qquad & h_{A}^{(1,1)} - h^{(1,1)}_{\mid A } + 2\,
h^{(1,1)}  \,, \label{??ff} \\
\Psi^{(1,1)} & : \qquad  &  \left[ r \left( k_{1 \, ,t}^{(1,1)} -
k_{0 \, ,r}^{(1,1)} \right) + 2 k_{0}^{(1,1)} \right] e^{-\left(\Phi +
\Lambda \right)} \,,   \label{ta} \\
L_A ^{(1,1)} & : \qquad &  \delta t_A ^{(1,1)} -
\bar{Q} \, h_A ^{(1,1)}  \\
L ^{(1,1)}  & : \qquad & \delta t ^{(1,1)} - \bar{Q} \, h ^{(1,1)}     \,.\label{??ext2}
\end{eqnarray}
and the quantities $ \sum_{\sigma} {\cal I}^{(1,0)}_\sigma {\cal
J}^{(0,1)}_\sigma$:
\begin{eqnarray}
k_A^{(1,1)} & : \qquad & 0  \, , \\
\Psi^{(1,1)} & : \qquad  &  -  \eta^{(1,0)} \Psi^{(0,1)}
 \, , \\
L_A ^{(1,1)} & : \qquad &  - Q^{(1,0)} \, h_A ^{(0,1)} \, , \\
L ^{(1,1)}  & : \qquad & - Q^{(1,0)} \, h
^{(0,1)} \, .
\end{eqnarray}

Now, we have to determine how the perturbative fields ${\cal
H}^{(1,1)}$ and ${\cal J}^{(0,1)}_\sigma$ change under a gauge
transformation given by the expressions~(\ref{ha11_B}) and
(\ref{J01}).
The covariant vector field $\xi_{(0,1)\alpha}$ that generates axial gauge
transformations can be expanded in odd-parity tensor harmonics,
\begin{equation}
\xi_{(0,1)\alpha}=( 0, 0 ,  r^2 V^{(0,1)} \, S_a )\,,
\label{xi01_ax}
\end{equation}
where $V^{(0,1)} \equiv V^{(0,1)} \left( x^A \right)$ is a scalar
function on the submanifold $M^2$.

\indent The first order metric components $h_A ^{(0,1)}$ and $h
^{(0,1)}$ change under the gauge transformation (\ref{gaf01}) as
follows:
\begin{eqnarray}
\tilde{h} _{A} ^{(0,1)} & = & h_{A} ^{(0,1)} + r^2
V^{(0,1)}_{\mid A} \, ,  \label{ha_01_tr}\\
\tilde{h}  ^{(0,1)} & = & h ^{(0,1)} + r^2  V^{(0,1)}  \label{h_01_tr}\, ,
\end{eqnarray}
while the master function $\Psi^{(0,1)}$ is gauge invariant by construction.

The transformation (\ref{ha11_B}) requires the evaluation of the Lie
derivatives of the radial metric and energy momentum tensor
perturbations with respect to the generator $\xi_{(0,1)\alpha}$.
Considering the expression of the radial metric in the gauge~(\ref{fixgauge})
we have:
\begin{eqnarray}
\pounds_{\xi_{(0,1)}} g ^{(1,0)}_{\alpha \beta} = 0 \, ,
\end{eqnarray}
and for the energy momentum tensor
\begin{eqnarray}
\pounds_{\xi_{(0,1)}} \delta t ^{(1,0)}_{AB} & = & 0 \, , \\
\pounds_{\xi_{(0,1)}} \delta t ^{(1,0)}_{Ab} & = & r^2 Q^{(1,0)}
V_{\mid A}^{(0,1)} S_b\, ,\\
\pounds_{\xi_{(0,1)}} \delta t ^{(1,0)}_{ab} & =
& r^2 Q^{(1,0)} V^{(0,1)} \left(S_{a:b} + S_{b:a} \right)\, .
\end{eqnarray}
Therefore the gauge transformation (\ref{ha11_B}) for the components
of the metric and energy momentum tensors leads to:
\begin{eqnarray}
\tilde{h} _{A} ^{(1,1)} & = & h_{A} ^{(1,1)}
\, , \label{ha_01_tr_red} \\
\tilde{h} ^{(1,1)} & = & h ^{(1,1)}  \, , \label{h_01_tr_red} \\
\delta \tilde{t} _{A} ^{(1,1)} &
= & \delta t _{A} ^{(1,1)} + r^2 Q^{(1,0)} V_{\mid A}^{(0,1)} \,
,\label{tA_11_tr_red}\\
\delta \tilde{t} ^{(1,1)} & = &\delta t_{A}
^{(0,1)} + r^2 Q^{(1,0)} V^{(0,1)} \label{t_11_tr_red} \, .
\end{eqnarray}
Finally, if we introduce equations
(\ref{ha_01_tr}),(\ref{h_01_tr}) and
(\ref{ha_01_tr_red})-(\ref{t_11_tr_red}) into the gauge transformation
(\ref{agi11_red}) of the axial variables (\ref{ka_11_exp}),
(\ref{LA_11_exp}) and (\ref{L_11_exp}), we find the gauge invariance
of the quantities $k_{A}^{(1,1)}$, $L_{A}^{(1,1)}$ and $L^{(1,1)}$
respectively.  The gauge invariant character of the master function
$\Psi^{(1,1)} $ derives directly from its expression~(\ref{Psi11_exp})
and from the fact that the radial gauge is fixed. In fact, the
equation~(\ref{Psi11_exp}) shows that it is a linear combination of
the gauge invariant tensor $k_A^{(1,1)}$ and the first order master
equation $\Psi^{(0,1)}$ and the linear radial perturbation
$\eta^{(1,0)}$.

The last variable of the axial sector to study is the axial component
$\beta$ of velocity perturbation (\ref{deltauax}).  For the first and
second order gauge transformations (\ref{gaf01}) and (\ref{ga_tran}),
the Lie derivatives of the background velocity and its radial
perturbations vanish:
\begin{equation}
\pounds_{\xi_{(i,j)}} \bar{u} _{\alpha} =
\pounds_{\xi_{(i,j)}}  \delta u^{(1,0)}  _{\alpha} =  0 \, ,
\end{equation}
where the indices $(i,1)$ with $i=0,1$ represent both the $\epsilon$
linear non-radial and the non-linear $\lambda \epsilon$ cases.
Therefore, the odd-parity velocity perturbations $\delta u^{(0,1)}
_{\alpha}$ and $\delta u^{(1,1)} _{\alpha}$ remain unchanged and
$\beta ^{(0,1)}$ and $\beta ^{(1,1)}$ are gauge invariant.

\subsection{Polar perturbations}
\label{sub_Pol}
Polar gauge invariant perturbations on a radially oscillating
spacetime are defined by the metric and fluid quantities
(\ref{pkab})-(\ref{pt2}).  The analysis of their gauge invariance will
start with the metric perturbations, then with the energy momentum
tensor and finally with the fluid variables. The details of the
calculations are given only for the metric variables. For the energy
momentum tensor and the fluid variables we determine the fields ${\cal
H}^{(1,1)}$, ${\cal I}^{(1,0)}_\sigma$ and ${\cal J}^{(0,1)}_\sigma$
obtained from the perturbative expansion (\ref{g11}) and the value of
the Lie derivatives of the gauge transformations (\ref{J01}) and
(\ref{ha11_B}). These quantities must be then introduced in
equation~(\ref{agi11_red}), where after some algebraic calculation the
two terms of the right hand side cancel each other.

\subsubsection{Metric polar perturbations}

The four first order gauge invariant metric variables in the GSGM
formalism are given by the symmetric tensor $k_{AB}^{(1)}$ and the
scalar function $k^{(1)}$ respectively defined in
equations~(\ref{pkab}) and (\ref{pk}). The perturbative expansion of
these quantities provides the following expressions at
first order in $\epsilon$:
\begin{eqnarray}
k_{AB}^{(0,1)}&=&h_{AB}^{(0,1)}-(p_{A|B}^{(0,1)}+p_{B|A}^{(0,1)})\,,
\qquad \qquad  \label{pkab01} \\
k^{(0,1)}&=&K^{(0,1)}-2 \bar v^A p_A^{(0,1)}  \,, \label{pk01}
\end{eqnarray}
where the vector $p_A^{(0,1)}$ is given by the following expression:
\begin{equation}
p_A^{(0,1)} = h^{(0,1)} _A - \frac{1}{2}r^2 G_{|A}^{(0,1)}\, , \label{ppa01_new}
\end{equation}
and at second order in $\lambda \epsilon$:
\begin{eqnarray}
k_{AB}^{(1,1)}&=&h_{AB}^{(1,1)}-(p_{A|B}^{(1,1)}+p_{B|A}^{(1,1)})
+ 2\, \Gamma^{(1,0)C}_{AB}  p^{(0,1)}_C \,,
\label{pkab11} \\
k^{(1,1)}&=&K^{(1,1)}-2 \bar v^A p_A^{(1,1)} + 2 \met^{(1,0) A B} \bar
v_{B} p^{(0,1)}_{A}\,, \label{pk11}
\end{eqnarray}
where $\Gamma^{(1,0)C}_{AB}$ are the radial perturbations of the
Christoffel symbols
\begin{equation}
\Gamma ^{(1,0) \ C} _{AB} = \frac{1}{2} \ \bar{\met}^{CD} \left[
h^{(1,0)} _{ D A \mid B} + h^{(1,0)} _{D B \mid A}- h^{(1,0)} _{AB
\mid D} \right]
\end{equation}
and the quantity  $p_A^{(1,1)}$ is given by the
following equation:
\begin{equation}
p_A^{(1,1)} = h^{(1,1)} _A - \frac{1}{2}r^2 G_{|A}^{(1,1)}\,. \label{ppa11}
\end{equation}

In these second order perturbations ${\cal G}^{(1,1)}$, it is easy to
identify the quantities ${\cal H}^{(1,1)}$:
\begin{eqnarray}
k_{AB}^{(1,1)}& : \qquad &  h_{AB}^{(1,1)}-(p_{A|B}^{(1,1)}+p_{B|A}^{(1,1)})
 \,, \label{pkab11_H} \\
k^{(1,1)} & : \qquad & K^{(1,1)}-2 \bar v^A p_A^{(1,1)}
\,, \label{pk11_H}
\end{eqnarray}
and the contribution brought by the first order perturbations
$\sum_{\sigma} {\cal I}^{(1,0)}_\sigma {\cal J}^{(0,1)}_\sigma$
\begin{eqnarray}
k_{AB}^{(1,1)}& : \qquad &   2\, \Gamma^{(1,0)C}_{AB}  p^{(0,1)}_C \,,
\label{pkab11_J} \\
k^{(1,1)} & : \qquad & 2\met^{(1,0) A B} \, \bar v_{B} \, p^{(0,1)}_{A}\, .
\label{pk11_J}
\end{eqnarray}
Now, we consider the gauge transformations (\ref{J01})
and (\ref{ha11_B}). The generator $\xi_{(0,1)\alpha}$
of the gauge transformations associated with the non-radial
perturbations can be expanded in polar tensor harmonics,
\begin{equation}
\xi_{(0,1)\alpha}=(\hat\xi_A^{(0,1)} \, Y\,,\, r^2 \xi^{(0,1)} \, Y_a )\,.
\label{xi01_pol}
\end{equation}
Under the transformation (\ref{gaf01}) the components of
the non-radial metric tensor $\met _{\alpha \beta }^{(0,1)}$
change as follows:
\begin{eqnarray}
\tilde{h}_{AB}^{(0,1)} & = & h_{AB}^{(0,1)} + \hat{\xi}_{A\mid B}^{(0,1)}
+ \hat{\xi}_{B\mid A}^{(0,1)} \, \label{hAB01_tr}\\
\tilde{h}_{A}^{(0,1)} & = & h_{A}^{(0,1)} + \hat{\xi}_{A}^{(0,1)} +
r^2  \xi_{\mid A} \, \\
\tilde{K}^{(0,1)} & = & K^{(0,1)} + 2 \bar v ^A \hat{\xi}_{A}^{(0,1)}
\, , \\
\tilde{G}^{(0,1)} & = & G^{(0,1)} + 2 \xi^{(0,1)}  \label{G01_tr} \, .
\end{eqnarray}
Thus, from the definition~(\ref{ppa01_new}) and the
expressions~(\ref{hAB01_tr})-(\ref{G01_tr}) the quantity
$p_A^{(0,1)}$, which constitutes the term ${\cal
J}^{(0,1)}_\sigma$ in equations (\ref{pkab11_J}) and
(\ref{pk11_J}) changes as:
\begin{equation}
\tilde{p}_A^{(0,1)} = p_A^{(0,1)} + \hat \xi _{A}^{(0,1)} \, . \label{pA01_tr}
\end{equation}
On the other hand, in the transformation (\ref{ha11_B}) the Lie
derivative of the linear radial metric $\met_{\alpha \beta } ^{(1,0)}$
provides the following results:
\begin{eqnarray}
\pounds_{\xi_{(0,1)}}\met_{AB}^{(1,0)}& = &  \left( \hat{\xi} ^{(0,1)~C} h_{AB
\mid C}^{(1,0)} + h_{C B}^{(1,0)} \hat{\xi} ^{(0,1)~C} _{\mid A }
+ h_{A C}^{(1,0)} \hat{\xi} ^{(0,1)~C} _{\mid B} \right) Y \, , \\
\quad \pounds_{\xi_{(0,1)}}\met_{Aa}^{(1,0)} &=& h_{A C}^{(1,0)}
\,\hat{\xi}^{C} \,Y_{:a} \,,  \\
\pounds_{\xi_{(0,1)}}\met^{(1,0)}_{ab} & = & 0\, .\label{Lieg10}
\end{eqnarray}
Therefore, the second order metric components changes with respect to
the transformation (\ref{ha11_B}) as follows:
\begin{eqnarray}
\tilde{h}_{AB}^{(1,1)} & = & h_{AB}^{(1,1)}
+ \hat{\!\!\pounds}_{\hat\xi}h_{AB}^{(1,0)} \, , \\
\tilde{h}_{A}^{(1,1)} & = & h_{A}^{(1,1)} + h_{A C}^{(1,0)}
\,\hat{\xi}^{C} \, , \\
\tilde{K}^{(1,1)} & = & K^{(1,1)} \, , \\
\tilde{G}^{(1,1)} & = & G^{(1,1)} \, ,
\end{eqnarray}
and then the perturbation $p_A^{(1,1)}$ as:
\begin{equation}
\tilde{p}_A^{(1,1)} = p_A^{(1,1)}  +  h_{A C}^{(1,0)}
\,\hat{\xi}^{C} \, .
\end{equation}

Finally, we have all the elements for checking the invariance of the
tensors $k_{AB}^{(1,1)}$ and $k^{(1,1)}$ under the gauge
transformation~(\ref{agi11_red}).  When we introduce in the
definitions~(\ref{pkab11}) and (\ref{pk11}) the relative values we find:
\begin{eqnarray}
\tilde{k}_{AB}^{(1,1)} & = & k_{AB}^{(1,1)} +
\hat{\xi} ^{(0,1)~C} h_{AB
\mid C}^{(1,0)} + h_{C B}^{(1,0)} \hat{\xi} ^{(0,1)~C} _{\mid A }
+ h_{A C}^{(1,0)} \hat{\xi} ^{(0,1)~C} _{\mid B} \nn \\
& - &  \left( h_{A C}^{(1,0)} \,
\hat{\xi}^{C} \right)_{\mid B } - \left( h_{B C}^{(1,0)} \,
\hat{\xi}^{C} \right)_{\mid A } + 2\, \Gamma^{(1,0)C}_{AB}
\hat{\xi}^{(0,1)}_C \, \\
\tilde{k}^{(1,1)} & = & k ^{(1,1)} - 2 \bar v^A h_{A C}^{(1,0)} \,
\hat{\xi}^{C} + 2 h^{(1,0) A B} \bar
v_{B} \hat{\xi}^{(0,1)}_{A} \, ,
\end{eqnarray}
and after some trivial calculations we get
\begin{eqnarray}
\tilde{k}_{AB}^{(1,1)} & = & k_{AB}^{(1,1)}  \, , \\
\tilde{k}^{(1,1)} & = & k ^{(1,1)} \, ,
\end{eqnarray}
thus the gauge invariance of the non-linear
perturbations.

\subsubsection{Energy momentum tensor}
The components of the energy momentum tensor $\delta t_{\alpha \beta}
^{(0,1)}$ can be combined to define the seven quantities
(\ref{pTAB})-(\ref{pt2}). These perturbations that are gauge invariant
on a spherical and time dependent spacetime time can be expanded as in
equation~(\ref{g11}). We can then write the following expressions:
\begin{eqnarray}
T_{AB}^{(1,1)} & :  \qquad &  \delta
t_{AB}^{(1,1)}- \bar t_{AB|C}\, p^{(1,1)~C}- \bar t_{AC} \,
p^{(1,1)~C}_{|B} - \bar t_{BC} \, p^{(1,1)~C}_{|A} \, ,
\label{pTAB11_H} \\
T^{(1,1)~3} & :  \qquad & \delta t^{(1,1)~3}- p^{(1,1)~C} (\bar Q_{|C}+2 \bar Q
\bar v_C) +\frac{l(l+1)}{2} \, \bar Q \, G^{(1,1)} \,, \\
T_A^{(1,1)}  & :  \qquad & \delta t_A^{(1,1)} - \bar t_{AC} \,
p^{(1,1)~C} - \frac{r^2}{2} \, \bar Q \, G_{|A}^{(1,1)}
\,, \\
T^{(1,1)~2} & :  \qquad &  \delta t^{(1,1)~2} -r^2\, \bar Q\, G^{(1,1)}
\,, \label{pt211_H}
\end{eqnarray}
and for the contribution carried by the first order perturbations
$\sum_{\sigma} {\cal I}^{(1,0)}_\sigma {\cal J}^{(0,1)}_\sigma$:
\begin{eqnarray}
T_{AB}^{(1,1)}  & :  \qquad &  p^{(0,1)C}    t^{(1,0)}_{A B \mid C} +
t^{(1,0)}_{AC}  p^{(0,1)  C}_{\mid B} + t^{(1,0)}_{CB}
 p^{(0,1)  C}_{\mid A}      - \bar{t}_{A B \mid C}
 \met^{(1,0) C D}  p^{(0,1)}_D    \nonumber \\
& & {} - \bar{t}_{AC} \left( \met^{(1,0)  C D}_{\mid B} \
p^{(0,1)}_D + \met^{(1,0)  C D}  p^{(0,1)}_{D  \mid B}
\right) \nn \\
&& - \bar{t}_{BC}  \left( \met^{(1,0)  C D}_{\mid A}
 p^{(0,1)}_D + \met^{(1,0)  C D}  p^{(0,1)}_{D  \mid A} \right)\,, \label{extab} \\
T^{(1,1)~3} & : \qquad &  \left( \bar{Q}_{\mid A} +  2 \bar{Q}   v_A
\right) \met^{(1,0) AB}  p^{(0,1)}_{B}   - \left( Q^{(1,0)}_{\mid A}
+ 2 Q^{(1,0)} v_A \right)\bar{\met}^{AB} p^{(0,1)}_{B} \nn \\
&& + \frac{l(l+1)}{2}   Q^{(1,0)}   G^{(0,1)}\,, \label{ext3} \\
T_A^{(1,1)} & : \qquad  &  \left( t^{(1,0)  C}_A - \bar{t}_{AB}
\met^{(1,0) BC} \right) p^{(0,1)}_{C} - \frac{r^2}{2}
Q^{(1,0)} G^{(0,1)}_{\mid A}   \,,   \label{exta} \\
T^{(1,1)~2} & : \qquad &  r^2   Q^{(1,0)}   G^{(0,1)}   \,.\label{ext2}
\end{eqnarray}
where $\bar{Q}$ and $Q^{(1,0)}$ are a static scalar function and its
radial perturbations, which corresponds for a perfect fluid to the
pressure. This notation has been introduced in the GSGM formalism in
equation (\ref{tblock}), and we keep it in this section in order
to distinguish the pressure from the metric vector $p_{A}$.

The perturbative fields ${\cal H}^{(1,1)}$ transform accordingly to
equation~(\ref{ha11_B}). We must then determine the Lie derivative
$\pounds_{\xi (0,1)} t_{\alpha\beta}^{(1,0)}$, where the energy
momentum tensor for the radial perturbations has this block diagonal
form
\begin{equation}
t_{\alpha\beta}^{(1,0)} = \textrm{diag} \left( t_{AB}^{(1,0)}
; r^2 Q^{(1,0)} \gamma_{ab} \right) \,. \label{tAB10}
\end{equation}
One then gets
\begin{eqnarray}
\pounds_{\xi_{(0,1)}} t_{AB}^{(1,0)} & = & \left( \hat{\xi} ^{(0,1)~C}
t_{AB \mid C}^{(1,0)} + t_{C B}^{(1,0)} \hat{\xi} ^{(0,1)~C} _{\mid A
} + t_{A C}^{(1,0)} \hat{\xi} ^{(0,1)~C} _{\mid B} \right) Y \, , \\
\pounds_{\xi_{(0,1)}} t_{Aa}^{(1,0)} & = & \left( t_{AC}^{(1,0)}
\hat{\xi}^{(0,1)~C} + r^2 Q^{(1,0)} \xi_{\mid A}^{(0,1)} \right) Y_a
\,,\\ \pounds_{\xi_{(0,1)}} t^{(1,0)}_{ab} & = & r^2 \left(
Q^{(1,0)}_{\mid C} \hat{\xi}^{(0,1)~C} - l(l+1) Q^{(1,0)} \xi^{(0,1)}
+ 2 v_C Q^{(1,0)} \hat{\xi}^{(0,1)~C} \right) Y \gamma_{a b} \nn \\ &&
+ \left(2 r^2 Q^{(1,0)}\xi^{(0,1)} \right) Z_{ab} \,.
\end{eqnarray}
Therefore, the gauge transformation (\ref{ha11_B})
for the ${\cal H}^{(1,1)}$ terms give the following expressions:
\begin{eqnarray}
\widetilde{T}_{AB}^{(1,1)} & = & T_{AB}^{(1,1)} + \hat{\xi}^{(0,1)~C}
 t^{(1,0)}_{A B \mid C} + t^{(1,0)}_{CB} \hat{\xi}^{(0,1)~C}_{\mid A}
 + t^{(1,0)}_{AC} \hat{\xi}^{(0,1)~C}_{\mid B} - \bar{t}_{A B \mid C}
 \met^{(1,0)C}_D \hat{\xi}^{(0,1)~D} \nn \\
&& {}
- \bar{t}^C{}_A \left(
 \met^{(1,0)}_{C D \mid B} \hat{\xi}^{(0,1)~D} + \met^{(1,0)}_{C D}
 \hat{\xi}^{(0,1)~D}_{\mid B} \right) \nn \\
&& {}
- \bar{t}^C{}_B
 \left( \met^{(1,0)}_{C D \mid A} \ \hat{\xi}^{(0,1)~D} +
 \met^{(1,0)}_{C D} \hat{\xi}^{(0,1)~D}_{\mid A} \right) \,,
\label{gttab} \\
\widetilde{T}^{(1,1)~3} & = & T^{(1,1)~3} + \left( Q^{(1,0)}_{\mid A}
+ 2 Q^{(1,0)} v_A \right) \hat{\xi}^{(0,1)~A} - l(l+1) Q^{(1,0)}
\xi^{(0,1)}  \nn \\
&& - \left( \bar{Q}_{\mid A} + 2\bar{Q} v_A \right)
\met^{(1,0) A}_B \hat{\xi}^{(0,1)~B}\,, \label{gtt3} \\
\widetilde{T}_A^{(1,1)} & = & T_A^{(1,1)} + t^{(1,0)}_{AB}
\hat{\xi}^{(0,1)~B} + r^2 Q^{(1,0)} \hat{\xi}^{(0,1)}_{\mid A} -
\bar{t}_{AB} \met^{(1,0) BC} p_{C}^{(0,1)}\,,
\label{gtta} \\
\widetilde{T}^{(1,1)~2} & = & T^{(1,1)~2} + 2 r^2 Q^{(1,0)}
\xi^{(0,1)} \,. \label{gtt2}
\end{eqnarray}

The gauge invariant character of the energy momentum perturbations
(\ref{pTAB})-(\ref{pt2}) at order $\lambda \epsilon$ can be now
proved by collecting the previous information. We must introduce
into the gauge transformation (\ref{agi11_red}): \emph{i)} the
expressions (\ref{gttab})-(\ref{gtt2}) due to the transformation
of the tensor fields ${\cal H}^{(1,1)}$, and \emph{ii)} the
changes~(\ref{G01_tr}) and (\ref{pA01_tr}) carried by the first
order non-radial fields ${\cal J}^{(0,1)}_\sigma$ of equations
(\ref{extab})-(\ref{ext2}).

\subsubsection{Fluid perturbations}
In the polar sector the fluid perturbations on the GSGM spacetime
$M_{G}$ are given by the two velocity perturbations (\ref{algi}) and
(\ref{gamgi}) and energy density (\ref{omgi}).
The gauge invariant character of the fluid perturbation
$H^{(1,1)}$, which is defined in equation~(\ref{En_11_exp}), will be
discussed at the end of this section.

The perturbative expansion~(\ref{g11}) of these variables produces the
following expression for the tensor fields~${\cal H}^{(1,1)}$:
\begin{eqnarray}
\alpha ^{(1,1)} & : \quad & \tilde\alpha^{(1,1)} -p^{(1,1)~B} \bar u_B
\,, \label{algi_H}  \\
\gamma ^{(1,1)} & : \quad & \tilde\gamma^{(1,1)} - \bar n^A \left[p^{(1,1)~B}
\bar u_{A|B}
+\frac{1}{2} \bar u^B \left(p^{(1,1)}_{B|A}-p^{(1,1)}_{A|B}\right)
\right]   \,, \label{gamgi_H}  \\
\omega ^{(1,1)} & : \quad & \tilde \omega^{(1,1)} -p^{(1,1)~A} \, \bar{\Omega}_{|A}
\, ,   \label{omgi_H}
\end{eqnarray}
and for the part related to the nonlinear contribution of the first
order perturbations~$\sum_\sigma {\cal I}^{(1,0)}_\sigma {\cal}
J^{(0,1)}_\sigma$:
\begin{eqnarray}
\alpha ^{(1,1)}& : \quad &   \met^{(1,0)}_{A B}  p^{(0,1)  A}
 \bar{u}^B -  p^{(0,1)}_{A}  u^{(1,0)  A} \,, \label{exal} \\
\gamma ^{(1,1)}& : \quad &   \left( n^{(1,0)  A} -
\met^{(1,0)  AC}  \bar{n}_C \right) \left( \delta u^{(0,1)}_A
- \frac{1}{2}  h^{(0,1)}_{AB}  \bar{u}^B  - \bar{u}_{A\mid B}
 p^{(0,1)  B} - p^{(0,1)}_{[B \mid A]}  \bar{u}^B\right)  \nn \\
& & {} - \bar{n}^A \left[ \left( u^{(1,0)  B} - \met^{(1,0)BC}
\bar{u}_C \right)  \left( p^{(0,1)}_{[B \mid A]}
+ \frac{1}{2}  h^{(0,1)}_{AB} \right) +
\left( u^{(1,0)}_{A \mid B} - \Gamma^{(1,0)  D}_{AB}
\bar{u}_{D} \right)  p^{(0,1)  B} \right.  \nn \\
& &  \left. - \met^{(1,0) BD}  \bar{u}_{A \mid B}  p^{(0,1)}_D \right] \,,
\label{exga}  \\
\omega ^{(1,1)}& : \quad &  \met^{(1,0)  A B}  p^{(0,1)}_B
 \bar\Omega_{\mid A} - p^{(0,1)  A} \Omega^{(1,0)}_{\mid A}\,.
\label{exom}
\end{eqnarray}

The gauge transformations (\ref{ha11_B}) for ${\cal H}^{(1,1)}$
are now determined by the Lie derivative of the radial perturbations
of the fluid velocity and energy density
\begin{equation}
u^{(1,0)}_{\alpha} \equiv  \left( u^{(1,0)}_{A},  0 \right)\,,~~~~
\Omega^{(1,0)} \equiv \Omega^{(1,0)} (t,r)\,,
\end{equation}
where $\Omega \equiv \ln \rho $. We obtain
\begin{eqnarray}
\pounds_{\xi (0,1)} u_A^{(1,0)} & = & \hat{\xi}^{(0,1)~B} u_{A \mid
 B}^{(1,0)} + u_B^{(1,0)} \hat{\xi}^{(0,1)~B}_{\mid A} \,, \\
 \pounds_{\xi (0,1)} u_{a}^{(1,0)} & = & u_A^{(1,0)}
 \hat{\xi}^{(0,1)~A} Y_a \,, \\
\pounds_{\xi (0,1)} \Omega^{(1,0)} & =  &\hat{\xi}^{(0,1)~A}
 \Omega_{\mid A} \,.
\end{eqnarray}
Therefore the gauge transformation (\ref{ha11_B})
for the ${\cal H}^{(1,1)}$ are
\begin{eqnarray}
\widetilde{\alpha} ^{(1,1)} & = & \alpha ^{(1,1)} +  u^{(1,0)}_A
\hat{\xi}^{A} - \met^{(1,0)}_{A B}\hat{\xi}^{A}  \bar{u}^B \,, \label{gtal}  \\
\widetilde{\gamma} ^{(1,1)} & = &  \gamma ^{(1,1)} + \bar{n}^A
\left\{\hat{\xi}^{B} \
u^{(1,0)}_{A \mid B} + \left( u^{(1,0)  B} -  \met^{(1,0)  B C}
 \bar{u}_C \right) \hat{\xi}_{B \mid A} - \met^{(1,0)  B C}
 \bar{u}_{A \mid B}  \hat{\xi}_{C}  \right. \nn \\
&& \left. - \Gamma^{(1,0)  B}_{AC}
 \hat{\xi}^{C}  \bar{u}_B \right\} \,, \label{gtga}  \\
\widetilde{\omega} ^{(1,1)} & = & \omega ^{(1,1)} + \hat{\xi}^A
\Omega^{(1,0)}_{\mid A} - \met^{(1,0)  A B}   \hat{\xi}_A
\bar\Omega_B \,, \label{gtom}
\end{eqnarray}
In addition, we can notice that in the
equations~(\ref{exal})-(\ref{exom}) the first order non-radial terms
$J^{(0,1)}_\sigma$ are given by the metric quantities
$h_{AB}^{(0,1)}$,$p_{A}^{(0,1)}$ and the fluid velocity $\delta
u_{A}^{(0,1)}$. The behaviour of $h_{AB}^{(0,1)}$ and $p_{A}^{(0,1)}$
under the gauge transformation (\ref{gaf01}) is expressed by the
equations~(\ref{hAB01_tr}) and~(\ref{pA01_tr}), while the velocity
perturbation changes as follows:
\begin{equation}
\widetilde{\delta} u_{A}^{(0,1)} = \delta u_{A}^{(0,1)} + \pounds_{\xi
(0,1)} u_A^{(1,0)} = \delta u_{A}^{(0,1)} +
\hat{\xi}^{(0,1)~B} \bar{u}_{A \mid B} +
\bar{u}_B \hat{\xi}^{(0,1)~B}_{\mid A} \, . \label{u01_tr}
\end{equation}

We can now bring the transformations~(\ref{gtal})-(\ref{gtom}),
(\ref{hAB01_tr}), (\ref{pA01_tr}) and~(\ref{u01_tr}) into the gauge
transformation~(\ref{agi11_red}), then find the gauge invariance of
these non-linear quantities.

Finally, we discuss the fluid perturbation $H^{(1)}$~(\ref{En_11}).
Its perturbative expansion provides the following expressions at first
and second order:
\begin{eqnarray}
H^{(0,1)} & = & {\cal H}^{(0,1)} = \frac{\bar{c}_s^2
\bar{\rho}}{\bar \rho + \bar p}\omega^{(0,1)} \,, \\
H^{(1,1)} & = & {\cal H}^{(1,1)} + \sum_\sigma {\cal I}^{(1,0)}_\sigma
{\cal} J^{(0,1)}_\sigma \,,
\end{eqnarray}
where we identify
\begin{eqnarray}
{\cal H}^{(1,1)} & = & \frac{\bar{c}_s^2   \bar{\rho}}{\bar \rho + \bar p}
\omega^{(1,1)} \,, \\
{\cal I}^{(1,0)}_\sigma & = & \left[ \bar c_s^2 + \bar \rho
\left( \frac{d\bar c_s^2}{d\bar\rho} - \left(1 + \bar{c}_s^2 \right)
\frac{\bar{c}_s^2}{\bar \rho + \bar p}\right)\right]
\frac{\bar{\rho}}{\bar \rho + \bar p}     \omega^{(1,0)} \,, \\
J^{(0,1)} & = & \omega^{(0,1)} \,.
\end{eqnarray}
Hence, the gauge-invariant character of $H^{(0,1)}$ and $H^{(1,1)}$
for a gauge transformation~(\ref{agi11_red}) with a fixed radial gauge
follows from the gauge invariance of $\omega^{(0,1)}$ and
$\omega^{(1,1)}$, previously proved (see equation~(\ref{exom})).

\chapter{Numerical Simulations}
\label{sec:NumInt}

This chapter is dedicated to the description of the numerical code
that simulates the non-linear dynamics arising from the coupling
between radial and non-radial perturbations.  In particular, this
numerical analysis is focused on the axial non-radial oscillations, as
for the polar non-radial perturbations the implementation of the code is still
under way.  The linear and non-linear perturbative equations form a
hierarchical boundary initial value problem, where initial values for
the linear radial and non-radial perturbations can be independently
set up.  The two independent initial configurations that we can
investigate are given by: \emph{i)} a radially pulsating and
differentially rotating star, and \emph{ii)} the scattering of an
axial gravitational wave on a radially pulsating star.  The former
configuration is the more interesting. In fact, at first perturbative
order the axial non-radial perturbations of the fluid quantities do
not have any dynamical properties. The only matter perturbation is
given by the axial velocity that describes a stationary differential
rotation.  As a result, the star is not a source of gravitational
radiation.  This aspect changes radically at second order, when the
radial oscillations couple with the differential rotation.  Now, the
presence of this stationary axial velocity and the related frame
dragging allows the radial perturbations to exhibit their pulsating
character and drive the oscillations in the source terms of the second
order axial master equation. This non-linear oscillating dynamics then
produces an axial gravitational signal that as we will see in
section~\ref{sec:diff_rot} precisely mirrors at coupling order the
spectral properties of the radial perturbations. In addition, it is
worthwhile to remark that this is a first order effect with respect to
the axial velocity perturbation and that it is strictly related to its
differential character. In fact, it is well known that for a uniform
rotating star $\Omega=const$ the quasi-radial modes appear at second
perturbative order in~$\Omega$, when the deformation of the star is
taken into account by the perturbative analysis.
In section~\ref{sec:7Numer_Frame} we introduce the structure of the
numerical code, focusing on its hierarchy, implementation of the
numerical grids and analysis of characteristic curves. In addition, we
describe also the introduction of the ``tortoise fluid coordinate'',
which provides a more accurate description of the radial pulsations
near the stellar surface. The background solution is illustrated in
section~\ref{sec:Num_Bac}, while the radial and axial non-radial
perturbations are described respectively in sections~\ref{Lin_Rad_Num}
and section~\ref{Lin_Ax_Num}. In these sections, we write the
numerical methods for simulating their evolution and discuss in detail
the setting up of the initial configurations.  Finally,
section~\ref{sec:Simul-NLAx11} is dedicated to the non-linear
simulations for the description of the coupling between the radial and
axial non-radial perturbations, where we provide the numerical methods
and discuss the results.

\section{Numerical framework}
\label{sec:7Numer_Frame}

\begin{figure}[h]
\begin{center}
\includegraphics[scale=0.7]{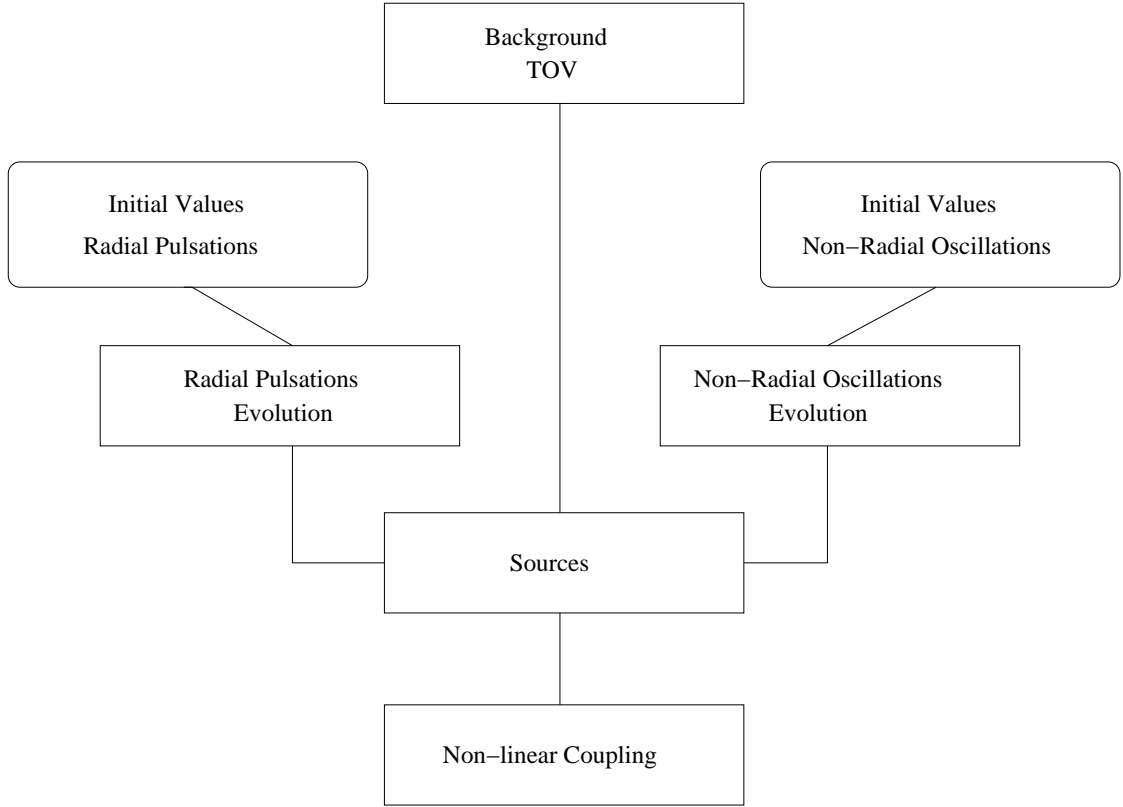}
\vspace{0.4cm} \caption{\label{Int_scheme} \small{Code hierarchy for
the time evolution of the non-linear axial perturbations arising
from the coupling between the radial and non-radial oscillations. The
initial configurations of the first order radial and non-radial
perturbations are independent.}}
\end{center}
\end{figure}

The numerical code for the evolution of non-linear oscillations in the
time domain reflects the hierarchy structure of the non-linear
perturbative theory. Starting from a solution of the TOV equations
that describes the equilibrium configuration of a nonrotating
spherical star, we must solve independently the two classes of first
order perturbations, i.e. the radial and the axial non-radial for an
arbitrary harmonic index $l$. These first order values are necessary
at any time step for updating the sources that drive the non-linear
oscillations. Eventually, the perturbative equations for the second
order perturbations can be integrated (see figure~\ref{Int_scheme}).

The separation in the perturbative fields of the angular dependence
from the other two coordinate, which derives from the 2+2 splitting of
the four dimensional spherical spacetime and the existence of a tensor
harmonics basis for the 2-sphere, leads to a 1+1 dimensional
problem.  One dimension is given by the spatial coordinate $r$ and the
other one by the coordinate time $t$.

The numerical simulations in this code are based on the ``finite
differencing method'', where derivatives are replaced with finite
difference approximations (see Appendix~\ref{sec:finit_appr}). As a
result, the differential equations become a set of algebraic equations
that can be solved with standard numerical methods for partial and
ordinary differential equations~\cite{1992nrfa.book.....P}.  In
particular, we adopt explicit numerical schemes with second order
approximations for the spatial derivatives and first or second order
approximations for the time derivatives.  The use of a first or second
order numerical algorithm in time will depend on the particular
properties of the perturbative equations.

The numerical stability and dissipation of the simulations will be
monitored with the $L2$-norms of the numerical solutions and its
accuracy with a convergence test, see Appendices~(\ref{sec:Norms})
and~(\ref{sec:Convtest}) respectively.

\subsubsection{Fluid tortoise coordinate transformation}
The radial perturbations can manifest a reduction in accuracy when the
speed of sound tends to vanish, near the stellar surface, where the
characteristic curves of the sound wave equations (\ref{ga10_WvEq})
and (\ref{eqzeta}) and of the radial perturbative system of equations
(\ref{eq:H10_ev}) and (\ref{eq:gam_t}), which are given by the
following expression:
\begin{equation}
  \xi_1 = r \pm v_s t \qquad \textrm{where} \qquad v_s = \bar c_s
  e^{\Phi-\Lambda}  \, ,
\end{equation}
lose their propagation character~(see figure \ref{fig:tort_coor})).
In practice, the numerical simulations carried out with a second order
scheme show that the convergence rate of the radial perturbative
solutions falls to one immediately after the wave packet touches the
surface. This behaviour has been well described by Sperhake in his PhD
thesis \cite{Sperhake:2001si}, here we report only the important
aspects necessary for our work.  A method for solving this problem is
given by a refinement of the numerical grid. However, since the
accuracy issues arises near the surface it is not necessary to increase
the resolution homogeneously on the whole grid. It would be more
appropriate to perform a coordinate transformation that, close to the
surface, simulates a high refined grid for the $r$ coordinate.  This
characteristics are satisfied by the ``tortoise fluid coordinate'',
which have been introduced by Ruoff for the analysis of stellar
non-radial perturbations where the matter is described by realistic
equation of state \cite{Ruoff:2000nj}.

\indent The new spatial variable $x$ in the  ``tortoise fluid coordinate'' is defined as follows:
\begin{equation}
dr = c_s \, dx \label{tort_fluid}
\end{equation}
where $c_s$ is the speed of sound.  The decreasing character of this
velocity~(figure~\ref{fig:TOV_variables}) and the definition
(\ref{tort_fluid}) imply that an evenly spaced grid with respect to
the new coordinate $x$ is able to simulate a grid for $r$ 
which will be more and more refined toward the surface.

This aspect is particularly important at second perturbative order,
where the perturbative equations contains source terms in the interior
spacetime. A coarse resolution in the low density regions near the
surface could produce some spurious oscillations in the radial
pulsations, which could be propagated into the exterior spacetime
through the junction conditions and then affect the gravitational
signal.

Therefore, we introduce the coordinate
transformation~(\ref{tort_fluid}) in the part of the code dedicated to
the radial pulsations and consequently to the TOV equations that
describe the hydrostatic equilibrium.  In practice, we must replace
the spatial derivatives with respect to the coordinate $r$ with the
following expression:
\begin{equation}
\partial _r = \frac{1}{c_s} \, \partial _x  \, .\label{torto_der}
\end{equation}
The new equations for the radial perturbations and for the background
are given in sections~\ref{sec:Num_Bac} and~\ref{Lin_Rad_Num}
respectively.  From the radial perturbative equations~(\ref{H10_ev_x})
and~(\ref{gam_t_x}) we can notice that with the tortoise coordinate
the velocity of the characteristic curves
\begin{equation}
  \tilde{\xi}_1 = x \pm \tilde{v}_s t \qquad \quad \textrm{where}  \qquad
  \tilde{v}_s = e^{\Phi-\Lambda} \,
\end{equation}
does not vanish on the surface.

\subsection{Numerical grids}

The fluid tortoise coordinate transformation concerns the radial
perturbative fields which are present in the interior spacetime only.
On the other hand, the linear and non-linear axial perturbations
are well described by the Schwarzshild coordinate $r$ on the
entire spacetime.
As a result these two perturbative families are defined on two distinct
integration domains that must be maintained separate in the
construction of the numerical code. These domains are both 1+1
dimensional, where one dimension is given by the time coordinate $t$
and the other by the spatial coordinate $x$ for the radial
perturbations and $r$ for the non-radial perturbations.

\indent The 2-dimensional and continuous evolution domain
$\mathcal{D}_r \subseteq \mathbb{R}^2$ for the non-radial
perturbations is discretized along the two dimensions $(t,r)$ with an
evenly spaced mesh that we call the $r$\emph{-grid}:
\begin{eqnarray}
r_{j} & = & r_{1} + j \Delta r \qquad \textrm{with} \qquad j =
0,1,..,J_r \, , \\    \label{r-grid}
t_{n} & = & t_{1} + n \Delta t \qquad \textrm{with}
\qquad n = 0,1,..,N_r \, ,
\end{eqnarray}
where the quantities $\Delta r$ and $\Delta t$ are the constant
spatial and time increments, and $J_r$ and $N_r$ denote the number of
grid points.  The background and perturbative fields of the
linear~$\mathcal{T}^{(0,1)}$ and non-linear~$\mathcal{T}^{(1,1)}$
axial oscillations are then approximated by a set of discrete
quantities evaluated on the points of the numerical grid
\begin{equation}
\left( \mathcal{T}^{(0,1)} \right)_{j}^{n} \equiv \mathcal{T}^{(0,1)} \left( t_{n} , r_{j} \right) \, , \qquad \qquad
\left( \mathcal{T}^{(1,1)} \right)_{j}^{n} \equiv \mathcal{T}^{(1,1)} \left( t_{n} , r_{j} \right)  \, ,  \label{disc_rapr_01}
\end{equation}
where the upper index $n$ denotes the time level and the lower index
$j$ the spatial mesh point. We have shown in
equation~(\ref{disc_rapr_01}) only the perturbative fields, for
background quantities the discrete approximation is obviously similar.

\noindent The 2-dimensional and evolution domain $\mathcal{D}_x
\subseteq \mathbb{R}^2$ for the radial perturbations is instead
discretized along the two dimensions $(t,x)$ with an evenly spaced
mesh, the~$x$\emph{-grid}:
\begin{eqnarray}
x_{j} & = & x_{1} + j \Delta x \qquad \textrm{with} \qquad j =
0,1,..,J_x \, , \\
\tilde{t}_{n} & = & \tilde{t}_{1} + n \Delta \tilde{t} \qquad \textrm{with}
\qquad n = 0,1,..,N_x \, ,
\end{eqnarray}
where $\Delta x$ and $\Delta \tilde{t}$ are now the constant
increments for the tortoise fluid coordinate and the time, which is in
general different from the representation given in
the~$r$\emph{-grid}. As before, the integers $J_x$ and $N_x$ denote
the grid dimensions. The radial perturbative fields~$\mathcal{T}^{(1,0)}$ are then
discretized on the~$x$\emph{-grid} as follows:
\begin{equation}
\left( \mathcal{T}^{(1,0)} \right)_{j}^{n} \equiv \mathcal{T}^{(1,0)} \left( t_{n} , x_{j} \right) \, .
\end{equation}
From the definition~(\ref{tort_fluid}) of the tortoise coordinate,
we can see that the radial simulations carried out on the~$x$\emph{-grid},
can also be mapped on this new representation of the coordinate
$\tilde{r}$:
\begin{eqnarray}
\tilde{r}_{j} & = & \tilde{r}_{1} + j \Delta
\tilde{r}
\quad \textrm{where} \qquad \Delta \tilde{r} = \bar c_s \Delta x
\qquad \textrm{and} \qquad j \in \mathbb{N} \, ,  \label{tilde_r}
\end{eqnarray}
which for a polytropic equation of state has an increasing resolution
torwards the surface.

\subsubsection{Interpolation}
\begin{figure}[ht]
\begin{center}
\includegraphics[scale=0.7]{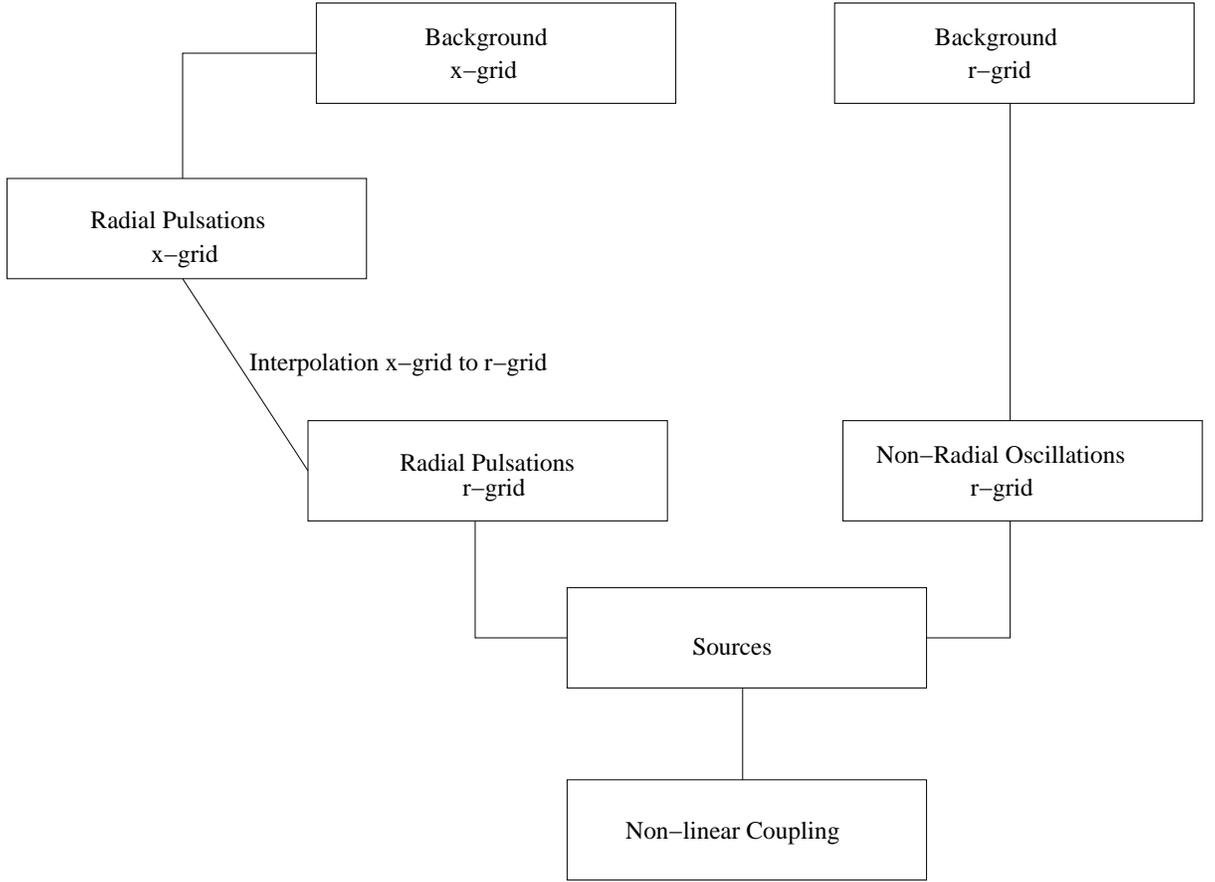}
\vspace{0.4cm}
\caption{\label{Int_scheme_new} Integration scheme of the numerical
code which simulates the coupling between the radial and non-radial
perturbations by using two different one-dimensional grids and an
interpolation procedure.}
\end{center}
\end{figure}

The implementation of the two grids in the numerical code is shown in
figure~\ref{Int_scheme_new}. The TOV solutions for the equilibrium
configuration are discretized on both the $x$- and $r$\emph{-grids}.
The simulations of linear radial perturbations are carried out on the
$x$\emph{-grid} while the linear non-radial on the $r$\emph{-grid}.
Therefore, in order to provide the source terms quantities
evaluated at the same spatial mesh points, the radial perturbations are
interpolated on the evenly spaced~$r$\emph{-grid}. In particular, this
procedure is constructed between the radial quantities determined in
the non-homegeous representation $\tilde{r}$~(\ref{tilde_r}) of the
spatial coordinate $r$ and the $r$\emph{-grid}. With the updated
values of the source terms the non-linear simulations can be then
carried out. More precisely, due to the implementation of explicit
numerical schemes the evaluation of the non-linear perturbations at
the $n+1$ time slice will rely on the source determined at the
previous time slice $n$. This property can be seen directly from the
discretization schemes given in the following sections.  In addition,
the interpolation displays more accurate results when the $x$\emph{-grid}
dimension is twice the~$r$\emph{-grid}, namely $J_x = 2 J_r$.

In the next section, we discuss the Courant-Friedrichs-Levy condition
and show how to set up numerical simulations on the two grids adopting
the same Eulerian time.

\subsection{Characteristic curves and Courant-Friedrichs-Levy condition}
\label{subsec:CFL}

A discrete representation of the spacetime must satisfy the
Courant-Friedrichs-Levy (CFL) condition which is a necessary but not
sufficient condition for the stability of the code. In order to well
describe the evolution of a physical system and its causal structure,
the numerical domain of dependence must include the physical domain of
dependence. This implies that the physical velocity $v$ of the system
has to be lower than the numerical velocity, i.e.
\begin{equation}
v \le \frac{\Delta r}{\Delta t} \, . \label{CFLcond}
\end{equation}
This equation can be also used to determine the maximum time increment
allowed for a given velocity and a spatial resolution,
\begin{equation}
\Delta t _{max} = \frac{\Delta r}{v} \, .
\end{equation}
In this code, we have set up the time and space grid increments in
order to: \emph{i)} satisfy the CFL condition~(\ref{CFLcond}) and
\emph{ii)} have the same discrete representation of the Eulerian time
coordinate, namely $\Delta t = \Delta \tilde{t}$. This latter
requirement allows us to couple, in the source terms, the first order
perturbations which are evaluated at the same instants of time.

The physical velocity~$v_{gw}$ of the axial gravitational waves can be
determined by the characteristic curves of the master
equations~(\ref{Psi01maseq}) and~(\ref{Psi11maseq}). For the radial
perturbations, which are evolved on the $x$\emph{-grid}, the sound
velocity~ $v_{s}$ is given by the system of
equations~(\ref{H10_ev_x})-(\ref{gam_t_x}).  They read
\begin{eqnarray}
\tilde v_{s} & = & e^{\Phi - \Lambda} \qquad \textrm{for} \quad x \in [\, 0,R_{s}^{x}]
\, ,\\
v_{gw} & = & e^{\Phi - \Lambda} \qquad \textrm{for} \quad r \in
[ \, 0,\infty ) \, ,
\end{eqnarray}
where the radius of the star with respect to the $x$ coordinate has
been defined by the quantity~$R_{s}^{x}$.  The two numerical
velocities are formally the same, but $v_{s}$ is defined on the
$x$\emph{-grid} and is present only in the interior spacetime, while
$v_{gw}$ is the velocity of the gravitational wave in the Schwarzshild
coordinate and is defined in the whole spacetime.
\begin{figure}[t]
\begin{center}
\includegraphics[width=145 mm, height=95 mm]{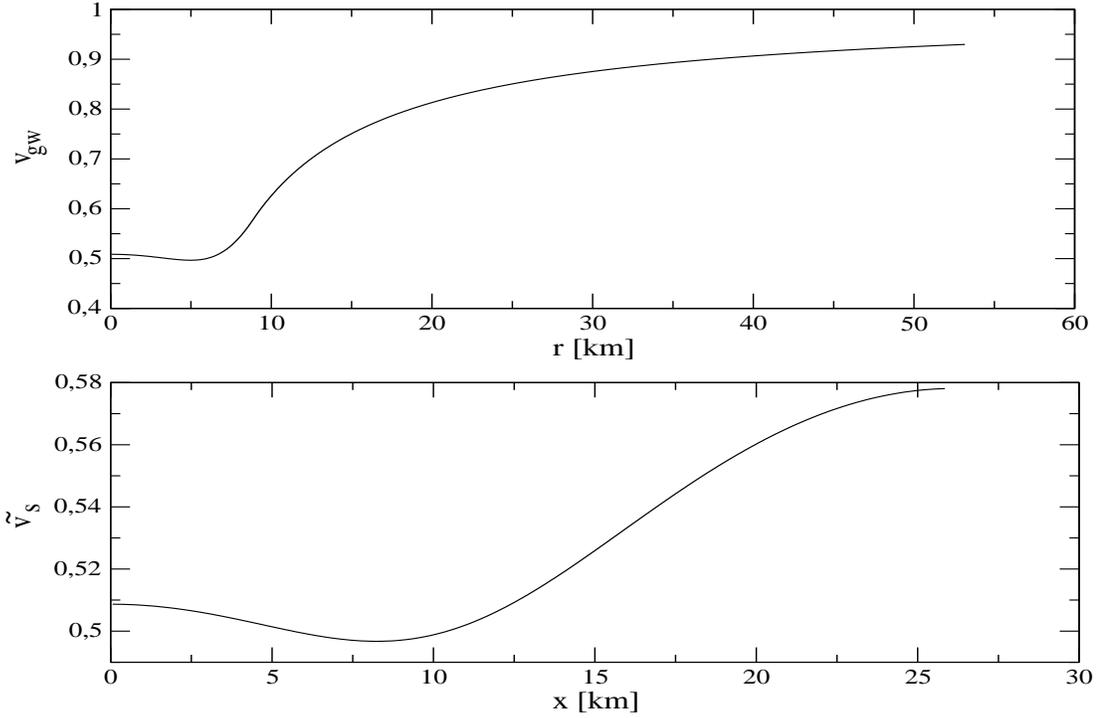}
\caption{\label{fig:Vch_s} \small{Characteristic curve velocity profiles
$v_{gw}$ for the master axial equations~(\ref{Psi01maseq})
and~(\ref{Psi11maseq}), and~$\tilde v_{s}$ for the radial
hyperbolic system of equations~(\ref{H10_ev_x})-(\ref{gam_t_x}).
The radial propagation velocity is plotted in $x$-coordinate.
The stellar radius is at~$R_s = 8.862~km$ in the~$r$-coordinate
and is at~$R_s^x = 25.80~km$ in the~$x$-coordinate.}}
\end{center}
\end{figure}
The profile of these velocities for our equilibrium stellar model is
plotted in figure~\ref{fig:Vch_s}, where the stellar radius is
at~$R_s = 8.862~km$ in the $r$\emph{-grid} and is mapped by
the tortoise coordinate transformation at~$R_s^x = 25.80~km$
in the $x$\emph{-grid},~(see section~\ref{sec:Num_Bac}).  If we
define the numerical velocities of the~$r$- and~$x$\emph{-grids}
respectively as follows:
\begin{equation}
v_{r} = \frac{\Delta r}{\Delta t} \, , \qquad \qquad \qquad v_{x} = \frac{\Delta
x}{\Delta \tilde{t}} \, ,
\end{equation}
the CFL condition~(\ref{CFLcond}) implies that
\begin{equation}
v_{s} < v_{x}     \qquad \qquad \qquad v_{gw} < v_{r}   \, .
\end{equation}
Therefore, in order to have the same Eulerian description of the time
coordinate in the two grids the following relation between the
numerical velocities must be imposed:
\begin{equation}
v_{x} = v_{r} \frac{\Delta r}{\Delta x} \, . \label{CfX}
\end{equation}

We set up in this code the two numerical $r$- and~$x$\emph{-grids} for
the interior spacetime, where the~$x$\emph{-grid} dimension is twice
the~$r$\emph{-grid}, $J_x = 2 J_r$.  The coarse dimension for
the~$r$\emph{-grid} is $J_r = 200$, which for the stellar model
adopted in this thesis leads to the spatial increment ~$\D r
= 0.04431~km$. This choice leads to an~$x$\emph{-grid} with a dimension
$J_x = 400$ and spatial increment~$\D x = 0.0646~km$.  The CFL
condition will certainly be satisfied if we fix the numerical velocity
of the ~$r$\emph{-grid} as $v_r = 0.99$. In fact, the corresponding
value for the $x$\emph{-grid} velocity is~$v_{x} = 0.67905$, which has
been determined from equation~(\ref{CfX}).  This value is higher
than the physical velocity~$\tilde v_{s}$ shown in
figure~\ref{fig:Vch_s} for the stellar model adopted in this
work. The same properties are valid also for higher resolution for
the~$x$\emph{-grid} as long as the relation $J_x = 2 J_r$ is
maintained between the two meshes.

In the exterior spacetime only the $r$\emph{-grid} is present which
conserves the spatial and time steps of the internal mesh.

\section{Background}  \label{sec:Num_Bac}

The background spacetime is represented by a perfect fluid,
spherically symmetric relativistic star in hydrostatic equilibrium.
The TOV equations~(\ref{Phi_r})-(\ref{M_r}) and an equation of state,
which describes the properties of the stellar matter, form a closed
system of equations, which can be integrated by specifying the central
density of the star.  As explained in section~\ref{sec:4_Back}, in
this work we will consider a polytropic equation of
state~(\ref{Poli_EOS}),
\begin{equation}
p = k \, \rho ^{\Gamma} \label{polyEOS2} \, .
\end{equation}
During the numerical integration of the TOV equations, we have
found a slight improvement in the rate of convergence by
adopting the metric function~$\Lambda$ as a new independent variable
instead of the mass function~$M$.
Furthermore, the presence of two numerical meshes in the code requires
an integration of the TOV equations in both the $r$- and $x$\emph{-grids}.

The equilibrium configuration is then the solution of the
following system of equations:
\begin{eqnarray}
\widetilde{\mathbf{D}} \Lambda  & = & \left( 4 \pi r \, \rho \, e^{2 \,
             \Lambda} + \frac{1- e^{2 \, \Lambda} }{2r} \right) \, , \label{Lam_r_tov}
             \\
\widetilde{\mathbf{D}} p & = & - \left( \rho + p \right) \left( 4 \pi r \,
             p \, e^{2 \, \Lambda} - \frac{1- e^{2 \, \Lambda} }{2r} \right)
             \, , \\
\widetilde{\mathbf{D}} \Phi & = & - \frac{ \widetilde{\mathbf{D}} p }{ \rho
             + p } \, , \label{Phi_x}\\
\widetilde{\mathbf{D}} r & = & 1  \, , \label{r_x} \\
p & = & K \, \rho ^{\,
\Gamma} \ , \label{polyEOS2_new}
\end{eqnarray}
where we have introduce the differential operator
$\widetilde{\mathbf{D}}$, which acts on a generic scalar function $f$
as follows:
\begin{equation}
 \widetilde{\mathbf{D}} f = \left\{ \begin{array}{ll} f_{, \, r} & \,
\, \textrm{~~for the}~~$r$\textrm{\emph{-grid} } \, ,
\\ \bar c_s ^{\, -1}
f_{, \, x} & \, \, \textrm{~~for the}~~$x$\textrm{\emph{-grid} } \, .
\label{D_ope}
\end{array} \right.
\end{equation}
This definition allows us to write the same expressions for the TOV
equations~(\ref{Lam_r_tov})-(\ref{r_x}) for the ``tortoise fluid
coordinate''~(\ref{tort_fluid}) as well as for the (area)
coordinate~$r$. By adopting this new system of equations the mass
function~$M$ can be determined in terms of the metric function
$\Lambda$ with its definition~(\ref{M_def}).  In case of an
integration carried out in the ``fluid tortoise coordinate'' frame, the
equation~(\ref{r_x}) shows that the radial coordinate $r$ is a scalar
field which is an unknown of the system as well as the other metric
and fluid variables.

\indent For specific values of the polytropic parameters $K$ and
$\Gamma$ in the EOS~(\ref{polyEOS2_new}), the numerical integration of
equations~(\ref{Lam_r_tov})-(\ref{polyEOS2_new}) can proceed from the stellar
origin $r=0$ outward as follows~\cite{Misner:1973cw}:
\begin{itemize}
\item[1)] Specification of the central mass-energy density $\rho_c$
          and consequently the corresponding value of pressure
          $p_{c}$ determined trough the EOS (\ref{polyEOS2_new}).

\item[2)] Imposition of the boundary conditions at the origin of the coordinates $r=0$
          for the metric variables $\Phi$ and $\Lambda$:
      \begin{equation}
            \Phi(0) = \Phi_{c}  \, , \qquad \qquad \Lambda(0) = 0 \,  . \label{Bac_Ori}
      \end{equation}
      The second condition in equation~(\ref{Bac_Ori}) can be
          derived from the definition~(\ref{M_def}), and from the
          behaviour $M \sim O(r^3)$ given by equation
          (\ref{M_r}). On the other hand, the choice of $\Phi_{c}$ is
          completely arbitrary. Due to the linearity of the
          ODE~(\ref{Phi_x}), its value can be later rescaled to the
          correct one in order to satisfy the matching condition on
          the stellar surface.

\item[4)] Integration of the
          equations~(\ref{Lam_r_tov})-(\ref{polyEOS2_new}) with a standard
          numerical method for ODEs~\cite{1992nrfa.book.....P}. The integration starts
          from the origin onward and ends when the pressure
          vanishes~$p(r) = 0$. This position of null pressure
          identifies the stellar surface and then the stellar radius
          $R_s$. Thus the value of the mass function in this
          point $M=M(R_s)$ is the total gravitational mass of the
          star.

\item[5)] Eventually, the function $\Phi$ can be rescaled with a
          constant value in order to match the Schwarzschild solution
          on the surface
      \begin{equation}
        \Phi(R_S) = - \Lambda(R_s) = \frac{1}{2}
            \ln \left( 1-\frac{2M}{R_s}\right)
        \label{Back_Srf_cond2} \, .
      \end{equation}
\end{itemize}
%
\begin{figure}[ht]
\begin{center}
\includegraphics[width=150 mm, height=95 mm]{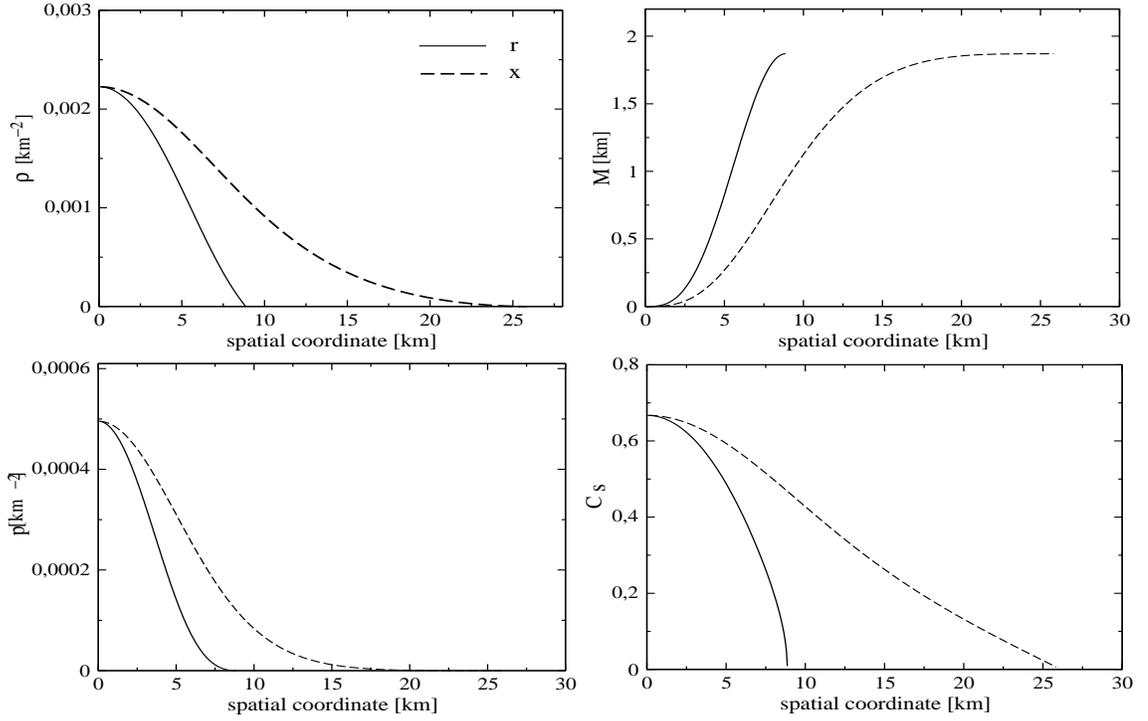}
\caption{\label{fig:TOV_variables} \small{For a polytropic
non-rotating star with indices $\Gamma = 2$, $k = 100$ km$^2$ and
with a central density~$\rho_c = 3\times 10^{15}~g cm^{-3}$, we plot
the spatial profile of (\emph{clockwise from top left}): the mass
energy density~$\rho$, mass function~$M$, speed of sound~$c _s$ and
the pressure~$p$ respectively.
The plots are against the $r$-coordinate (\emph{solid line}) and
$x$-coordinate (\emph{dashed line}). }}
\end{center}
\end{figure}
%
\noindent We consider a stellar model which is described by the
polytropic EOS~(\ref{polyEOS2_new}) with the following parameters
\begin{equation}
K = 100~km^{2} \qquad \qquad \Gamma = 2 \, .  \label{Pol_para}
\end{equation}
This choice allows us to determine a star with a mass and radius
similar to those obtained with realistic equation of state of average
stiffness. For a supranuclear central density given by $\rho_c =
3\times 10^{15}~g cm^{-3}$, the integration procedure carried out with
a fourth order Runge-Kutta (RK4) method~\cite{1992nrfa.book.....P}
provides a star with a radius of $R_s=8.862~km$ and a mass of
$M=1.869~km = 1.266~M_{\odot}$. When we use the $x$-fluid tortoise
coordinate, the stellar radius is mapped at $R_{x}=25.840~km$.  In
figure~\ref{fig:tort_coor}, one can notice the relation between the
$\tilde r$ and $x$ representation of the spatial coordinate $r$ and
the higher point density near the stellar surface. The equilibrium
configuration for all the fluid and metric quantities are plotted in
figures~(\ref{fig:TOV_variables}) and~(\ref{fig:tort_coor}).  In the
exterior the Schwarzschild solution is described only by the metric
functions~$\Phi$ and~$\Lambda$ that are written in terms of the total
gravitational mass of the star as follows:
      \begin{equation}
        \Phi(r) = - \Lambda(r) = \frac{1}{2}
            \ln \left( 1-\frac{2M}{r}\right)
        \label{Extern} \, ,
      \end{equation}
where their values on the stellar surface are given by the junction
condition~(\ref{Back_Srf_cond2}).
%
\begin{figure}[t]
\begin{center}
\includegraphics[width=150 mm, height=100 mm]{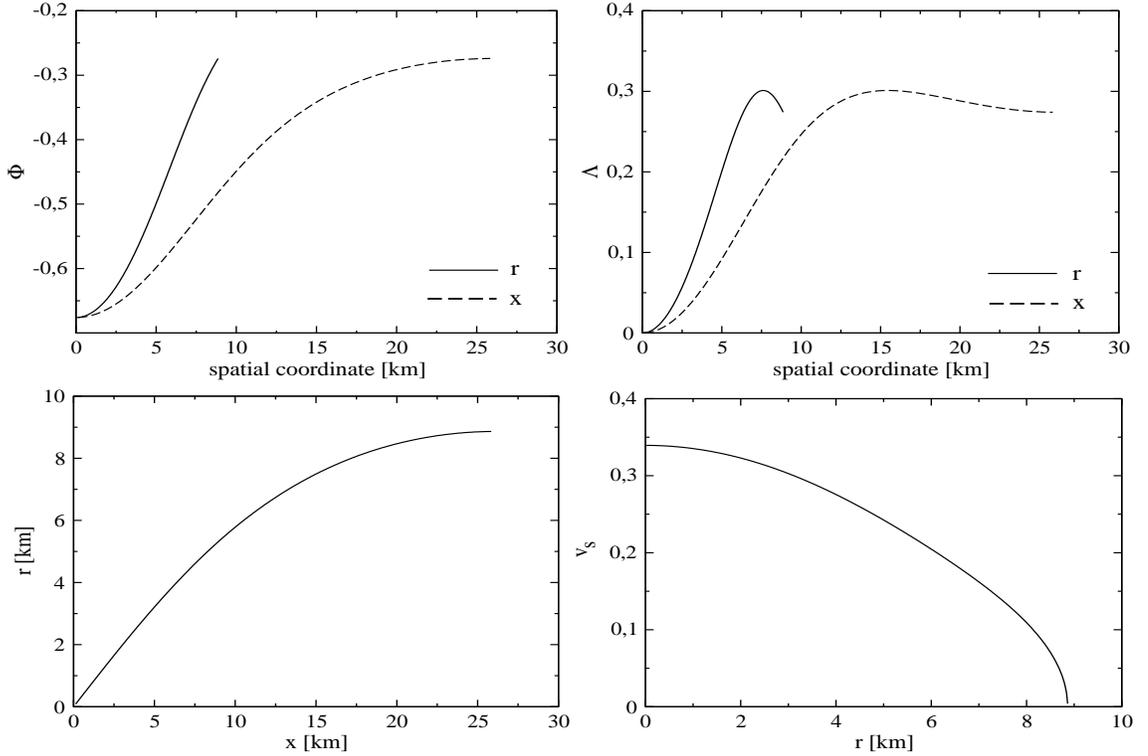}
\caption{\label{fig:tort_coor} \small{For the same stellar model
adopted in figure~\ref{fig:TOV_variables}, the metric
quantities~$\Phi$ and~$\Lambda$ are shown in the two upper plots,
where the solutions determined in the $r$\emph{-grid} are denoted with
the \emph{solid lines} while the \emph{dashed lines} are relative to
the~$x$\emph{-grid} solutions.  In the lower figures, the relation
between the Schwarzschild $r$-coordinate and the fluid tortoise
$x$-coordinate is shown on the left, while the sound velocity~$v_s$
is on the right.}}
\end{center}
\end{figure}
%

Finally, we have determined the rate of convergence of the TOV
solutions with the method described in appendix~\ref{sec:Convtest} by
using three grids of 200,400, and 800 points. The results are shown
in table~\ref{tab:TOV_conv}, where $\sigma_r$ and $\sigma_x$ are the
convergence factor for solutions determined in
the~$r$-~and~$x$-coordinates respectively.  The fourth order convergence
is found in accordance with the numerical scheme used (RK4).
%
\begin{table}[!t]
\begin{center}
\begin{tabular}{|c|c c c c c c|}
\hline
 Convergence & m       & p       & $\rho$  & $\bar c_s $ & $\Phi $  &  $\Lambda $  \\
\hline \hline
$\sigma_x$   & 4.017   & 4.007   & 4.007   & 4.007       & 4.046    & 4.031         \\
$\sigma_r$   & 4.055   & 4.000   & 3.997   & 3.995       & 3.987    & 4.055         \\
\hline
\end{tabular}
\caption[]{\label{tab:TOV_conv} \small{Convergence rate of the TOV
solutions for the stellar model considered in this thesis, where
$\sigma_x$ and $\sigma_r$ are calculated for the $x$- and
$r$\emph{-grids} respectively. The convergence factor has been
calculated for a set of three grids of dimension~(200,400,800).
The results confirm the fourth order convergence expected
by the accuracy of the RK4 method.}}
\end{center}
\end{table}

\section{Linear radial pulsations}
\label{Lin_Rad_Num}

The time-evolution of the linear radial pulsations of a static star
can be studied with the system of equations
(\ref{eq:H10_ev})-(\ref{eq:S10_cn}) for the four variables
$S^{(1,0)},\eta^{(1,0)}, \gamma^{(1,0)} , H^{(1,0)}$, which has been
presented in section (\ref{sec:Linear_Rad_pert_anal_Eqs}).  The
presence of hyperbolic and elliptic partial differential equations in
this system allows us to choose two integration approaches:
\begin{itemize}
\item[1)] a purely hyperbolic formulation (PHF), where we solve the three
evolution equations (\ref{eq:H10_ev}), (\ref{eq:gam_t}) and
(\ref{chi_t}) and we monitor numerical errors in time by looking at
the Hamiltonian constraint (\ref{eq:S10_cn}).
\item[2)] A hyperbolic-elliptic system of equations (HEF), where we
integrate Eqs.~(\ref{eq:H10_ev}) and (\ref{eq:gam_t}) and then 
solve the Hamiltonian constraint (\ref{eq:S10_cn}) for the metric
variables $S^{(1,0)}$. In this case the constraint will be satisfied 
by construction.
\end{itemize}
As shown in figure~\ref{Int_scheme_new}, the radial perturbations
are first integrated on the $x$\emph{-grid}. These solutions are then
interpolated on the $r$\emph{-grid} at any time slice.  In terms of
the fluid tortoise coordinate $x$ defined in
equation~(\ref{tort_fluid}) the system of perturbative
equations~(\ref{eq:H10_ev})-(\ref{eq:eta_cn}) becomes the following:
\begin{eqnarray}
\! \! \! \! \! \! \! \!  H_{, \, t}^{(1,0)}  & = & - \bar{c}_s\, e^{\Phi -\La} \,
          \ga^{(1,0)}_{, \, x} - \bar{c}_s^2 \,
      \left[\left(1-\frac{1}{\bar{c}_s^2}\right) \left(4 \pi \bar p \, r
        + \frac{m}{r^2}  \right) + \frac{2}{r} \, e^{-2 \La}  \right. {} \nn \\
        {} & & -  \left.  4\pi
        \left(\bar\rho + \bar p \right) r \, \right]\, e^{\Phi + \La} \,
      \ga^{(1,0)} \,,   \label{H10_ev_x}  \\
\! \! \! \! \! \! \! \! \ga_{, \, t}^{(1,0)} & = & - \frac{e^{\Phi-\La}}{\bar c_s}
          \, H^{(1,0)}_{,\,x} -
      4\pi \left(\bar\rho + \bar p\right)\,r \, e^{\Phi + \La}  \,H^{(1,0)} -
      \left( 4 \pi \bar p \, r^2 + \frac{1}{2} \right)  \, e^{\Phi + \La} S^{(1,0)} ,
      \label{gam_t_x} \\
\! \! \! \! \! \! \! \! S_{, \, t}^{(1,0)}  & = &  -8 \pi \left(\bar\rho + \bar p
          \right)\, e ^{\Phi + \La} \, \ga^{(1,0)}\,, \label{S_t_x} \\
\! \! \! \! \! \! \! \! \eta^{(1,0)}_{, \, x} & = & \bar c_s \, 4\pi (\bar\rho+\bar p)
         \, r \, \left[r\,S^{(1,0)}+\left(1+ \frac{1}{\bar{c}^2_s}\right)H^{(1,0)}
       \right]  \, e^{2 \La} \, , \label{eta_cn_x}
\end{eqnarray}
 and the Hamiltonian constraint~(\ref{eq:S10_cn}) is given by:
\begin{equation}
\label{S10_cn_x}
S^{(1,0)}_{, \, x} =  \bar{c}_s \left( 8 \pi \bar\rho r -\frac{2}{r} +
\frac{2 m}{r^2} \right)e^{2 \La} \, S^{(1,0)}+8 \pi
\frac{\bar\rho + \bar p}{\bar c_s} \,
 e^{2 \La} \,  H^{(1,0)}   \, . \label{Ham10_x}
\end{equation}

\subsection{Numerical algorithm}

The two evolution equations~(\ref{H10_ev_x}) and~(\ref{gam_t_x}) for
the enthalpy $H^{(1,0)}$ and the radial velocity $\ga ^{(1,0)}$
perturbations can be solved with various PDE numerical
algorithms~\cite{1992nrfa.book.....P}. We have used a two-step
McCormack algorithm (see appendix~\ref{sec:Num_meth}), which
implements a predictor and corrector step and provides second order
approximation in space and time.  For the \emph{predictor} step the finite
difference approximation of equations~(\ref{eq:H10_ev}) and
(\ref{eq:gam_t}) reads:
\begin{eqnarray}
\widetilde H_{j}^{n+1} & = &  H_{j}^n - e^{\Phi_j-\Lambda_j} \, (\bar c_{s})_{ \, j} \, \frac{\D t}{\D x} \,
\left(  \ga_{j}^n -  \ga_{j-1}^n \right)
         + \D t  \, (b_{1})_{j} \,  \frac{ \ga_{j-1}^n  + \ga_{j}^n }{2}     \, ,   \\ \nn \\
\widetilde \ga_{j}^{n+1} & = &  \ga_{j}^n - e^{\Phi_j-\Lambda_j}  \, \frac{\D t}{\D x} \,
         \frac{H_{j}^n -  H_{j-1}^n }{(\bar c_{s})_{j} }
         + \D t  \, (b_{2})_{ j} \,  \frac{ H_{j-1}^n  + H_{j}^n }{2}    \nn \\
         && + \D t  \, (b_{3})_{j}  \, \frac{ S_{j-1}^n  + S_{j}^n }{2}  \, .
\end{eqnarray}
The values found, which have been denoted with a tilde, are now used
in the \emph{corrector} step as follows:
\begin{eqnarray}
H_{j}^{n+1} & = &  H_{j}^n - e^{\Phi_j-\Lambda_j} \, (\bar c_{s})_{j} \, \frac{\D t}{\D x} \,
          \frac{  \ga_{ j}^n -  \ga_{ j-1}^n +
          \widetilde \ga_{ j+1}^{n+1} -  \widetilde \ga_{ j}^{n+1} }{2 }  \nn \\
        &&  + \D t  \, (b_{1})_{j} \,
            \frac{  \ga_{ j}^n  + \ga_{ j-1}^n
            + \widetilde \ga_{ j+1}^{n+1}   + \widetilde \ga_{ j}^{n+1} }{4} \, ,\\  \nn \\
\ga_{ j}^{n+1} & = & \ga_{ j}^n - \frac{e^{\Phi_j-\Lambda_j}}{ (\bar c_{s})_{j} } \,
         \frac{\D t}{\D x } \,
             \frac{  H_{j}^n - H_{j-1}^n
             + \widetilde H_{j+1}^n - \widetilde H_{j}^n  }{2}   \nn \\
        &&  + \D t  \, (b_{2})_{j} \, \frac{ H_{j}^n  + H_{j-1}^n
            + \widetilde H_{j+1}^{n+1}   + \widetilde H_{j}^{n+1} }{4}        \\
        &&  + \D t  \, (b_{3})_{j} \, \frac{ S_{j}^n  + S_{j-1}^n
            + \widetilde S_{j+1}^{n+1}   + \widetilde S_{j}^{n+1} }{4}    \, .
\end{eqnarray}
The coefficients $(b_{1})_{j}, (b_{2})_{j}$ and $(b_{3})_{j}$ are
the discrete approximations in $r_j$ of the following background
quantities:
\begin{eqnarray}
b_1 & = & - \bar{c}_s^2 \,
\left[\left(1-\frac{1}{\bar{c}_s^2}\right) \left(4 \pi \bar p \, r
+ \frac{m}{r^2}  \right) + \frac{2}{r} \, e^{-2 \La}   -   4\pi
\left(\bar\rho + \bar p \right) \right]\, e^{\Phi + \La} \, ,\\
b_2 & = & -
      4\pi \left(\bar\rho + \bar p\right)\,r \, e^{\Phi + \La} \, , \\
b_3 & = & -
      \left( 4 \pi \bar p \, r^2 + \frac{1}{2} \right)  \, e^{\Phi + \La} \, .
\end{eqnarray}

The metric variable $S^{(1,0)}$ can be updated at every time--step
with one of the two equations~(\ref{S_t_x}) and~(\ref{Ham10_x}).  In a
purely hyperbolic formulation (PHF) the variable is evolved by the
equation~(\ref{S_t_x}), which can be solved with an up--wind
method~\cite{1992nrfa.book.....P} where we have introduced a second
order approximation in space.  The numerical algorithm is then given
by
\begin{equation}
S^{n+1}_j  = S^n_j + \D t \, (b_4)_{j} \, \frac{\ga ^n_j + \ga ^{n}_{j+1} }{2} \, ,
\end{equation}
where the coefficient $b_{4}$ is
\begin{equation}
b_{4} = -8 \pi \left(\bar\rho + \bar p
          \right)\, e ^{\Phi + \La}  \, .
\end{equation}
Alternatively, in the hyperbolic--elliptic formulation (HEF)
discussed above we can integrate the Hamiltonian
constraint~(\ref{Ham10_x}) as an ODE at any time step.  The
equation~(\ref{Ham10_x}) can be discretized by using a second
order finite approximations in space and then written as a
tridiagonal linear system.  We can then use a standard
$\mathbf{LU}$ decomposition and a ``tridiagonal
subroutine''~\cite{1992nrfa.book.....P} for getting the value of
$S^{(1,0)}$.
The components of the
$\mathbf{LU}$ decomposition of equation~(\ref{Ham10_x}) are
given by the following expressions:
\begin{equation}
a_j S_{j-1}^n + b_j S_{j}^n + c_j S_{j+1}^n = f_j  \,
\end{equation}
where the coefficients $a, b, c$, and~$f$ are
\begin{eqnarray}
a & = &   - \frac{e^{-2 \Lambda}}{ 2 } \frac{1}{\D x } \, , \\
b & = &   - \bar{c}_s \left( 8 \pi \bar\rho r -\frac{2}{r} +
\frac{2 m}{r^2} \right)e^{2 \La} \, , \\
c & = &   \frac{e^{-2 \Lambda}}{ 2 } \frac{1}{\D x }  \, ,\\
f & = & 8 \pi\frac{ \bar\rho + \bar p }{c_s}  \, H^{(1,0)} \, .
\end{eqnarray}

The other metric perturbation $\eta^{(1,0)}$ can be solved by
integrating equation~(\ref{eta_cn_x}). At any time slice, this
equation is a two point boundary value problem that must satisfy the
condition~(\ref{eta_bc}) at the origin and vanishes on the surface. We
use a \emph{shooting method}, where the integration is carried out
from the origin to the surface by specifying an initial guess
for~$\eta^{(1,0)}$ at origin. The algorithm will shoot the solution up
to the surface and control whether the surface condition is
satisfied. If not, the routine will correct the value of the variable
at the origin and repeats the operation until the surface condition is
fulfilled.

\subsection{Boundary conditions}

The condition at the origin for the radial perturbations are given by 
equations~(\ref{bc_rad_in})-(\ref{bc_rad_ori}).  We implement a
grid where the first point of the spatial coordinate $r_1$ is not at
the origin $r_0 = 0$ but at $r_1 = r_0 + \D r$, where $\D r$ is the
spatial grid step. The boundary condition at the origin of the
perturbative variables is then implemented at $r_1$ by using the
behaviours~(\ref{bc_rad_in})-(\ref{bc_rad_ori}), which have been
obtained with a Taylor expansion. This procedure is particularly
useful in the axial sector~\cite{Ruoff:2000nj} for avoiding numerical
instability which can be generated by the presence of the term
$l(l+1)r^{-2}$ in the potential of the axial master
equation~\cite{Ruoff:2000nj}.  This method for the radial quantities
$\gamma ^{(1,0)}$ and $S^{(1,0)}$ gives
\begin{eqnarray}
\gamma _{1}^{n}  =   \frac{r_1}{r_2} \, \gamma _{2}^{n}  \, , \qquad \qquad \qquad
S _{1}^{n} =  \frac{r_1}{r_2} \, S _{2}^{n} \, ,
\end{eqnarray}
where we have assumed that the
behaviours~(\ref{bc_rad_in})-(\ref{bc_rad_ori}) are valid for both
the first and second grid points $r_1$ and $r_2$.
For the enthalpy $H^{(1,0)}$ and the other metric perturbation $\eta ^{(1,0)}$
we implemented the following conditions:
\begin{eqnarray}
H_1^n  = \frac{4 H_2^n -H_3^n }{3} \, , \qquad \qquad \qquad
\eta_1^n  = \frac{4 \eta_2^n -\eta_3^n }{3} \, ,
\end{eqnarray}
which have been determined by spatial differentiation of 
equations~(\ref{H10_bc}) and~(\ref{eta_bc}), and using the second
order finite one-sided approximation~(\ref{2Dplus}) for the first
order derivative. This condition for~$\eta^{(1,0)}$ can be also
derived by the constraint~(\ref{eta_cn_x}).

At the surface the vanishing of the Lagrangian perturbation of the
pressure leads to equation~(\ref{Surf_bc_gam10}) for the
variable~$\gamma ^{(1,0)}$.  The stellar matter in our model is
described by a polytropic EOS, where pressure, mass energy density and
speed of sound vanish on the surface. Therefore, the
condition~(\ref{Surf_bc_gam10}) is certainly satisfied if the
velocity~$\gamma ^{(1,0)}$ and its spatial derivative~$\gamma
^{(1,0)}_{, \, r}$ are finite. According to the fluid
tortoise transformation~(\ref{torto_der}) the spatial derivative
in the tortoise coordinate is related to the derivative in $r$ by the
following expression:
\begin{equation}
\gamma ^{(1,0)}_{, \, x} = \bar c _{s} \gamma ^{(1,0)}_{, \, r} \, .
\end{equation}
Hence, on the stellar surface we can impose the vanishing of $\gamma
^{(1,0)}_{, \, x}$ which also ensures the finiteness of $\gamma
^{(1,0)}$ and $\gamma ^{(1,0)}_{, \, r}$~\cite{Ruoff:2000nj}.  The
second order finite one-sided approximation~(\ref{2Dminus}) of the
first order spatial derivative leads to the following expression:
\begin{equation}
\gamma _{J_x}^{n}  = \frac{4 \gamma_{Jx-1}^n -\gamma_{J_x-2}^n }{3} \, .
\end{equation}
The other values for the enthalpy and the metric variable $S^{(1,0)}$ can
be directly determined by the perturbative equations~(\ref{H10_ev_x}) and~(\ref{Ham10_x}), while
for $\eta^{(1,0)}$ we can easily implement its trivial condition, i.e. $\eta^{(1,0)} = 0$.
The resulting expressions are the following:
\begin{eqnarray}
S _{J_x}^{n}  & = &  \frac{4 S_{J_x-1}^{n} - S_{J_x-2}^{n} }{3}    \, ,  \label{S10_sf_bc}\\
H_{J_x}^{n}  & = &  H_{J_x}^{n-1} + \D t \, (b_5)_{J_x}  \ga ^{n}_{J_s} \, , \\
\eta_{J_s}^n   & = &  0 \, .
\end{eqnarray}
where we have used the finite difference approximation~(\ref{2Dminus})
also for the condition~(\ref{S10_sf_bc}). The coefficient $b_5$ is
so defined:
\begin{equation}
b_5 = - \left( 4 \pi \bar p r + \frac{m}{r^2} \right) e^{\Phi+\Lambda}  \, .
\end{equation}

%
\begin{figure}[t]
\begin{center}
\includegraphics[width=150 mm, height=100 mm]{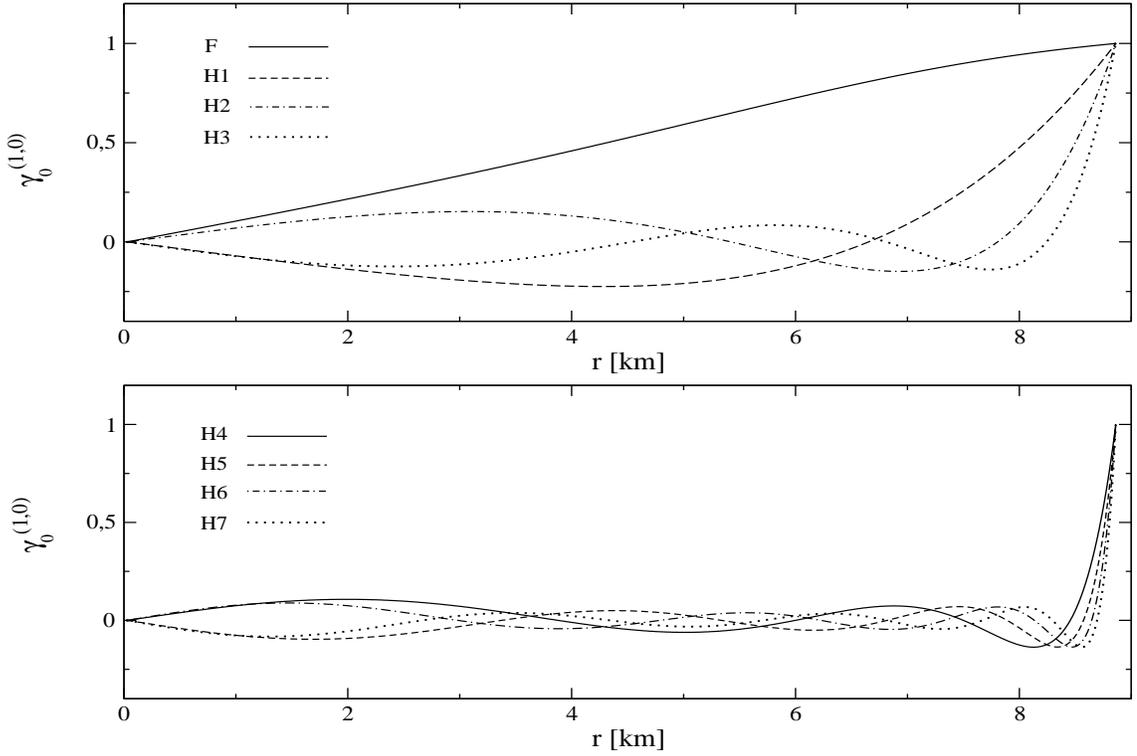} 
\caption{\label{fig:rad_Eigenf_1} \small{Eigenfunctions of the radial
perturbation~$\gamma ^{(1,0)}$.  The \emph{upper panel} displays the
eigenfunctions of the fundamental mode and the first three
overtones, while the \emph{lower panel} from the fourth to the seventh
overtones.}}
\end{center}
\end{figure}
%

\subsection{Initial configuration for radial pulsations}
\label{sec:IV_Rad}
The time--domain integration of the radial perturbative equations
(\ref{H10_ev_x})-(\ref{eta_cn_x}) requires the choice of initial
values for the variables $\gamma^{(0,1)}$, $H^{(0,1)}$, $S^{(0,1)}$
and $\eta^{(0,1)}$ on a Cauchy surface. However, the presence of two
constraint equations implies that we can independently specify two
perturbations.  In addition, the radial perturbations must satisfy the
boundary conditions on the initial slice as well as along the time
evolution.
%
These conditions are certainly satisfied if we specify a profile only
for the radial velocity perturbation $\gamma^{(1,0)}$ and set to zero
the enthalpy perturbation $H^{(0,1)}$, as the constraint equations
(\ref{eta_cn_x}) and (\ref{Ham10_x}) and boundary conditions imply the
vanishing also of the two metric perturbations $S^{(0,1)}$ and
$\eta^{(0,1)}$.  In the case of a single eigenfunction, this
corresponds to a choice of the origin of time. Indeed, as we can see
from the radial perturbative
equations~(\ref{H10_ev_x})-(\ref{eta_cn_x}), we can consistently
choose a normal mode oscillation with eigenfunction $\omega_n$ to have
the form:
\begin{eqnarray}
\gamma^{(1,0)} & = & \gamma^{(1,0)}_{n} (r) \cos \omega_n t \, , \qquad \qquad 
H^{(1,0)}  =  H^{(1,0)}_{n} (r) \sin \omega_n t \, , \nn \\  \label{Rad_Ic} \\
S^{(1,0)} & = & S^{(1,0)}_{n} (r) \sin \omega_n t  \, , \, \,  \, \qquad \qquad 
\eta^{(1,0)}   =   \eta ^{(1,0)}_{n} (r) \sin \omega_n t \, , \nn 
\end{eqnarray}
thus at $t=0$ only $\gamma^{(1,0)}$ is nonvanishing.
With this choice the Hamiltonian constraint~(\ref{Ham10_x}) is
initially satisfied by construction. This choice is also consistent in
the case of an initial Gaussian pulse, which can be considered as a
particular linear conbination of the eigenmodes~(\ref{Rad_Ic}).

The time evolution of radial pulsations of a spherically symmetric
perfect fluid star is completely determined by a complete set of
radial eigenfunctions.  In this work, we consider two different
configurations for the dynamical evolution, respectively described by:
\emph{i)} the presence of a broad range of normal mode frequencies,
\emph{ii)} the excitation of a few selected radial modes. The first
configuration can be accomplished by imposing for $\gamma^{(1,0)}$ an
initial Gaussian pulse so that many radial modes are excited at the
same time. This kind of initial data is mainly used to test and
calibrate the reliability of the code. On the other hand, the second
configuration can be determined by imposing as initial data the
eigenfunctions of~$\gamma^{(1,0)}$ in order to excite the desired
associated eigenfrequencies. \\
\indent The eigenvalue problem for the radial perturbation~$\gamma^{(1,0)}$
has been described in section~\ref{sec:Rad_freq}.  However, since the
simulations for the radial perturbations are carried out on the
$x$\emph{-grid} we have to introduce the tortoise fluid coordinate
also in the Sturm-Liouville problem~(\ref{eqyy_fr}). The system of two
ordinary differential equations~(\ref{yeq})-(\ref{zeq}) transforms as
follows:
\begin{eqnarray}
y_{,x}^{(1,0)}  & = & \bar{c}_{s} \frac{z}{\hat P}^{(1,0)} \, ,\label{yeq_x} \\
z_{,x}^{(1,0)}  & = & -  \bar{c}_{s} \left( \omega^2 \hat W + \hat Q \right) y ^{(1,0)}
\label{zeq_x}\, ,
\end{eqnarray}
and the functions $\hat W,\hat P,\hat Q$ are given by:
\begin{eqnarray}
 r^2  \hat W  & \equiv & \left( \bar \rho + \bar p  \right) \,
e^{3 \Lambda  + \Phi }, \\
 r^2   \hat P &  \equiv & \left( \bar \rho + \bar p  \right) \, \bar c _{s}^{2}
\, \bar p  \, e^{ \Lambda + 3 \Phi} \, , \\
 r^2  \hat Q  & \equiv  & \left( \bar \rho + \bar p  \right) \,
  \left[ \frac{1}{ \bar{c}_{s}^{2}} \Phi _{,x} ^{2} - \frac{4}{r \bar{c}_{s}} \,
 \Phi_{,x} - 8 \pi
\, \bar p \, e^{2 \Lambda  } \right] \, e^{ \Lambda + 3 \Phi }  \, .
\end{eqnarray}
Now, the boundary conditions at the origin and surface are:
\begin{equation}
y_{0}  = \bar{c}_{s} \frac{z_{0}}{3 P} \, , \qquad \qquad
\qquad \left. (\bar\rho + \bar p)\,\bar{c}_s\, e^{-\Phi}
y^{(1,0)} _{,x} \right|_{r=R_x}= 0\,.
\end{equation}
The method used to integrate the radial eigenvalue system of equations
(\ref{yeq_x})-(\ref{zeq_x}) is the ``relaxation
method''~\cite{1992nrfa.book.....P}.
The first eight eigenfunctions for the radial velocity $\gamma
^{(1,0)}$ are plotted in figure~\ref{fig:rad_Eigenf_1} with respect to
the $r$ coordinate and after a normalization with the norm of their
maximum value.  These profiles have the characteristic node numbers
expected by the theory, i.e. absence of nodes for the F-mode, and a
node number equal to the order of the overtone.  The associated
eigenfrequencies are written in table~\ref{tab:Rad_modes}.  The first
three have been compared with published values obtained by Kokkotas
and Ruoff~\cite{Kokkotas:2000up} for our stellar model.  The results
are accurate to better than $0.2$ percent (see
table~\ref{tab:Rad_modes}).  The rate of convergence for the
eigenfrequencies and eigenfunctions is of second order, as expected 
by the accuracy of the numerical method.  
%
\begin{table}[!t]
\begin{center}
\begin{tabular}{ |c|c|c| c| c| }
\hline
 Normal mode  & Frequecy domain   & Time domain  & Kokkotas and Ruoff & Relative error \\
              &       [kHz]       & [kHz]        & [kHz]              &
\\
\hline \hline
    F        & 2.138              &  2.145       & 2.141              & 0.10 \%        \\
    H1       & 6.862              &  6.867       & 6.871              & 0.13 \%         \\
    H2       & 10.302             & 10.299       & 10.319             & 0.16 \%         \\
    H3       & 13.545             & 13.590       &                    & 0.33 \%         \\
    H4       & 16.706             & 16.737       &                    & 0.18 \%          \\
    H5       & 19.823             & 19.813       &                    & 0.05 \%          \\
    H6       & 22.914             & 22.889       &                    & 0.11 \%          \\
    H7       & 25.986             & 25.964       &                    & 0.08 \%          \\
\hline
\end{tabular}
\caption[]{\label{tab:Rad_modes} \small{The table shows the
eigenfrequencies of the first eight normal modes of the radial
perturbations, which have been determined with the Sturm-Liouville
problem (\emph{second column}) and with an FFT of the time evolution
(\emph{third column}).  The first three normal modes obtained in
frequency domain have been compared with those published by Kokkotas
and Ruoff~\cite{Kokkotas:2000up} (\emph{fourth column}), and the
relative error is given in the first three rows of the last
column. The remaining rows display the relative errors between the
frequency determined in the time and frequency domains.}}
\end{center}
\end{table}

\begin{figure}[t]
\begin{flushleft}
\includegraphics[width=150 mm, height=100 mm]{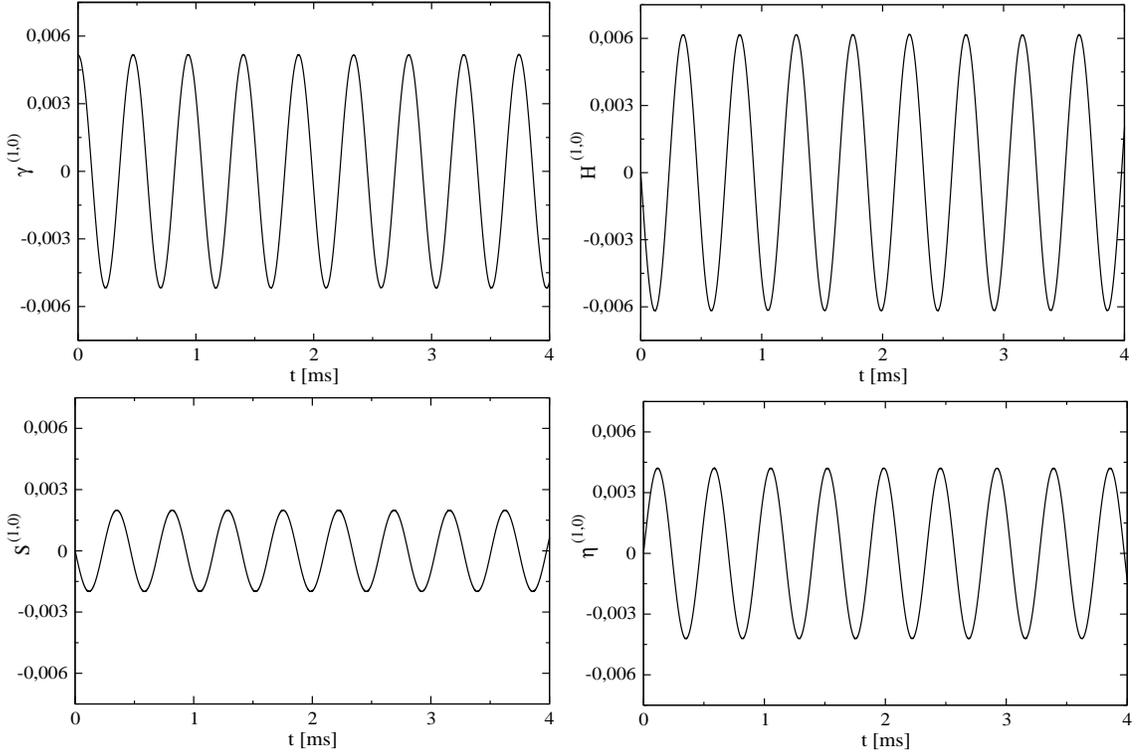}
\vspace{0.1cm}
\caption{\label{fig:rad_ev_FF} \small{Time evolution of the four
radial perturbations, where the oscillations have been excited with
the F-mode eigenfunction of the variable~$\gamma ^{(1,0)}$.
The quantities have been averaged in the interior spacetime by using
the definition~(\ref{aver_f}).
}}
\end{flushleft}
\end{figure}
%
%
\begin{figure}[t]
\begin{flushleft}
\includegraphics[width=150 mm, height=120 mm]{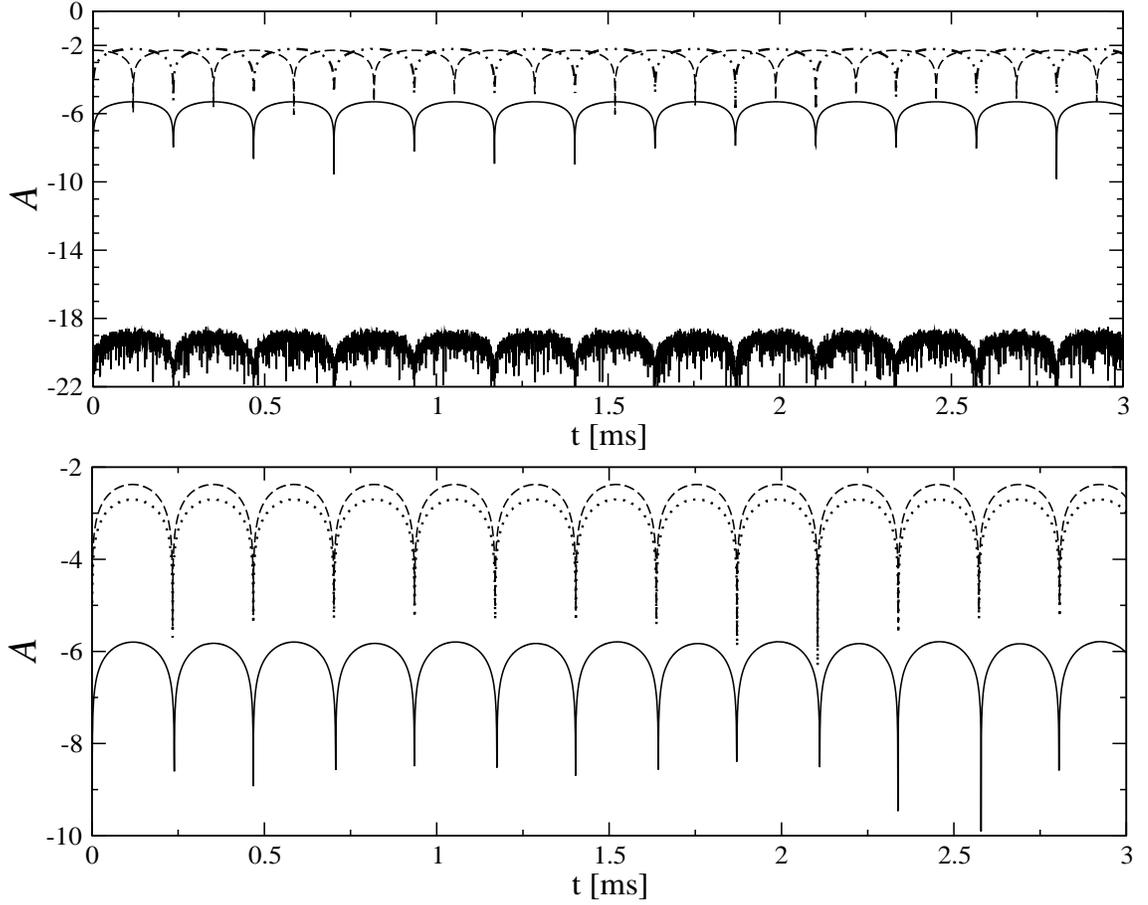}
\caption{\label{fig:rad_Ham10} \small{The fundamental mode
oscillations of the four radial perturbations are compared with the
numerical errors due to the violation of the Hamiltonian constraint.
The quantities are plotted in semilogarithmic scale after having
performed the spatial average as defined in equation~(\ref{aver_f}).
In the \emph{upper} and \emph{lower} panels are shown the results for a
HEF and PHF formulation respectively.  In both figures the curves relative to the
Hamiltonian constraints are represented in (\emph{solid lines}) while
the radial perturbations in \emph{dashed} and \emph{point-dashed
line}. For the details of the results, see the discussion in
section~\ref{sec:Ch7_rad_sim}. }}
\end{flushleft}
\end{figure}
%
\begin{figure}[t]
\begin{flushleft}
\includegraphics[width=150 mm, height=120 mm]{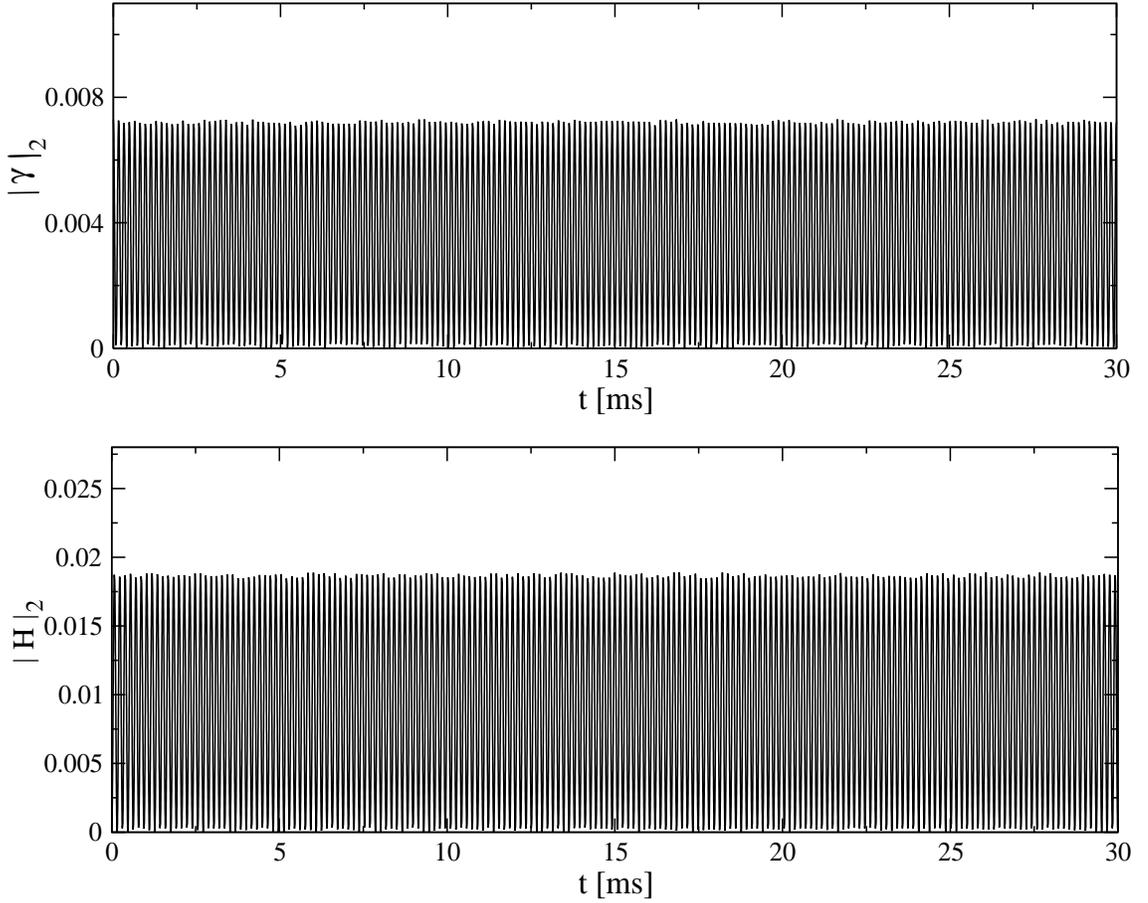}
\vspace{0.1cm}
\caption{\label{fig:rad_Norms} \small{Norms of the radial perturbations
$\ga^{(1,0)}$ and $H^{(1,0)}$ for a $30~ms$ simulation where
only the F-mode has been excited.}}
\end{flushleft}
\end{figure}

\begin{figure}[t]
\begin{flushleft}
\includegraphics[width=150 mm, height=100 mm]{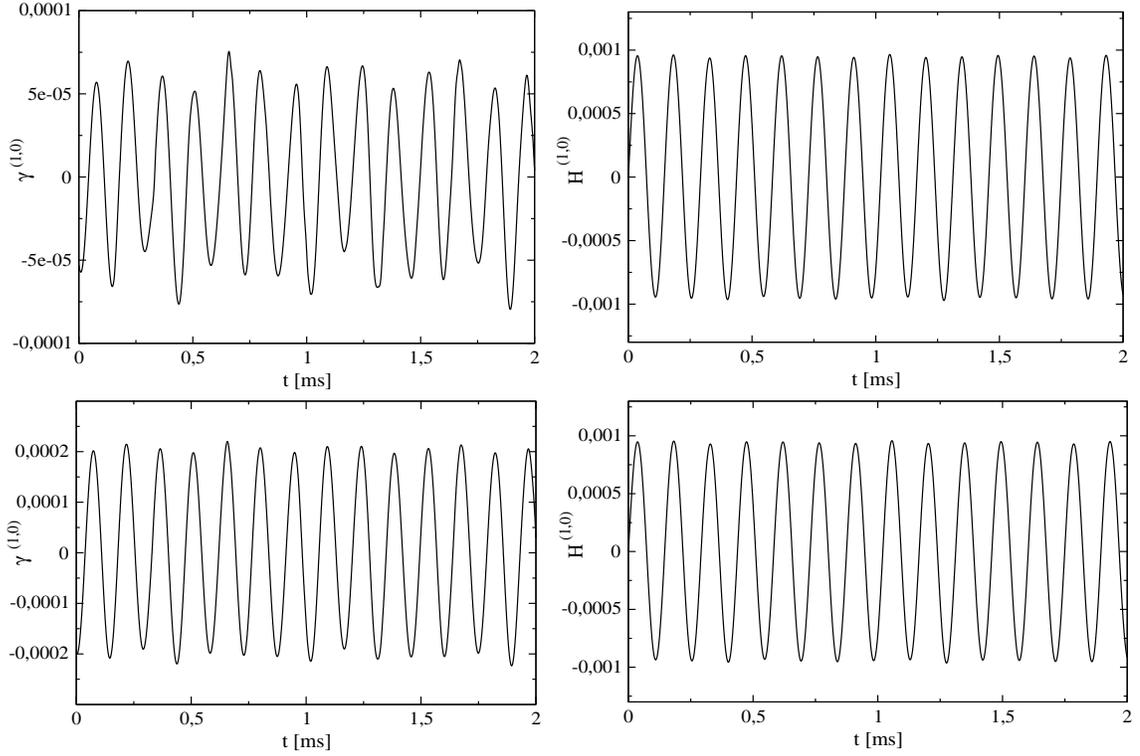} \qquad
\vspace{0.1cm}
\caption{\label{fig:rad_var_H1} \small{Time evolution of the spatial
average~(\ref{aver_f}) for the radial variables~$\ga^{(1,0)}$ and
$H^{(1,0)}$.  The pulsations have been excited with the eigenfunction of~$\ga^{(1,0)}$
associated with the first overtone.  On the top, the averages
have been calculated in the whole interior spacetime. On the bottom,
the three grid points near the surface have been neglected.}}
\end{flushleft}
\end{figure}
%
\begin{figure}[t]
\begin{flushleft}
\includegraphics[width=150 mm, height=120 mm]{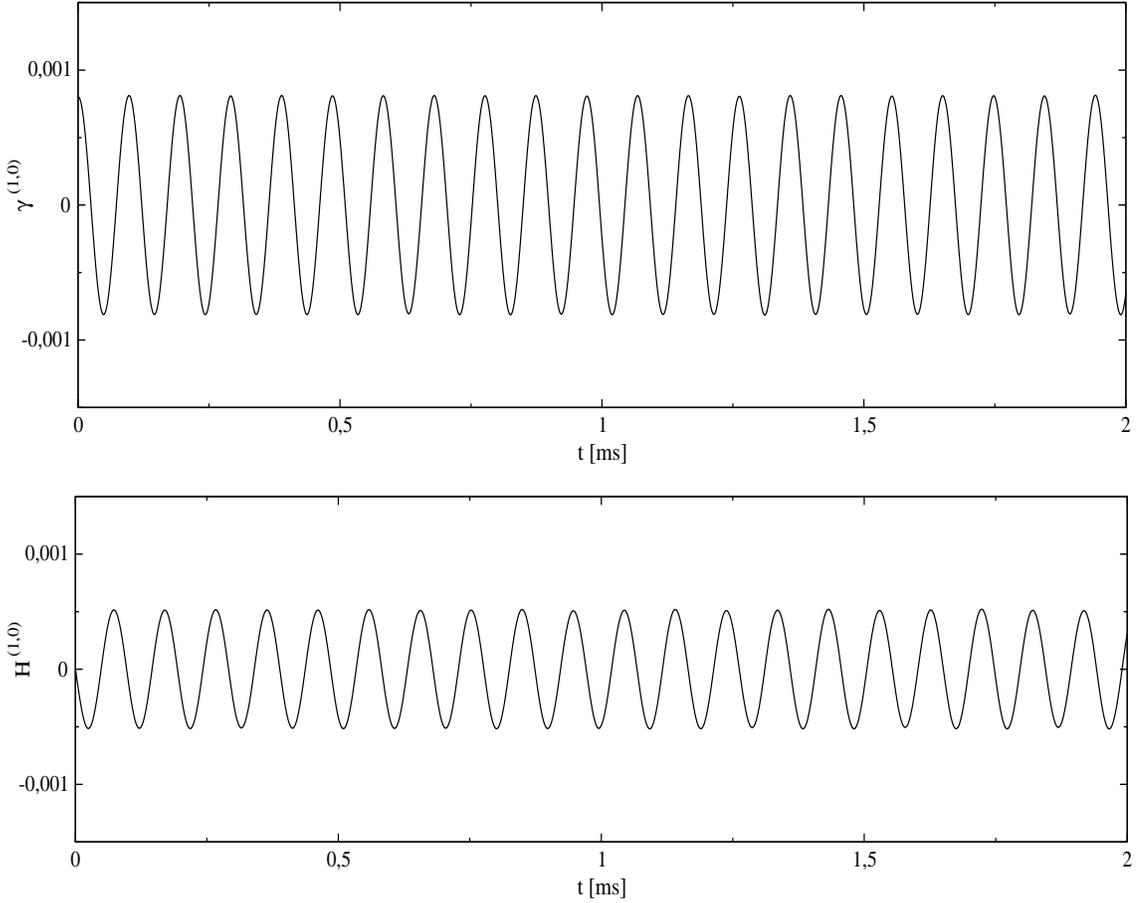}
\vspace{0.1cm}
\caption{\label{fig:rad_var_H2} \small{Time evolution of the spatial
average profiles of the radial variables~$\ga^{(1,0)}$ and
$H^{(1,0)}$. The radial pulsations have been excited with the eigenfunction of $\gamma^{(1,0)}$
associated with the second overtone.}}
\end{flushleft}
\end{figure}
%
\begin{figure}[t]
\begin{center}
\includegraphics[width=150 mm, height=120 mm]{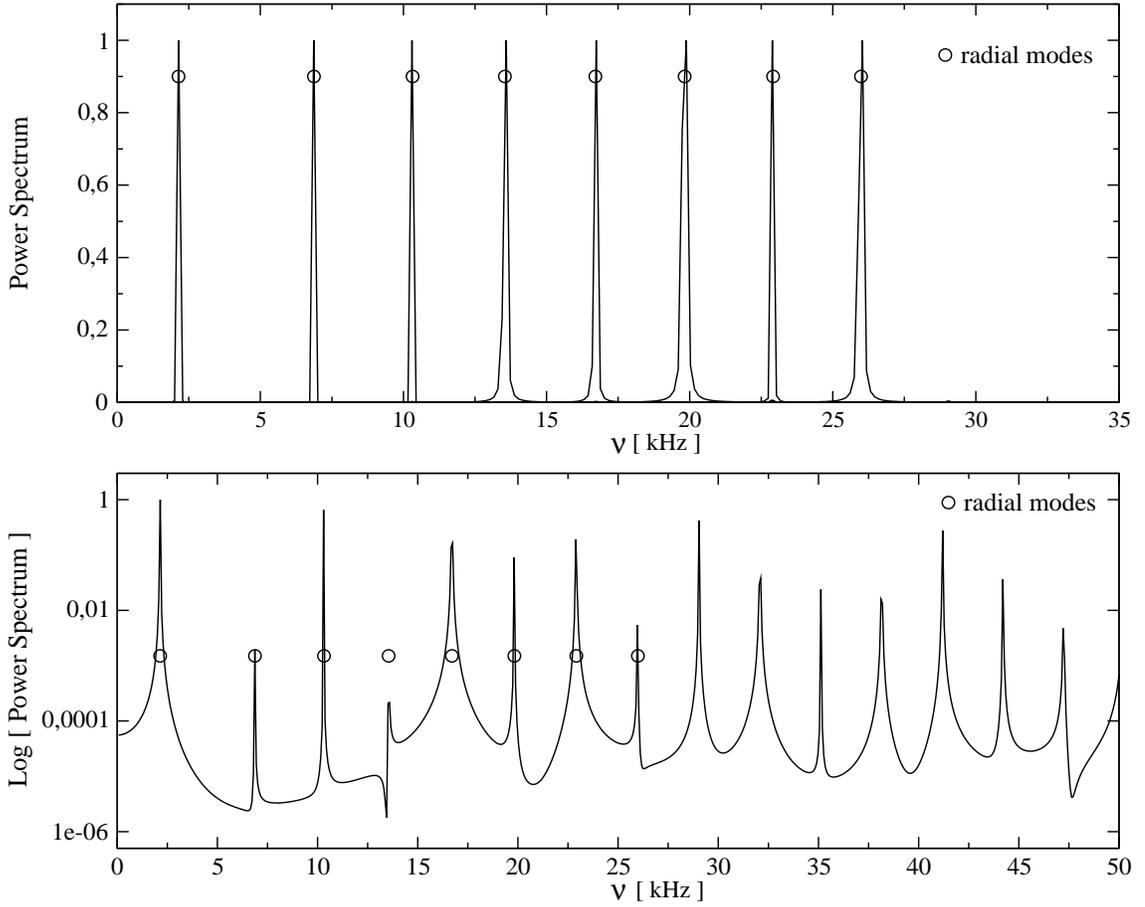}
\caption{\label{fig:Rad_Spec} \small{Power spectrum of the radial
perturbation $\gamma ^{(1,0)}$, which has been determined by an FFT of
the time profiles. In the \emph{upper panel} we show in the same plot
the spectrum of eight different time evolutions, where everytime a
single radial mode has been excited. In the \emph{lower panel},we show
the spectrum of a time evolution excited by an initial Gaussian pulse.
The excitation of the radial modes is evident in both cases. The first
eight frequencies are compared to the values determined with a code in
the frequency domain and are shown with a~\emph{circle}. }}
\end{center}
\end{figure}

\subsection{Simulations of radial perturbations}
\label{sec:Ch7_rad_sim}

The analysis of the radial oscillation part of the code starts with a pulsating
configuration described by selected radial modes.  The desired
oscillation frequency is excited by introducing the initial
condition (\ref{Rad_Ic}), where at $t=0$ the eigenfunction of the
velocity perturbation $\gamma ^{(1,0)}$ is given by
\begin{equation}
\gamma ^{(1,0)}_{in} = A^{(1,0)} \gamma ^{(1,0)}_{n} \label{gam_IC_nm}
\end{equation}
where we have introduced a constant factor~$A^{(1,0)}$ to control the
amplitude of the oscillations and~$\gamma ^{(1,0)}_{n}$ is one of the
normalized eigenfunctions plotted in figure~\ref{fig:rad_Eigenf_1}.
The first simulation is carried out for the fundamental mode with an
initial amplitude of $A^{(1,0)}=0.01$ and for an evolution time of
$4~ms$.  The $x$\emph{-grid} has dimension~$J_x = 400$, while the time
step is chosen as explained in section~\ref{subsec:CFL} in order to
satisfy the CFL condition and have the same Eulerian time discrete
representation of the $r$\emph{-grid}.  The evolution of the four
radial perturbations $\ga^{(1,0)},~H^{(1,0)},~S^{(1,0)}$
and~$\eta^{(1,0)}$ is shown in figure~\ref{fig:rad_ev_FF}, where
we have plotted the following average values determined at any time
step:
\begin{equation}
  <f> = \frac{1}{R_s} \int_0^{R_{s}} f dr \, , \label{aver_f}
\end{equation}
in order to have global information about the oscillation 
properties of the variables studied. The function~$f$ in
equation~(\ref{aver_f}) obviously represents one of the four radial
perturbations cited above. The results show the typical periodic
character of the adiabatic radial pulsations. This monochromatic
character is also confirmed by their spectra 
(figure~\ref{fig:Rad_Spec}) which have been determined with a Fast
Fourier Transformation (FFT) of the simulations.

In order to test the stability and possible numerical dissipative
effects, we perform a longer simulation of $30~ms$ and we monitor
the Hamiltonian constraint and determine the $L_2$ norms of the
variables under consideration.
In figure~\ref{fig:rad_Ham10}, we show the oscillation amplitude on
a logarithmic scale of the four radial perturbations, i.e
$\ga^{(1,0)},H^{(1,0)}, S^{(1,0)}, \eta^{(1,0)}$, and the
numerical errors due to the violation of the Hamiltonian constraint
for a hyperbolic-elliptic formulation~(HEF) (\emph{upper panel}) and a
purely hyperbolic formulation~(PHF) (\emph{lower panel}).  In the HEF,
the Hamiltonian constraint (\emph{solid lines}) remained bounded and
is three orders of magnitude lower than the amplitude of the radial
perturbations $\ga^{(1,0)}$ and~$H^{(1,0)}$, which are denoted with a
\emph{dashed} and \emph{point-dashed line} respectively.  In addition,
when we perform the average~(\ref{aver_f}) of the Hamiltonian
constraint by neglecting the first two grid points $r_1$ and $r_2$, we
find an appreciable reduction of the numerical errors.  As shown by
the \emph{lowest solid line} the numerical errors are less
than~$10^{-18}$. This accuracy is expected as the Hamiltonian
constraint in the HEF is solved by updating one of the radial
variables. However, the results show that the implementation of
boundary conditions at the centre of the star introduce some numerical
errors which is anyway three orders of magnitude lower than the radial
physical oscillations and is not propagated along the star. We perform
a similar analysis for the PHF~(lower panel of
figure~\ref{fig:rad_Ham10}) by comparing the two metric radial
perturbations $S^{(1,0)}$ and~$\eta^{(1,0)}$ (\emph{dashed} and
\emph{point-dashed line} respectively) with the numerical oscillations
due to the violations of the Hamiltonian constraints~(\emph{solid
line}). We can conclude that even for this case the constraint is well
satisfied.

The stability of the radial simulations is confirmed also by the
analysis of the $L_2$ norms. For a $30~ms$ evolution excited by the
eigenfunction of $\ga^{(1,0)}$ associated with the F-mode, the
$L_2$-norms for $\ga^{(1,0)}$ and $H^{(1,0)}$ show a constant
oscillatory character without any presence of dissipative effects (see
figure~\ref{fig:rad_Norms}).  The properties of the Hamiltonian
constraint violation and $L_2$-norms illustrated for the case of an
evolution dominated by the F-mode, remains valid also when the other
overtones are excited and the simulations preserve their stability
and absence of dissipation also for longer evolutions.

The simulations carried out with the excitation of the first overtone
(H1) show a behaviour near the surface that deserves some
attention. In figure~\ref{fig:rad_var_H1}, we plot the time profile of
the two fluid radial perturbations~$\ga^{(1,0)}$ and $H^{(1,0)}$. The
two upper panels have been obtained by performing the
average~(\ref{aver_f}) for the interior spacetime, while on the bottom
the last three grid points that are near the surface have been
neglected from the average operation. We can first notice that the
enthalpy~$H^{(1,0)}$ preserves the same oscillating properties while
the velocity~$\ga^{(1,0)}$ has a higher amplitude and a smoother
oscillating dynamics when the points near the surface are
neglected. By investigating the movie of these simulations for the
variable~$\ga^{(1,0)}$, we have noticed the presence of small spurious
oscillations near the surface. This numerical noise is not continuous
but has a random character, which is sufficient to modify the
evolution of this variable.  In order to have a better understanding
we have also analyzed the simulations for the second (H2) and third
overtones (H3). Similarly to the case of the F-mode, the spatial
average profiles of these time evolutions do not present any spurious
oscillations near the surface. For the second overtone, this behaviour
is shown in figure~\ref{fig:rad_var_H2}.  However, when we study the
corresponding movies we notice also in these cases the presence of
random and very small spurious oscillations near the surface that
decrease by increasing the resolution of the meshes.  From the
analysis of the movies and the eigenfunction profiles, we can argue
that for the H2 and H3 normal modes the presence of a node near the
surface seems to prevent the propagation of these oscillations along
the star and actually reduce their effects.  On the other hand for the
F-mode, the absence of micro-oscillations in the time profiles of
figure~\ref{fig:rad_ev_FF} seems more due to the smallness of these
numerical oscillations with respect to the average value of the
physical pulsations, about two order of magnitude less.  This
motivation is confirmed by the analysis of the non-linear simulations
when the F-mode will display this noise.
As a result,  the presence of these oscillations has to be taken into
account during the implementation of the matching conditions of the
non-linear perturbations.

The spectral properties of the oscillations are studied by performing
an FFT of the time profiles. In figure~\ref{fig:Rad_Spec} we show
the power spectrum of seven simulations where every time only one of
the first seven normal modes has been excited by the initial
configuration. Furthermore, we have performed a $12~ms$ simulation for
radial pulsations excited by an initial Gaussian pulse. The relative
spectrum is shown in the lower panel of figure~\ref{fig:Rad_Spec},
where the frequencies of the normal modes are labelled with a circle.
The radial modes are determined in time domain with an accuracy to
better than $0.3\%$, see table~\ref{tab:Rad_modes}.

Eventually, we determine the convergence rate for the radial evolution
which is given in table~\ref{tab:conv_rad} for the four radial
perturbations.
\begin{table}
\begin{center}
\begin{tabular}{|c|c c c c|}
\hline 
Convergence~ & $~\gamma^{(1,0)}~$ & $~H^{(1,0)}~$  & $~S^{(1,0)}~$  
& $~\eta^{(1,0)}~$   \\
\hline 
 $\sigma^{(1,0)}$   & 2.04   & 2.06   & 2.07   & 1.53 \\
\hline
\end{tabular}
\caption[]{Convergence test for radial perturbations:
         $\sigma^{(1,0)}$ denotes the convergence rate in the
         $L_2$ norm. \label{tab:conv_rad}}
\end{center}
\end{table}

The amount of pulsation energy contained in the radial
perturbations can be determined with the expressions of the
relativistic kinetic and potential energy derived with a variational
analysis \cite{Chandrasekhar:1964pr, Chandrasekhar:1964tc,
Meltzer:1966mt}, and~\cite{Bardeen:1966tm}.  In particular, according
to the initial conditions that we have set up for the radial
perturbations we can derive the pulsation energy introduced on the
initial Cauchy surface.  For an initial vanishing Lagrangian
displacement and a specific eigenfunction for the radial velocity
$\ga^{(1,0)}$, the initial pulsation energy is given by the kinetic
energy~\cite{Bardeen:1966tm}:
\begin{equation}
E^{(1,0)}_{n} = 2 \pi W \left( A^{(1,0)} \right)^2 \int_{0}^{R_s} dr \left( r^2
e^{-\Lambda} \gamma^{(1,0)}_{n} \right) ^2   \label{Kin_ener} \, ,
\end{equation}
where we have used the relation~(\ref{yydef}), and
$\gamma^{(1,0)}_{n}$ is an eigenfunction of $\ga^{(1,0)}$ and
$A^{(1,0)}$ its amplitude. In table~\ref{tab:TOV_Ener}, the
oscillation energy for the first seven normal modes has been
calculated with the expression~(\ref{Kin_ener}) for an amplitude
$A^{(1,0)}=0.001$.

\section{Linear, axial non-radial oscillations}
\label{Lin_Ax_Num}

The dynamics of axial perturbations on a static star is described by
the system of two differential equations: the odd-parity master
equation~(\ref{Psi01maseq}) for the axial master variable $\Psi
^{(0,1)}$ and the conservation equation~(\ref{traseq01}), which is
satisfied by the redefined axial velocity perturbation $\hat \beta
^{(0,1)}$.  In section (\ref{sec:Lin_Axial_per}) we have already
shown that the functional form of the conservation equation
~(\ref{traseq01}) allows us to discern the dynamical degree of freedom
of the axial gravitational spacetime from the stationary frame
dragging profile produced by the presence of an axial differential
rotation. Hence, the solution of the master
equation~(\ref{Psi01maseq}) can be decomposed in two
parts~(\ref{sol_dec}),
\begin{equation}
\Psi^{(0,1)}  = \Psi^{(0,1)}_{hom} + \Psi^{(0,1)}_{p} \, , \label{sol_dec_num}
\end{equation}
namely the solution $\Psi^{(0,1)}_{hom}$ of the homogeneous
equation~(\ref{Psi01_hom_eq}) plus a particular static solution
$\Psi^{(0,1)}_{p}$ of equation~(\ref{Psi01part}).

\subsection{Numerical algorithm}
\label{sec:Numer_alg_01}

The homogeneous equation~(\ref{Psi01_hom_eq}) is integrated by means
of a standard leapfrog method, which is an explicit second order and
three level method~\cite{1992nrfa.book.....P}.  By using a centred
finite differential approximation in space and time we can write the
discrete approximation of equation~(\ref{Psi01_hom_eq}) as follows:
\begin{equation}
- \frac{\psi^{n+1}_j - 2 \psi^{n}_j + \psi^{n-1}_j }{\D t^2} +  v^2_j \, \frac{\psi^{n}_{j+1} - 2 \psi^{n}_{j} + \psi^{n}_{j-1} }{\D r^2}
+ \alpha _{j} \, \frac{\psi^{n}_{j+1} -  \psi^{n}_{j-1} }{2 \D r}  + V_{j} \, \psi^n_j  = 0  \, .
\end{equation}
Thus, the value of~$\psi_{hom}^{(0,1)}$ is updated in time with the
following expression:
\begin{eqnarray}
\psi^{n+1}_j & = & 2 \psi^{n}_j - \psi^{n-1}_j  + \frac{\D t^2}{\D r^2} v^2_j \,
             \left( \psi^{n}_{j+1} - 2 \psi^{n}_{j} + \psi^{n}_{j-1} \right)
         + \alpha_j \, \D t^2  \, \frac{\psi^{n}_{j+1}
           -   \psi^{n}_{j-1} }{2 \D r}
         + V_j\, \D t^2  \, \psi^n_j   \, . \nn  \label{psi01_alg}\\
\end{eqnarray}
The coefficient~$v$ is the propagation speed of the wave and
$\alpha, V$ are coefficients depending only on the coordinate $r$:
\begin{eqnarray}
v & = & e^{\Phi-\Lambda}  \, , \\
\alpha & = & e^{2 \Phi} \left(\frac{2 M}{r^2} - 4 \pi \left( \bar \rho - \bar p \right) r \right)  \, , \\
V & = & e^{2 \Phi} \left( \frac{6 M}{r^3} - 4 \pi \left( \bar \rho - \bar p \right)
 - \frac{l(l+1)}{r^2} \right) \, .
\end{eqnarray}

The particular solution is obtained from 
equation~(\ref{Psi01part}), which is second-order discretized in space
and written as a tridiagonal linear system, which is then solved using
a standard $\mathbf{LU}$ decomposition~\cite{1992nrfa.book.....P}. The
components of the $\mathbf{LU}$ decomposition are given by
\begin{equation}
\hat a_j \psi_{j-1}^n + \hat b_j \psi_{j}^n + \hat c_j \psi_{j+1}^n = \hat f_j
\end{equation}
where the coefficients $\hat a_j, \hat b_j$, and~$\hat c_j$ are
\begin{eqnarray}
\hat a & = & \frac{1}{\D r^2 } - \left(\frac{2
       m}{r^2} - 4 \pi \left( \bar \rho - \bar p \right) r
       \right)\frac{e^{2 \Lambda}}{2 \D r } \, , \\
\hat b & = & V e^{-2 \Phi} - \frac{2}{\D r^2 } \, , \\
\hat c & = &
       \frac{1}{\D r^2 } + \left(\frac{2 m}{r^2} - 4
       \pi \left( \bar \rho - \bar p \right) r \right)
       \frac{e^{2 \Lambda}}{\D r } \, , \\
\hat f & = & 16 \pi \left( 4 \pi p r^2
       + \frac{m}{r} \right) e^{3 \Lambda} \hat \beta^{(0,1)} +
       16 \pi r e^{\Lambda} \hat \beta^{(0,1)} _{, \, r} \, .
\end{eqnarray}
The dragging of the inertial frame can be determined by using 
expression (\ref{frmdr-rel}) and the relation (\ref{mtr_form}), which
connects the particular solution $\Psi^{(0,1)}_{p}$ with the metric
perturbation components $k_0^{lm}$ and $h_{0} ^{lm}$.  Alternatively,
the metric variable $k_0^{lm}$ as well as the related frame dragging
$\omega^{lm}$ can also be determined directly by the ordinary 
differential equation (\ref{k0-fradr}). We can apply the same
procedure used for solving the particular solution
$\Psi^{(0,1)}_{p}$. In this case the $\mathbf{LU}$ decomposition
reads:
\begin{equation}
\tilde a_j k_{0, \, j-1}^n + \tilde b_j k_{0, \, j}^n + \tilde c_j k_{0, \, j+1}^n = \tilde f_j
\end{equation}
where the coefficients $\tilde a_j, \tilde b_j$, and~$\tilde c_j$ are
\begin{eqnarray}
\tilde a & = &  \frac{e^{- 2 \Lambda}}{\D r ^2 }
               + \frac{2 \pi r }{\D r } \left( \bar \rho + \bar p \right)  \, , \\
\tilde b & = &  8 \pi \left( \bar \rho + \bar p \right) - \frac{l (l+1)}{r^2}
               + \frac{4 m }{r^3} - \frac{2 e^{-2 \Lambda}}{\D r^2} \, , \\
\tilde c & = & \frac{e^{- 2 \Lambda}}{\D r ^2 }
               - \frac{2 \pi r }{\D r } \left( \bar \rho + \bar p \right)  \, , \\
\tilde f & = & 16 \pi e^{\Phi} \hat \beta^{(0,1)} \, .
\end{eqnarray}

\subsection{Boundary conditions and initial configuration}
\label{sec:BC-IV_Axial01}
The boundary conditions at the origin, stellar surface and at
infinity are implemented in accordance with the discussion given
in section~\ref{sec:Lin_Axial_per}. At the origin we impose the
condition~(\ref{BC01_orig}) at the first grid point $r_1 = \Delta
r$ as follows:
\begin{equation}
\Psi ^{(0,1)} _1 = \Psi ^{(0,1)} _2 \, \left( \frac{r_1}{r_2} \right) ^{l+1}\, .
\end{equation}
At infinity we use a standard outgoing wave condition.  We carry out
the integration of the wave equation~(\ref{Psi01_hom_eq}) for the
homogeneous solution and the ordinary equation~(\ref{Psi01part}) for
the particular solutions on the whole numerical grid without imposing
any condition at the stellar surface. We have found that the numerical
solutions satisfy the continuity of the function $\psi^{(0,1)}$ and
its first spatial derivatives $\psi^{(0,1)}$, which are the
requirements prescribed by the junction conditions.  For the two point
boundary value problems~(\ref{Psi01part}) and (\ref{k0-fradr}), we
have imposed at the centre and at infinity the conditions discussed in
section~\ref{sec:Lin_Axial_per}.

The system of equations~(\ref{Psi01maseq}) and~(\ref{traseq01})
requires the specification of three functions on the initial Cauchy
surface, namely
\begin{equation}
\label{IC_01}
{\cal I} ^{(0,1)}= \left( \Psi ^{(0,1)}, \Psi_{, \,t} ^{(0,1)},
\beta ^{(0,1)} \right) \ .
\end{equation}
In the initial data we specify the original axial perturbations
$\beta^{(0,1)}$ and we determine the function $\hat \beta ^{(0,1)}$
according to  equation~(\ref{hatbeta_def}).

We can set up two independent classes of initial conditions
for the first--order fields:
\begin{itemize}
\item[1)] a stationary differentially rotating star, where the amount of
differential rotation can be determined by specifying the profile
of the axial velocity $\beta^{(0,1)}$. The functional form of this
perturbation can be derived as we see later from one of the
theoretical rotation laws used in the literature. The profile of the
master function $\Psi^{(0,1)}$ is then obtained by numerically 
solving equation~(\ref{Psi01part}). Therefore, the set of
initial data is given by the following expression:
\begin{equation}
\label{IC_01_frmdr}
{\cal I} ^{(0,1)}_{1} = \left( 0,0,\beta ^{(0,1)}\right) \ ,
\end{equation}
\item[2)] In the second case, we may consider a non-rotating
star, $\beta^{(0,1)}=0$, and arbitrarily fix the metric perturbation
$\Psi_{{\rm hom}}$ and its time derivative $ \Psi_{{\rm hom},\, t}$,
\begin{equation}
\label{IC_01_sct}
{\cal I}^{(0,1)}_{2} =
\left( \Psi_{{\rm hom}, } ^{(0,1)}, \Psi_{{\rm hom},t} ^{(0,1)},0 \right) \ .
\end{equation}
This initial condition can be used to understand how the second-order
metric perturbations can be affected by the radial pulsations of the
star.  In fact, it is well known that the odd-parity grativational
wave signal arising from a non-rotating compact star can only contain
the imprint of the spacetime $w$-modes. The question is then what this
particular signal looks at second order for a star that is also
pulsating radially. In practice, we generate $w$-modes ringing at
first order and look at the corresponding signal at second order and
at its dependence on the radial pulsation of the star. The $w$-mode
ringing is induced in a standard way, i.e. by means of a Gaussian pulse
of GWs impinging on the star.
\end{itemize}
Let us describe in more detail the choice of the initial conditions
for a differentially rotating star (\ref{IC_01_frmdr}).  The
differential rotation law of a neutron star is unknown.  An accurate
description of a differentially rotating configuration for newly born
neutron stars should come out from the numerical simulations of core
collapse, references~\cite{1997A&A...320..209Z, Dimmelmeier:2002bk,
Dimmelmeier:2002bm}. However, a set of rotation laws has been
introduced in Newtonian analysis, whose main motivations are: 
mathematical simplicity and the satisfaction of Rayleigh's
stability criterion for rotating inviscid fluids: $d\left( \varpi ^2
\Omega \right) / d\varpi > 0 $, where $\varpi = r \sin \theta $ is the
cylindrical radial coordinate. The ``j-constant law'', one of these
Newtonian laws, has also been extended to the general relativistic
approach where the dragging of the inertial frame must be taken into
account \cite{1989MNRAS.239..153K, 1989MNRAS.237..355K}.

\indent In this thesis, we provide the initial profile for the
odd-parity velocity perturbation $\beta^{(0,1)} _{lm}$ by using
an expansion in vector harmonics of the velocity
perturbation of a slowly differentially rotating star.
In slow rotation approximation the covariant
velocity perturbation is the following:
\begin{equation}
\delta u_{\mu}^{(0,1)} = \left( 0, \delta u_{a}^{(0,1)} \right) =
e^{-\Phi} \left( 0,0,0, r^2 \sin^2 \theta \, \left( \Omega - \omega
\right) \right) \, , \label{u_sl_rot}
\end{equation}
where $\Omega = \Omega(r,\theta)$ is the angular velocity measured by
an observer at infinity which describes the stellar differential
rotation, while the function $\omega = \omega(r,\theta)$ denotes the
dragging of inertial frame associated with the stellar rotation.  In
barotropic rotating stars, the integrability condition of the
hydrostatic equilibrium equation requires that the specific angular
momentum measured by the proper time of the matter is a function of
$\Omega $ only \cite{1989MNRAS.237..355K}, i.e., $u^{t} u_{\phi} =
j(\Omega)$. In the slow rotation case this condition leads to the
following expression:
\begin{equation}
\delta j ^{(0,1)} (\Omega) = u^{t} \delta u_{\phi}^{(0,1)}
= e^{- 2 \Phi} r^2 \sin^2 \theta \, \left( \Omega - \omega \right) \, ,  \label{dj10_sp}
\end{equation}
which is valid up to first order in $\Omega$.  The choice of the
functional form of $j (\Omega)$, and  hence $\delta j ^{(0,1)} (\Omega)$, 
must satisfy the Rayleigh's stability criterion
against axisymmetric disturbances for inviscid fluids:
\begin{equation}
d j \tilde{} /d\Omega < 0 \, ,
\end{equation}
where $j \tilde{}$ is the specific angular momentum
\begin{equation}
j \tilde{} = ( \bar \rho + \bar p) u_{\phi} / \bar \rho_{0} \, ,
\end{equation}
and $ \bar \rho_{0}$ the rest mass density.  The specific angular
momentum $j \tilde{}$ is locally conserved during an axisymmetric
collapse of perfect fluids \cite{Stergioulas:2003ep}.  A common choice
for $j (\Omega)$ that satisfies these 
conditions~\cite{1989MNRAS.237..355K, 1989MNRAS.239..153K} is the following:
\begin{equation}
\delta j ^{(0,1)}(\Omega)   = A^2 \left( \Omega_c - \Omega \right)
\label{jOmega}  \, ,
\end{equation}
where $\Omega _{c} $ is the angular velocity at the rotation axis
$\Omega(r=0)$, and $A$ is a constant parameter that governs the amount
of differential rotation.  Equations~(\ref{dj10_sp})
and~(\ref{jOmega}) give the following expression for the rotation law:
\begin{equation}
\Omega (r,\theta) = \frac{A^2 \Omega _{c} + e^{- 2 \Phi} r^2 \sin^2
\theta \, \omega (r,\theta) }{A^2 + e^{- 2 \Phi} r^2 \sin^2 \theta }
\, . \label{j-cons}
\end{equation}
This equation in the Newtonian limit reduces to the j-constant
rotation law used in Newtonian analysis \cite{1986ApJS...61..479H}:
\begin{equation}
\Omega (r,\theta) = \frac{A^2 \Omega _{c} }{A^2 + r^2 \sin^2 \theta }
\, . \label{j-cons_New}
\end{equation}
A uniformly rotating configuration with $\Omega = \Omega_c$ is
attained for high values of $A$, namely for $A \rightarrow \infty$. On
the other hand for small values of $A$, the law (\ref{j-cons})
describes in the Newtonian limit a configuration with constant angular
momentum.
The only non-null vector harmonic components of $\delta u_{a}$ are
given by the following expansion:
\begin{equation}
\delta u_{\phi} ^{(0,1)}  = \sum_{lm}
\beta _{lm} ^{(0,1)} S_{\phi}^{lm}  \, ,
\end{equation}
where $\beta ^{(0,1)}$ is the scalar function of the coordinate $r$ that
has been defined in Eq.~(\ref{deltauax}) and that can be derived
by the inner product with the basis of axial vector harmonics:
\begin{equation}
\beta _{lm}^{(0,1)} = \left( \delta u_{a}^{(0,1)}, S_{b}^{lm} \right) =
 \frac{1}{ l \left( l + 1  \right) } \,
\int _{S^2} \sin \theta d\theta d\phi \, \delta u_{a} ^{(0,1)}
S_{b}^{lm} \gamma^{ab} \, , \label{innpr}
\end{equation}
where $\gamma ^{ab} $ is the contravariant unit metric tensor of the
sphere $S^2$.

\begin{figure}[t]
\begin{center}
\includegraphics[width=150 mm, height=120 mm]{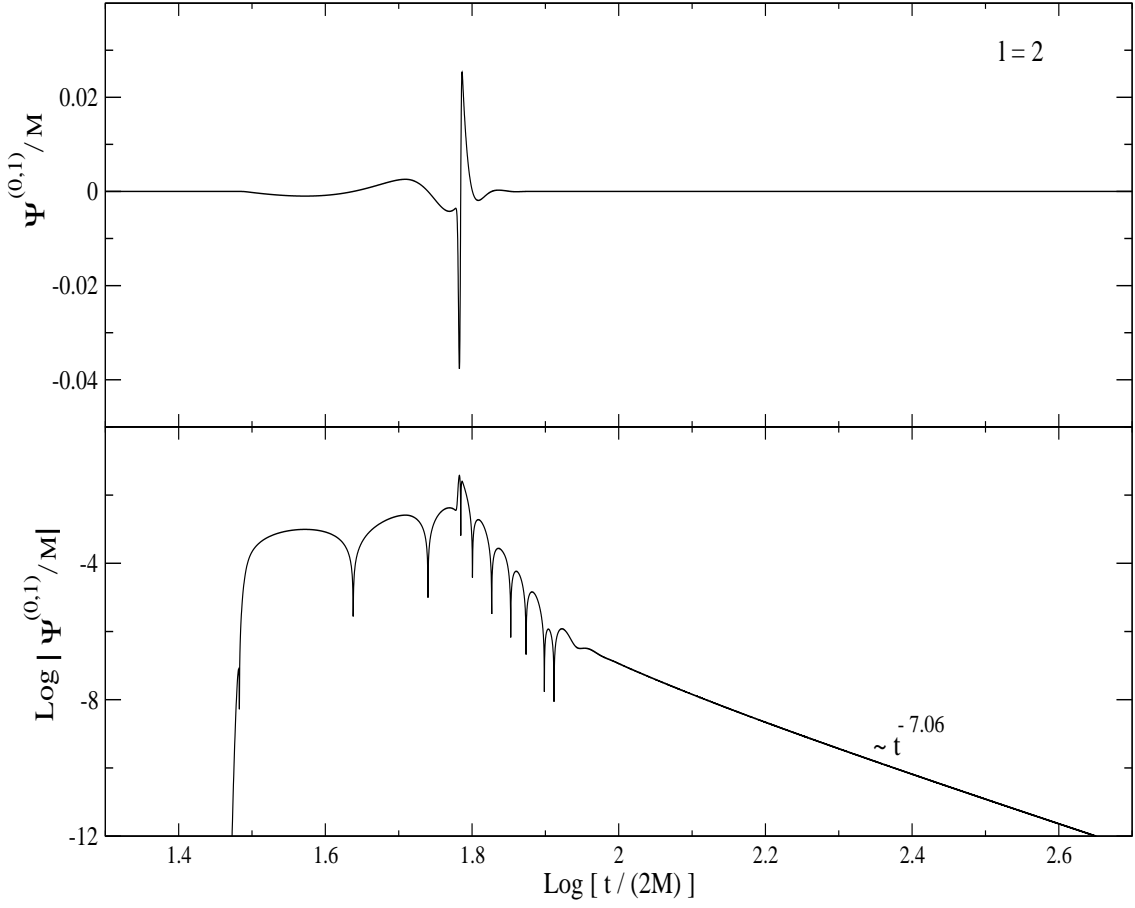}
\vspace{0.1cm}
\caption{\label{fig:Non_rad_L2} \small{Wave form in semi-logarithmic
    and logarithmic scale of the quadrupolar component ($l=2$) of the
    axial master function~$\Psi^{(0,1)}$ scaled by the stellar mass
    $M$. The excitation of the first $w$-mode and its strongly damped
    ringing phase are evident for $1.78 \leq \log \left[ t/(2M)
    \right] \leq 1.95$. The late time power-law tail is also in
    accordance to the theoretical results.}}
\end{center}
\end{figure}
%
\begin{figure}[t]
\begin{center}
\includegraphics[width=150 mm, height=120 mm]{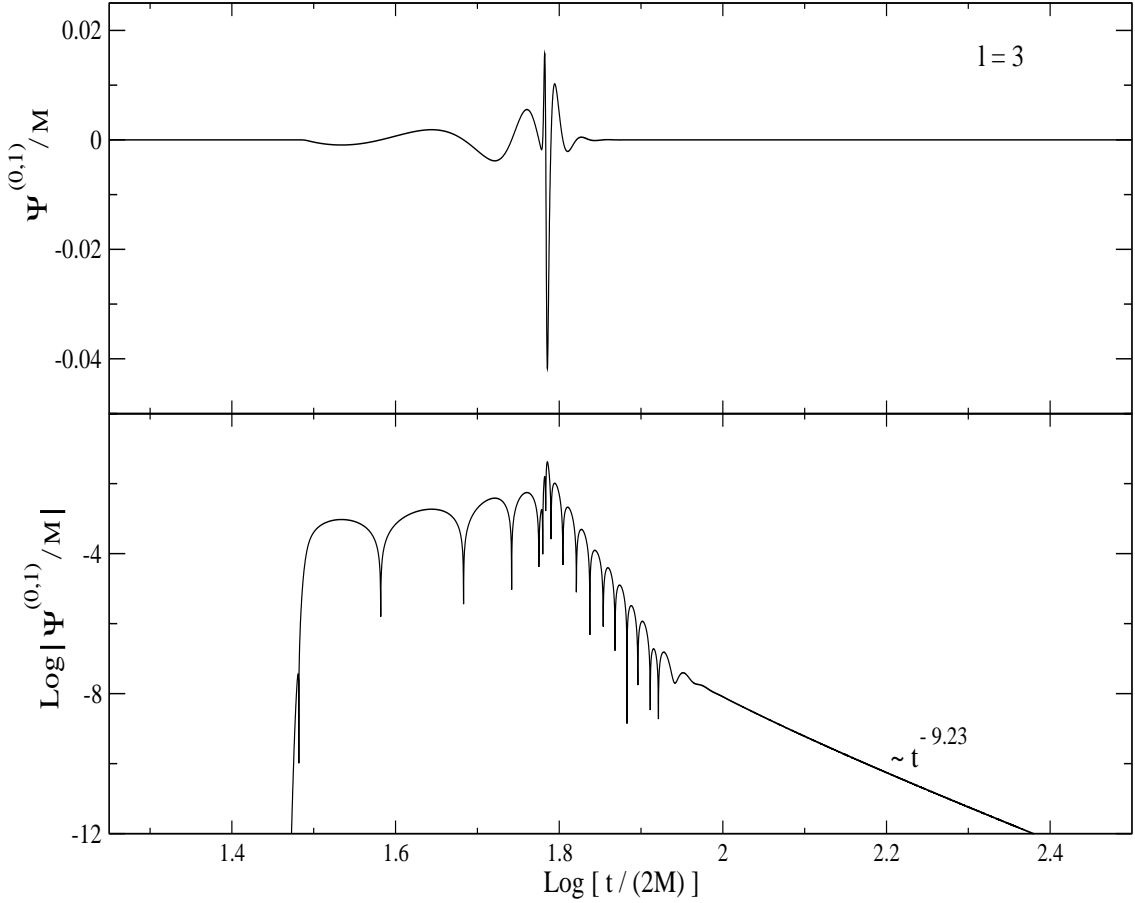}
\vspace{0.1cm}
\caption{\label{fig:Non_rad_L3} \small{Wave form in semi-logarithmic
    and logarithmic scale of the component $l=3$ of the axial master
    function~$\Psi^{(0,1)}$ scaled by the stellar mass $M$. As for
    the quadrupolar case, the $w$-mode excitation, the ringing phase
    and the long term time decay is clearly present.}}
\end{center}
\end{figure}

\begin{figure}[t]
\begin{center}
\vspace{0.8cm}
\includegraphics[width=120 mm, height=60 mm]{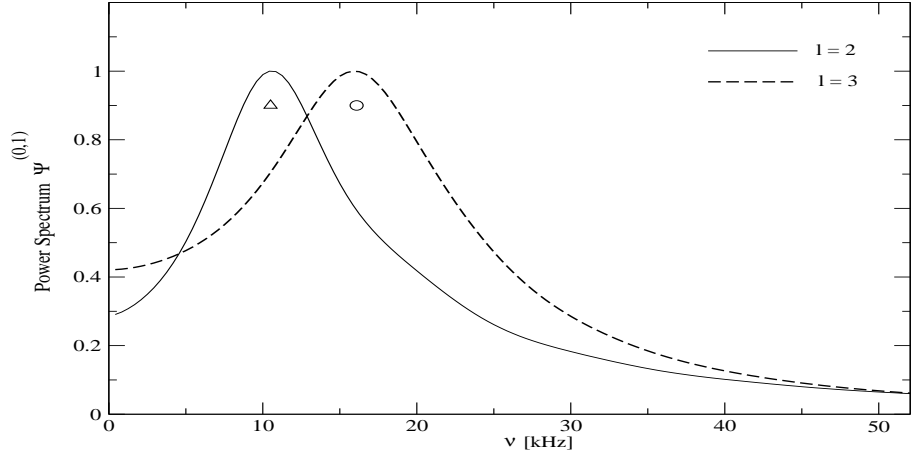}
\vspace{0.1cm}
\caption{\label{fig:Spectrum01} \small{Power spectrum of the
gravitational signal produced by the scattering of an axial
gravitational wave on a spherical non-rotating star. The \emph{solid
line} refers to the quadrupolar term $l=2$ while the \emph{dashed
line} to the $l=3$ case.  The frequencies of the first $w$-mode are
shown with a triangle for $l=2$ and a circle for $l=3$. The curves
are normalized with their maximum values assumed at the peak. }}
\end{center}
\end{figure}

In order to determine the initial profile of the axial velocity we can
introduce equation~(\ref{j-cons}) into the velocity
perturbation~(\ref{u_sl_rot}), and obtain:
\begin{equation}
  \delta u_{\phi} ^{(0,1)} =  \frac{ A^2 e^{-\Phi} }{A^2 + e^{-2\Phi} r^2 \sin^2 \theta }
 \left( r^2 \sin^2 \theta \, \Omega _{c} + \sum_{lm} k_0 ^{lm} S_{\phi}^{lm}
 \right)   \, , \label{betjlaw}
\end{equation}
where we have used the relation~(\ref{frmdr-rel}) that connects the
frame dragging function with the metric perturbations.  It is worth
noticing that the axial velocity (\ref{betjlaw}) contains as first
term the Newtonian j-rotation law up to an exponential factor
(hereafter for simplicity, we will call this term the nearly Newtonian
j-rotation law), while the second part accounts for the frame dragging.
When we introduce equation~(\ref{betjlaw}) into the inner product
(\ref{innpr}), we can easily determine the expression for the nearly
Newtonian term while the relativistic correction requires more
attention. In fact the metric variable $k_0 ^{lm}$ is itself the
unknown of the differential equation (\ref{k0-fradr}) and the inner
product (\ref{innpr}) of this term produces quantities which are
products of different harmonic indices $(l,m)$. In order to decouple
the various terms of the relativistic corrections we will assume that
the dominant contributions will be provided by the $k_0 ^{lm}$ which
has the same harmonic index as the nearly Newtonian rotation law.
When we have determined the harmonic expansion of the nearly
Newtonian part, we have found a well posed solution only for $A > e^{-
\Phi(R_s)} R_{s}$.
This relation and the values assumed by the metric field $\Phi$ in the
stellar model considered in this thesis allow us to give the following
estimation:
\begin{equation}
 \frac{1}{2}  < \frac{ A^2 }{A^2 + e^{-2 \Phi} r^2 \sin^2 \theta }
\le 1 \, ,
\end{equation}
which can be introduced in the relativistic term of
equation~(\ref{betjlaw}) to give:
\begin{equation}
\delta u_{\phi} ^{(0,1)} = \frac{ A^2 r^2 \sin^2 \theta \,  e^{-\Phi} \,  {}
 }{A^2 + e^{-2 \Phi} r^2 \sin^2 \theta }  \, \Omega _{c}  + \alpha_0  e^{-\Phi}
 \sum_{lm} k_0 ^{lm} S_{\phi}^{lm}  \, , \label{betjlawred}
\end{equation}
where $\alpha_0 \in (0.5,1]$.
The expansion of this law in
odd-parity vector harmonic provides the following result:
\begin{equation}
 \delta u_{\phi, l0}^{(0,1) } = \left\{ \begin{array}{ll}
   \alpha_0 e^{-\Phi}   k_0 ^{l0}
   S_{\phi}^{l0} & \quad \textrm{for} \quad l \quad \textrm{even} \, ,
   \\ \nonumber\\ \beta^{(0,1)}_{l0} S_{\phi}^{l0} & \quad
   \textrm{for} \quad l \quad \textrm{odd} \, ,
\end{array} \right.
\end{equation}
where the components with odd $l$ are given by the following
expression:
\begin{equation}
\beta^{(0,1)}_{l0} = e^{\Phi} \Omega_c  f_{l0} \left(x,A\right)
 + \alpha_0 e^{-\Phi}  \, k_0 ^{l0}    \, , \label{bet01_IC}
\end{equation}
where $x=r e^{-\Phi}$ (not to be confused with the fluid tortoise
coordinate).  The functions $f_{l0}$ for $l=1$ and $l=3$ are given by
the following expressions:
\begin{eqnarray}
\! \! \! \! \! \! \! \! \! \! \! \!  f ^{(0,1)} _{10} & = & - 3.069 A^2
\left[  1- \frac{A^2}{x \sqrt{A^2 + x^2} }
\ln \frac{ \sqrt{ x + \sqrt{A^2 + x^2}} }{\sqrt{ \sqrt{A^2 + x^2} - x} } \right] \, ,  \label{bet01_j_law_l1}\\
\! \! \! \! \! \! \! \!\! \! \! \!  f ^{(0,1)} _{30} & = & - 0.781 A^2 \left[  1 + 7.5 \frac{A^2}{x^2}
- \frac{6 A^2}{x  \sqrt{A^2 + x^2}} \left( 1 + \frac{5}{4} \frac{A^2}{x^2} \right)
\ln \frac{ \sqrt{ x + \sqrt{A^2 + x^2}} }{\sqrt{ \sqrt{A^2 + x^2} - x} } \label{bet03_j_law_l1}
\right] ,
\end{eqnarray}
which have been derived by imposing the condition $A > R_s
e^{-\Phi(R_s)}$.
%
\begin{figure}[t]
\begin{center}
  \includegraphics[width=150 mm, height=70
  mm]{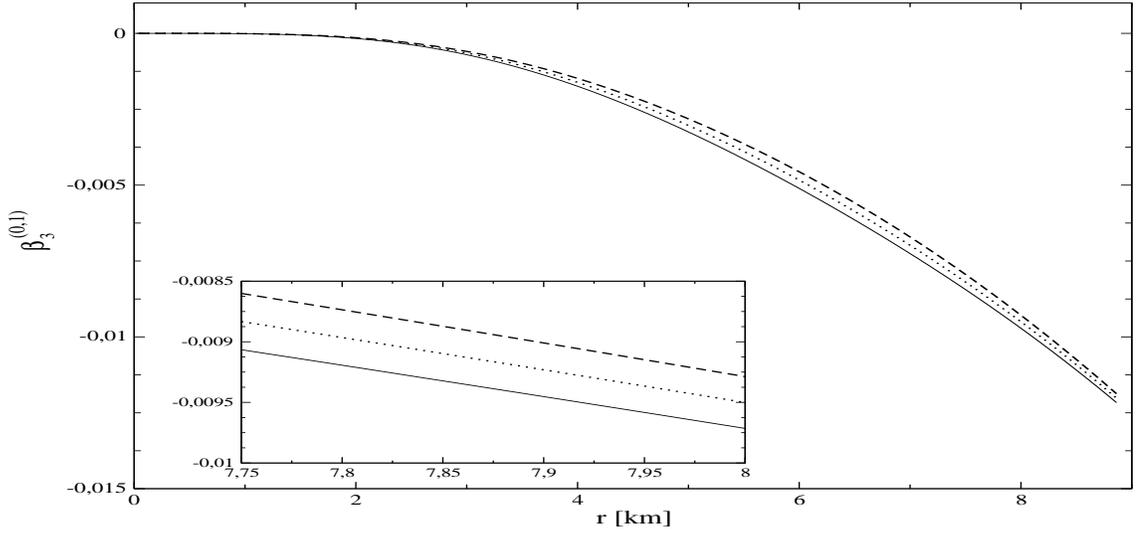}
  \vspace{0.4cm}\caption{\label{fig:beta01} \small{Profiles of the
      $l=3$ component of the axial velocity perturbation
      $\beta^{(0,1)}_{30}$ in $km$, determined from a j-constant 
      rotation law with $A=15~km$ and a $T=10~ms$ rotation period at
      the axis. The axial velocity associated with the nearly
      Newtonian j-law is shown with a \emph{solid line}, while the
      \emph{dotted} and \emph{dashed lines} denote the
      velocity~(\ref{bet01_IC}) with $\alpha = 0.5$ and $\alpha = 1$
      respectively.  }}
\end{center}
\end{figure}
%
\begin{figure}[t]
\begin{center}
  \includegraphics[width=150 mm,
  height=70mm]{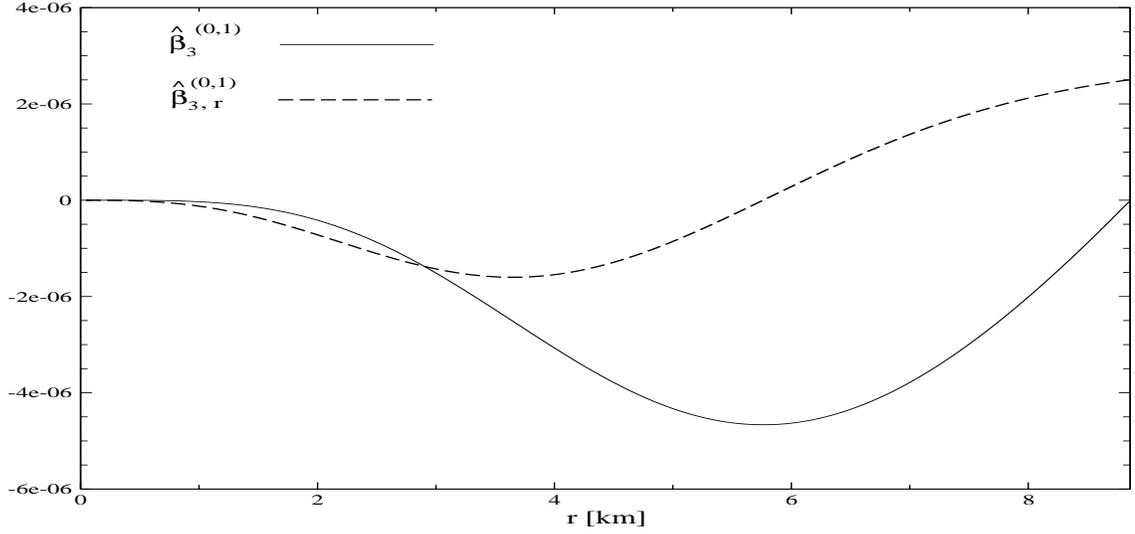}
  \vspace{0.4cm}\caption{\label{fig:bren01} \small{Profiles of the
      $l=3$ components of the axial fluid perturbation $\hat
      \beta^{(0,1)}_{30}$ in $km^{-1}$ (\emph{solid line}) and its
      spatial derivative $\hat \beta^{(0,1)}_{30 , \, r}$ in $km^{-2}$
      (\emph{dashed lines}), determined for a j-constant rotation law
      with $A=15~km$ and a $T=10~ms$ rotation period at the axis, and
      with $\alpha = 0$.}}
\end{center}
\end{figure}
%
\begin{figure}[t]
\begin{center}
\includegraphics[width=150 mm, height=100
mm]{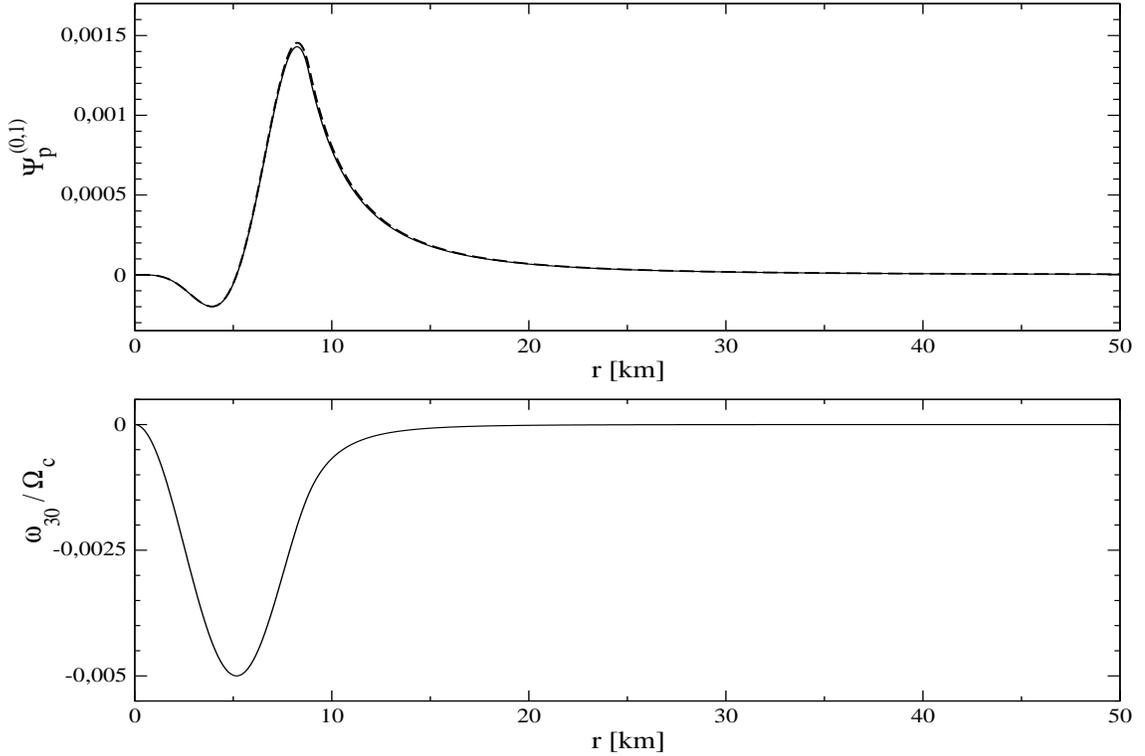} 
\vspace{0.1cm}
\caption{\label{fig:Non_rad_Psip} \small{The upper panel displays the
    stationary axial master function $\psi^{(0,1)}_p$, in $km$, for a
    nearly Newtonian j-constant rotation law with $A=15~km$ and a
    period $T=10~ms$ at the rotation axis. The solution of
    equation~(\ref{Psi01part}) is shown as a \emph{solid line} while
    the \emph{dashed line} is the solution found indirectly by first
    solving equation~(\ref{k0-fradr}) for the variable~$k_0^{(0,1)}$
    and then using the definition~(\ref{Psi01def_stat}).}}
\end{center}
\end{figure}
%
%
\begin{figure}[t]
\begin{center}
\includegraphics[width=150 mm, height=120 mm]{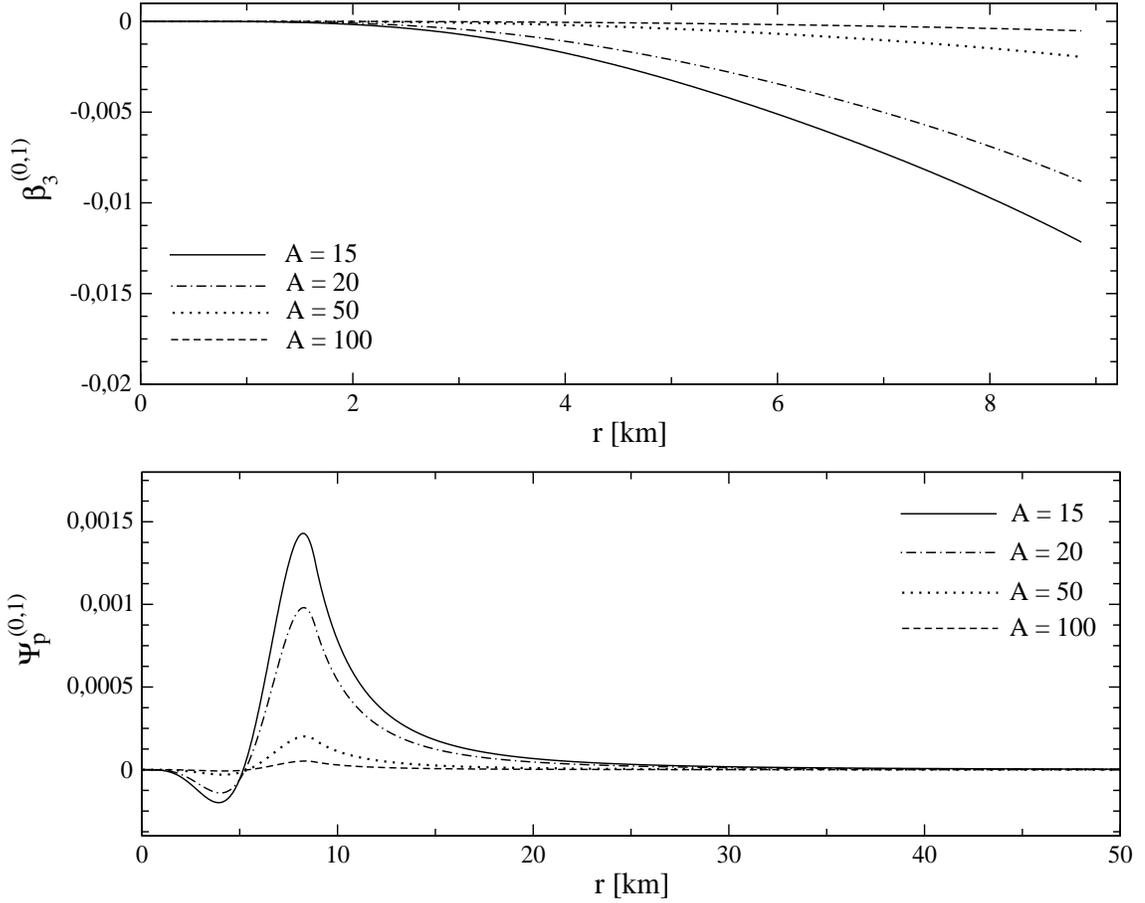}
\vspace{0.1cm}
\caption{\label{fig:Non_rad_beta01_diff} \small{The upper panel
    displays the $l=3$ component of the axial velocity
    perturbation~$\beta^{(0,1)}$, in $km$, for a nearly Newtonian
    j-constant rotation law with a period $T=10~ms$ at the rotation
    axis and for different values of the differential parameter~$A$ (in $km$).
    The associated solutions of equation~(\ref{Psi01part}) of the
    stationary master function $\psi^{(0,1)}_p$, in $km$, are shown in
    the lower panel.}}
\end{center}
\end{figure}

In the limit of $A \rightarrow \infty$, the function $f_{l0}
\left(r,A\right)$ behaves correctly, i.e. the nearly Newtonian part
assumes the following expressions:
\begin{equation}
 \lim _{A \rightarrow \infty}  e^{\Phi}
 \Omega_c   f_{l0} S_{\phi}^{l0} = \left\{
 \begin{array}{ll}
e^{- \Phi}  \Omega_c  r^2 \sin^2 \theta & \quad  \textrm{for} \quad l = 1 \, ,
\\ \nonumber\\
0 & \quad \textrm{for}  \quad l \geq 3 \, ,
\end{array} \right.
\end{equation}
which corresponds to a uniform rotating configuration, where $\Omega =
\Omega_c$ is the angular velocity measured at infinity.
The aim in this work is the analysis of the gravitational signal
produced by the non-linear coupling. Therefore, we will
consider only the case $l=3$ as the dipolar term $l=1$
does not produce gravitational waves.

In figure~\ref{fig:beta01} is plotted the axial velocity
perturbation $\beta_{30} ^{(0,1)}$. The figure shows that in the two
extreme values of the constant $\alpha_0$, the solutions disagree up
to ten percent.  In the following simulations we will use for
simplicity only the nearly Newtonian term and we obtain results which
are correct to better than ten percent.

\subsection{Simulations for axial non-radial perturbations}
\label{sec:Simul-Ax01}

In this section, we carried out numerical simulations of linear
axial perturbations for the two initial configurations described in
section~\ref{sec:BC-IV_Axial01}.

There are no fluid oscillations associated with linear axial
perturbations, therefore these are characterized by pure spacetime
oscillations represented in terms of a set of high frequency and
strongly damped \emph{w-modes}.  These spacetime modes can be excited
dynamically by studying the scattering of an axial gravitational wave
on a compact star. This is a standard procedure extensively used in
the literature~\cite{Andersson:1996ak, Ferrari:2000fk}. The gravitational
wave can be modelled by an impinging initial Gaussian profile,
\begin{equation}
\left. \Psi_{hom}^{(0,1)} \right|_{t=t_0} = A^{(0,1)} e^{-q
(r-r_{0})^2} \, , \label{Imp_psi01}
\end{equation}
where $A^{(0,1)}$ designs the initial wave amplitude, while the
constants~$q$ and~$r_0$ control the wide and initial position of the
Gaussian respectively.  We set up two simulations for the $l=2$ and
$l=3$ components of the axial perturbative fields. The initial
Gaussian pulse is at $40~km$ far from the centre of the star, has an
amplitude $A^{(0,1)}=0.1~km$ and $q=1.25~km^{-2}$.  The wave forms determined by
an observer at $150~km$ are shown in figures~(\ref{fig:Non_rad_L2})
and (\ref{fig:Non_rad_L3}).  The scattered signal in both cases
exhibits the characteristic excitation of the first $w$-mode at about
$\log[t/(2M)]=1.8$ evolution time, followed by the ringing phase which
is strongly damped by the emission of gravitational radiation. The
ringing of the QNM is more evident in the lower panels of
figures~(\ref{fig:Non_rad_L2}) and (\ref{fig:Non_rad_L3}) where the
signal is plotted on a  logarithmic scale.  After this phase, the master
function $\Psi_{hom}^{(0,1)}$ shows the typical decreasing behaviour
predicted by the theoretical power law $\Psi_{hom}^{(0,1)} \sim
t^{-\left(2l+3\right)}$~\cite{1995PhRvD..52.2118C}. For an evolution
time of $8~ms$ a linear regression of the tail of the signal provides 
the following values:
\begin{equation}
 \Psi_{hom}^{(0,1)} = \left\{ \begin{array}{ll}
t^{- 7.06 } & \mbox{\qquad for~~$l=2$\,,} \\
t^{-  9.23 } &  \mbox{\qquad for~~$l=3$\,.}
\end{array} \right.
\end{equation}
The power law gets closer to the theoretical value for longer
evolutions.

The excitation of the first \emph{w-modes} for the two harmonic
indices $l=2,3$ is confirmed by the analysis of the spectral
properties of the signal. We perform an FFT of the signal part that
starts with the excitation of the spacetime mode.  The results in
figure~\ref{fig:Spectrum01} show the characteristic broad shape of
\emph{w-modes} which is due to the short values of their damping time,
respectively $\tau = 29.5~\mu s$ and $\tau = 25.2~\mu s$ for the $l=2$
and $l=3$ cases.  The curve peaks at the $\nu = 10.501~kHz$ for the
quadrupolar case and at $\nu = 16.092~kHz$ for the $l=3$ case, which
are denoted in figure~\ref{fig:Spectrum01} with a \emph{triangle}
and \emph{circle} respectively. For these two harmonic indices these
values are the frequencies of the first curvature $w$-mode, which have
been determined by Gualtieri's numerical code in a frequency
domain approach~\cite{Gualtieri:2001cm}.

The other initial configuration for the linear axial perturbation
describes a stationary differential rotation induced by the axial
velocity perturbation. In particular, the structure of this
perturbative framework allows us to investigate the $(l,m)$ component
of the axial perturbation $\psi^{(0,1)}$, which is induced by the
$(l,m)$ component of the axial velocity perturbation $\beta^{(0,1)}$.
As illustrated in section~\ref{sec:BC-IV_Axial01}, we are going to
implement an axial velocity perturbation described by the
function~(\ref{bet01_IC}), which was derived from the relativistic
j-constant rotation law.  The first point to clarify in
equation~(\ref{bet01_IC}) is the amount of the relativistic correction
due to its second term with respect to the nearly Newtonian j-constant
law. This issue can be studied with equation~(\ref{k0-fradr}) for the
metric gauge invariant quantity $k_0^{(0,1)}$. We have introduced the
functional dependence~(\ref{bet01_IC}) for $\beta^{(0,1)}$ into the
source term of equation~(\ref{k0-fradr}) and solved the equation first
for a nearly Newtonian rotation law, i.e. $\alpha_0 = 0$, and secondly
for $\alpha_0 = 0.5$ and $\alpha_0 = 1$, which are the two extreme
values of the relativistic correction to the $l=3$ velocity component.
In addition, we choose the following value of the differential
rotation parameter $A=15~km$ and a $T=10~ms$ rotation period at the
rotation axis of the star. The choice of $A$ is motivated by the
regularity of the $l=3$ component of the axial velocity (see
section~\ref{sec:BC-IV_Axial01}), that leads to the condition $A>
e^{\Phi(R_s)}~R_s$. For higher values of $A$, the harmonic components
$\beta^{(0,1)}$ with $l > 1$ will decrease, as the rotation law
approaches the uniform rotation configuration.
On the other hand, the angular velocity $\Omega_c$ is the physical
perturbative parameter that controls the strength of the axial
perturbations. Its values must satisfy the requirement imposed by the
``slow rotation approximation'', i.e. the dimensionless perturbative
parameter $\epsilon = \Omega_c / \Omega_K \ll 1$, where $\Omega_K $ is
the Keplerian angular rotation which describes the mass-sheeding limit
of the stellar model under consideration. 
In order to have a simple extimate, we consider the $\Omega_K$ of a
uniformly rotating star, which can be extimated with the
\emph{empirical formula} \cite{1989Natur.340..617H,
1987ApJ...314..594F} which is accurate to better than $10$
percent. This formula is based on the classical value up to a
corrrective factor and is given by the following expression:
\begin{equation}
\Omega_K \equiv 0.625 \, \sqrt{M/R_s^3} \, ,
\end{equation}
where $M$ and $R_s$ are the mass and radius of the non-rotating star
in hydrostatic equilibrium. For our stellar model we get $\Omega_K =
0.0324~km^{-1}$ which corresponds to a rotational period $T_{K} =
193.996~km = 0.64~ms$.  Therefore, the dimensionless
parameter is $\epsilon = 6.45 \times 10^{-2}$.

The numerical integrations of equation~(\ref{k0-fradr}) with the
method explained in section~\ref{sec:Numer_alg_01} are shown in
figure~\ref{fig:beta01}. The two solutions obtained for $\alpha_0 =
0.5$ and $\alpha_0 = 1$ disagree up to $5$ percent, while the
Newtonian differential rotation law is accurate with respect to the
relativistic $\alpha_0 = 1$ case to better than $10$ percent.  We will
use the nearly Newtonian rotation law in this and the next sections, being
aware that the results could be accurate within ten percent due to
the relativistic corrections of the dragging of inertial frame.

The particular solution of the axial master function $\Psi_p^{(0,1)}$
can then be determined by equation~(\ref{Psi01part}), which is
shown in figure~\ref{fig:Non_rad_Psip} for the same parameter of
the j-constant rotation law used above. This solution can also be determined
indirectly by first solving equation~(\ref{k0-fradr}) for the
metric variable $k_0$ and then getting $\Psi_p^{(0,1)}$ through the
definition~(\ref{Psi01def}) for the stationary case:
\begin{equation}
\Psi^{(0,1)}_p = \left( 2 k_{0}^{(0,1)} - r k_{0 \, , \, r}^{(0,1)}
\right) e^{-\left(\Phi + \Lambda \right)} \, . \label{Psi01def_stat}
\end{equation}
The indirect solution, which is shown in
figure~\ref{fig:Non_rad_Psip} as a \emph{dashed line}, reproduces
the solution obtained with the direct method with a maximum error less
than $2.3$ percent.  In the lower panel of
figure~\ref{fig:Non_rad_Psip}, we show the $l=3$ component of the
frame dragging function $\omega_{30}$, which has been found from 
equation~(\ref{frmdr-rel}).

So far, we have used a rotation law with a rotation period at the axis
fixed at $T=10~ms$.  However, the linearity of
equation~(\ref{Psi01part}) allows us to determine the
solutions~$\Psi_p^{(0,1)}$ with a simple rescaling. Let
$\left. \Psi_p^{(0,1)} \right| _{T_1}$ be a solution related to a
differential rotation period $T_1$.  The solution corresponding to a
rotational period $T_2$ is given by:
\begin{equation}
\left. \Psi_p^{(0,1)} \right| _{T_2} = \frac{T_1}{T_2} \, \left. \Psi_p^{(0,1)} \right| _{T_1} \, .
\end{equation}

In figure~\ref{fig:Non_rad_beta01_diff}, we show the effects of the
differential parameter $A$ on the profile of the axial velocity
$\beta^{(0,1)}$ for $\alpha_{0} = 0$~(\ref{bet01_IC}) and the
particular solution of the master function$\Psi_p^{(0,1)}$, determined
by equation~(\ref{Psi01part}). The $l=3$ axial velocity and master
function decrease for higher values of $A$, when the star tends to a
uniform rotational configuration and the only non-vanishing component
is the $l=1$.  In addition, we have noticed that for $A> 100~km$ the
axial velocity is not accurately described by the
expressions~(\ref{bet01_j_law_l1}) and~(\ref{bet03_j_law_l1}), as it
has an irregular behaviour near the origin due to appearence of high
peaks.

The rotation energy associated with the differential rotation can be
determined with the following equation~\cite{Hartle:1970ha}:
\begin{equation}
E^{(0,1)} = \frac{1}{2} \int_{0}^{R_s}  dr \int _{0}^{\pi} d \theta \, 2 \pi r^4 \sin \theta^3
\left( \bar \rho + \bar p \right) e^{\Lambda - \Phi}
\Omega \left( \Omega - \omega \right) \, .  \label{E01_rot}
\end{equation}
By neglecting  the dragging of the inertial frame $\omega$ we can determine an upper limit of the
rotational energy~\cite{Hartle:1970ha}:
\begin{equation}
E^{(0,1)}_{rot} = \frac{1}{2} \int_{0}^{R_s}  dr \int _{0}^{\pi} d \theta \, 2 \pi r^4 \sin \theta^3
\left( \bar \rho + \bar p \right) e^{\Lambda - \Phi } \Omega ^2 \, ,   \label{E01_rot2}
\end{equation}
where $E^{(0,1)} \leq E^{(0,1)}_{rot}$. Furthermore, this expression can be expanded in tensor harmonic
as follows:
\begin{equation}
E^{(0,1)}_{rot} = \sum_{l\geq 1} \frac{l \left( l +1
\right)}{2\left( 2 l +1 \right)} \int_{0}^{R_s} dr \, 4 \pi \left(
\bar \rho + \bar p \right) e^{\left( \Lambda + \Phi \right) } \left(
\beta^{(0,1)}_{lm} \right) ^2  \,
\end{equation}
where we have applied the following relation $\Omega _{lm} =
r^{-2}e^{\Phi}\beta^{(0,1)}_{lm}$, which is valid when the frame
dragging function $\omega$ is neglected. The quantities $\Omega _{lm}$
are the harmonic components of the angular velocity,
\begin{equation}
\Omega \left( r, \theta \right) = \sum _{l \geq 1} \Omega_{lm} S_{lm}
^{\phi} \, .
\end{equation}
The upper limit of the $l=1,3$ component of the rotational energy for
a differential rotation with $A = 15~km$ and $T=10~ms$ is
$E^{(0,1)}_{rot, 10} = 7.90 \times 10^{-5}~km$ and $E^{(0,1)}_{rot,
30} = 8.62\times 10^{-7}~km$ respectively.

The solutions $\Psi^{(0,1)}_{hom}$ and $ k_0$ exhibit a second order
convergence, while $\Psi^{(0,1)}_{p}$ manifests a convergence of first
order. This lower convergence rate is due to the discontinuity of
$\Psi^{(0,1)}_{p, \, rr}$ at the stellar surface.

\section{Non-linear axial oscillations}
\label{sec:Simul-NLAx11}

The dynamical properties of the coupling between the radial and axial
non-radial oscillations are described by the solutions of the two
inhomogeneous partial differential equations~(\ref{Psi11maseq})
and~(\ref{traseq11}), where the source terms are present only in the
interior spacetime.  The initial values adopted for the linear
perturbations are able to describe the non-linear coupling for the
following configurations: \emph{i)} a radially pulsating and
differentially rotating star, \emph{ii)} scattering of a gravitational
wave on a radially oscillating star. In particular in the former case,
we will see that  the coupling between the stationary axial velocity with the
radial pulsations produces an oscillating $\lambda \epsilon$ corrective
term of the redefined axial velocity $\hat \beta^{(1,1)}$.

\subsection{Numerical algorithms}

The conservation equation~(\ref{traseq11}) for the
perturbation~$\hat \beta ^{(1,1)}$ exists only inside the star and
is integrated with an up-wind algorithm. The numerical
discretization is given by the following expression:
\begin{equation}
\hat \beta ^{n+1}_j  = \hat \beta ^n_j + \D t \, e^{\Phi _j} \, \left( \Sigma _{\beta} \right)^n _{j}
\end{equation}
where $\left( \Sigma _{\beta} \right) _{j}$ is the second order
discrete approximation of the source term given in
appendix~(\ref{AppSources_axial}). Thus, the first and second order
derivatives that are present in~$\left( \Sigma _{\beta} \right) _{j}$
are approximated by second order centered finite difference
approximations in the internal grid points and by second order
one-sided finite approximations at the origin and stellar surface.

The integration domain of the axial master equation~(\ref{Psi11maseq})
is the entire spacetime, where the source~$\Sigma _{\psi}$ is present
only in the interior spacetime. After several tests, we have found
that the simulations of the master function~$\Psi^{(1,1)}$ are more
accurate when we implement two different numerical methods.  An
Up-wind algorithm for the interior and a Leapfrog for the exterior,
for studying the coupling between the radial pulsations and
differential rotation, and a Leapfrog on the whole spacetime for the
scattering of an axial gravitational wave on a radially oscillating
star.  The implementation of these two methods is due to the different
properties of the source terms~$\Sigma _{\psi}$ and the junction
conditions~(\ref{Sfbc11}) for the two cases mentioned above.

In case of coupling between the radial pulsations and axial
differential rotation, we prefer to separate the numerical integration
of the axial master equation in the interior and exterior
spacetime. Furthermore, in order to reduce the numerical noise caused
by the discontinuity of the sources at the stellar surface, we
implement a first order accurate numerical scheme in the interior
spacetime.  Therefore, the axial master equation is simulated with two
different numerical schemes inside and ouside the star. In the
interior, the second order PDE~(\ref{Psi11maseq}) is transformed in a
system of first order PDEs which will be integrated with a
generalization of the Up-wind method~\cite{Leveque_mio}.  In the
exterior, the Regge-Wheeler equation at order $\lambda \epsilon$ is
instead updated in time by using a second order Leapfrog method, where
we extract the values of $\Psi^{(1,1)}$ and $\Psi^{(1,1)}_{, \, r}$ on
the surface with the matching conditions.
Let us first describe the interior spacetime.
We can define two new quantities as follows:
\begin{eqnarray}
\W  = \psi^{(1,1)}_{int} \, , \qquad \qquad  \U _1  =  \W _{\, t}    \, ,\qquad  \qquad
\U _2  =  \W _{, \, r}  \, ,  \label{UpW_def}
\end{eqnarray}
where the variable $\W$ has been introduce to reduce the number of
indices in the discrete equations.
With the definition of the vector $\U \equiv \left[ \, \U _1, \, \U _2
\,\right] ^{T}$ we can write the master wave
equation~(\ref{Psi11maseq}) in the interior spacetime in the following
conservative form:
\begin{equation}
\U_{, \, t} + \F_{, \, r} = \S \label{eq_cons}
\end{equation}
where the flux is given by
\begin{equation}
\F = \AA \U \, ,
\end{equation}
and $\AA$ is the following 2x2 matrix:
\begin{displaymath}
  \AA = - \left(  \begin{array}{cc}
            0 &  v_{gw}^{2}  \\
        1 & 0
\end{array} \right)  \, .
\end{displaymath}
The quantity $v_{gw}= e^{\Phi- \Lambda}$ is the propagation velocity
of the gravitational signal, and the source $\S$ is a two dimensional
vector with the following components:
\begin{eqnarray}
\S _1 & = & -e^{2 \Phi} \left( 4 \pi \left(\bar p - \bar \rho \right)
          r + \frac{2 m }{r} \right) \U _{2} + V \W +16 \pi \left( 4 \pi
          \rho r^2 - \frac{m}{r} \right) e^{2 \Phi+\Lambda} \hat
          \beta^{(1,1)} \nn \\ && + 16 \pi r e^{2 \Phi - \Lambda} \hat
          \beta^{(1,1)}_{,\, r} + \Sigma_{\psi} \, ,\\ \S _2 & = & 0 \,
          ,
\end{eqnarray}
where $\Sigma_{\psi}$ are the sources expressed in
equation~(\ref{Spsi11}) and are second order accurate in space.  The
variable $\U$ is updated at every time step by the following
differentiating scheme,
\begin{equation}
\U_{j}^{n+1}  =  \U_{j}^{n}  - \frac{\D t}{\D x} \, \AA
^{+}  \left( \U_{j}^{n} - \U_{j-1}^{n} \right) -  \frac{\D t}{\D x}
\, \AA ^{-}  \left( \U_{j+1}^{n} - \U_{j}^{n} \right) + \D t  \, \S ^n_j \, ,
\label{Up_scheme}
\end{equation}
where the matrices
$\mathbf{A^{+}}$ and $\mathbf{A^{-}}$ are given by,
\begin{displaymath}
\mathbf{A^{+}} = \frac{1}{2} \, \left(
\begin{array}{cc}
v & -v^{2}  \\
-1 & v
\end{array} \right) \, , \qquad \qquad \mathbf{A^{-}} = - \frac{1}{2}  \, \left(
\begin{array}{cc}
v & v^{2}  \\
1 & v
\end{array} \right) \, .
\end{displaymath}
The value of the $\psi^{(1,1)}_{int}$ is then obtained from the definition~(\ref{UpW_def}):
\begin{equation}
\W ^{n+1}_{j} = \W ^{n}_{j}
+ \frac{\Delta t}{2} \, \left[ \left(\U _1\right)^{n+1}_{j} + \left(\U _1\right)^{n}_{j} \right]
\end{equation}
In the exterior spacetime the system of equation reduces to the
Regge-Wheeler equation for the master function $\Psi^{(1,1)}$. This
axial wave-like equation is solved with a Leapfrog algorithm as well
as we did for the first order axial master equation. We do not write
the explicit discrete expression for the variable $\Psi^{(1,1)}$ here as it
can be determined directly by equation~(\ref{psi01_alg}), where 
the coefficients are now given by
\begin{eqnarray}
v & = & e^{\Phi-\Lambda}  \, , \\
\alpha & = & e^{2 \Phi}  \frac{2 M}{r^2}  \, ,  \\
V & = & e^{2 \Phi} \left( \frac{6 M}{r^3}  - \frac{l(l+1)}{r^2} \right) \, .
\end{eqnarray}

The scattering of a gravitational wave on a radially pulsating star is
instead studied with a Lepfrog algorithm in the whole spacetime.  When
we discuss the numerical implementation of the junction conditions for
non-linear perturbations (section~\ref{sec:BIC11_Num}), we will see
that the movement of the stellar surface in a Eulerian gauge leads us
to adopt some approximate treatments of the matching conditions. As
already explained in section~\ref{sec:BC01_ana}, we have chosen in
this work to impose the junction conditions on the surface
$\Sigma_{jc}$ which is always well inside the star, even during the
phases of maximal contraction. Consequently, the axial junction
conditions reduce to the continuity of $\Psi^{(1,1)}$ and its
derivatives~(\ref{Sfbc11}). We can then integrate  
equation~(\ref{Psi11maseq}) by using the explicit discrete
scheme~(\ref{psi01_alg}), where we introduce in addition the second
order finite approximation of the source~$\Sigma _{\psi}$ in the
interior spacetime. We have noticed that the junction conditions are
automaticaly satisfied by the simulations.

\subsection{Boundary and initial conditions}
\label{sec:BIC11_Num}

The three non-linear coupling perturbations $\Psi^{(1,1)}$,
$\Psi^{(1,1)}_{,t}$ and $\beta^{(1,1)}$ are not all independent on the
initial Cauchy surface, because of the presence of not vanishing
source terms in the non-linear perturbative equations. A correct
initial value for the axial master function $\Psi^{(1,1)}$ must come
out from the solution of the coupling axial constraints, as in case of
first order axial and polar perturbations in presence of
sources~\cite{Nagar:2005ea}. However, since we are more interested in
the part of the gravitational wave signal which is driven by the
stationary radially oscillating sources, we set a vanishing value for
all three coupling perturbations at $t=0$ without solving the
constraints. As a result, we will see later that after an initial
burst of $w$-mode gravitational waves due to the violation of the
constraints the solution relax to the correct periodic solution.

Let us now discuss the boundary conditions. At the \emph{origin} we
implement the conditions~(\ref{BC11_orig}) by applying the same
procedure described for the linear perturbations. Hence, we have for
the metric variable:
\begin{equation}
\W_1 = \W _2 \left( \frac{r_1}{r_2} \right) ^{l+1} \, , \qquad \qquad
\left( \U_1 \right) _1  = \left( \U_1 \right) _2 \left( \frac{r_1}{r_2} \right) ^{l+1} \, , \qquad \qquad
\left( \U_2 \right) _1  = \left( \U_2 \right) _2 \left( \frac{r_1}{r_2} \right) ^{l} \, , \qquad \qquad
\end{equation}
and for the fluid quantity
\begin{equation}
\hat \beta^{(1,1)} _1 = \hat \beta^{(1,1)} _2  \left( \frac{r_1}{r_2} \right) ^{l+1}\, .
\end{equation}
At the \emph{outermost grid point} we implement the outgoing
Sommerfeld boundary condition, where the outermost grid point is set
far enough from the star in order to not corrupt the gravitational
signal extraction with the numerical noise coming from the numerical
reflection at the grid buondary.      \\
\indent The internal and external spacetimes are separated by the
stellar surface, where the matching conditions have to be imposed in
order to connect the physical and geometrical descriptions between the
interior and exterior spacetime. In an Eulerian gauge, the perturbed
surface could not coincide with the surface of the equilibrium
configuration.  Therefore, when we compare near the surface the
perturbations and background tensor fields related to a same physical
quantities some problems can arise.  Let us for instance consider
radial pulsations of a static star, and a point $x_P$, which is in the
region between the static and perturbed surface during a contraction
phase of the star $x(t) < x_P < R_s$, where $x(t) = R_s + \lambda
\xi^{(1,0)}$ and $\xi^{(1,0)}$ is the Lagrangian displacement of
radial perturbations. During this dynamical phase $x_P$ is outside the
star, then the mass energy density $\rho$ should vanish there.
However, when we determine at $x_P$ the Eulerian perturbation
\begin{equation}
\rho^{(1,0)} = \rho - \bar \rho
\end{equation}
where $\bar \rho$ is the background density, we find this unphysical
value $\rho^{(1,0)} = \bar \rho < 0$.  At any perturbative order, the
dynamical properties of the perturbation fields depend on the
quantities which have been determined at previous perturbative
orders. Therefore, the introduction of this negative values for the
density into the $\lambda \epsilon$ perturbative equations could leads
to an unphysical description of the non-linear perturbations.  This
issue can be solved with different methods~\cite{Sperhake:2001si}.
For instance, one may add to the stellar model an atmosphere of low
density, which extends beyond the size of the neutron star. Otherwise,
one can adopt a Lagrangian gauge for the perturbative description. In
this case, the perturbed and background surfaces concide and the
energy density would get the correct value
\begin{equation}
\Delta \rho^{(1,0)} = \rho^{(1,0)} + \pounds _{\xi^{(1,0)}} \, \rho \geq 0 \, ,
\end{equation}
where the vanishing value is taken at the background stellar surface.
Alternatively, we can impose the junction conditions on the
hypersurface~(\ref{hyp_sigjc}), which remains always inside the star.
With this method we remove the outer layers of the star, but since the
density in this region is extremely low the mass neglected is less
than one percent. Therefore, the wave forms and spectra of the
gravitational signal are still well approximated by this procedure.
The position of the junction hypersurface $\Sigma_{jc}$ is evaluated
as follows: \emph{i)} we evolve the radial perturbations and determine
the maximum movement of the surface with the Lagrangian displacement
$\xi^{(1,0)}_{sf}$. The values for the particular initial conditions
used in our evolution are written in table~\ref{tab:TOV_Ener}.
\emph{ii)} Then, we place the surface $\Sigma_{jc}$ at the first grid
point that remains inside the star during the evolution.  It is worth
remarking that for an $r$\emph{-grid} of dimension $J_r = 200$, the
values of $\xi^{(1,0)}_{sf}$ in table~\ref{tab:TOV_Ener} leads to
neglect only one point, i.e. the junction conditions can be imposed at
grid point $J_{jc} = 199$. However, since the source of the axial
master equation~(\ref{Psi11maseq}) has its maximal amplitude at the
stellar surface, this removal of a grid point induces a error of about
five percent in the gravitational signal.

\noindent For the coupling between radial pulsation and differential
rotation, we impose the continuity of the metric variable~$\Psi
^{(1,1)}$ and the relation~(\ref{Sfbc11}) for~$\Psi ^{(1,1)}_{, \,
r}$. For the latter condition we discretize the sided external
derivative with a first order approximation in order to get the
appropriate value of the master function~$\Psi ^{(1,1)}$ in the first
point outside the junction surface $\Sigma_{jc}$:
\begin{eqnarray}
\psi^{ext} _{J_{jc}} & = &\W _{J_{jc}} \, , \\
\psi^{ext} _{J_{jc}+1} & = &
\psi^{ext} _{J_{jc}} + \D r \left[ \left(\U_2 \right) _{J_{jc}} + 16 \pi \, r \, \left( \hat \beta
^{(1,1)} \right) _{J_{jc}} \right] \, , \\
\end{eqnarray}
%
\begin{table}[!t]
\begin{center}
\begin{tabular}{l  c c c c c c c  }
\hline  \hline \\
                             &  F      & H1   & H2     & H3       & H4      & H5       & H6  \\ \\
\hline \hline \\
$  E^{(1,0)}_{n}~[10^{-8}~km]$   & 35.9    & 4.2  & 1.37   & 0.62     & 0.34    & 0.21     & 0.14     \\  \\
$  \xi^{(1,0)}_{sf}~ [m] $   & 12.65   & 4.02 & 2.66   & 2.02     & 1.64    & 1.38     & 1.19      \\ \\
\hline
\end{tabular}
\caption[]{\label{tab:TOV_Ener} \small{Energy $E^{(1,0)}_{0}$ and
maximum surface displacement $\xi^{(1,0)}_{sf}$ of radial pulsations.
These values correspond to the initial conditions~(\ref{Rad_Ic})
and~(\ref{gam_IC_nm}) with $A^{(1,0)}=0.001$.  Every evolution has
been excited with a single radial mode. }}
\end{center}
\end{table}
%

\begin{figure}[t]
\begin{center}
\includegraphics[width=150 mm, height=100 mm]{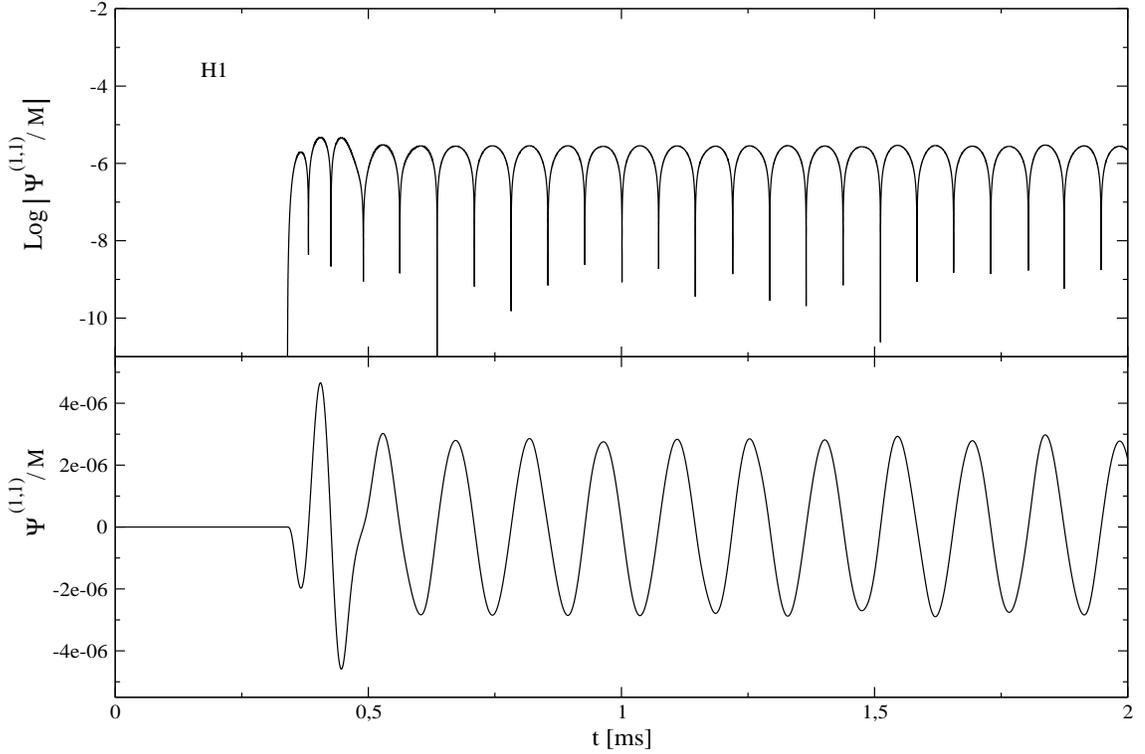} 
\vspace{0.1cm}
\caption{\label{fig:Coupl_Psi11_H1} \small{Wave form of the axial
    master function~$\psi^{(1,1)}$ scaled by the stellar mass $M$,
    which describes the coupling between the linear radial pulsations
    and the axial differential rotation of a neutron star. The
    rotation is given by a nearly Newtonian j-constant rotation law
    with a $T=10~ms$ period at the rotation axis and $A=15~km$, while
    the radial pulsating dynamics has been excited with the overtone
    H1. The signal manifests an initial excitation of the first $l=3$
    $w$-mode, which is followed by a periodic oscillation driven by
    the pulsating source terms.}}
\end{center}
\end{figure}
%
\begin{figure}[t]
\begin{center}
\includegraphics[width=140 mm, height=80 mm]{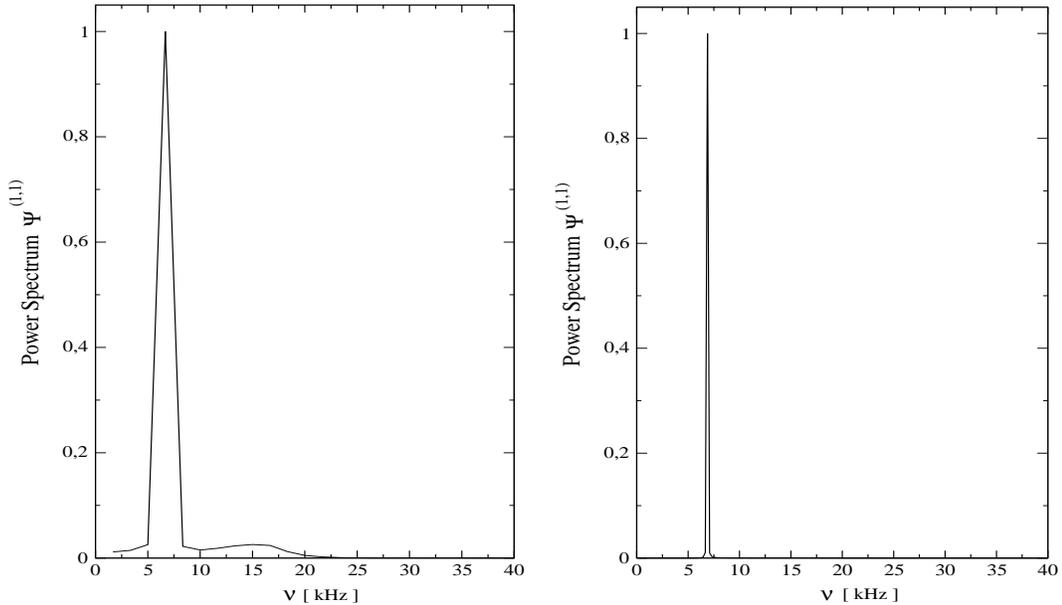}
\caption{\label{fig:Spec_Psi11_H1} \small{Normalized power spectrum of
the wave form shown in figure~\ref{fig:Coupl_Psi11_H1}. The
\emph{left panel} displays the presence of the H1 radial mode and the
$l=3$ $w$-mode in the gravitational signal. In the \emph{right panel}
the FFT has been performed for a longer evolution time, thus the
energy contained in the H1 peak becomes dominant and the spacetime
mode is not visible.}}
\end{center}
\end{figure}
%
\begin{figure}[t]
\begin{center}
\includegraphics[width=150 mm, height=80 mm]{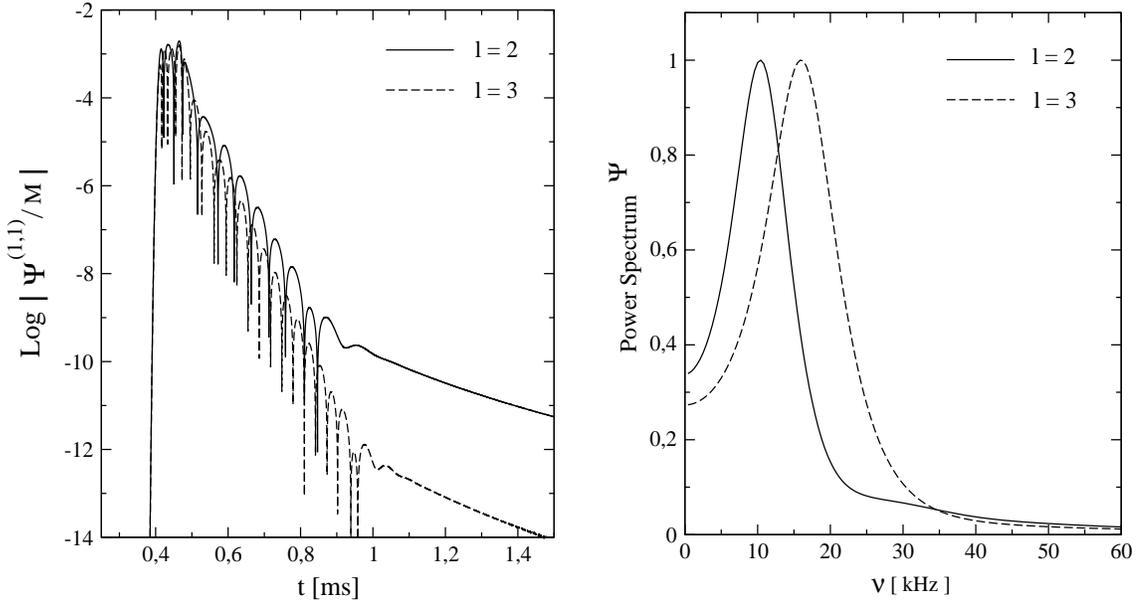}
\vspace{0.1cm}
\caption{\label{fig:Coupl_toymodel} \small{The wave forms (\emph{left
panel}) and spectra (\emph{right panel}) of the master function
$\psi$, which corresponds to the toy model studied in
section~\ref{sec:diff_rot}.  The presence of an arbitrary transient in
the source of equation~(\ref{sec:diff_rot}), which has been induced by
a delta function, excites the first $w$-mode.}}
\end{center}
\end{figure}
%
\begin{figure}[t]
\begin{center}
\includegraphics[width=150 mm, height=120 mm]{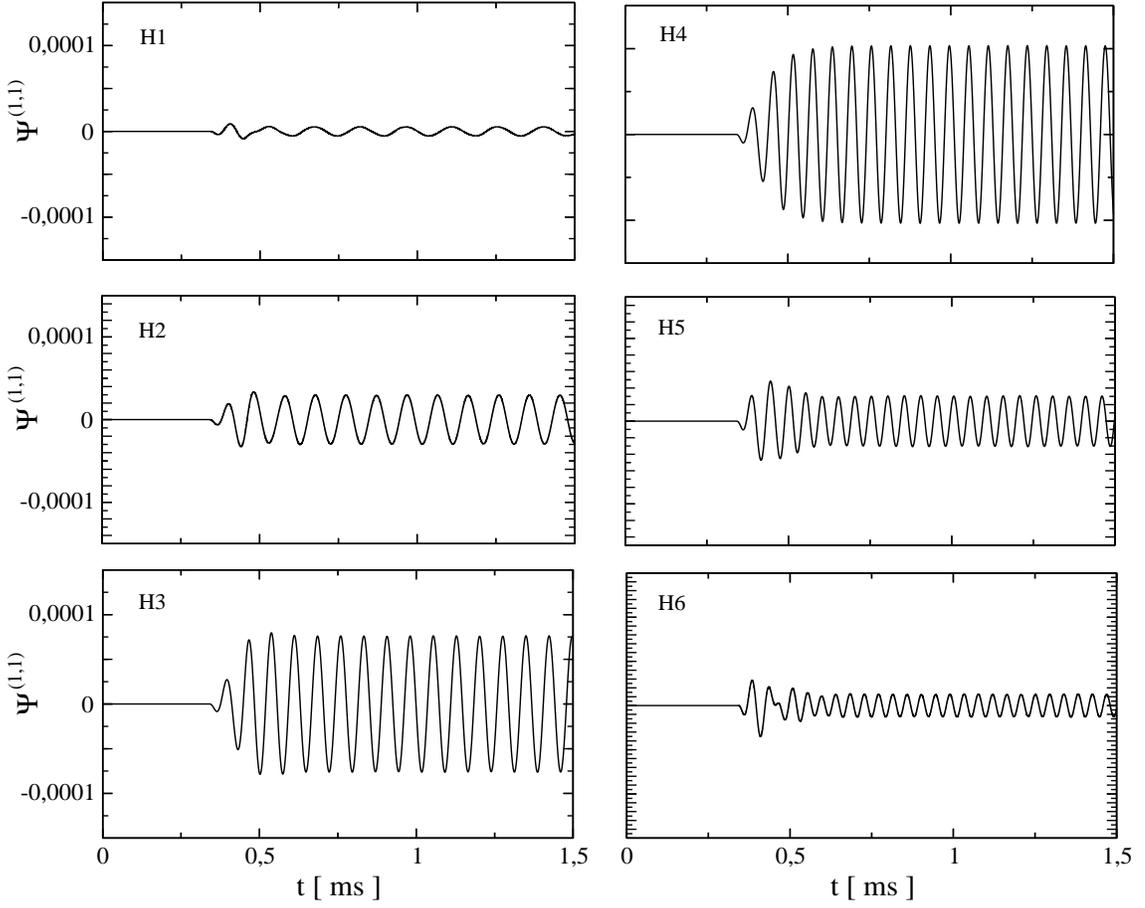} 
\caption{\label{fig:Coupl_Psi11_H16} \small{Comparison of six wave
    forms of the axial master function $\psi^{(1,1)}$ given in $km$,
    where the axial perturbations are described by the same
    differentially rotating configuration illustrated in
    figure~\ref{fig:Coupl_Psi11_H1}.  On the other hand, the radial
    pulsations have been excited each time with one of the six radial
    overtone (H1-6). The function $\psi^{(1,1)}$, which is plotted on
    the same scale, shows a resonance effect when the radial
    perturbations pulsate at H4 overtone, whose frequency is close to
    the first $w$-mode. }}
\end{center}
\end{figure}
%
\begin{figure}[t]
\begin{center}
\includegraphics[width=150 mm, height=120 mm]{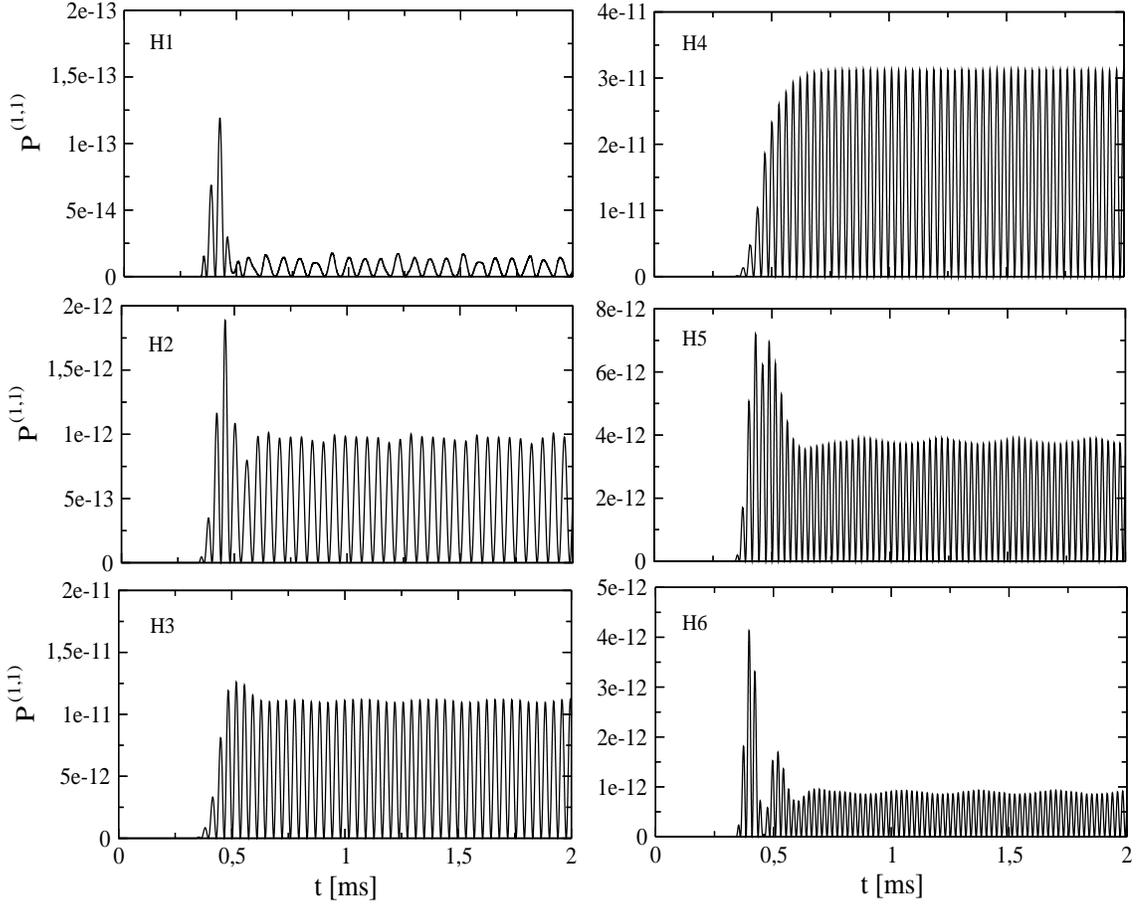} 
\caption{\label{fig:Coupl_Power11_H1H6} \small{For the same six
    simulations described in figure~\ref{fig:Coupl_Psi11_H16}, we plot
    the power emitted in gravitational waves at infinity. The six
    panels do not have the same scale.  }}
\end{center}
\end{figure}
%
\begin{figure}[t]
\begin{center}
\includegraphics[width=150 mm, height=120 mm]{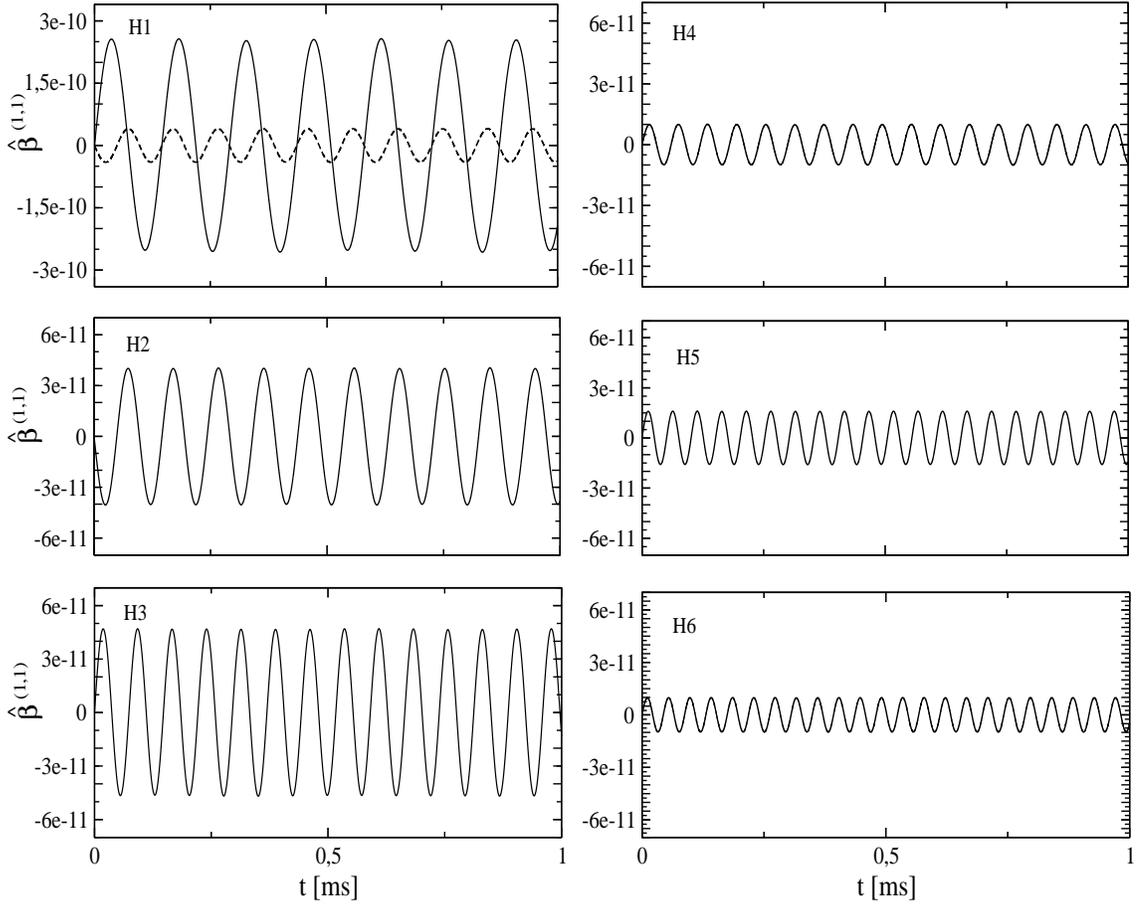} 
\vspace{0.1cm}
\caption{\label{fig:Coupl_bre11} \small{Time evolution of the second
    order fluid perturbation $\hat \beta^{(1,1)}$ in $km^{-1}$,
    which is related to the axial velocity through the
    definition~(\ref{hatbet11}). The perturbation $\hat \beta^{(1,1)}$
    has been averaged at each time step on the interior spacetime with
    the formula~(\ref{aver_f}).  The curves plotted in the six
    \emph{panels} refer to six evolutions where the radial pulsations
    have been excited every time with a single overtone of the radial
    modes. In the \emph{upper panel} of the \emph{first column}, we
    compare the $\hat \beta^{(1,1)}$ arising from the H1 (\emph{solid
      line}) and H2 (\emph{dashed line}) radial pulsations.  With the
    exception of the top panel of the first column, all the figures
    have the same scale.}}
\end{center}
\end{figure}
%
%
\begin{figure}[t]
\begin{center}
\includegraphics[width=150 mm, height=100 mm]{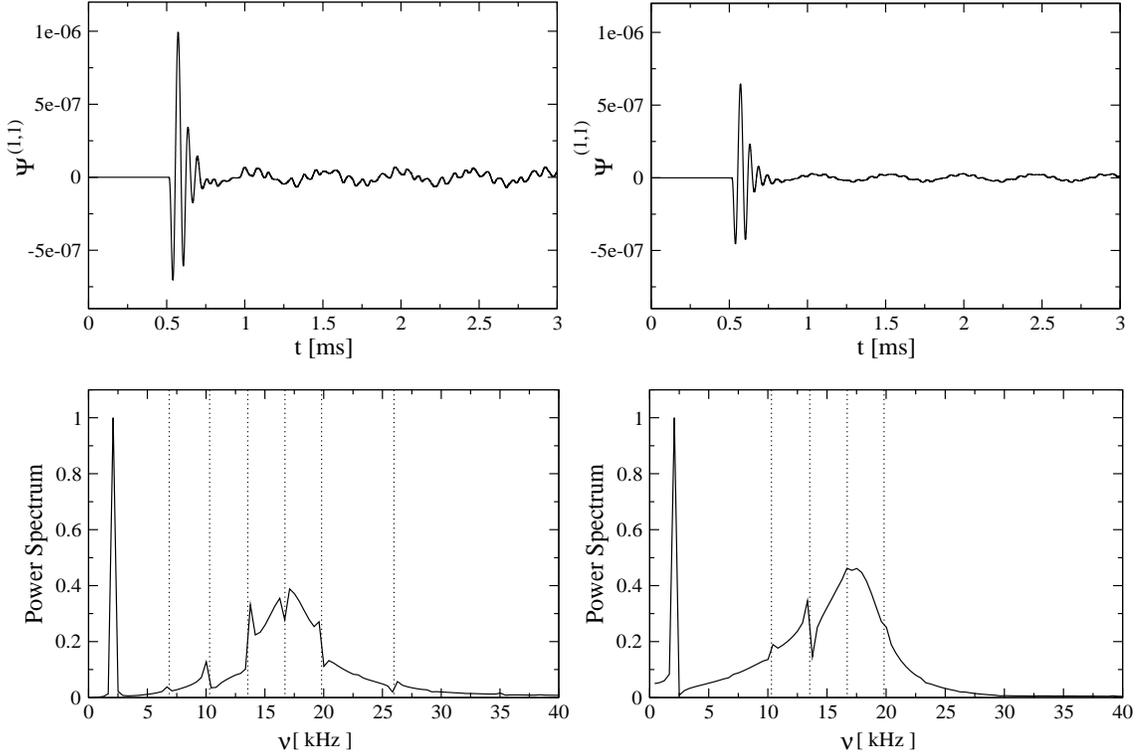}
\vspace{0.1cm}
\caption{\label{fig:Coupl_Psi11_FF} \small{Wave forms and spectra of
    the axial master function $\Psi ^{(1,1)}$, in $km$, for the
    coupling between differential rotation and radial pulsations
    excited by the F-mode.  The \emph{left column} displays the wave
    form (top) and spectrum (bottom) for a simulation where the
    junction conditions have been imposed on a hypersurface at $r
    =8.64~km$.  The \emph{right column} shows the same quantities, but
    now the matching surface is at $r=7.75~km$. In accordance to the
    results shown in figure~\ref{fig:Coupl_Psi11_H16}, the excitation
    of the F-mode and $w$-mode for $l=3$ are evident. In addition, the
    curves manifest the presence of spurious micro-oscillations that
    appear at the frequencies of the radial higher overtones
    (\emph{vertical dotted lines}). This noise is reduced when the
    matching is imposed on a more internal hypersurface.  }}
\end{center}
\end{figure}
\begin{figure}[t]
\begin{center}
\includegraphics[width=150 mm, height=100 mm]{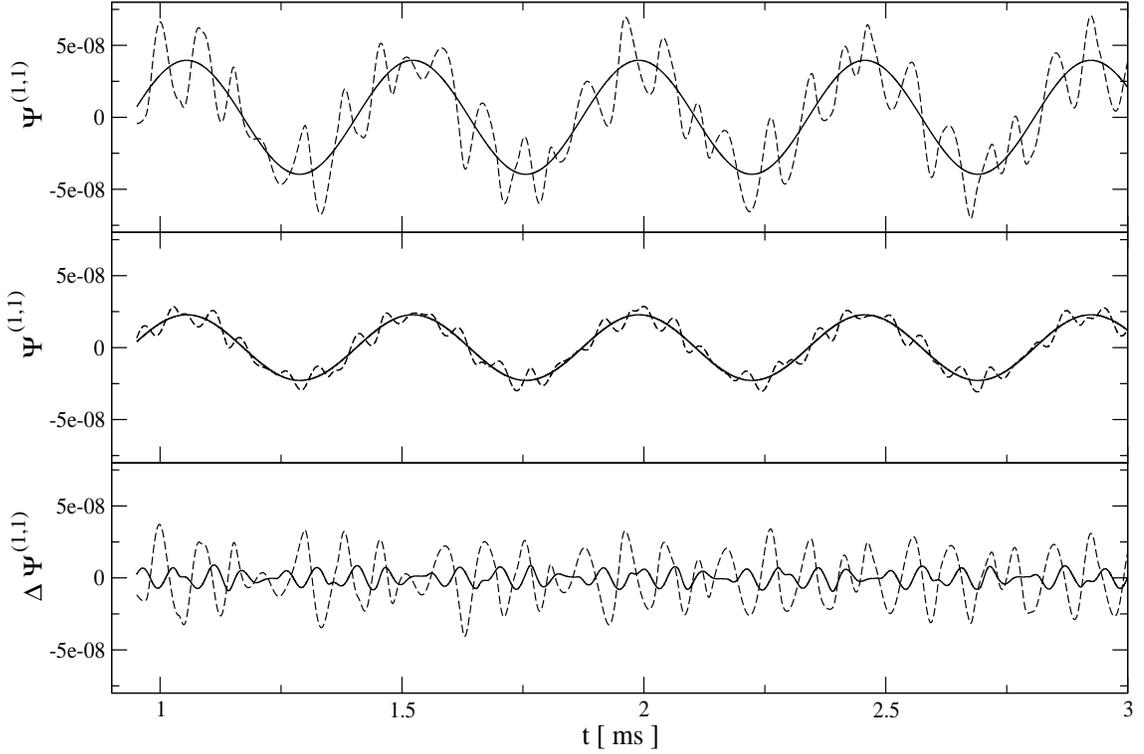}
\vspace{0.1cm}
\caption{\label{fig:Coupl_Fnoise} \small{The axial master function
    $\Psi ^{(1,1)}$, in $km$, (\emph{dashed line}) and interpolation
    function $f^{int}$ (\emph{solid line}) are plotted for the
    coupling between differential rotation and radial pulsations
    excited by the F-mode.  In the \emph{upper panel} the curves are
    obtained for a simulation where the junction conditions have been
    imposed on a hypersurface at $r =8.64~km$.  In the \emph{middle
    panel}, the same quantities are now determined by matching at
    $r=7.75~km$.  In the \emph{lower panel}, the difference $\Delta
    \Psi^{(1,1)} \equiv \Psi^{(1,1)} - f^{int}$ is plotted for $r
    =8.64~km$ (\emph{dashed line}) and $r=7.75~km$ (\emph{solid
    line}). }}
\end{center}
\end{figure}

\subsection{Coupling between radial pulsations and axial differential
 rotation}
\label{sec:diff_rot}

In this section, we present the results of the non-linear
perturbations which describe the coupling between a stationary
differential rotation and radial pulsations.  For the particular
choice of the initial rotating configuration the non-linear
perturbations studied in this section are relative to the harmonic
index $l=3$.  The axial rotation of the fluid is described with a good
approximation by the nearly Newtonian j-constant rotation law which
has been derived in section~\ref{sec:BC-IV_Axial01}.
As in section~\ref{sec:BC-IV_Axial01} and~\ref{sec:Simul-Ax01}, we
have specified the differential parameter ($A = 15~km$) and the
angular velocity at the rotation axis ($\Omega_c =
2.09~10^{-3}~km^{-1}$). This value, which corresponds to a $10~ms$
rotation period at the axis, is small with respect to the mass
shedding limit rotation rate $\Omega_c/ \Omega_{K} = 6.45 10^{-2}$.
The stationary profile of the axial velocity perturbation
$\beta^{(0,1)}$ for the harmonic index $l=3$ is given by the
expression~(\ref{bet01_IC}), and has been plotted in
the figures~(\ref{fig:beta01}) and (\ref{fig:Non_rad_beta01_diff}). \\
The radial pulsations are instead excited for any evolution at $t=0$
by selecting a single normal mode given by equation~(\ref{Rad_Ic}),
where the initial radial velocity perturbation $\gamma^{(1,0)}$ has
the form of equation~(\ref{gam_IC_nm}) with an initial amplitude
$A^{(1,0)}=0.001$. The radial simulations related to these initial
conditions provide the maximum Lagrangian displacements shown in
table~\ref{tab:TOV_Ener}, which have been determined at the static
surface of the star.
From these values, we can argue that the region of the spacetime where
the perturbed surface of the star moves is really confined in a narrow
region around the static equilibrium configuration. Therefore, the
issues related to the negative values of the mass-energy density in a
Eulerian description can be approximatively described by matching the
non-linear perturbations on a surface always contained into the star.
For more details see section~\ref{sec:BIC11_Num}.

We start by investigating the evolution of the axial master function
$\Psi^{(1,1)}$, when the radial perturbations oscillate at the
frequency of the first overtone H1. In
figure~\ref{fig:Coupl_Psi11_H1}  we show the wave form for an
observer at $100~km$ from the stellar centre. This curve presents
a first excitation of a typical $w$-mode followed by a monochromatic
periodic signal. This interpretation is confirmed by the spectrum
associated to this signal. Indeed the \emph{left panel} of
figure~\ref{fig:Spec_Psi11_H1} displays the  broad peak of the
$w$-mode, which is centered around the frequency $\nu = 16.092~kHz$,
and a narrow peak, which has the same frequency of the first overtone
H1 that has been mirrored in the gravitational signal at second
order. In the \emph{right panel}, the FFT of the signal has been
performed for a higher number of time cycles where the signal is
dominated by the periodic evolution of the source terms. As a result,
the energy contained at the frequency of H1 is much higher than that
of the curvature mode.
The presence of a periodic dynamics which reflects the radial
pulsations into these non-linear coupling arises from the structure of
the source terms. In fact for a stationary axial perturbation, we can
schematically write the sources of equations~(\ref{Psi11maseq})
and~(\ref{traseq11}) as
\begin{equation}
{ \cal S} = \sum_\sigma {\cal I}^{NR}_{n} \left( t,r \right)
{\cal J}^{R}_n \left( t,r \right) =
\sum_n {\cal I}^{NR}_{n} \left(r \right)
{\cal J}^{R}_n \left( r \right) e^{i\omega_n t}  \label{Source_scheme}
\end{equation}
where the superscripts $R$ and $NR$ denote respectively the linear
radial and non-radial perturbations and $\omega_\sigma$ are the
discrete set of radial modes. It is then evident that the dynamics is
entirely sustained by the normal modes of the radial perturbations and
that the $\lambda \epsilon$ gravitational signal $\Psi^{(1,1)}$ and
the axial velocity perturbation $\hat \beta^{(1,1)}$ oscillate at the
frequencies of the radial modes.  This behaviour is expected from the
symmetries of the stellar model. In fact, the radial frequencies could
be corrected by rotation starting from the perturbative order $\lambda
\epsilon ^2$, as they must be invariant for an inversion of the
rotation direction $\Omega \to - \Omega$. \\
On the other hand, the presence of the first $w$-mode seems to have a
different origin. This spacetime mode is only excited during the early
phases of the dynamical evolution.  As explained in
section~\ref{sec:BIC11_Num}, this gravitational burst is due to the
initial transient in $\Psi^{(1,1)}$, which is produced by the initial
constraint violation.  In order to confirm that a transient can
generate a burst of gravitational wave characterized by the excitation
of a $w$-mode, we modify the source of the axial master
equation~(\ref{Psi11maseq}) with an artificial transient described by
a delta function placed in a point inside the star at a given time.
This delta function is then immediatly removed after a time step. The
wave form and the spectrum of the resulting simulation carried out for
the harmonic indices $l=2,3$ are shown in
figure~\ref{fig:Coupl_toymodel}, where the excitation of the first
$w$-mode is evident.
%

When we extend this analysis to simulations where the radial
pulsations are excited with higher overtones and with the fundamental
mode, the wave forms and the spectra have the same features described
for the H1 case.  However, an interesting amplification is noticed in
the gravitational signal when the radial perturbations pulsate at
frequencies close to the axial $w$-mode.  In
figure~\ref{fig:Coupl_Psi11_H16}, we show with the same scale the
axial master function $\Psi^{(1,1)}$ for six simulations, where the
initial data for the radial perturbations are provided each time by
one of the first six overtones.  It is evident that the function
$\Psi^{(1,1)}$ increases in amplitude when the order of the radial
normal mode increases from the first to the fourth overtone, while the
amplitude decreases for the fifth and sixth overtone. The amplitude of
$\Psi^{(1,1)}$ related to the radial H4 evolution results is about
sixteen times the amplitude of $\Psi^{(1,1)}$ for a H1 evolution. For
this stellar model, the spacetime mode for the harmonic index $l=3$
has frequency $16.092~kHz$ which is between the frequencies of the
third and fourth overtones, $13.545~kHz$ and $16.706~kHz$
respectively.  In addition, we can notice that this effect is present
although the energy contained in the radial pulsations and the maximum
displacement of the surface decreases proportionally for higher radial
modes~(see table~\ref{tab:TOV_Ener}).  Considering the structure of
the axial master equation~(\ref{Psi11maseq}), this amplification is
certainly due to a resonance between the axial potential $V$, which
contains the properties of the QNM of the spacetime, and the source
that is pulsating at the radial eigenfrequencies.  In fact, we have
noticed that this effect disappears when the axial potential is
arbitrarily removed from the equation. In addition, the fluid variable
$\hat \beta ^{(1,1)}$ that obeys the conservation 
equation~(\ref{traseq11}) does not show amplification and decreases
proportionally with the order of the radial mode.  In analogy with a
forced oscillator, the amplification of the signal appears when one of
the natural frequencies, determined by the form of the axial potential
$V$, is sufficiently close to the frequencies associated
with the external force term. 
Moreover, it is interesting to notice in
figure~\ref{fig:Coupl_Psi11_H16} that in accordance with the broad
spectral peak of the $w$-modes (figure~\ref{fig:Spectrum01}), this
resonance affects a wide frequency band around the $w$-mode. \\
Consistently with the previous results, the power emitted at infinity
by gravitational radiation exhibits the amplification present in the
axial master function. In figure~\ref{fig:Coupl_Power11_H1H6}, we
determine the rate of the radiate energy for the six simulations which
have been described before and were shown in
figure~\ref{fig:Coupl_Psi11_H16}.  The power displays two distinct
contributions: \emph{i)} a first peak which is due to the energy
emitted by the first $w$-mode which comes out from the initial 
constraint violation, \emph{ii)} the periodic emission
due to the pulsations of the source.  However, the relative strengh of
these two effects changes when we excite the radial oscillations with
different eigenmodes.  In fact, the periodic emission first increases
proportionally with the order of the radial overtone excited (H1, H2
,H3). In correspondence with the H4 radial overtone it reaches its
maximum and dominate completely the radiation, and eventually it
decreases for the H5 and H6 overtones.

Now, we can study the coupling between differential rotation and
radial pulsations that oscillate at the fundamental frequency F-mode.
From the previous analyses, we can expect a gravitational signal with
at initial excitation of the axial $w$-mode, which is immediately
followed by periodic pulsations with the F-mode frequency which are
driven by the source. In addition, the axial potential $V$ should
lower the second part of the signal as the F-mode frequency is far
from the first $w$-mode.  This expected behaviour is present in the
wave forms shown in the upper panels of
figure~\ref{fig:Coupl_Psi11_FF}. Unfortunately, the periodic part of
the signal exhibits also some micro-oscillations.  They are the
spurious numerical oscillations discussed in
section~\ref{sec:Ch7_rad_sim}, which arise from the regions at low
density near the stellar surface and that seem to depend on the
profile of the eigenfunction.  The signal can be partially improved by
neglecting the outer layers of the star. In the \emph{left column} of
figure~\ref{fig:Coupl_Psi11_FF} the junction conditions have been
imposed on the hypersurface at radial coordinate $r=8.64~km$, which
corresponds to neglect the $0.3~\%$ of the stellar gravitational
mass. The waveform (\emph{upper panel}) and the spectrum (\emph{lower
panel}) clearly diplay the corruption of the signal due to the
numerical oscillations, which appear at the frequencies of the radial
mode overtones.  In the \emph{right column}, the matching surface is 
at $r=7.75~km$ and the mass neglected is the $6.3~\%$ of the total
mass of the star. Despite the lower amplitude of the oscillations we
can now notice that the waveform and spectrum are closer to the
expected form and the numerical noise is reduced. In the spectrum, we
can distinguish the excitation of the F-mode and $l=3$ $w$-mode as
well as some small peaks due to the numerical noise, especially at the
frequency $\nu = 13.4~kHz$.
In order to estimate the error introduced by the micro-oscillations in
the ``expected'' wave forms, we first proceed to eliminate the first
part of the signal characterized by the initial transient.  Then we
interpolate the periodic part of the wave form with a trigonometric
function,
 \begin{equation}
  f^{int} = c_0 \sin \left( 2 \pi ~\nu_{F} t + c_1 \right) \, 
 \end{equation}
 where $\nu_{F}$ is the F mode radial frequency and $c_0$ and $c_1$
 are two free parameters, which are determined by the interpolation
 algorithm. It is worth noticing that the interpolation functions for
 the two signals shown in figure~\ref{fig:Coupl_Psi11_FF} will be in
 general different, since the amplitude of the wave form depends on
 the position of the matching surface. Thus, we separately estimate
 the deviation between the evolved wave form and its interpolation
 function $f^{int}$ by using the following expression:
 \begin{equation}
    < \Delta \Psi >   \equiv   
\sqrt{ \frac{1}{J}  \sum _{j=1}^{J} \left(
  \Psi^{(1,1)}_j - f^{int}_j \right) ^2 }  \label{dev11} \, ,
 \end{equation}
 where $J$ is the number of grid points.  Equation~(\ref{dev11})
 provides $< \Delta \Psi > = 1.76\times10^{-8}~km$ and $< \Delta \Psi
 > = 4.45\times10^{-9}~km$ for the wave forms extracted on the
 junction surface at $r=8.64~km$ and $r=7.75~km$ respectively.  The
 corruption of the signal is then reduced in average by a factor of
 four at $r=7.75~km$. This improvement can be also noticed in
 figure~\ref{fig:Coupl_Fnoise}, where the wave forms and the
 interpolation functions are shown for the junction conditions at
 $r=8.64~km$ (\emph{top panel}) and $r=7.75~km$ (\emph{middle panel}),
 while the deviation $\Delta \Psi^{(1,1)} \equiv \Psi^{(1,1)} -
 f^{int}$ is shown in the \emph{lower panel}.

The perturbative approach used in this thesis does not include the
back reaction analysis, i.e. the damping of the radial oscillations or
the slowing of the stellar rotation due to the energy emitted by the
$\lambda \epsilon$ gravitational radiation.
This effect could be studied by investigating the third perturbative
order, which is beyond the aims of this thesis.  However, we can
provide a rough extimate of the damping time of the radial pulsations.
We can assume that the energy emitted in gravitational waves is
completely supplied by the first order radial oscillations, and that
the power radiated in gravitational waves is constant in time.  Thus
the damping time is given by the following expression:
\begin{equation}
\tau^{(1,1)}_{lm} \equiv \frac{E^{(1,0)}_n}{ < \dot E^{(1,1)}_{lm} > }  \label{dam_time} \, ,
\end{equation}
where $E^{(1,0)}_n$ is the energy of the radial pulsations relative to
the specific initial eigenfunction~(see table~\ref{tab:TOV_Ener}),
while $< \dot E^{(1,1)}_{lm} >$ is the time averaged value of the
power emitted by the $\lambda \epsilon$ axial gravitational wave,
which has been calculated by averaging equation~(\ref{Power11ax}).
The calculations carried out with the initial data described before
for the linear perturbations provide the results reported in
table~\ref{tab:damping_time}. From the definition (\ref{dam_time}) and
the bilinear dependence of the non-linear perturbations $\lambda
\epsilon$, the damping time $\tau ^{(1,1)}_{30}$ depends only on the
non-radial parameter $\epsilon$, as $\tau ^{(1,1)}_{lm} \sim
\epsilon^{-2}$.  In the third row of table~\ref{tab:damping_time}, we
write the damping of the radial pulsations due to the coupling in
terms of oscillation periods for any radial mode, namely
\begin{equation}
N_{osc} = \frac{ \tau ^{(1,1)}_{lm} }{ P_{n} } \, ,
\end{equation}
where $P_{n} = \nu_{n}^{-1}$ and $\nu _{n}$ is the eigenfrequency of
the radial normal mode.

\begin{figure}[t]
\begin{center}
\includegraphics[width=150 mm, height=100 mm]{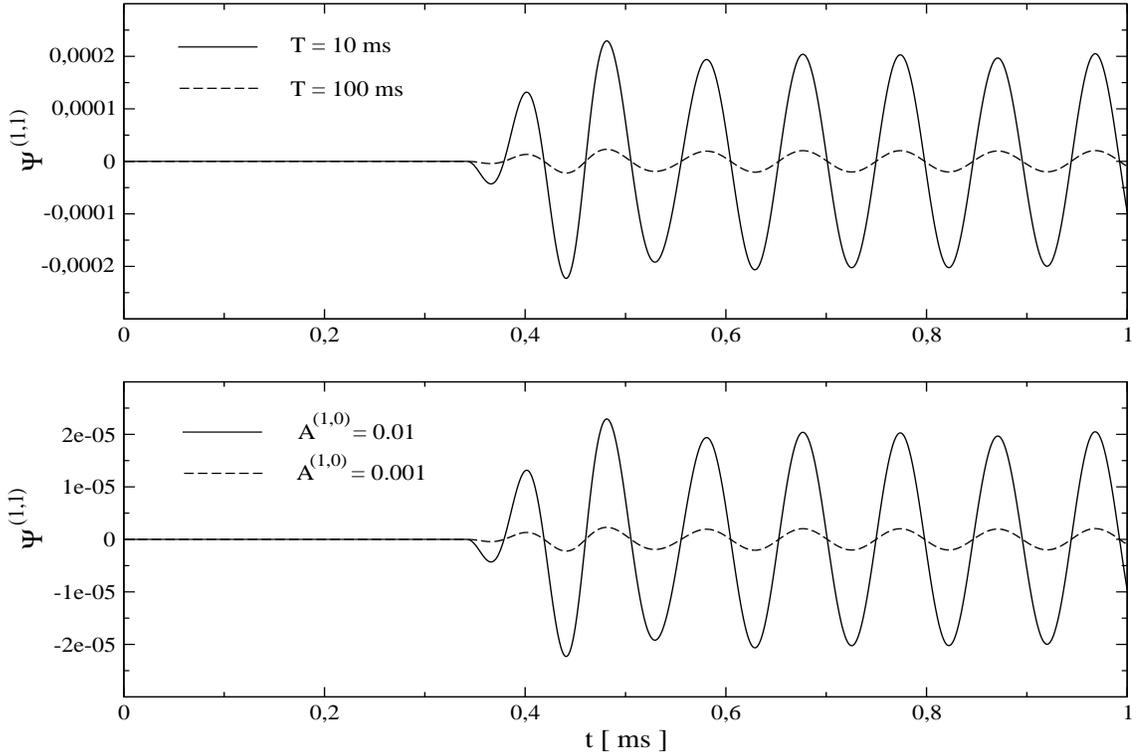}
\vspace{0.1cm}
\caption{\label{fig:Cou_consPsi} \small{Axial master function
    $\Psi^{(1,1)}$, in $km$, for four different initial values of the
    first order perturbative parameters. The \emph{upper panel}
    displays the wave forms obtained for radial perturbations excited
    with a H2 overtone of amplitude $A^{(1,0)} = 0.01$ and for
    axial perturbations with rotation periods $T_c = 10~ms$
    (\emph{solid line}) and $T_c = 100~ms$ (\emph{dashed line}). In
    the \emph{lower panel} the rotation period is fixed at $T_c =
    10~ms$ and the amplitude of the radial pulsations is changed by an
    order of magnitude, $A^{(1,0)} = 0.01$ (\emph{solid line}) and
    $A^{(1,0)} = 0.001$ (\emph{dashed line}).}}
\end{center}
\end{figure}

The $\lambda \epsilon$ non-linear perturbations are bilinear with
respect to their perturbative parameters. In order to test this
property in the numerical simulations, we have performed a first
simulation by fixing a basic rotation period for the axial
perturbations $T_0 = 10~ms$ and an initial amplitude of the velocity
perturbation $A_0 = 0.001$, and we have determined the averaged values
of the non-linear perturbations $\Psi^{(1,1)}_{0}$ and
$\beta^{(1,1)}_{0}$. Then, we have carried out several simulations
with different periods $T_n = n T_0$, and radial amplitude $A_m = m
A_0$, where $n,m \in \mathbb{N}$. The associated values for
$\Psi^{(1,1)}_k = k \Psi^{(1,1)}_0$ and $\beta^{(1,1)}_k = k
\beta^{(1,1)}_0$, manifest the bilinear character expected as their
values scale almost perfectly with the coupling constant $k = m n$.
An example is given in figure~\ref{fig:Cou_consPsi}, where we have
carried out four simulations with different values of the perturbative
parameters. In the \emph{upper panel}, we have maintained fixed the
amplitude of the radial pulsations $A^{(1,0)} = 0.01$ for the H2
overtone, and we have changed the rotation period of the differential
rotation by an order of magnitude. In the \emph{lower panel}, the
rotation rate is fixed at $T_c = 100~ms$ and the radial initial
amplitude is modified by an order of magnitude.  In both cases, the
axial master function $\Psi^{(1,1)}$ scales by a factor of ten.
Therefore, the results obtained in this section can be used to
extrapolate the values of the non-linear perturbations for
different values of the initial perturbative parameters associated with
the radial and non-radial perturbations. Of course, this procedure is
correct as long as the strength of the parameters is within the range
of validity of the perturbative approach.

The simulations for the non-linear perturbations are stable for very
long time evolutions and have a first rate order of convergence, as
expected from the implementation of the first order accurate Up-wind
numerical scheme.

%
%
\begin{table}[!t]
\begin{center}
\begin{tabular}{l  c c c c c c c  }
\hline \hline \\ & F & H1 & H2 & H3 & H4 & H5 & H6 \\ \\ \hline \hline
                             \\ $ \dot E^{(1,1)}~[10^{-13}]$ & $1.85
                             \times 10^{-6}$ & $6.83 \times 10^{-2}$ &
                             $4.85$ & $56.03 $ & $157.01 $ & $19.08 $
                             & $4.45 $ \\ \\ $ \tau^{(1,1)}_{30}~[ms]$
                             & $6.49 \times 10^{9}$ & $20.49 \times
                             10^{3}$ & $94.15$ & 3.69 & 0.72 & 3.67 &
                             10.51 \\ \\ $ N_{osc} $ & $1.39 \times
                             10^{10}$ & $1.408 \times 10^{5}$ & $
                             971.58 $ & $49.99$ & 12.07 & 72.78 &
                             240.77 \\ \\ \hline
\end{tabular}
\caption[]{\label{tab:damping_time} \small{Rate of energy $\dot
E^{(1,1)}$ emitted in gravitational waves at infinity with the
coupling between radial pulsations and axial differential
rotation. Estimated values of the damping times $\tau ^{(1,1)}_{30}$
and number of oscillation periods $N_{osc}$ necessary for the
non-linear oscillations to radiate the whole energy initially
contained by the radial modes.}}
\end{center}
\end{table}
%
%
\begin{figure}[t]
\begin{center}
\includegraphics[width=150 mm, height=100 mm]{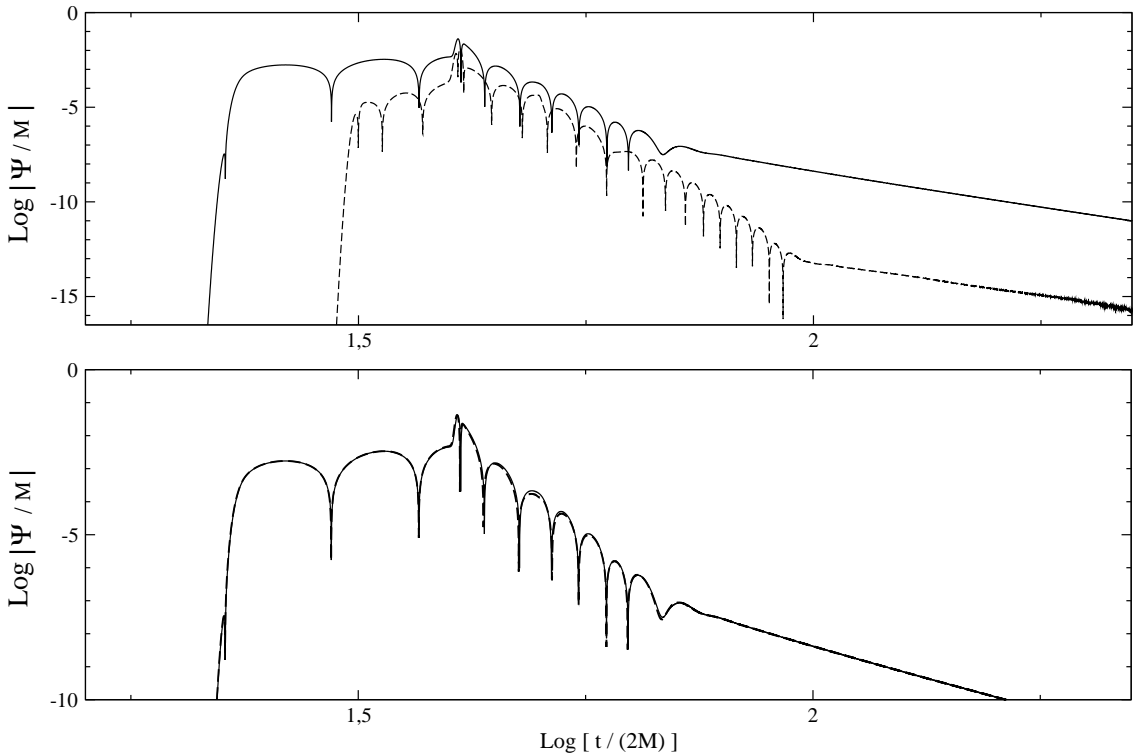}
\vspace{0.1cm}
\caption{\label{fig:Cou_ScaF} \small{Scattering of an axial
gravitational wave on a radially oscillating star which is pulsating 
in the fundamental radial mode.  In logarithmic scale, the \emph{upper
panel} displays the wave forms of the first order axial function
$\Psi^{(0,1)}$ (\emph{solid line}) and the second order axial function
$\Psi^{(1,1)}$ (\emph{dashed line}). The \emph{lower panel} shows
again with the \emph{solid line} the wave forms of $\Psi^{(0,1)}$ and
with the \emph{dashed line} the function $\Psi^{(0,1)} +
\Psi^{(1,1)}$.}}
\end{center}
\end{figure}
\begin{figure}[t]
\begin{center}
\includegraphics[width=150 mm, height=100 mm]{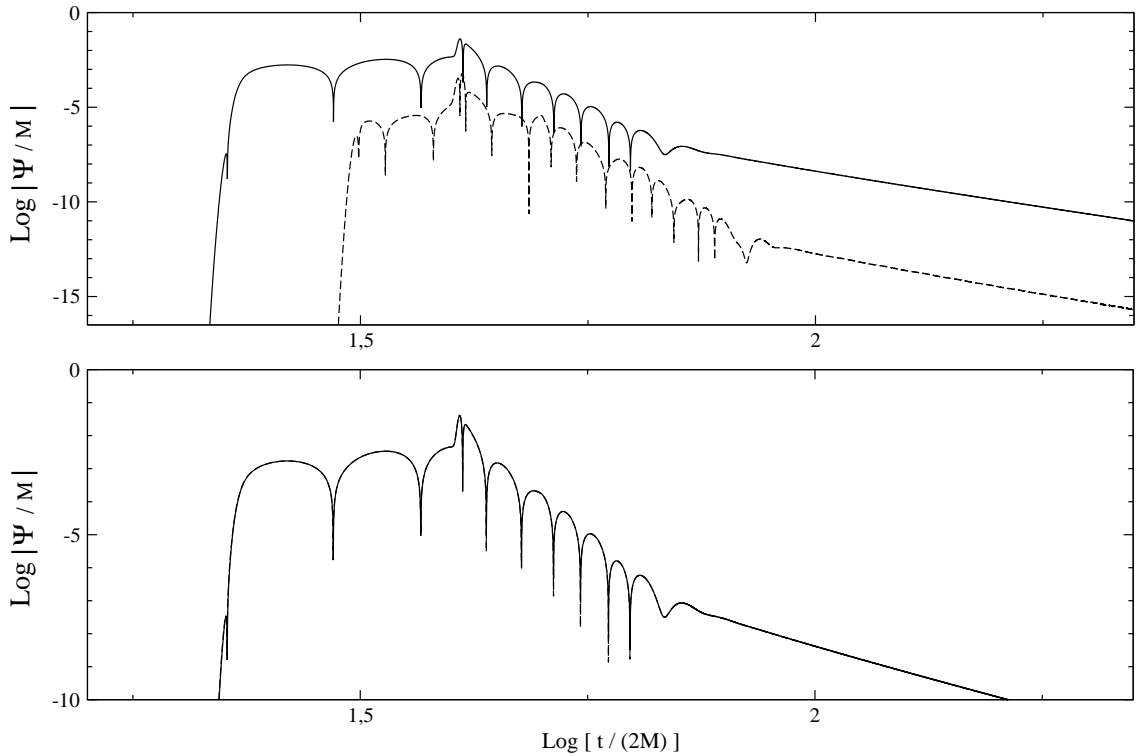}
\vspace{0.1cm}
\caption{\label{fig:Cou_ScaH1} \small{With the same notation of
figure~\ref{fig:Cou_ScaF}, the wave forms are related to the
scattering of the gravitational waves on a radially pulsating star
where only the first overtone H1 is excited.}}
\end{center}
\end{figure}
\begin{figure}[t]
\begin{center}
\includegraphics[width=150 mm, height=100 mm]{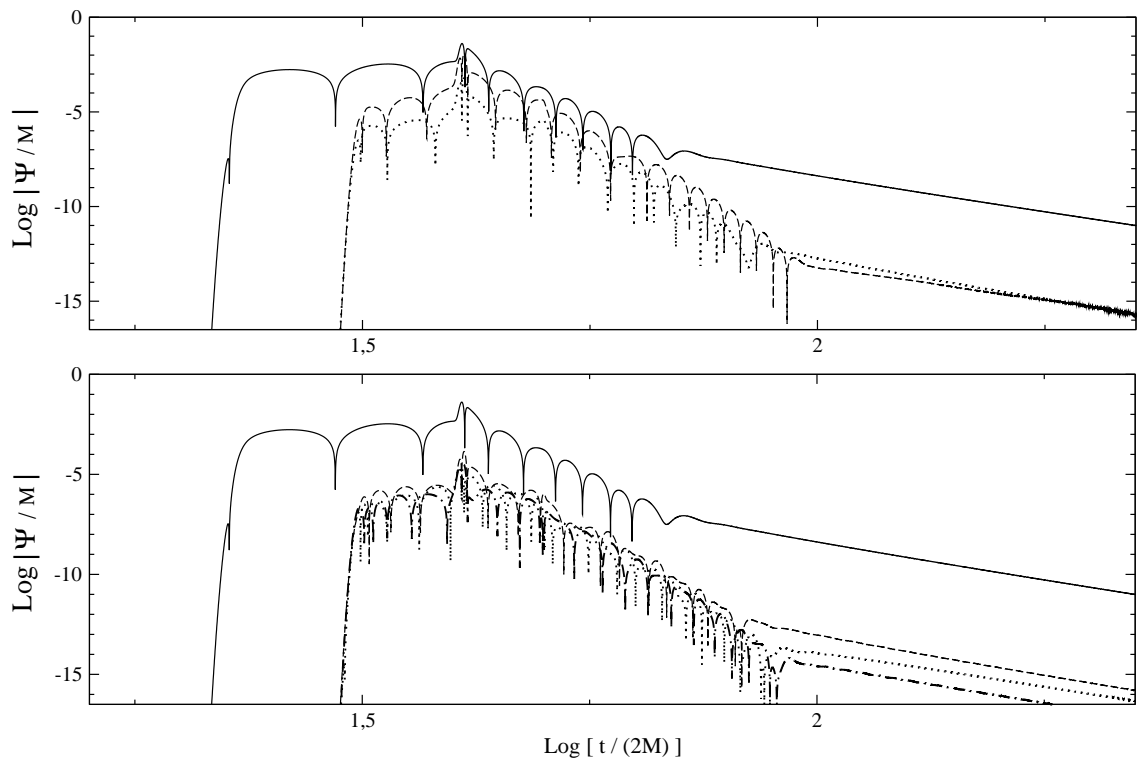}
\vspace{0.1cm}
\caption{\label{fig:Cou_Scacomp} \small{Scattering of the axial
gravitational wave on a radially pulsating star. The \emph{solid line}
in both panels refers to the wave form of $\Psi^{(0,1)}$. In the
\emph{upper panel} the \emph{dashed} and \emph{dotted lines} denote 
the wave forms of $\Psi^{(1,1)}$ for radial pulsations excited by a
F-mode and H1-mode respectively. In the \emph{lower panel} instead,
the \emph{dashed, dotted} and \emph{ dot-dashed lines} describe the
wave forms of $\Psi^{(1,1)}$ for radial pulsations excited by H2, H3,
and H4 overtones respectively. }}
\end{center}
\end{figure}
%
\begin{figure}[t]
\begin{center}
\includegraphics[width=150 mm, height=100 mm]{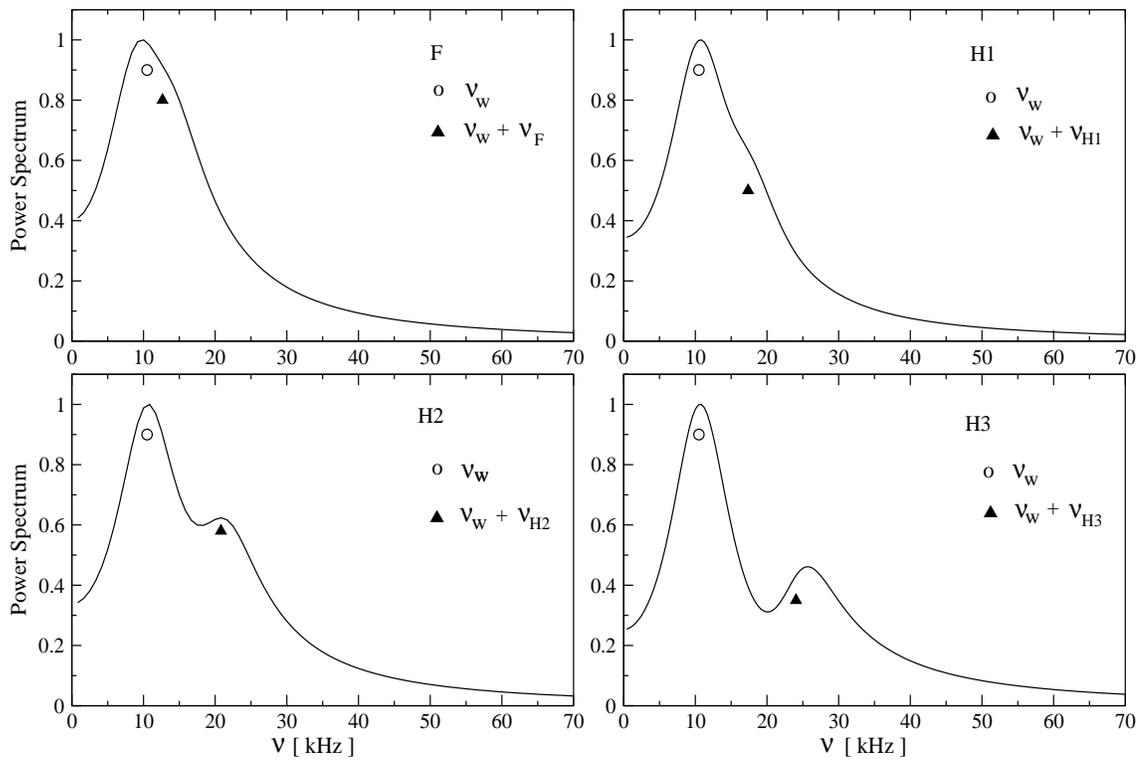}
\vspace{0.1cm}
\caption{\label{fig:Cou_Scacomp} \small{Power Spectrum of the wave
form $\Psi^{(1,1)}$, which is obtained from the scattering of a axial
gravitational wave on a radially pulsating star. The radial pulsations
have been excited by selecting in each simulation a particular radial
mode, i.e. the F mode and its first three overtones (see labels in the
four panels).}}
\end{center}
\end{figure}

\subsection{Effects of radial pulsations on the scattering of a gravitational
wave}

In this section, we summarise our study of the scattering of an axial
gravitational waves on a radially pulsating spherically symmetric
star. The aim is to understand whether the outcoming signal contains
some signature of the radially oscillating dynamics of the star.  At
first order in $\epsilon$, the results are well known and have been
re-proposed and briefly discussed in section~\ref{sec:Simul-Ax01}.
The scattered gravitational wave displays an excitation of the first
$w$-mode and the associated strongly damped ringing phase.

The initial values for the first order axial perturbations are set up
for the harmonic index $l=2$ with the condition~(\ref{IC_01_sct}).
The velocity perturbation $\beta^{(0,1)}$ then vanishes and the axial
master function $\Psi^{(0,1)}$ is given by the Gaussian
pulse~(\ref{Imp_psi01}) centered at $r_0 = 20~km$ with amplitude
$A^{(0,1)} = 0.1~km$ and width parameter $q=1.25~km^{-2}$. The radially
oscillating phase is excited with the radial eigenfunctions as we have
already done in the previous section~\ref{sec:diff_rot}. However, in
order to have a stronger coupling we have increased the initial
amplitude of $\gamma^{(1,0)}$ by an order of magnitude,
i.e. $A^{(1,0)} = 0.01$.

In figures~(\ref{fig:Cou_ScaF})-(\ref{fig:Cou_Scacomp}), we have
plotted the first order gravitational master function $\Psi^{(0,1)}$
and the second order correction due to the coupling $\Psi^{(1,1)}$ on
a logarithmic scale.  The second order signal provides small
corrections, less than $2$ percent when the radial pulsations are
excited by the F-modes, and less than $0.1$ percent for the higher
radial overtones. These corrections do not modify the properties of
the wave forms and spectra of the linear analyses. Furthermore, we
can notice that the amplification of the signal found in the
differential rotating configuration is not present in this case.
The fundamental difference between this and the previous case, when
the axial perturbation was related to differential rotation, is
essentially due to the difference in the properties of the source
terms.  For the scattering, the source mainly acts for a relatively
short period ot time, which is given by the travel time of the linear
gravitational wave across the star. After the wave has been scattered,
the first order signal still present in the star decreases according
to the long time decay power law~(section~\ref{sec:Simul-Ax01}). As a
result at second order a longer ringing phase appears which is anyway
well below the first order signal.
On the other hand, in the coupling between radial pulsations and
differential rotation the source terms act periodically into the star
forever, as this model does not contain the back-reaction. As a
result, the source has more time to couple with the non-radial
perturbations.

Although the non-linear coupling does not change the linear results,
we can study whether the $\lambda \epsilon$ wave forms and spectra
contain some signatures of the radial pulsating dynamics of the star.
From figures~\ref{fig:Cou_ScaF} and \ref{fig:Cou_ScaH1}, we can notice
that for dynamical times $\log \left[ t/(2M) \right] \sim 1.70$ the
ringing phase presents a small bump, which is due to a small numerical
reflection at the stellar surface that appears when the $\lambda
\epsilon$ gravitational wave is moving out the star.  In particular,
this effect is caused by the discontinity of the source terms, which 
vanish in the exterior spacetime. In addition, the ringing phase of
$\Psi^{(1,1)}$ seems to contain some interference effects (see
figure~\ref{fig:Cou_Scacomp}). This suggests the presence of at least
two oscillation frequencies.  Therefore, we perform an FFT of the
signal by selecting the part of the wave form after the numerical
reflection. In figure~\ref{fig:Cou_Scacomp}, the spectra show the
presence of the $l=2$ $w$-mode and a non-linear harmonics, whose
frequency is the sum of the radial and $w$-mode frequencies.  It is
worth noticing that this non-linear harmonic exhibits a strong damping
and then the characteristic broad shape of the spacetime modes.
However, as we have already specified these effects are much lower than
the linear results and do not modify the spectra of the linear
analysis (see e.g.figure~\ref{fig:Spectrum01}).

The code for the analysis of the non-linear scattering of a
gravitational wave on a radial pulsating star manifests a first order
of convergence. This is expected from the discontinuity of
$\Psi^{(1,1)}_{, \, rr}$ at the stellar surface, due to the presence
of the source terms only in the interior spacetime.

\chapter{Conclusions}
\label{ch:conclusions}

With the aim of investigating the dynamics of non-linear oscillations
of compact stars and the related gravitational radiation, we have
presented in this thesis a formalism and a numerical code which enable
us to study in the time domain the coupling between the radial and
non-radial perturbations of perfect fluid non-rotating compact stars.
The formalism for the polar perturbations has been worked out in a
first paper~\cite{Passamonti:2004je}, while the formalism and the
applications to axial perturbations are presented
in~\cite{Passamonti:2005axial}. The applications of the polar
perturbations will be presented in a future work.\\
\indent In order to have a well-defined framework to consider the
gauge dependence of linear and non-linear perturbations, we have found
it very convenient to address this topic by using the multi- parameter
relativistic perturbation theory introduced
in~\cite{Bruni:2002sm,sopuerta-2004-70}.  We have then carried out an
expansion of the metric, the energy-momentum tensor and the Einstein
and conservation equations in terms of two parameters $\lambda$ and
$\epsilon$, where $\lambda$ parameterizes the radial pulsations,
$\epsilon$ the non-radial oscillations and the $\lambda\epsilon$ terms
describe their coupling.  The spherical symmetry of the stellar
background allows us to separate in the perturbative tensor fields the
time and radial dependence from the angular variables, by using the
standard tensor harmonic basis.  Therefore, for any harmonic indices
$(l,m)$ the linear and the $\lambda\epsilon$ non-linear perturbations
have been described by a system of perturbative equations which forms
a 1+1 problem, where one dimension is given by the time coordinate and
the other by the radial (area) coordinate $r$. The tensor harmonic
expansion separates all the perturbative fields in two independent
classes, the axial (odd-parity) and polar (even-parity) perturbations,
which are defined by their behaviour under a parity transformation.
The non-linear perturbative equations and the definition of the gauge
invariant non-linear quantities have been derived by using the
2-parameter perturbative theory in connection with the formalism of
Gerlach and Sengupta~\cite{Gerlach:1979rw}, and Gundlach and
Mart\'{\i}n Garc\'{\i}a~\cite{Gundlach:1999bt,Martin-Garcia:2000ze}.
This formalism describes generic one-parameter non-radial
perturbations of a time-dependent and spherical symmetric spacetime in
terms of a set of gauge invariant perturbations.  Our approach
consists in exploiting the spherical symmetry of the radial
perturbations in order to separate the spherical and time-dependent
spacetime of the GSGM formalism in a static background, which
describes the equilibrium configuration, and a first order radially
pulsating spacetime.  We have then fixed the gauge of the radial
perturbations and have {\it i)} used the GSGM gauge-invariant
non-radial $\epsilon$ variables on the static background, and {\it
  ii)} defined new second order $\lambda\epsilon$ variables,
describing the non- linear coupling of the radial and non-radial
linear perturbations, which are also gauge-invariant for general
$\lambda\epsilon$ gauge transformations with the radial gauge fixed.
We have then derived the evolution and constraint equations for the
non-linear $\lambda\epsilon$ perturbations both for the polar and
axial sectors.  As expected, in the interior the $\lambda\epsilon$
variables satisfy inhomogeneous linear equations where the homogeneous
part is governed by the same linear operator acting on the first order
$\epsilon$ non-radial perturbations, while the source terms are
quadratic and made of products of $\lambda$ and $\epsilon$ terms. In
the exterior the sources vanish and there is no direct coupling, and
the whole dynamics is described by the $\lambda\epsilon$ order Zerilli
and Regge-Wheeler functions, respectively for the polar and axial
perturbations. Thus the effect of the coupling is transmitted from the
interior to the exterior through the junction conditions at the
surface of the star. \\
We have given a brief discussion of the boundary conditions, focusing
on those at the stellar surface. In order to avoid negative values
of the mass energy density near the surface due to the Eulerian
description of the radial and non-radial polar perturbations, we have
adopted an approximation already used in
literature~\cite{Sperhake:2001si, Sperhake:2001xi,
2003PhRvD..68b4002H}, i.e. we have removed the outer layers of the
neutron star, which corresponds to neglecting less than one percent of
the total gravitational mass of the star. This approximation leads to
a description of the second order $\lambda\epsilon$ gravitational
radiation which is accurate to better than five per cent.

The perturbations at first order have been studied with the GSGM
quantities. However, in some cases we have redefined these variables
or changed them for computational purposes.  The \emph{radial
perturbations} have been described by four perturbative fields, two
metric and two fluid, which obey three first order in time evolution
equations and two constraints, as there is a single radial degree of
freedom. This allows us to set up either a hyperbolic-elliptic
formulation (HEF) or a purely hyperbolic formulation (PHF).  The
numerical simulations have shown a good accuracy in both cases.  The
perturbative equations for the \emph{polar} $\epsilon$
\emph{non-radial perturbations} and for the \emph{polar} $\lambda
\epsilon$ \emph{coupling} rely on the hyperbolic-elliptic system of
three equations already used in ~\cite{Nagar:2004ns}, where the two
hyperbolic equations describe respectively the gravitational wave and
the sound wave, while the elliptic equation is the Hamiltonian
constraint.  We have found that this system of equations is more
suitable for numerical integration than others available in
literature, as the Hamiltonian constraint is used for updating at any
time step one of the unknowns of the problem.  Therefore, the errors
associated with the violation of this constraint are automatically
corrected. However, the sound wave equation is only known for first
order polar non-radial perturbations. Thus, in order to determine it
at the $\lambda \epsilon$ perturbative order, we have first obtained
this wave equation for generic non-radial perturbations on a
barotropic, time dependent and spherically symmetric background.
Then, we have used the 2-parameter expansion and determined the source
terms relative to the $\lambda \epsilon$ perturbations.  The
\emph{axial non-radial perturbations} for the linear $\epsilon$ and
non-linear $\lambda \epsilon$ orders have been studied with a system
of two equations, the axial master wave equation and a conservation
equation.  The former describes the only gauge invariant metric
variable of the axial sector, while the latter the axial velocity
perturbation.  At first perturbative order, the stationary character
of the axial velocity allowed us to study separately the dynamical
degree of freedom of the spacetime and its stationary part.  Linear
axial metric perturbations describe the dragging of the inertial
frame, the linear axial velocity perturbation represents a stationary
differentially rotation.

Since the linear axial velocity induces a stationary differential
rotation, the related metric solution describes the dragging of the
inertial frame. In the axial sector, we have redefined the axial
velocity in order to have a vanishing quantity on the stellar surface.
This definition leads to simpler perturbative equations and a better
behaviour of the $\lambda \epsilon$ second order source term near the
surface of the star.

We have then presented the numerical code for investigating in the time
domain the evolution of the axial $\lambda\epsilon$ perturbations.
This code is based on finite differencing methods and on standard
explicit numerical schemes for integrating partial and ordinary
differential equations.  The structure of the code reflects the
hierarchy of the perturbative frameworks. Starting from the TOV
solutions and two independent initial conditions for the radial and
non-radial perturbations, the code evolves at any time step the linear
radial $\lambda$ and axial non-radial $\epsilon$ perturbations, thus
updates the source terms of the $\lambda\epsilon$ perturbative
equations, and eventually simulates the evolution of the non-linear
perturbations.
In order to have more accuracy in the description of the radial
pulsations near the surface, we have increased the resolution in this
region by adopting the fluid tortoise coordinate, which is necessary
only for the radial perturbative equations. Therefore, we have
introduced in the numerical code two meshes for describing the radial
and non-radial perturbations separately at first order. In order to
have source terms calculated at the same grid point, we have also
introduced an interpolation that at any time step connects the
simulations of the radial pulsations between the two meshes.

In this thesis the static equilibrium configuration is determined by
solving the TOV equations for a polytropic equation of state with
adiabatic index $\Gamma=2$ and $k = 100~km^2$, and for a central mass
energy density $\rho_c = 3 \times 10^{15}~g~cm^{-3}$. The star then
has $M = 1.26 M_{\odot}$ and $R=8.862~km$.  \\ The structure of the
radial and axial perturbative equations enables us to set up two
independent initial configurations: \emph{i)} scattering of an axial
gravitational wave on a radially pulsating star, \emph{ii)} a first
order differentially rotating and radially pulsating star.  \\ 
\indent The initial configuration for the radial pulsations has been
excited by selecting specific radial eigenmodes. We have chosen an
origin of time such that the radial eigenmodes are described only by
the eigenfunctions associated with the radial velocity perturbation
$\gamma^{(1,0)}$.  To this end we have first determined the wave
equation for this variable, then we have managed it in order to set up
a Sturm-Liouville problem.  This eigenvalue problem has been solved
with a numerical code based on the shooting method, where the
eigenfrequencies of radial modes are accurate to better than 0.2
percent with respect to published values.  We have performed some
tests for the part of the code dedicated to the radial pulsations. The
simulations for any initial radial mode satisfy with high accuracy the
Hamiltonian constraint and are stable for very long evolutions. The
radial spectrum, which has been determined with a FFT of the time
evolution, reproduce the published results with an accuracy to better
than 0.2 percent. \\
\indent For the first order axial perturbations we have excited the dynamical
degree of freedom of the spacetime with a standard gravitational wave
scattering on the star, where the impinging axial gravitational wave
is described by a Gaussian pulse.  The tests carried out on the
numerical code show that we are able to reproduce the wave forms and
the spectra known in literature for the harmonic indices $l = 2,
3$. The axial differential rotation instead has been described by
expanding in vector harmonics the relativistic j-constant rotation
law, then taking the first component which is related to the
gravitational wave emission, that is $l=3$.
We can specify two parameters in the initial profile for the axial
velocity perturbations, i.e. the differential parameter $A$ and the
angular velocity at the rotation axis $\Omega_c$.  The value for $A$
has been chosen in order to have a smooth profile and a relatively
high degree of differential rotation, as for high $A$ the rotation
tends to be uniform and then the $l=3$ component vanishes.  We have
chosen an angular velocity that corresponds to a $10~ms$ rotation
period at the axis of the star. For other values of the angular
velocity, the linearity of the perturbative equations allows us to get
the respective gravitational signal with a simple re-scaling.
The relativistic j-constant rotation law contains a term which is
related to the dragging of the inertial frame. In order to simplify
the expression of our initial condition we have estimated the amount
of correction associated with the $l=3$ component of this relativistic
term. We have then neglected it, thereby introducing an error of less
than ten percent.

For the \emph{first initial configuration}, i.e. the scattering of an
axial gravitational wave on a radially pulsating star, the second
order signal provides small corrections, less than $2$ percent when
the radial pulsations are excited by the F-modes, and less than $0.1$
percent for the higher radial overtones. These corrections do not
modify the properties of the wave forms and spectra of the linear
analyses.  The \emph{second configuration}, where the linear axial
perturbations describe a stationary differential rotation and the
associated frame dragging, produces an new interesting gravitational
signal. The wave forms have these properties: i) an excitation of the
first $w$-mode at the early stage of the evolution ii) a periodic
signal which is driven by the radial pulsations through the source
terms.  The spectra confirm this picture by showing that the radial
normal modes are mirrored in the gravitational signal at non-linear
perturbative order.  
However, the excitation of the $w$-mode at the early stages of the
numerical simulations is an unphysical response of the system to the
initial violation of the axial constraint equations for the coupling
perturbations. 
Moreover, a resonance effect is present when the
frequencies of the radial pulsations are close to the first $w$-mode.
For the stellar model considered in this thesis the amplitude of the
gravitational wave signal related to the fourth radial overtone is
about three orders of magnitude higher than that associated with the
fundamental mode.  We have also roughly estimated the damping times of
the radial pulsations due to the non-linear gravitational emission.
Their values radically depend on the presence of resonances.  For a
$10~ms$ rotation period at the axis and $15~km$ differential
parameter, the fundamental mode damps after about ten billion
oscillation periods, while the fourth overtones after ten only.  This
is not surprising, and shows that the coupling near resonances is a
very effective mechanism for extracting energy from the radial
oscillations.

It is worthwhile to remark that a possible detection of this
gravitational signal could provide new information of the stellar
parameters, as the second order gravitational spectra reproduce those
of the radial modes of a non-rotating star, which can be determined
easily for a large class of equations of state.  The numerical code
manifests a first order convergence for the simulations of the
non-linear axial perturbations arising from the coupling between the
first order radial and axial non-radial perturbations.

\subsubsection{Future extensions}
The implementation of the numerical code for studying the coupling
between the radial pulsations and the polar non-radial perturbations
is currently under way.  The spectrum of the polar non-radial
perturbations is richer than the axial case due to the presence of the
fluid modes, which may have frequencies lower than the spacetime modes
and a longer gravitational damping. As a result, we may expect a more
effective coupling with the radial pulsation modes.

Future extension of this work certainly must consider more realistic
models of the star, which take into account the effects of rotation,
composition gradients, magnetic fields, dissipative effects, etc.  In
particular, it would be interesting in a protoneutron star to compare
the damping rate due to the gravitational radiation produced by the
coupling with the strong damping induced by the presence of a
high-entropy envelope, which surrounds the newly created neutron
star. \\
\indent New interesting non-linear effects could come out when the star is
rotating, mainly due to the different behaviour of the radial and
non-radial modes in a rotating configuration. While the radial modes
are only weakly affected and their spectrum can be considered almost
the same as that of a radially pulsating non-rotating star (when
scaled by the central density), the non-axisymmetric modes manifest a
splitting similar to that observed in the atomic energy levels due to
the Zeeman effect. The rotation has the effect of removing the mode
degeneracy of the azimuthal quantum number, which is present in the
non-rotating case. The amount and the details of this frequency
separation depend on the stellar compactness and rotation rate. We
might then expect that for a given stellar rotation rate and
compactness the non-radial frequencies should cross the sequence of
the radial modes~\cite{Dimmelmeier:2004prep}.  In this case the
frequencies of these two kinds of modes are comparable, and possible
resonances or instabilities could influence the spectrum and wave
profile of the gravitational wave radiated.  The possibility of
identifying new resonances in the high frequency gravitational wave
spectrum could provide new relations, which can be used to determine
the stellar parameters through asteroseismology.

\appendix

\chapter{Gundlach-Garcia Source terms}
\label{full-equations}

In this section we re-write the source terms of the equations of polar
non-radial perturbations on a time dependent and spherically symmetric
background, which have been found by Gundlach and M. Garc\'{\i}a
\cite{Gundlach:1999bt}. These equations are valid for a perfect fluid
star with constant entropy along each element fluid trajectory.
The equations for a barotropic fluid can be easily
determined by neglecting the terms relative to the
entropy $s$, its perturbation $\sigma$ and the quantity
$C = \left( \frac{\partial p}{\partial s} \right)_{\rho} \, .$
\begin{eqnarray}
\label{S_chi}
\nonumber
S_\chi  & = &
- 2 \left[ 2 \nu^2 + 8 \pi \rho - {6m\over r^3}
         - 2 U ( \mu -U )
    \right] ( \chi + k )
+ \frac{(l-1)(l+2)}{r^2} \chi
\\ \nonumber &&
+ 3 \mu \dot\chi
+ 4 ( \mu - U ) \dot k
- ( 5 \nu - 2 W ) \chi'
- 2 [ 2 \mu \nu - 2 ( \mu - U ) W + \mu' - \dot\nu ] \psi
\\ &&
+2 \eta''
- 2 ( \mu - U ) \dot\eta
+ ( 8 \nu - 6 W ) \eta'
- \left[ - 4 \nu^2 + \frac{l(l+1)+8}{r^2} + 8 \nu W \right.
\nn \\
 && +  \left. 4 ( 2\mu U + U^2 - 4W^2 - 8 \pi \rho )
  \right] \eta,
\\
\label{S_k}
\nonumber S_k
& = &
  ( 1 + c_s^2 ) U \dot\chi
+ [ 4 U + c_s^2 ( \mu + 2 U ) ] \dot k
- W ( 1 - c_s^2 ) \chi'
- ( \nu + 2 W c_s^2 ) k'
\\ \nonumber &&
- \left[ 2 \left( \frac{1}{r^2} - W^2 \right) + 8 \pi p
       - c_s^2 \left( \frac{l(l+1)}{r^2} + 2 U ( 2\mu + U ) - 8 \pi \rho
               \right)
  \right] ( \chi + k )
\\ \nonumber &&
- \frac{(l-1)(l+2)}{2 r^2} ( 1 + c_s^2) \chi
+ 2 [ - \mu W ( 1 - c_s^2 ) + ( \nu + W ) U ( 1 + c_s^2 ) ] \psi
+ 8\pi \cee \rho \sigma
\\ &&
- 2 U \dot\eta + 2 W \eta'
+ \left[ \frac{l(l+1)+2}{r^2} - 6 W^2 + 16 \pi p - 2 U (2\mu+U) c_s^2
  \right] \eta,
\\
S_\psi
& = &
2 \nu ( \chi + k ) + 2 \mu \psi + \chi' -2 \eta (\nu-W) - 2 \eta',
\\
C_\gamma
& = &
- W \dot\chi + U \chi' - ( \mu - 2 U ) k'
+ \frac{1}{2} \left[ \frac{l(l+1)+2}{r^2} + 2 U ( 2 \mu + U )
                     - 2 W ( 2 \nu +  W ) \right. \nn \\
&& \left. + 8 \pi (p-\rho)
  \right] \psi
- 2 U \eta',
\\
C_\omega
\nonumber & = &
  \left[ \frac{l(l+1)}{r^2} + 2 U ( 2\mu + U ) - 8 \pi \rho
  \right] ( \chi + k )
- \frac{(l-1)(l+2)}{2 r^2} \chi
+ 2 [ \nu U + ( \mu + U ) W ] \psi
\\ &&
+ U \dot\chi + ( \mu + 2 U ) \dot k
+ W \chi' - 2 W k'
- 2 \eta U ( 2 \mu + U ),
\\
C_\alpha
& = &
2 \mu ( \chi + k ) + 2 \nu \psi + \dot\chi + 2 \dot k
- 2 \eta ( \mu + U ),
\\
S_\omega
& = &
\left( 1 + \frac{p}{\rho} \right)
\left[ - \frac{l(l+1)}{r^2} \alpha
       + \frac{\dot\chi + 3\dot k}{2}
       + \left( \nu + 2 W - \frac{\nu}{c_s^2} \right)
         \left( \gamma + \frac{\psi}{2} \right)
\right]   \nn \\
&& + ( \mu + 2 U ) \left( c_s^2 - \frac{p}{\rho} \right) \omega
- \cee \left[ \left(\gamma+\frac{\psi}{2}\right)\frac{s'}{c_s^2}
           - \sigma (\mu+2U)
    \right],
\\
S_\gamma
& = &
\left( 1 + \frac{p}{\rho} \right)
\left[ \frac{\chi' + k' - 2 \eta'}{2}
       + \left[ c_s^2 (\mu + 2 U ) - \mu \right]
         \left( \gamma - \frac{\psi}{2} \right)
\right]
- \nu \left( c_s^2 - \frac{p}{\rho} - \frac{\rho + p}{c_s^2}
               \frac{\partial c_s^2}{\partial \rho}
      \right) \omega
\nonumber \\ &&
- \cee \sigma'
-\sigma\left[ \cee \left(\nu-\frac{s'}{c_s^2}
                         \frac{\partial c_s^2}{\partial s}\right)
            + s' \frac{\partial \cee}{\partial s}
            - \nu \left(1+\frac{p}{\rho}\right)
              \frac{1}{c_s^2}\frac{\partial c_s^2}{\partial s}
       \right]   \nn \\ &&
- \omega s' \left[ \frac{\partial c_s^2}{\partial s}
                 - \cee \left(1+\frac{\rho}{c_s^2}
                   \frac{\partial c_s^2}{\partial\rho}\right)
            \right],
\\
S_\alpha
& = &
- \frac{k+\chi}{2} + \eta - c_s^2 ( \mu + 2 U ) \alpha
+ \frac{c_s^2 \omega + \cee \sigma}{1+\frac{p}{\rho}}, \\
\nonumber
\bar S_\omega & = &
\left( 1 + \frac{p}{\rho} \right)
\left[ \left( - \frac{l(l+1)}{r^2} + 8\pi(\rho+p) \right) \alpha
       + \frac{\dot k}{2}
       + (\mu+U) \eta - \mu (\chi+k)
\right]
\nn \\
&& + ( \mu + 2 U ) \left( c_s^2 - \frac{p}{\rho} \right) \omega
\cee ( \mu+2U ) \sigma
- \frac{1}{c_s^2}
  \left[ s' \cee + \left(1+{p\over\rho}\right) ( \nu-2W\kbar ) \right]
 \nn \\
&& \times   \left( \gamma+\frac{\psi}{2} \right)
+ \nu \left(1+{p\over\rho}\right) \left( \gamma-\frac{\psi}{2} \right),
\\
\nonumber
\bar S_\gamma & = &
\left( 1 + \frac{p}{\rho} \right)
\left[ \frac{k'}{2}
       + \left( c_s^2 (\mu + 2 U ) - \mu \right)
         \left( \gamma - \frac{\psi}{2} \right)
       - \mu \psi - \nu (\chi+k) + (\nu-W) \eta
\right]
\\ \nonumber &&
- \cee \sigma'
- \sigma \cee \left[ \nu
                   + \frac{s'}{\cee} \frac{\partial\cee}{\partial s}
                   - \left( \frac{\nu}{\cee}
                            \left( 1 + \frac{p}{\rho} \right)
                          + s'
                     \right)
                     \frac{1}{c_s^2} \frac{\partial c_s^2}{\partial s}
              \right]
\\ &&
+ \omega
  \left[ \nu \left(\frac{p}{\rho}-c_s^2\right)
       + s' \left( \cee - \frac{\partial c_s^2}{\partial s} \right)
       + \left[ \nu (\rho+p)+ \rho \cee s' \right]
         \frac{1}{c_s^2} \frac{\partial c_s^2}{\partial\rho}
  \right] .
\end{eqnarray}
In this section, we report also the equations relative to the radial
perturbations of a time dependent and spherically symmetric star,
which have been determined in the radial gauge by Gundlach and
M. Garc\'{\i}a \cite{Gundlach:1999bt}. The two fluid evolution equations are given by
\begin{eqnarray}
-\dot\omega - \eor \gamma' &=&
  (\mu+2U) \cee \sigma
- \omega \left[ 4\pi \frac{U}{|v|^2} (\rho+p)
              + (\mu+2U) \left( \frac{p}{\rho} - c_s^2 \right) \right]
\nonumber \\ &-&  \gamma \left[ \frac{\cee s'}{c_s^2}
              - \eor \left( -4\pi \frac{W}{|v|^2} (\rho+p)
                          + \nu + 2W - \frac{\nu}{c_s^2} \right)
         \right]
\nonumber \\ &-&  ( \chi - \eta )
  \left[ -\frac{UW}{U^2+W^2} \frac{\cee s'}{c_s^2} \right.
\nonumber \\ &+&
       \left.  \eor \left( \mu + U
                   - \nu \frac{UW}{U^2+W^2} \frac{1+c_s^2}{c_s^2}
                   + \frac{U}{2} \frac{|v|^2}{U^2+W^2}  \right. \right. \nn \\
    &+&   \left. \left. 4\pi \frac{U W^2}{-U^4+W^4} (\rho+p) \right)
  \right]
- \eta \frac{U}{|v|^2} \eor \left(-\frac{1}{2r^2}
+ 4\pi \rho\right)
, \\
\eor \dot\gamma + c_s^2 \omega' &=&
  \gamma \eor \left( -4\pi \frac{U}{|v|^2} (\rho+p)
                   - \mu + (\mu+2U) c_s^2 \right)
\nonumber \\ &-&  \sigma \left[ \nu \left( \cee - \eor
\frac{1}{c_s^2}
                                  \frac{\partial c_s^2}{\partial s}
                    \right)
              + 4\pi \frac{W\cee}{|v|^2} (\rho+p)
              + s' \left( \frac{\partial\cee}{\partial s}
                        - \frac{\cee}{c_s^2}
                          \frac{\partial c_s^2}{\partial s} \right)
         \right]
\nonumber \\ &-&  \omega \left[ \nu \left( c_s^2 - \frac{p}{\rho}
                         - (\rho+p) \frac{1}{c_s^2}
                           \frac{\partial c_s^2}{\partial\rho} \right)
              + 4\pi \frac{W c_s^2}{|v|^2} (\rho+p)
              - s' \left( \cee - \frac{\partial c_s^2}{\partial s}
                        + \frac{\rho \cee}{c_s^2}
                          \frac{\partial c_s^2}{\partial\rho} \right)
         \right]
\nonumber \\ &-&  (\chi-\eta) \eor
  \left( \nu - \mu \frac{UW}{U^2+W^2}(1+c_s^2)
       - \frac{2U^2W}{U^2+W^2} c_s^2
       + \frac{W}{2} \frac{|v|^2}{U^2+W^2}
       \right. \nn \\
       & + &   \left.
4\pi \frac{U^2 W}{U^4-W^4} (\rho+p)
  \right)
-\eta \frac{W}{|v|^2} \eor \left( \frac{1}{2r^2}
+ 4\pi p \right)
- \cee \sigma', \\
\dot{\sigma}&=&-s'\left(\gamma+\frac{UW}{U^2+W^2}(\eta-\chi)\right).
\end{eqnarray}
The remaining two metric perturbations $\eta$ and $\chi$ can be
obtained from the following constraints:
\begin{eqnarray}
r|v|^2D\eta &=&
  4\pi(\rho+p) \left( \chi + \frac{2U^2}{|v|^2} \eta \right)
+ 8\pi (\rho+p) \frac{2UW}{|v|^2} \gamma \nn \\
&+&  \, 4\pi \rho
\frac{U^2+W^2}{|v|^2} (\cee\sigma + (1+c_s^2)\omega)
, \\
\nonumber r|v|^2D\chi &=& \frac{4UW}{U^2+W^2} \left( \mu W - \nu U
+ 4\pi \frac{UW}{|v|^2} (\rho+p) \right) ( \chi - \eta )
\nn \\ &+&
 \left(
-\frac{1}{r^2} + 8\pi \rho \right)
  \left( \chi + \frac{2U^2}{|v|^2} \eta \right)
+ 8\pi (\rho+p) \frac{2UW}{|v|^2} \gamma
\nn \\ &+ &
 8\pi \rho \frac{U^2+W^2}{|v|^2} \omega.
\end{eqnarray}

\chapter{Sound wave equation}
 \label{AppSW11MG}
\noindent Here, we give the complete form of the sound wave equation
~(\ref{En_eq_GG}) for non-radial perturbations on a barotropic, time
dependent and spherically symmetric background.  This equation is
written in terms of the quantities and frame derivatives introduced in
the GSGM formalism~(\ref{sec:GSGM}).
\begin{eqnarray}
& - & \ddot{H} + c_s^2 H'' + \left( \mu + 2 U \right) \left( c_s^2
- \frac{p}{\rho} - \frac{2}{c_s^2} \left(\rho+p\right) \frac{d
c_s^2}{d \rho} \right) \dot{H} + \left( \left( 2  c_s^2 -1
\right) \nu + 2  c_s^2W \right) H'  \nn \\
& + & \left\{\left(\rho + p \right) \left[ \left(\rho + p \right)
\left(\mu + 2 U \right)^2 \frac{1}{c_s^2} \left( \frac{d^2
c_s^2}{d \rho^2}   - \frac{2}{c_s^2} \frac{d c_s^2}{d \rho}^2
\right)  + \left[ \left(\mu+ 2 U \right)^2
\left( 2+ \frac{\rho- p}{\rho  c_s^2}   \right) \right.  \right. \right. \nn \\
& + & \left. \left. \left.   \frac{1}{c_s^2} \left( 3 U^2 -
\dot{\mu}  - \left(2 \nu + W \right) W + 8 \pi p + \frac{1}{r^2}
\right) \right] \frac{d c_s^2}{d \rho} + 4 \pi \left(1 + 3 c_s^2
\right) \right]   - \frac{l \left( l+1
\right)}{r^2} c_s^2\right\} H \nn \\
& + & \frac{1}{2} \left( c_s^2 -1 \right) \nu \left( \chi' - k'
\right) + c_s^2  \mu  \dot{\chi} + \frac{c_s^2}{2} \left( \mu + 2
U \right) \left( 1 + c_s^2 - \frac{p}{\rho} \right) \ \dot{k} +
\left\{c_s^2  \left[ 2 \left( 2 \nu + W \right) W
\right. \right.  \nn \\
& + & \left.  \left.
 4
\pi \left( \rho - p \right)
+ 2 \dot{\mu} - \frac{2}{r^2} + \left( 1 +
\frac{p}{\rho} -  c_s^2 \right) \mu^2 - 2 \left( 1 + c_s^2 -
\frac{p}{\rho} \right)  \mu  U - 2 U^2  \right] -2 \nu^2 \right\}
\left( \chi + k \right)  \nn \\
& + & \frac{c_s^2}{2} \left( \left(1 +c_s^2 \right) \mu - 2 \left(
1 -c_s^2\right) U \right)  \psi' +   \left\{ \frac{1}{2} \left(1
+c_s^2 \right) \left( c_s^2 \mu' + \dot{\nu} \right) - \frac{1}{2}
\left( \rho + p\right) \frac{d c_s^2}{d \rho}  \left( 1 -
\frac{1}{c_s^2} \right)  \right. \nn \\
& \times & \left.
\left(\mu + 2 U\right) \nu
\left[  c_s^2 \left( 1 - \frac{2  p}{\rho} +3 c_s^2
\right) W + \left[ c_s^2 \left( c_s^2 + \frac{p}{\rho} -3 \right)
+ \frac{p}{\rho}  \right] \nu \right]  U +
\left[  c_s^2 \left( 1 - \frac{p}{\rho} + 3 c_s^2  \right) W \right. \right. \nn \\
& + & \left.  \left. \frac{1}{2} \left[ c_s^2 \left( c_s^2 - 2
\right) + \frac{p}{\rho} \left( 1 + c_s^2 \right) -3 \right] \nu
\right]  \mu \right\} \psi + c_s^2 \left[ 2 \mu - \frac{p}{\rho}
\left( \mu + U \right) \right] \ga' + \left\{\left( 1 -c_s^2
\right)  \left( c_s^2 \mu' + \dot{\nu} \right)
\right. \nn \\
& + & \left. \left(\rho + p \right) \left( 1 +  \frac{1}{c_s^2}
\right) \left( \mu + 2 U \right) \nu \frac{d c_s^2}{d \rho} +
\left[  2   c_s^2   \left( 1 + c_s^2 - 2
 \frac{p}{\rho} \right) W - 2 \left[ c_s^2 \left( 1 + c_s^2
+ \frac{p}{\rho} \right)  - \frac{p}{\rho}  \right] \nu   \right] U \right. \nn \\
& + & \left. \left[  2   c_s^2   \left( 1 - c_s^2 - \frac{p}{\rho}
\right) W +  \left[ c_s^2 \left( 2 - c_s^2 \right) +
\frac{p}{\rho} \left( 1-c_s^2 \right)  -1   \right] \nu \right]
\mu
\right\} \ga  \nn \\
& + &  \left\{\left[ 8 \pi \left(\rho + p \right)  c_s^4 + \left(
\frac{l\left( l+1 \right)}{r^2} \frac{p}{\rho} + 8 \pi
\frac{\rho^2 - p^2}{\rho}  \right)   c_s^2 \right] \left( \mu + 2
U \right)  - 2  \frac{l\left( l+1 \right)}{r^3}  c_s^2 \ \dot{r}
\right\} \alpha = 0 \,.
\end{eqnarray}

\chapter{Source terms for the $\lambda \epsilon$ polar perturbative equations}
\label{AppSources}

\noindent The perturbative equations which describe the coupling
between the radial and non-radial polar
perturbations~(\ref{GW11})-(\ref{Sham11}) have long terms that are
written in this appendix.
\noindent For the \emph{gravitational wave}
equation~(\ref{Sgw11}) the source has the following form:
\begin{eqnarray}
{\cal S}_{S} & = & a_1 S_{,rr}^{(0,1)} + a_2 S_{,r}^{(0,1)} + a_3
S_{,t}^{(0,1)} + a_4 S^{(0,1)} + a_5 \left( \psi_{,r}^{(0,1)} - 2
e^{\Lambda -\Phi} k_{,t}^{(0,1)} \right) + a_6 k^{(0,1)} \nn \\
&& + a_7
\psi^{(0,1)} \,,
\end{eqnarray}
where the coefficients $a_i$ are given by
\begin{eqnarray}
a_1 & = & 2\left( r
S^{(1,0)}- \eta^{(1,0)}\right) e^{-2\Lambda} \,,  \\
a_2 & = &
\left[ 2\left(\Lambda_{,r}-5\Phi_{,r}\right)\eta^{(1,0)} -
\left(\left( \Lambda_{,r}-5\Phi_{,r} \right) r+3 \right) S^{(1,0)} -
\left( \Lambda_{,r}+\Phi_{,r} \right)\left(5-\bar{c}_s^{\,-2} \right)
H^{(1,0)}\right]{e^{-2\,\Lambda}} \nn \\
&& - 4\ga_{,t}^{(1,0)}
e^{-\Phi-\Lambda} \,,  \\
a_3 & = & - 4\left(
\Lambda_{,r}+\Phi_{,r}\right)\ga^{(1,0)} e^{-\Lambda-\Phi} -
e^{-2\Phi} \eta^{(1,0)}_{,t} + \frac{2}{r} \left( r\ga^{(1,0)}
e^{-\Phi} \right)_{, \, r} e^{-\Lambda} \,, \\
a_4 & = &
-\left\{\frac{4}{r}\left(1+2r\Phi_{,r}\right)\ga^{(1,0)}_{,t}
e^{-\Phi+\Lambda} + 2\left[2\Phi_{,r} \left( \frac{1}{r} +
2\Phi_{,r} \right)
+\frac{2-\left(l(l+1)+2\right)e^{2\Lambda}}{r^2} \right. \right. \nn \\
&&\left. \left.  +3\frac{\Lambda_{,r}
+\Phi_{,r}}{r} \right]\eta^{(1,0)} -
\left[\Phi_{,r}+3\Lambda_{,r}+\frac{3-\left(l(l+1)+2\right)
e^{2\Lambda}}{r}\right] S^{(1,0)} \nn \right. \\
&& \left. + \left( \Lambda_{,r}+\Phi_{,r}
\right) \left[\left(5+\frac{3}{\bar{c}_s^2}\right) \frac{1}{r} +
8\Phi_{,r}\right]H^{(1,0)}\right\}e^{-2\Lambda} \,,  \\
a_5 & = & - 2\left[\left(\frac{e^\Phi}{r}\ga^{(1,0)}\right)_{,r}
-\left(\Lambda_{,r}+\Phi_{,r}\right)\left(\frac{e^\Phi}{r}\ga^{(1,0)}\right)
\right]e^{-2\Lambda-\Phi}\,,  \\
a_6 & = &
-\left[\frac{2}{r^2}\left(-5+2r(\Phi_{,r}-\Lambda_{,r})+2e^{2\Lambda}
\right) S^{(1,0)}
+\frac{8}{r^3}\left(1+r^2\Phi_{,r}^2+r\Lambda_{,r}-
e^{2\Lambda}\right)\eta^{(1,0)} \right. \nn \\
+  & & \left. \frac{4}{r^2} \left(\Lambda_{,r}+\Phi_{,r}\right)
\left(2r\Phi_{,r}+\frac{1}{\bar{c}_s^2}\right)H^{(1,0)}
+\frac{8}{r}\Phi_{,r}\ga^{(1,0)}_{,t}e^{-\Phi+\Lambda}\right] e^{-
2\Lambda} \,,  \\
a_7 & = &
-\frac{2}{r}\left\{\left(1-\bar{c}_s^2\right)r \left(
\frac{e^\Phi}{r}\ga^{(1,0)}\right)_{,rr}\!\!\! + \left(
\frac{e^{\Phi}}{r}\ga^{(1,0)}\right)_{,r}
\left[r\left(\Phi_{,r}-2\Lambda_{,r}\right) + \left(2\Lambda_{,r}r
+\Phi_{,r}r-4\right)\bar{c}_s^2 \right.
\right. \nn \\
{} & + & \left. \frac{\Phi_{,r}}{4\pi} \frac{\Lambda_{,r}+
\Phi_{,r}}{\bar{c}_s^2}
\frac{d\bar{c}_s^2}{d\bar\rho}e^{-2\Lambda} \right] +
\left[\left(2-2r\Phi_{,r}
-3r\Lambda_{,r}\right)\Phi_{,r}-\left(1+r\Lambda_{,r}\right)\Lambda_{,r}
+ \left(r\left(\Lambda_{,r}^2- \Phi_{,r}^2\right)
\right. \right.  \nn \\
& + &  \left. \left. \left.
   2\Phi_{,r} + 5\Lambda_{,r} \right) \, \bar{c}_s^2
+ \frac{\Lambda_{,r}+\Phi_{,r}}
{\bar{c}_s^2}\Phi_{,r}\left[r+\frac{e^{-2\Lambda}}{4\pi r}
\left(3-\left(\Lambda_{,r}+\Phi_{,r}\right)r
\right)\frac{d\bar{c}_s^2} {d\bar\rho} \right]
\frac{e^{\Phi}}{r}\ga^{(1,0)}\right]
 \right\} e^{-\Phi -2\Lambda} \,. \nn \\
\end{eqnarray}

The source of the \emph{sound wave} equation~(\ref{SW11}) is given by
(\ref{Ssw11}):
\begin{eqnarray}
{\cal S}_H & = & b_1 H^{(0,1)}_{,rr}+b_2
H^{(0,1)}_{,tr} + b_3 H^{(0,1)}_{,t} + b_4 H^{(0,1)}_{,r} + b_5
H^{(0,1)} + b_6 k^{(0,1)}_{,t} + b_7 r S^{(0,1)}_{,t} \nn \\
& + &  b_8\left[k^{(0,1)}_{,r}
-\left(r S^{(0,1)}\right)_{,r}\right]
+ b_9\left(r S^{(0,1)} + k^{(0,1)}\right) + b_{10}
\ga^{(0,1)}_{,r} + b_{11} \ga^{(0,1)} + b_{12} \psi^{(0,1)}_{,r}
\nn \\ & + &
b_{13}\psi^{(0,1)} + b_{14}\alpha^{(0,1)} \,,
\end{eqnarray}
and the expression of the coefficients $b_i$ is the following
\begin{eqnarray}
b_1 & = &-\left[2\left(\eta^{(1,0)}-rS^{(1,0)}\right)\bar{c}_s^2+
\frac{e^{-2\Lambda}}{4\pi r}
\frac{\Lambda_{,r}+\Phi_{,r}}{\bar{c}_s^2}
\frac{d\bar{c}_s^2}{d\bar\rho}H^{(1,0)} \right]e^{-2\Lambda}\,,  \\
b_2 & = &2\left(1-\bar{c}_s^2\right) e^{-\Phi-\Lambda}\ga^{(1,0)}
\,, \\
b_3 & = & - \left\{\frac{e^{-\Lambda}}{r^3}
\left(-\frac{e^{-2\Lambda}}{2\pi}
\frac{\Lambda_{,r}+\Phi_{,r}}{\bar{c}_s^2} \frac{d\bar{c}_s^2}
{d\bar\rho} + \frac{2\bar{c}_s^2\bar\rho - \bar{p}}{\bar\rho}r\right )
\left(r^2\ga^{(1,0)} e^{\Phi}\right)_{,r} \right. \nn \\ & + & \left.
\left(\Lambda_{,r}+\Phi_{,r}\right) \left(\frac{e^{-2\Lambda}}{2\pi
r}\frac{\Lambda_{,r}+\Phi_{,r}}{\bar{c}_s^2}
\frac{d\bar{c}_s^2}{d\bar\rho}-\bar{c}_s^2+1+
\frac{\bar{p}}{\bar\rho}\right)\ga^{(1,0)}
e^{\Phi-\Lambda}+\eta^{(1,0)}_{,t}\right\} e^{-2\Phi}\,, \\
b_4
& = & \frac{1}{4\pi r} \left[\Lambda_{,r}\left(\Lambda_{,r}-
\frac{1}{r}\right)-\left(\frac{1}{r}+\Phi_{,r}\right)\left(2\Phi_{,r}+
\Lambda_{,r}\right)\right] \frac{1}{\bar{c}_s^2}
\frac{d\bar{c}_s^2}{d\bar\rho}H^{(1,0)} e^{-4\Lambda} \nn \\ & - &
\left\{2\left(1-\bar{c}_s^2\right ) H^{(1,0)}_{,r} +2\left[\left(
2\Phi_{,r}-\Lambda_{,r}+\frac{2}{r}\right) \bar{c}_s^2
-\Phi_{,r}\right] \eta^{(1,0)} \right. \nn \\ & + & \left. \left[
3\Phi_{,r}+\frac{1}{2r}+\left(\Lambda_{,r} - 4\Phi_{,r} -
\frac{7}{2r}\right)\bar{c}_s^2\right] r S^{(1,0)}\right\}
{e^{-2\Lambda}} \,,  \\
b_5 & = &
-\left\{\left[\frac{e^{-2\Lambda}}{4\pi r^2}
\left(1-{e^{2\Lambda}}+\Phi_{,r}\Lambda_{,r}r^2+ \left(\Lambda_{,r}
+\frac{5}{2}\Phi_{,r}\right)r\right)
\frac{\Lambda_{,r}+\Phi_{,r}}{\bar{c}_s^2}\frac{d\bar{c}_s^2}
{d\bar\rho} - \left(\Lambda_{,r}+\Phi_{,r}\right)
\left(1+3\bar{c}_s^2\right) \right. \right. \nn \\ & + &
\left. \left. \frac{l(l+1)}{r}\bar{c}_s^2 e^{2\Lambda}\right]
S^{(1,0)} + \frac{\Lambda_{,r}+\Phi_{,r}}{\bar{c}_s^2
r}\left[\frac{1}{4\pi r} \left( \left(6\bar{c}_s^2+1+
r\Phi_{,r}\right)
\frac{\Lambda_{,r}+\Phi_{,r}}{\bar{c}_s^2}{e^{-2\Lambda}}
-\frac{l(l+1)}{r}\right )\frac{d\bar{c}_s^2}{d\bar\rho} \nn
\right.\right.\\ & + &
\left. \left. 3\bar{c}_s^4+4\bar{c}_s^2+1\right] H^{(1,0)} +
\frac{e^{-2\Lambda}}{4\pi r}
\frac{\Lambda_{,r}+\Phi_{,r}}{\bar{c}_s^2}
\frac{d\bar{c}_s^2}{d\bar\rho}\left[ \left( \frac{2}{r} -
\Lambda_{,r}+ 2\Phi_{,r} \right) H^{(1,0)}_{,r} + H^{(1,0)}_{,rr}
\right] \right. \nn \\ & + & \left. \frac{2}{r}\left[
\left(\Lambda_{,r}+\Phi_{,r}\right ) \left(1+3\bar{c}_s^2\right)
-\frac{l(l+1)}{r}\bar{c}_s^2 e^{2\Lambda} \right]
\eta^{(1,0)}\right\}{e^{-2\Lambda}} \,,  \\
b_6 & = & \left\{\left[
\left(\Lambda_{,r}+\Phi_{,r} \right)\left(1+\bar{c}_s^2
-{\frac{\bar{p}}{\bar\rho}}
\right)\bar{c}_s^2-\left(1-\bar{c}_s^2\right ) \Phi_{,r}\right]
\ga^{(1,0)} e^{\Phi}  \right. \nn \\
&&  - \left. \frac{\bar{c}_s^2}{r^2} \left(1+\bar{c}_s^2-
\frac{\bar{p}}{\bar\rho}\right) \left(r^2\ga^{(1,0)} e^{\Phi}
\right)_{,r} \right\}e^{-2\Phi-\Lambda} \,, \\
b_7 & = &
\left\{\left[ \left( \Lambda_{,r}+\frac{2}{r} \right )\bar{c}_s^2 +
\frac{\Phi_{,r}}{2}\left(1+\bar{c}_s^2\right)
\right]\ga^{(1,0)}e^{\Phi} - \frac{\bar{c}_s^2}{r^2} \left( r^2
\ga^{(1,0)}e^{\Phi} \right)_{,r} \right\} {e^{-2\Phi-\Lambda}} \,, \\
b_8 & = &  -\left[ \frac{\Phi_{,r}}{8\pi
r}\frac{\Lambda_{,r}+\Phi_{,r}}
{\bar{c}_s^2}\frac{d\bar{c}_s^2}{d\bar\rho} H^{(1,0)} e^{-2\Lambda} +
\left(1-\bar{c}_s^2 \right) \left( \Phi_{,r} \eta^{(1,0)} +
\frac{1}{2}H^{(1,0)}_{,r} \right )\right] e^{-2\Lambda} \,, \\
b_9 & = & -\left\{\left[
\frac{2}{r^2}\left(1-{e^{2\Lambda}}+3r\Phi_{,r}
+r\Lambda_{,r}\right)\bar{c}_s^2 -4{\Phi_{,r}}^2 \right] \eta^{(1,0)}
\nn \right.  \\ & + & \left. \left[\left[3\frac{e^{2\Lambda} -1}{r} -
2\Phi_{,r} \left( 4 + r \Lambda_{,r} \right) - 3\Lambda_{,r}
\right]\bar{c}_s^2 +4 r \Phi_{,r}^2\right] S^{(1,0)} \nn \right.  \\ &
- & \left. \frac{\Lambda_{,r}+\Phi_{,r}}{r} \left[ 1+ 2\Phi_{,r} r
+3\bar{c}_s^2 - \frac{e^{-2\Lambda}}{4\pi r^2} \frac{1}{\bar{c}_s^2}
\frac{d\bar{c}_s^2}{d\bar\rho} \left(\left(3 \Phi_{,r}
+\Lambda_{,r}\right)r+ 1 - e^{2\Lambda} \right)\right] H^{(1,0)} \nn
\right.  \\
& + & \left. 2\left[ 2\Phi_{,r}+\left(
\Lambda_{,r}-2\Phi_{,r}- \frac{2}{r}\right ) \bar{c}_s^2 \right]
H^{(1,0)}_{,r} - 2\bar{c}_s^2 H^{(1,0)}_{,rr}\right\} e^{-2\,\Lambda}
\,,  \\
b_{10} & = & \frac{2 \bar{c}_s^2}{r^2} \left\{\left[
\left( 1-{\frac{\bar{p}}{2\bar\rho}} \right) \left
(\Lambda_{,r}+\Phi_{,r}\right ) + \frac{2}{r} \right] \left( r^2
\ga^{(1,0)} e^{\Phi} \right) \right. \nn \\
& - & \left.
\left( 1 -{\frac{p}{2 \rho}}\right)
\left( r^2 \ga^{(1,0)} e^{\Phi} \right)_{,r} \right\} e^{-\Phi-2
\Lambda} \,,  \\
b_{11} & = & - \frac{1}{r^3} \left\{2 r
\bar{c}_s^2 \left(1-\bar{c}_s^2\right ) \left( r^2 \ga^{(1,0)}
e^{\Phi} \right)_{,rr} + \left[ \frac{\Phi_{,r}}{2 \pi}
\left(\Lambda_{,r}+\Phi_{,r}\right ) \frac{d\bar{c}_s^2}{d\bar\rho}
{e^{-2 \Lambda}} +2 \left(\Phi_{,r}+2 \Lambda_{,r}+\frac{2}{r} \right
)r \bar{c}_s^4 \right. \right. \nn \\ & - & \left. \left. \left( 4
\left(1+r \Lambda_{,r} \right) + {\frac{\bar{p}}{\bar\rho} \left(2+ r
\Phi_{,r} \right ) }\right )\bar{c}_s^2 -
\left(2-{\frac{\bar{p}}{\bar\rho}}\right ) r \Phi_{,r} \right] \left(
r^2 \ga^{(1,0)} e^{\Phi} \right)_{,r}  \right. \nn \\
& + &
 \left. \left[ -2 \left
(\Lambda_{,r}+\Phi_{,r}\right )\left(1+\left(\Phi_{,r}-
\Lambda_{,r}\right )r\right ) \bar{c}_s^4
+ \left( \frac{5}{r} -\frac{\Phi_{,r}}{2} \left( 2
\Phi_{,r}r+7+8 r\Lambda_{,r}\right)
\right.  \right. \right.  \nn \\ & - &
 \left. \left. \left.
 2 \Lambda_{,r}
\left(r\Lambda_{,r}-1\right )+\frac{\bar{p}}{\bar\rho}\frac{\Phi_{,r}
r \left(2 \Phi_{,r}r+5\right )+2}{2 r} -\frac{e^{2 \Lambda}}{2 r}
\left(1+{\frac{\bar{p}}{\bar\rho}}\right) \left( 2+\Phi_{,r}r\right )
\right )\bar{c}_s^2 \right.  \right. \nn \\ & + &
\left. \left. \left(\left(4-{\frac{\bar{p}}{\bar\rho}}\right
)\left(\Lambda_{,r}+\Phi_{,r}\right )r+4\right )\Phi_{,r} -
\frac{e^{-2\Lambda}}{2 \pi} \left(\Lambda_{,r}+\Phi_{,r}\right )^2
\Phi_{,r} \frac{d\bar{c}_s^2}{d\bar\rho}\right]  \right. \nn \\
& \times & \left. \left( r^2 \ga^{(1,0)}
e^{\Phi} \right) \right\} e^{-\Phi-2 \Lambda}\,,  \\
b_{12} & = &
\frac{\bar{c}_s^2}{2 r^2} \left\{\left[ \left
(\Lambda_{,r}+\Phi_{,r}\right )\left (1+\bar{c}_s^2\right
)+\frac{4}{r} \right] \left( r^2 \ga^{(1,0)} e^{\Phi} \right) - \left
(1+\bar{c}_s^2\right ) \left( r^2 \ga^{(1,0)} e^{\Phi} \right)_{,r}
\right\} e^{-\Phi-2 \Lambda} \,,  \\
b_{13} & = & -
\frac{1}{r^2} \left\{\bar{c}_s^2 \left( 1 + \bar{c}_s^2 \right) \left(
r^2 \ga^{(1,0)} e^{\Phi} \right)_{,rr} \!\!+ \frac{e^{-2 \Lambda}}{4
\pi r} \left(\Lambda_{,r}+\Phi_{,r} \right ) \Phi_{,r} \left[
\left(\Lambda_{,r}+\Phi_{,r} \right ) \left( r^2\ga^{(1,0)} e^{\Phi}
\right) \nn \right. \right. \\
& - & \left. \left.
\left( r^2\ga^{(1,0)} e^{\Phi} \right)_{,r} \right]
\frac{d\bar{c}_s^2}{d\bar\rho} \right. \nn \\
& - & \left. \left[
\left( 2 \Lambda_{,r}+\Phi_{,r} \right) \bar{c}_s^4 +
\left(\frac{2}{r} + 2\Lambda_{,r} + 3 \Phi_{,r} +
\frac{\bar{p}}{\bar\rho} \left( \frac{1}{r} -
\frac{\Phi_{,r}}{2}\right) \right) \bar{c}_s^2 + \left(2
-\frac{\bar{p}}{2\bar\rho} \right)\Phi_{,r}\right] \left( r^2
\ga^{(1,0)} e^{\Phi} \right)_{,r} \right. \nn \\ & + & \left. \left[
\left( \frac{4}{r} + \frac{\Lambda_{,r}+\Phi_{,r}}{2} \left( 6 -
\frac{\bar{p}}{\bar\rho} \right) \right)\Phi_{,r} - \left(
\Lambda_{,r}+\Phi_{,r} \right) \left( \Lambda_{,r} - \Phi_{,r} +
\frac{1}{r} \right) \bar{c}_s^4 \right. \right. \nn \\ & + &
\left. \left. \left[ \frac{1}{2} \left( \Phi_{,r} + \frac{1}{2 r}
\right) \left( 6 \Lambda_{,r} + 7 \Phi_{,r} - \frac{\bar{p}}{\bar\rho}
\left( \Phi_{,r} -\frac{2}{r} \right) \right) + \frac{2}{r^2} - \left(
\Lambda_{,r} - \frac{1}{2 r} \right) \left( \Lambda_{,r} + \frac{1}{r}
\right) \right. \right. \right. \nn \\ & + &
\left. \left. \left. \frac{1}{4 r^2} \left( 1 +
\frac{\bar{p}}{\bar\rho} \right) \left( r \Phi_{,r} -2 \right)
e^{2\Lambda}\right] \bar{c}_s^2\right]r^2 \ga^{(1,0)} e^{\Phi}\right\}
e^{-\Phi - 2\Lambda}\,,  \\
b_{14} & = & -
\frac{\bar{c}_s^2}{r^3} \left\{\left[ 2 \left(\Lambda_{,r}+\Phi_{,r}
\right) \left(1+\bar{c}_s^2-{\frac{\bar{p}}{\bar\rho}} \right ) +
\frac{\bar{p}}{\bar\rho} \frac{l \left( l + 1\right)}{r} e^{2
\Lambda}\right] \left[ \left( r^2 \ga^{(1,0)} e^{\Phi} \right)_{,r}
\right. \right. \nn \\
& - & \left. \left.
\left(\Lambda_{,r}+\Phi_{,r}\right)r^2 \ga^{(1,0)} e^{\Phi}
\right]e^{-2 \Lambda} \right. \nn \\ & - & \left. 2 l \left( l +
1\right) \ga^{(1,0)} e^{\Phi} \right\} e^{-\Lambda- \Phi}\,.
\end{eqnarray}

The third equation of the system is the \emph{Hamiltonian constraint},
whose source term has the following form~(\ref{Sham11}):
\begin{eqnarray}
{\cal S}_{Hamil} & = & c_1\left(
k^{(0,1)}_{,rr} - S^{(0,1)}_{,r}\right) + c_2 k^{(0,1)}_{,r} + c_3
k^{(0,1)}_{,t} + c_4 S^{(0,1)}
+ c_5 k^{(0,1)} + c_6 H^{(0,1)}
 \nn \\
& + &
 c_7 \psi^{(0,1)}_{,r}
+ c_8 \psi^{(0,1)} + c_9 \ga^{(0,1)}  \,,
\end{eqnarray}
where the coefficients $c_i$ are given by
\begin{eqnarray}
c_1 & = &  r  S^{(1,0)} \,, \\
c_2 & = & \left( \frac{3}{2}  S^{(1,0)} +  \frac{\Lambda_{,r}+
\Phi_{,r}}{\bar{c}_s^2}    H^{(1,0)} \right) \,, \\
c_3 & = &  - \left(\Lambda_{,r}+\Phi_{,r}\right)e^{\Lambda-\Phi}\ga^{(1,0)}  \,, \\
c_4 & = &  - \left(S^{(1,0)} +
2\frac{\Lambda_{,r}+\Phi_{,r}}{\bar{c}_s^2}
H^{(1,0)} \right)\,,  \\
c_5 & = &  -  \frac{2}{r} \frac{\Lambda_{,r}+\Phi_{,r}}{\bar{c}_s^2}H^{(1,0)}  \,, \\
c_6 & = &  - \frac{2}{r} \frac{\Lambda_{,r}+
\Phi_{,r}}{\bar{c}_s^2}\left[ 1+ \frac{1}{\bar{c}_s^2} -
\frac{e^{-2 \Lambda}}{4\pi r}\frac{\Lambda_{,r}+
\Phi_{,r}}{\bar{c}_s^4}\frac{d\bar{c}_s^2}{d\bar\rho}\right] H^{(1,0)} \,,   \\
c_7 & = & \frac{2}{r}\ga^{(1,0)}  \,,  \\
c_8 & = & \frac{1}{r^2}\left[\left(2-4\Lambda_{,r}r +
l\left(l+1\right)
{e^{2\Lambda}}\right)\ga^{(1,0)} + 2 r\ga_{,r}^{(1,0)} \right] \,,  \\
c_9 & = &
-\frac{4}{r}\left(\Lambda_{,r}+\Phi_{,r}\right)\ga^{(1,0)} \,.
\label{ccc}
\end{eqnarray}

\chapter{Source terms for the $\lambda \epsilon$ axial perturbative equations}
\label{AppSources_axial}

\noindent The equations which describe the coupling between the radial
and non-radial axial perturbations are given by the two
equations~(\ref{Psi11maseq})-(\ref{traseq11}). Their source terms
$\Sigma_{\Psi}$ and $\Sigma_{\beta}$ have the following form:
\begin{eqnarray}
\Sigma_{\Psi} & = &   2 \left( r \Suo - \etuo \right) \,
\emLL \Psi^{(0,1)}_{, \, rr} + \left\{ 4\pi \left( \bar \rho +
\bar p \right) r \, \frac{1-c_s^2}{ c_s^2} \, H^{(1,0)}  \right. \nn \\
&+ & \left.  \left[ \pmr r^2 -1 + \frac{4M}{r}  \right]  \Suo
+ 2 \, \left[ 4 \pi \, \left( \bar \rho - \bar p \right) r  - \frac{2M}{r^2}
\right] \etuo \right\}  \, \Psi^{(0,1)}_{, \, r}  \nn \\ 
 & - &\left[ \eta^{(1,0)}_{,t} +  8 \pi (\bar \rho + \bar p ) r e^{\Lambda + \Phi }
  \gamma^{(1,0)} \right]  \Psi^{(0,1)}_{,t}  \nn \\
&+ &
 \left\{ 4\pi \left( \bar \rho + \bar p \right) \frac{1-c_s^2}{
c_s^2} \, H^{(1,0)} + \left[ 4 \pi \left( \bar \rho - \bar p \right) r - \frac{\llcf + 3 }{r} +
\frac{12 M}{r^2}  \right]  \Suo
 {} \right. \nn \\
& + & {} \left. 2 \, \left[ \pmr  +   \frac{\llcf}{r^2} -
\frac{6M}{r^3}  \right] \etuo \right\}  \, \Psi^{(0,1)} - 8 \pi r
  \left( 4 \, \eta^{(1,0)}
 -  3 \, r S^{(1,0)} \right)
      \, e^{-\Lambda} \, \hat \beta ^{(0,1)}_{, \, r}  \nn   {} \\
& + & {} \left\{  24 \pi   \left(  4 \pi \, \bar p \,  r^3 + M
   \right) \, e^ { \La } \, S^{(1,0)}  -  32 \pi \,
    \left( 4 \pi \bar p  \, r^2 + \frac{M}{r} \right) \, e^{ \La }  \, \eta^{(1,0)}   {} \nn \right. \\
& + & {} \left.  16 \pi  \, r \, e^{-\La} \, H^{(1,0)}_{, \, r}
\right\}  \hat \beta^{(0,1)}  \label{Spsi11} \\ \nn \\
\Sigma_{\beta} & \equiv &  -
     \ga ^{(1,0)} \, e^{-\La } \, \hat \beta^{(0,1)}_{, \, r}
     + \left[
\left( \left(4 \pi \, \rho \, r - \frac{M}{r^2} \right) \, e^{2
\La}  - \frac{2}{r} \right) \ga ^{(1,0)}  - \ga _{, \, r} ^{(1,0)}
\right]  \, e^{-\La }  \, \hat \beta^{(0,1)}   \label{Sbe11}
\end{eqnarray}

\chapter{Tensor harmonics}
\label{sec:Tens_Harm}

In this section we write the even parity (polar) and the odd parity
(axial) tensor harmonics, which have been defined in a covariant way in
section~\ref{sec:GSGM_pert}.  The polar scalar spherical harmonics,
which satisfy the condition~(\ref{Sph_H_cond}), are given by the
following explicit expression:
\begin{equation}
Y^{lm} \left(\theta , \phi \right) = e^{i m \phi} \sqrt{
\frac{2l+1}{4\pi} \frac{\left( l -m \right)!}{\left( l + m \right)!}
}  P_{lm} \left( \cos \theta \right) \, ,
\end{equation}
where $P_{lm}$ are the associated Legendre functions with the harmonic 
indices~$(l,m)$, which are given by
\begin{equation}
 P_{lm} \left( x \right) = \left( -1 \right) ^m \left( 1 - x^2
 \right)^{\frac{m}{2}} \frac{d ^m }{d x^m} P_{l} \left(x \right) =
 \frac{ \left( -1 \right) ^m }{2^l ~l!} \left( 1 - x^2
 \right)^{\frac{m}{2}} \frac{d ^{l+m} }{d x^{l+m}} \left( x^2 -1
 \right) ^l \, ,
\end{equation}
where $x \in \left[ -1 , 1 \right]$, and in the last equality we have
used the ``Rodrigues representation'' for the Legendre function
$P_{l} \left(x \right)$:
\begin{equation}
P_{l} \left(x \right) = \frac{1}{2^l ~l!}  \frac{d ^{l} }{d x^{l}}
\left( x^2 -1 \right) ^l \, .
\end{equation}
It is worth noticing that for axisymmetric stellar perturbations
$m=0$, the associated Legendre polynomial $P_{lm}$ reduces to the
Legendre polynomial $P_{l}$. \\
\indent The polar and axial vector harmonic bases are then given by
\begin{eqnarray}
Y_{a}^{lm} & = & Y^{lm}_{:a} = \left( Y_{, \, \theta}^{lm} , \, Y_{,
\,\phi}^{lm} \right) \, , \\
S_{a}^{lm} & = & \epsilon_a^{~b} Y^{lm}_b = \left( - \frac{1}{\sin
\theta } Y_{, \, \phi}^{lm} , \, \sin \theta \, Y_{, \,\theta }^{lm}
\right) \, .
\end{eqnarray}
The tensor harmonics for the polar sector are given by
\begin{equation}
Y_{ab}^{lm}\equiv Y^{lm}\gamma_{ab}\,, \qquad
Z^{lm}_{ab}\equiv Y^{lm}_{:ab}+\frac{l(l+1)}{2}Y^{lm}\gamma_{ab}\, ,
\end{equation}
where the explicit expression of the second order covariant derivative of the spherical scalar
harmoncis is given by the following expressions:
\begin{eqnarray}
Y^{lm}_{: \, ab } & = & \left(\begin{array}{cc} Y^{lm}_{ , \,
  \theta \theta } & Y^{lm}_{\, ,\theta \phi} - \cot \theta \, Y^{lm}_{,
  \, \phi} \\ \\ Y^{lm}_{\, ,\theta \phi} - \cot \theta \, Y^{lm}_{, \,
  \phi} & Y_{, \, \phi \phi }^{lm} + \sin \theta \cos \theta \,
  Y^{lm}_{, \, \phi} \\ \label{Yab_pol}
\end{array}\right) \,,
\end{eqnarray}
The axial sector has the following tensor harmonics:
\begin{equation}
S^{lm}_{ab}\equiv S^{lm}_{a:b}+S^{lm}_{b:a} \,,
\end{equation}
where  $S^{lm}_{a:b}$ assumes the following form:
\begin{eqnarray}
S^{lm}_{a: \, b } & = & \left( \begin{array}{cc}
 - \frac{1}{\sin \theta } \left(  Y^{lm}_{ , \,
  \theta \phi }  - \cot \theta  \, Y^{lm}_{ , \, \phi }  \right)
            &  \sin \theta \, Y^{lm}_{\, ,\theta \theta}
 \\ \\ - \frac{1}{\sin \theta }  Y^{lm}_{\, ,\phi \phi} - \cos \theta \, Y^{lm}_{, \,  \theta}
              & \sin \theta \left( Y_{, \, \theta \phi }^{lm} - \cot \theta
                    \, Y^{lm}_{, \, \phi} \right) \\ \label{Sab_ax}
\end{array}\right) \, .
\end{eqnarray}
The harmonic tensors form an orthogonal basis of the 2-dimensional
sphere~$S^2$.  The orthogonality relations for the polar sector are
defined as follows:
\begin{eqnarray}
\int d \Omega \left( Y^{lm \,} \right) ^{ \ast} Y^{l'm'}   & = & \delta^{l l'} \delta^{m m '}  \, , \\
\int d \Omega \, \gamma ^{ab} \,
 \left( Y^{lm \,}_{a} \right) ^{ \ast} Y^{l'm'}_{b}   & = &
l \left( l + 1 \right) \delta^{l l'} \delta^{m m '}  \, , \\
\int d \Omega \, \gamma ^{ab}  \gamma ^{cd }\,
\left( Z^{lm \,}_{ac} \right) ^{ \ast} Z^{l'm'}_{bd}   & = &
l \left( l + 1 \right) \frac{l^2 + l - 2  }{2} \delta^{l l'} \delta^{m m '}  \, ,
\end{eqnarray}
where $\delta^{l l'}$ is the Kronecker delta and the asterisk denotes
complex conjugation. For the axial sector the orthogonality conditions
are given by
\begin{eqnarray}
\int d \Omega \, \gamma ^{ab} \, \left( S^{lm \,}_{a} \right) ^{ \ast} S^{l'm'}_{b}   & = & l \left( l + 1 \right) \delta^{l l'} \delta^{m m '}  \, , \\
\int d \Omega \, \gamma ^{ab}  \gamma ^{cd }\, \left( S^{lm \,}_{ (a : c) } \right) ^{ \ast} S^{l'm'}_{ (b:d)}   & = &
2 l \left( l + 1 \right) \left( l^2 + l - 1 \right) \delta^{l l'} \delta^{m m '}  \, ,
\end{eqnarray}
where we have denoted $(a:b) = a:b + b:a$.

\chapter{Finite difference approximations}
\label{sec:finit_appr}

In ``finite differencing method'' the derivative operators are
approximated by finite difference formulae.  The finite approximations
are based on an appropriate linear combination of the Taylor
expansions of the function of interest. In this section we illustrate
this procedure only for the first derivative $u'=u_{,\, x}$ of a
one-dimensional scalar function $u = u(x)$. For higher derivative
orders and variable numbers the method is similar.  Let $u = u(x)$ be
discretized on a one-dimensional mesh of dimension $J$ and increment
$\Delta x$,
\begin{equation}
u_j = u(x_j) \qquad \qquad  \textrm{for} \qquad   j=1,..,J \, .
\end{equation}
The Taylor expansion of $u_{j+1}$ and  $u_{j-1}$ around the grid point
$x_j$ reads:
\begin{eqnarray}
u_{j+1} & = & u_{j} + \Delta x u'_{j}  + \frac{\Delta x}{2} u''_{j}
+ O(\Delta x ^2)  \, ,  \label{Tayu+}\\
u_{j-1} & = & u_{j} - \Delta x u'_{j}  + \frac{\Delta x}{2} u''_{j}
+ O(\Delta x ^2) \, .\label{Tayu-}
\end{eqnarray}
Now, when we take their difference and isolate the term $u'_{j}$ we
obtain:
\begin{equation}
 D_{0} u \equiv \frac{u_{j+1}-u_{j-1}}{2\Delta x} + O(\Delta x ^2)  \, ,
\end{equation}
where $D_{0} u$ denotes the centered finite difference approximation of
the derivative $u'_{j}$, which shows a second order accuracy $O(\Delta
x ^2)$.  For grid points near the surface these centered formulae are
not able to approximate derivatives by using only internal points of
the mesh. For this aim it is more appropriate to use one-sided approximations
of $u'$. First order accurate approximation of $u'$ is given by the
following expressions:
\begin{eqnarray}
D_{+} u \equiv \frac{u_{j+1}-u_{j}}{\Delta x} + O(\Delta x )  \, , \\
D_{-} u \equiv \frac{u_{j}-u_{j-1}}{\Delta x} + O(\Delta x )  \, ,
\end{eqnarray}
which can be determined by the Taylor expansions (\ref{Tayu+}) and
(\ref{Tayu-}).

\noindent Second order accuracy is instead reached with the following finite difference
formulae:
\begin{eqnarray}
D_{+} u &\equiv & - \frac{3 u_{j} - 4u_{j+1} + u_{j+2}}{2\Delta x} +
O(\Delta x ^2 ) \, ,  \label{2Dplus}\\
D_{-} u &\equiv & \frac{3 u_{j} - 4u_{j-1} + u_{j-2}}{2\Delta x} +
O(\Delta x ^2 ) \, .   \label{2Dminus}
\end{eqnarray}

The finite expressions of the second derivative $u''$ can be determined on
the same line as the first order. The centered finite approximation is given
by the following expression:
\begin{equation}
 D^2_{0} u \equiv \frac{u_{j+1}-2 u_{j}+u_{j-1}}{\Delta x^2}
+ O(\Delta x ^2)  \, ,
\end{equation}
while the one-sided formulae for $u''$ are the following:
\begin{eqnarray}
D^2_{+} u &\equiv &  \frac{2 u_{j} - 5u_{j+1} + 4 u_{j+2} - u_{j+3}
}{\Delta x ^2} +
O(\Delta x ^2 ) \, , \\
D^2_{-} u &\equiv & \frac{2 u_{j} - 5u_{j-1} + 4u_{j-2} - u_{j-3}}{\Delta x ^2} +
O(\Delta x ^2 ) \, .
\end{eqnarray}
%
\chapter{Numerical methods}
\label{sec:Num_meth}
%
The numerical algorithms used in the simulations are all present
in the reference ``Numerical Recipes''~\cite{1992nrfa.book.....P}. Here
we briefly illustrate the McCormack scheme that is not mentioned in
this reference.  Furthermore, we also write the methods used for
determing the convergence rate and the norms of the perturbative
solutions.

\section{McCormack algorithm}
It is an explicit second order differencing algorithm and a 2-level
method.  The scheme consists of two computational steps, namely a predictor
and a corrector step.  For a given partial differential equation of a
function $f = f(t,x)$ that we indicate as follows:
\begin{equation}
f _{, \,t } = G \left( f, f_{, \, x}, t, x \right) \, ,
\end{equation}
the value of the function $f$ on the new time-slice is updated as
follows:  \\
\emph{i) Estimator step}
\begin{equation}
\tilde f _{j} ^{n+1} = \Delta
t \, G ^{n}_{j,j-1}
\end{equation}
where the source term $G ^{n}_{j,j-1}$ is evaluated at $x_{j-1/2}$ by
using the values $ f _{j-1} ^{n}$ and $f _{j} ^{n}$. \\
\emph{i) Corrector step}
\begin{equation}
\tilde f _{j} ^{n+1} = f _{j} ^{n} + \Delta t \, \frac{1}{2} \left( G
^{n}_{j,j-1} + \tilde G ^{n+1}_{j+1,j} \right) \, ,
\end{equation}
where now the term $\tilde G ^{n+1}_{j+1,j}$ is determined by the
preliminary values $ \tilde f _{j} ^{n+1}$ and $\tilde f _{j+1} ^{n+1}$.

\section{Convergence test}
\label{sec:Convtest}

The exact solution of a given analytical equation has to be more and
more accurately approximated by numerical solutions determined with an
increasing resolution of the numerical mesh. In the limit of an
infinite dimension of the grid the numerical solution must tend to the
exact solution. In addition, the rate of the convergence has to be
consistent with the degree of approximation of the numerical method
used.

Let $f^{ex}$ be an exact solution of a given equation, and $f^{c}$ and
$f^{m}$ the numerical solutions determined respectively on a coarse
grid of dimension~$J_c$ and a medium grid of~$J_m = 2 J_c$.  The
deviation of these solutions from the analytical one can be then
written as:
\begin{eqnarray}
\Delta f ^{c} & \equiv & \sqrt{ \frac{1}{J_c}  \sum _{j=1}^{J_c} \left(
  f^c_j - f^{ex}_j \right) ^2 } = E_0 \Delta r ^{\sigma} + O(r ^{\sigma +1 })\, , \\
\Delta f ^{m} & \equiv & \sqrt{ \frac{1}{J_c}  \sum _{j=1}^{J_c} \left(
  f^m_j - f^{ex}_j \right) ^2 } = E_0  \left( \frac{\Delta r}{2} \right) ^{\sigma} + O (r ^{\sigma +1})\, ,
\end{eqnarray}
where the differences in these two expressions are both evaluated at
the points of the coarse mesh.  The letter $\sigma$ denotes the
accuracy order of the numerical solution and $E_0$ is the unknown
error term.
The ratio of the previous two expressions leads to the following expression:
\begin{equation}
\frac{ \Delta f ^{c} }{\Delta f ^{m}} = 2^{\sigma} + O (r ^{\sigma +1}) \, ,
\end{equation}
thus the convergence factor~$\sigma$ is given by
\begin{equation}
\sigma = \frac{ \log \left[ \Delta f^c/\Delta f^m \right]}{ \log 2} \, .
\end{equation}

When the exact solution is unknown the convergence test requires three
numerical solutions determined on three different grids whose
resolution is in the following proportion 1:2:4.  Therefore, in the
previous expressions we can replace the exact solution $f^{ex}$ with
the numerical solution $f^{f}$ obtained on the fine grid of dimension
$J_f = 2 J_m = 4 J_c$.  It is important to remark that in this second
case the convergence test informs us of the correct scaling of the
numerical error but it does not imply that the numerical solution is
going to converge to the true solution.

\section{Numerical stability and dissipation}
\label{sec:Norms}
The numerical stability and dissipation of the simulations can be
monitored by the constancy of the norms, which can be determined for
numerical solutions during their time evolution.  Let $f_j^n$ be the
finite approximation of a quantity $f$, which has been determined on
a one-dimensional grid of dimension $J$ at the $n$ time slice. The
$\mathbf{L}_2$ norm can then be calculated at any time step as follows:
\begin{equation}
 || f ||_{2} = \frac{1}{J} \sum_{j=1}^{J} \left( f_j^n \right) ^2 \, ,
\end{equation}
where we have introduced the division by $J$ for having an expression
averaged on the  number of grid points.



\nocite*



\begin{thebibliography}{100}

\bibitem{bars}
{igec.lnl.infn.it}.

\bibitem{lisa}
{lisa.jpl.nasa.gov}.

\bibitem{virgoetal}
{www.virgo.infn.it; www.ligo.caltech.edu; www.geo600.uni-hannover.de;
  tamago.mtk.nao.ac.jp}.

\bibitem{2004PhRvD..69l2004A}
B.~{Abbott et al.}
\newblock {Analysis of first LIGO science data for stochastic gravitational
  waves}.
\newblock {\em \prd}, 69:122004, 2004.

\bibitem{2004PhRvD..69l2001A}
B.~{Abbott et al.}
\newblock {Analysis of LIGO data for gravitational waves from binary neutron
  stars}.
\newblock {\em \prd}, 69:122001, 2004.

\bibitem{2004PhRvD..69j2001A}
B.~{Abbott et al.}
\newblock {First upper limits from LIGO on gravitational wave bursts}.
\newblock {\em \prd}, 69:102001, 2004.

\bibitem{2004PhRvD..69h2004A}
B.~{Abbott et al.}
\newblock {Setting upper limits on the strength of periodic gravitational waves
  from PSR J1939+2134 using the first science data from the GEO 600 and LIGO
  detectors}.
\newblock {\em \prd}, 69:082004, 2004.

\bibitem{2005PhRvL..94r1103A}
B.~{Abbott et al.}
\newblock {Limits on Gravitational-Wave Emission from Selected Pulsars Using
  LIGO Data}.
\newblock {\em \prl}, 94:181103, 2005.

\bibitem{Abbott:2005ez}
B.~{Abbott et al.}
\newblock Upper limits on a stochastic background of gravitational waves.
\newblock {\em \prd}, 95:221101, 2005.

\bibitem{2005PhRvD..72f2001A}
B.~{Abbott et al.}
\newblock {Upper limits on gravitational wave bursts in LIGO's second science
  run}.
\newblock {\em \prd}, 72:062001, 2005.

\bibitem{allen-1998-58}
G.~{Allen}, N.~{Andersson}, K.~D. {Kokkotas}, and B.~F. {Schutz}.
\newblock Gravitational waves from pulsating stars: Evolving the perturbation
  equations for a relativistic star.
\newblock {\em \prd}, 58:124012, 1998.

\bibitem{Andersson:1996ak}
N.~{Andersson} and K.~D. {Kokkotas}.
\newblock {Gravitational Waves and Pulsating Stars: What Can We Learn from
  Future Observations?}
\newblock {\em \prl}, 77:4134, 1996.

\bibitem{Andersson:1998ak}
N.~{Andersson} and K.~D. {Kokkotas}.
\newblock {Pulsation modes for increasingly relativistic polytropes}.
\newblock {\em \mnras}, 297:493, 1998.

\bibitem{1998MNRAS.299.1059A}
N.~{Andersson} and K.~D. {Kokkotas}.
\newblock {Towards gravitational wave asteroseismology}.
\newblock {\em \mnras}, 299:1059, 1998.

\bibitem{2003ApJ...591.1129A}
P.~{Arras}, {\'E}.~{\'E}. {Flanagan}, S.~M. {Morsink}, A.~K. {Schenk}, S.~A.
  {Teukolsky}, and I.~{Wasserman}.
\newblock {Saturation of the r-Mode Instability}.
\newblock {\em \apj}, 591:1129, 2003.

\bibitem{1982MNRAS.200P..43B}
E.~{Balbinski} and B.~F. {Schutz}.
\newblock {A puzzle concerning the quadrupole formula for gravitational
  radiation}.
\newblock {\em \mnras}, 200:43P, 1982.

\bibitem{Bardeen:1966tm}
J.~M. Bardeen, K.~P. Thorne, and D.~W. Meltzer.
\newblock A catalogue of methods for studying the normal modes of radial
  pulsations of general-relativistic stellar models.
\newblock {\em \apj}, 145:505, 1966.

\bibitem{2005MNRAS.358..923B}
E.~{Berti}, F.~{White}, A.~{Maniopoulou}, and M.~{Bruni}.
\newblock {Rotating neutron stars: an invariant comparison of approximate and
  numerical space-time models}.
\newblock {\em \mnras}, 358:923, 2005.

\bibitem{Bruni:2002sm}
M.~{Bruni}, L.~{Gualtieri}, and C.~F. {Sopuerta}.
\newblock Two-parameter nonlinear space-time perturbations: Gauge
  transformations and gauge invariance.
\newblock {\em Class. Quant. Grav.}, 20:535, 2003.

\bibitem{Bruni:1996im}
M.~{Bruni}, S.~{Matarrese}, S.~{Mollerach}, and S.~{Sonego}.
\newblock Perturbations of spacetime: Gauge transformations and gauge
  invariance at second order and beyond.
\newblock {\em Class. Quant. Grav.}, 14:2585, 1997.

\bibitem{Cerdonio:2000bh}
M.~Cerdonio, L.~Conti, J.~A. Lobo, A.~Ortolan, and J.~P. Zendri.
\newblock Wideband dual sphere detector of gravitational waves.
\newblock {\em {\prd}}, 87:031101, 2001.

\bibitem{Chandrasekhar:1964pr}
S.~Chandrasekhar.
\newblock Dynamical instability of gaseous masses approaching the schwarzshild
  limit in general relativity.
\newblock {\em \prl}, 12:114, 1964.

\bibitem{Chandrasekhar:1964tc}
S.~{Chandrasekhar}.
\newblock {The Dynamical Instability of Gaseous Masses Approaching the
  Schwarzschild Limit in General Relativity.}
\newblock {\em \apj}, 140:417, 1964.

\bibitem{1970PhRvL..24..762C}
S.~{Chandrasekhar}.
\newblock {Solutions of Two Problems in the Theory of Gravitational Radiation}.
\newblock {\em \prl}, 24:762, 1970.

\bibitem{Chandrasekhar:1991fi}
S.~{Chandrasekhar} and V.~{Ferrari}.
\newblock {On the non-radial oscillations of a star}.
\newblock {\em Royal Society of London Proceedings Series A}, 432:247, 1991.

\bibitem{1995PhRvD..52.2118C}
E.~S.~C. {Ching}, P.~T. {Leung}, W.~M. {Suen}, and K.~{Young}.
\newblock {Wave propagation in gravitational systems: Late time behavior}.
\newblock {\em \prd}, 52:2118, 1995.

\bibitem{Cowling:1941co}
T.~G. {Cowling}.
\newblock {The non-radial oscillations of polytropic stars}.
\newblock {\em \mnras}, 101:367, 1941.

\bibitem{1980tsp..book.....C}
J.~P. {Cox}.
\newblock {\em {Theory of stellar pulsation}}.
\newblock Research supported by the National Science Foundation Princeton, NJ,
  Princeton University Press, 1980.

\bibitem{Cunningham:1978cp}
C.~T. {Cunningham}, R.~H. {Price}, and V.~{Moncrief}.
\newblock {Radiation from collapsing relativistic stars. I - Linearized
  odd-parity radiation}.
\newblock {\em \apj}, 224:643, 1978.

\bibitem{Cunningham:1980cp}
C.~T. {Cunningham}, R.~H. {Price}, and V.~{Moncrief}.
\newblock {Radiation from collapsing relativistic stars. III - Second order
  perturbations of collapse with rotation}.
\newblock {\em \apj}, 236:674, 1980.

\bibitem{Darmois:1927gd}
G.~Darmois.
\newblock Les \'equations de la gravitation einsteinienne.
\newblock In {\em M\'emorial des Sciences Math\'ematiques}, volume XXV,
  chapter~V. Gauthier-Villars, Paris, 1927.

\bibitem{Detweiler:1985dl}
S.~{Detweiler} and L.~{Lindblom}.
\newblock {On the nonradial pulsations of general relativistic stellar models}.
\newblock {\em \apj}, 292:12, 1985.

\bibitem{Dimmelmeier:2002bk}
H.~{Dimmelmeier}, J.~A. {Font}, and E.~{Muller}.
\newblock Relativistic simulations of rotational core collapse. \rm{I}.
  methods, initial models, and code tests.
\newblock {\em \aap}, 388:917, 2002.

\bibitem{Dimmelmeier:2002bm}
H.~{Dimmelmeier}, J.~A. {Font}, and E.~{Muller}.
\newblock Relativistic simulations of rotational core collapse. \rm{II}.
  collapse dynamics and gravitational radiation.
\newblock {\em \aap}, 393:523, 2002.

\bibitem{Dimmelmeier:2004prep}
H.~Dimmelmeier, N.~Stergioulas, and J.~A. Font.
\newblock Nonlinear axisymmetric pulsations of rotating relativistic stars in
  the conformal flatness approximation.
\newblock {\em astro-ph/0511394}, 2004.

\bibitem{Ferrari:2000fk}
V.~{Ferrari} and K.~D. {Kokkotas}.
\newblock {Scattering of particles by neutron stars: Time evolutions for axial
  perturbations}.
\newblock {\em \prd}, 62:107504, 2000.

\bibitem{2003LRR.....6....4F}
J.~A. {Font}.
\newblock {Numerical Hydrodynamics in General Relativity}.
\newblock {\em Living Reviews in Relativity}, 6:4, 2003.

\bibitem{1987ApJ...314..594F}
J.~L. {Friedman} and J.~R. {Ipser}.
\newblock {On the maximum mass of a uniformly rotating neutron star}.
\newblock {\em \apj}, 314:594, 1987.

\bibitem{1978ApJ...222..281F}
J.~L. {Friedman} and B.~F. {Schutz}.
\newblock {Secular instability of rotating Newtonian stars}.
\newblock {\em \apj}, 222:281, 1978.

\bibitem{1993ApJ...419..768F}
M.~Y. {Fujimoto}.
\newblock {The Evolution of Accreting Stars with Turbulent Mixing}.
\newblock {\em \apj}, 419:768, 1993.

\bibitem{Garat:2000gp}
A.~{Garat} and R.~H. {Price}.
\newblock {Gauge invariant formalism for second order perturbations of
  Schwarzschild spacetimes}.
\newblock {\em \prd}, 61:044006, 2000.

\bibitem{Gerlach:1979ih}
U.~H. {Gerlach} and U.~K. {Sengupta}.
\newblock Even parity junction conditions for perturbations on most general
  spherically symmetric space-times.
\newblock {\em J. Math. Phys.}, 20:2540, 1979.

\bibitem{Gerlach:1979rw}
U.~H. Gerlach and U.~K. Sengupta.
\newblock Gauge invariant perturbations on most general spherically symmetric
  space-times.
\newblock {\em \prd}, D19:2268, 1979.

\bibitem{Gerlach:1979ze}
U.~H. {Gerlach} and U.~K. {Sengupta}.
\newblock Junction conditions for odd parity perturbations on most general
  spherically symmetric space-times.
\newblock {\em \prd}, 20:3009, 1979.

\bibitem{Gerlach:1980tx}
U.~H. Gerlach and U.~K. Sengupta.
\newblock Gauge invariant coupled gravitational, acoustical, and
  electromagnetic modes on most general spherical space- times.
\newblock {\em \prd}, D22:1300, 1980.

\bibitem{Gleiser:1995gx}
R.~J. {Gleiser}, C.~O. {Nicasio}, R.~H. {Price}, and J.~{Pullin}.
\newblock Second order perturbations of a schwarzschild black hole.
\newblock {\em Class. Quant. Grav.}, 13:L117, 1996.

\bibitem{Gualtieri:2001cm}
L.~Gualtieri, E.~Berti, J.~A. Pons, G.~Miniutti, and V.~Ferrari.
\newblock Gravitational signals emitted by a point mass orbiting a neutron
  star: A perturbative approach.
\newblock {\em \prd}, D64:104007, 2001.

\bibitem{Gundlach:1999bt}
C.~{Gundlach} and J.~M. {Mart{\'\i}n-Garc{\'\i}a}.
\newblock Gauge-invariant and coordinate-independent perturbations of stellar
  collapse. \rm{I}: The interior.
\newblock {\em \prd}, D61:084024, 2000.

\bibitem{1986ApJS...61..479H}
I.~{Hachisu}.
\newblock {A versatile method for obtaining structures of rapidly rotating
  stars}.
\newblock {\em \apj SS}, 61:479, 1986.

\bibitem{1989Natur.340..617H}
P.~{Haensel} and J.~L. {Zdunik}.
\newblock {A submillisecond pulsar and the equation of state of dense matter}.
\newblock {\em \nat}, 340:617, 1989.

\bibitem{2003PhRvD..68b4002H}
T.~{Harada}, H.~{Iguchi}, and M.~{Shibata}.
\newblock {Computing gravitational waves from slightly nonspherical stellar
  collapse to a black hole: Odd-parity perturbation}.
\newblock {\em \prd}, 68:024002, 2003.

\bibitem{Hartle:1967ha}
J.~B. {Hartle}.
\newblock {Slowly Rotating Relativistic Stars. I. Equations of Structure}.
\newblock {\em \apj}, 150:1005, 1967.

\bibitem{Hartle:1970ha}
J.~B. {Hartle}.
\newblock {Slowly-Rotating Relativistic Stars.IV. Rotational Energy and Moment
  of Inertia for Stars in Differential Rotation}.
\newblock {\em \apj}, 161:111, 1970.

\bibitem{Hartle:1968ht}
J.~B. {Hartle} and K.~S. {Thorne}.
\newblock {Slowly Rotating Relativistic Stars. II. Models for Neutron Stars and
  Supermassive Stars}.
\newblock {\em \apj}, 153:807, 1968.

\bibitem{1975ApJ...195L..51H}
R.~A. {Hulse} and J.~H. {Taylor}.
\newblock {Discovery of a pulsar in a binary system}.
\newblock {\em \apjl}, 195:L51, 1975.

\bibitem{Ipser:1991ip}
J.~R. {Ipser} and R.~H. {Price}.
\newblock {Nonradial pulsations of stellar models in general relativity}.
\newblock {\em \prd}, 43:1768, 1991.

\bibitem{Israel:1966nc}
W.~Israel.
\newblock Singular hypersurfaces and thin shells in gr.
\newblock {\em Nuovo Cimento}, B44:1, 1966.

\bibitem{Kind:1993kn}
S.~{Kind}, J.~{Ehlers}, and B.~G. {Schmidt}.
\newblock Relativistic stellar oscillations treated as an initial value
  problem.
\newblock {\em Class. Quantum Grav.}, 10:2137, 1993.

\bibitem{Kokkotas:2000up}
K.~D. {Kokkotas} and J.~{Ruoff}.
\newblock Radial oscillations of relativistic stars.
\newblock {\em \aap}, 366:565, 2001.

\bibitem{kokkotas-1999-2}
K.~D. {Kokkotas} and B.~G {Schmidt}.
\newblock Quasi-normal modes of stars and black holes.
\newblock {\em Living Reviews in Relativity}, 2:2, 1999.

\bibitem{Kokkotas:1992ks}
K.~D. {Kokkotas} and B.~F. {Schutz}.
\newblock {W-modes - A new family of normal modes of pulsating relativistic
  stars}.
\newblock {\em \mnras}, 255:119, 1992.

\bibitem{kokkotas-2005-}
K.~D. {Kokkotas} and N.~{Stergioulas}.
\newblock Gravitational waves from compact sources.
\newblock {\em gr-qc/0506083}, 2005.

\bibitem{1989MNRAS.237..355K}
H.~{Komatsu}, Y.~{Eriguchi}, and I.~{Hachisu}.
\newblock {Rapidly rotating general relativistic stars. I - Numerical method
  and its application to uniformly rotating polytropes}.
\newblock {\em \mnras}, 237:355, 1989.

\bibitem{1989MNRAS.239..153K}
H.~{Komatsu}, Y.~{Eriguchi}, and I.~{Hachisu}.
\newblock {Rapidly rotating general relativistic stars. II - Differentially
  rotating polytropes}.
\newblock {\em \mnras}, 239:153, 1989.

\bibitem{1969mech.book.....L}
L.~D. {Landau} and E.~M. {Lifshitz}.
\newblock {\em {Mechanics}}.
\newblock Course of Theoretical Physics, Oxford: Pergamon Press, 1969, 2nd ed.

\bibitem{1993PhRvD..48.3467L}
M.~{Leins}, H.-P. {Nollert}, and M.~H. {Soffel}.
\newblock {Nonradial oscillations of neutron stars: A new branch of strongly
  damped normal modes}.
\newblock {\em \prd}, 48:3467, 1993.

\bibitem{Leveque_mio}
R.~J. {LeVeque}.
\newblock {\em Numerical Methods for Conservation Laws}.
\newblock Birkh\"auser Verlag, Basel, 1999.

\bibitem{2001MNRAS.322..515L}
Y.~{Levin} and G.~{Ushomirsky}.
\newblock {Non-linear r-modes in a spherical shell: issues of principle}.
\newblock {\em \mnras}, 322:515, 2001.

\bibitem{Lichnerowicz:1971al}
A.~Lichnerowicz.
\newblock Sur les ondes de choc gravitationnelles.
\newblock {\em C. R. Acad. Sci.}, 273:528, 1971.

\bibitem{1983ApJS...53...73L}
L.~{Lindblom} and S.~L. {Detweiler}.
\newblock {The quadrupole oscillations of neutron stars}.
\newblock {\em \apjs}, 53:73, 1983.

\bibitem{2001PhRvL..86.1152L}
L.~{Lindblom}, J.~E. {Tohline}, and M.~{Vallisneri}.
\newblock {Nonlinear Evolution of the r-Modes in Neutron Stars}.
\newblock {\em \prl}, 86:1152, 2001.

\bibitem{Lindblom_mio}
L.~{Lindblom}, J.~E. {Tohline}, and M.~{Vallisneri}.
\newblock {Numerical Evolutions of non-linear r-Modes in Neutron Stars}.
\newblock {\em \prd}, 65:084039, 2002.

\bibitem{Martin-Garcia:1998sk}
J.~M. {Mart{\'\i}n-Garc{\'\i}a} and C.~{Gundlach}.
\newblock All nonspherical perturbations of the choptuik spacetime decay.
\newblock {\em \prd}, D59:064031, 1999.

\bibitem{Martin-Garcia:2000ze}
J.~M. {Mart{\'\i}n-Garc{\'\i}a} and C.~{Gundlach}.
\newblock Gauge-invariant and coordinate-independent perturbations of stellar
  collapse. \rm{II}: Matching to the exterior.
\newblock {\em \prd}, D64:024012, 2001.

\bibitem{1988ApJ...325..725M}
P.~N. {McDermott}, H.~M. {van Horn}, and C.~J. {Hansen}.
\newblock {Nonradial oscillations of neutron stars}.
\newblock {\em \apj}, 325:725, 1988.

\bibitem{Meltzer:1966mt}
D.~W. {Meltzer} and K.~S. {Thorne}.
\newblock {Normal Modes of Radial Pulsation of Stars at the End Point of
  Thermonuclear Evolution}.
\newblock {\em \apj}, 145:514, 1966.

\bibitem{Misner:1973cw}
C.~W. Misner, K.S. Thorne, and J.~A. Wheeler.
\newblock {\em Gravitation}.
\newblock W. H. Freeman \& Co., San Francisco, 1973.

\bibitem{Moncrief:1974vm}
V.~Moncrief.
\newblock Gravitational perturbations of spherically symmetric systems. \rm{I}.
  \rm{The Exterior Problem.}
\newblock {\em Ann. Phys. (N.Y.)}, 88:323, 1974.

\bibitem{Moncrief:1974vmII}
V.~Moncrief.
\newblock Gravitational perturbations of spherically symmetric systems.
  \rm{II}. \rm{Perfect Fluid Interiors.}
\newblock {\em Ann. Phys. (N.Y.)}, 88:343, 1974.

\bibitem{Nagar:2004pr}
A.~{Nagar} and G.~{D{\'\i}az}.
\newblock Fluid accretion onto relativistic stars and gravitational radiation.
\newblock In {\em Proceedings of the 27th Spanish Relativity Meeting.
  Gravitational Radiation}, Alicante, 2004. Editorial Services of the
  University of Alicante.

\bibitem{Nagar:2004ns}
A.~{Nagar}, G.~{D{\'\i}az}, J.~A. {Pons}, and J.~A. {Font}.
\newblock Accretion driven gravitational radiation from nonrotating compact
  objects. infalling quadrupolar shells.
\newblock {\em \prd}, D69:124028, 2004.

\bibitem{Nagar:2005ea}
A.~{Nagar} and L.~{Rezzolla}.
\newblock Gauge-invariant non-spherical metric perturbations of schwarzschild
  black-hole spacetimes.
\newblock {\em Class. Quant. Grav.}, 22:R167, 2005.

\bibitem{Nakamura:2003wk}
K.~{Nakamura}.
\newblock Gauge invariant variables in two-parameter nonlinear perturbations.
\newblock {\em Prog. Theor. Phys.}, 110:723, 2003.

\bibitem{Nakamura:2004gi}
K.~{Nakamura}.
\newblock General framework of higher order gauge invariant perturbation
  theory.
\newblock {\em gr-qc/0402032}, 2004.

\bibitem{Nollert_mio}
H.~P. {Nollert}.
\newblock Quasinormal modes: the characteristic 'sound' of the black holes and
  neutron stars.
\newblock {\em Class. Quant. Grav.}, 16:R159, 1999.

\bibitem{Obrien:1952bs}
S.~O'Brien and J.~L. Singe.
\newblock {\em Proc. Dublin Inst. Adv. Stud.}, A9:1, 1952.

\bibitem{Oppenheimer:1939ne}
J.~R. Oppenheimer and G.~M. Volkoff.
\newblock On massive neutron cores.
\newblock {\em \prd}, 55:374, 1939.

\bibitem{1973ApJ...185..277O}
Y.~{Osaki} and C.~J. {Hansen}.
\newblock {Nonradial oscillations of cooling white dwarfs.}
\newblock {\em \apj}, 185:277, 1973.

\bibitem{Passamonti:2004je}
A.~{Passamonti}, M.~{Bruni}, L.~{Gualtieri}, and C.~F. {Sopuerta}.
\newblock Coupling of radial and non-radial oscillations of relativistic stars:
  gauge-invariant formalism.
\newblock {\em \prd}, D71:024022, 2005.

\bibitem{Passamonti:2005axial}
A.~Passamonti, A.~Nagar, M.~Bruni, L.~Gualtieri, and C.~F. Sopuerta.
\newblock Coupling of radial and axial non-radial oscillations of compact
  stars: Gravitational waves from first-order differential rotation.
\newblock {\em gr-qc/0601001}.

\bibitem{1992nrfa.book.....P}
W.~H. {Press}, S.~A. {Teukolsky}, W.~T. {Vetterling}, and B.~P. {Flannery}.
\newblock {\em {Numerical recipes in FORTRAN 77. The art of scientific
  computing}}.
\newblock Cambridge: University Press, 1999, 2nd ed.

\bibitem{1999bhgr.conf..351P}
R.~H. Price.
\newblock The two black hole problem: Beyond linear perturbations.
\newblock In {\em Black Holes, Gravitational Radiation, and the Universe:
  Essays in Honor of C.V. Vishveshwara}, page 351, Dordrecht, 1998. Kluwer
  Academic Publishers.

\bibitem{Regge:1957}
T.~Regge and J.~A. Wheeler.
\newblock Stability of a schwarzschild singularity.
\newblock {\em \prd}, 108:1063, 1957.

\bibitem{2000ApJ...531L.139R}
L.~{Rezzolla}, F.~K. {Lamb}, and S.~L. {Shapiro}.
\newblock {R-Mode Oscillations in Rotating Magnetic Neutron Stars}.
\newblock {\em \apj}, 531:L139, 2000.

\bibitem{Ruoff:2000nj}
J.~{Ruoff}.
\newblock {\em The Numerical Evolution of Neutron Star Oscillations}.
\newblock PhD thesis, Universitaet Tuebingen, 2000.

\bibitem{Ruoff:2001ux}
J.~{Ruoff}.
\newblock New approach to the evolution of neutron star oscillations.
\newblock {\em \prd}, D63:064018, 2001.

\bibitem{schnabel-2004-21}
R.~Schnabel, J.~Harms, K.~A. Strain, and K.~Danzmann.
\newblock Squeezed light for the interferometric detection of high frequency
  gravitational waves.
\newblock {\em Class. Quant. Grav.}, 21:S1045, 2004.

\bibitem{Seidel:1990xb}
E.~Seidel.
\newblock Gravitational radiation from even parity perturbations of stellar
  collapse: Mathematical formalism and numerical methods.
\newblock {\em \prd}, D42:1884, 1990.

\bibitem{Seidel:1987in}
E.~Seidel and T.~Moore.
\newblock Gravitational radiation from realistic relativistic stars: Odd parity
  fluid perturbations.
\newblock {\em \prd}, D35:2287, 1987.

\bibitem{sopuerta-2004-70}
C.~F. {Sopuerta}, M.~{Bruni}, and L.~{Gualtieri}.
\newblock Non-linear n-parameter spacetime perturbations: Gauge
  transformations.
\newblock {\em \prd}, 70:064002, 2004.

\bibitem{Sperhake:2001si}
U.~{Sperhake}.
\newblock Non-linear numerical schemes in general relativity.
\newblock {\em gr-qc/0201086}, 2001.

\bibitem{Sperhake:2001xi}
U.~{Sperhake}, P.~{Papadopoulos}, and N.~{Andersson}.
\newblock Non-linear radial oscillations of neutron stars: Mode- coupling
  results.
\newblock {\em astro-ph/0110487}, 2001.

\bibitem{2003LRR.....6....3S}
N.~{Stergioulas}.
\newblock {Rotating Stars in Relativity}.
\newblock {\em Living Reviews in Relativity}, 6:3, 2003.

\bibitem{Stergioulas:2003ep}
N.~{Stergioulas}, T.~A. {Apostolatos}, and J.~A. {Font}.
\newblock Nonlinear pulsations in differentially rotating neutron stars:
  Mass-shedding-induced damping and splitting of the fundamental mode.
\newblock {\em \mnras}, 352:1089, 2004.

\bibitem{Stergioulas:2000vs}
N.~{Stergioulas} and J.~A. {Font}.
\newblock Nonlinear r-modes in rapidly rotating relativistic stars.
\newblock {\em \prl}, 86:1148, 2001.

\bibitem{1974RSPSA.341...49S}
J.~M. {Stewart} and M.~{Walker}.
\newblock {Perturbations of space-times in general relativity}.
\newblock {\em Royal Society of London Proceedings Series A}, 341:49, 1974.

\bibitem{Thorne:1969to}
K.~S. {Thorne}.
\newblock {Nonradial Pulsation of General-Relativistic Stellar Models. III.
  Analytic and Numerical Results for Neutron Stars}.
\newblock {\em \apj}, 158:1, 1969.

\bibitem{Thorne:1969th}
K.~S. {Thorne}.
\newblock {Nonradial Pulsation of General-Relativistic Stellar Models.IV. The
  Weakfield Limit}.
\newblock {\em \apj}, 158:997, 1969.

\bibitem{Thorne:1967th}
K.~S. {Thorne} and A.~{Campolattaro}.
\newblock {Non-Radial Pulsation of General-Relativistic Stellar Models. I.
  Analytic Analysis for L $\ge$ 2}.
\newblock {\em \apj}, 149:591, 1967.

\bibitem{Thorne:1968tc}
K.~S. {Thorne} and A.~{Campolattaro}.
\newblock {Erratum: Non-Radial Pulsation of General-Relativistivc Stellar
  Models. I. Analytic Analysis for L $\ge$ 2}.
\newblock {\em \apj}, 152:673, 1968.

\bibitem{Tolman:1939jz}
R.~C. {Tolman}.
\newblock Static solutions of einstein's field equations for spheres of fluid.
\newblock {\em \prd}, 55:364, 1939.

\bibitem{1989nos..book.....U}
W.~{Unno}, Y.~{Osaki}, H.~{Ando}, H.~{Saio}, and H.~{Shibahashi}.
\newblock {\em {Nonradial oscillations of stars}}.
\newblock Nonradial oscillations of stars, Tokyo: University of Tokyo Press,
  1989, 2nd ed.

\bibitem{2002MNRAS.333..943W}
A.~L. {Watts} and N.~{Andersson}.
\newblock {The spin evolution of nascent neutron stars}.
\newblock {\em \mnras}, 333:943, 2002.

\bibitem{2005MNRAS.356.1371Z}
O.~{Zanotti}, J.~A. {Font}, L.~{Rezzolla}, and P.~J. {Montero}.
\newblock {Dynamics of oscillating relativistic tori around Kerr black holes}.
\newblock {\em \mnras}, 356:1371, 2005.

\bibitem{Zerilli:1970fj}
F.~J. Zerilli.
\newblock Effective potential for even-parity regge-wheeler gravitational
  perturbation equations.
\newblock {\em \prl}, 24:737, 1970.

\bibitem{Zerilli:1970la}
F.~J. Zerilli.
\newblock Gravitational field of a particle falling in a schwarzschild geometry
  analyzed in tensor harmonics.
\newblock {\em \prd}, 2:2141, 1970.

\bibitem{1997A&A...320..209Z}
T.~{Zwerger} and E.~{Mueller}.
\newblock {Dynamics and gravitational wave signature of axisymmetric rotational
  core collapse.}
\newblock {\em \aap}, 320:209, 1997.

\end{thebibliography}

\end{document}